\documentclass[oneside,12pt]{Classes/jiit-phd-thesis-1}
\usepackage{titlesec}
\usepackage[centertags]{amsmath}
\usepackage[chapter]{algorithm}
\usepackage{algorithmic}
\usepackage[acronym]{glossaries}
\usepackage{longtable}
\usepackage[normalem]{ulem}
\usepackage{multicol}
\usepackage{multirow}
\usepackage{amsfonts}
\usepackage{amssymb}
\usepackage{amsthm}
\usepackage{acronym}
\usepackage{tabu}
\usepackage{mathtools}
\usepackage{graphicx}
\usepackage{tabularx}
\usepackage{multirow}
\usepackage[center]{caption}
\usepackage{rotating}
\usepackage{amsmath}
\usepackage{epstopdf}
\usepackage{pdflscape}
\usepackage[T1]{fontenc}
\usepackage[utf8]{inputenc}
\usepackage{geometry}
\usepackage{amssymb} 
\usepackage{physics}
\usepackage{color,graphicx}
\usepackage{pdfpages}
\usepackage{subcaption}
\usepackage{float}
\usepackage{wrapfig}
\usepackage{array}
\usepackage{hyphenat}
\usepackage{ragged2e}
\usepackage{blindtext}
\usepackage[nopar]{lipsum}
\usepackage{blindtext}
\usepackage{hyperref}
\usepackage{csquotes}
\usepackage{afterpage}
\newacronym{BM}{BM}{Bell Measurements}
\newacronym{CJRIO}{CJRIO}{Controlled Joint Remote Implementation of Operators}
\newacronym{COW}{COW}{Coherent One Way}
\newacronym{CRIO}{CRIO}{Controlled Remote Implementation of Operators}
\newacronym{CR}{CR}{Compression Ratio}
\newacronym{DOF}{DOF}{Degree of Freedom}
\newacronym{DPR}{DPR}{Distributed Phase Reference}
\newacronym{DPTS}{DPTS}{Differential Phase Time Shifting}
\newacronym{DPS}{DPS}{Differential Phase Shift}
\newacronym{DR}{DR}{Disclose Rate}
\newacronym{DT}{DT}{Dead Time}
\newacronym{JRIO}{JRIO}{Joint Remote Implementation of Operators}
\newacronym{KR}{KR}{Key Rate}
\newacronym{LOCC}{LOCC}{Local Operation and Classical Communication}
\newacronym{MDI}{MDI}{Measurement Device Independent}
\newacronym{MQT}{MQT}{Multi-output Quantum Teleportation}
\newacronym{MZI}{MZI}{Mach Zehnder Interferometer}
\newacronym{NISQ}{NISQ}{Noisy Intermediate-Scale Quantum}
\newacronym{P-DOF}{P-DOF}{Polarization Degree of Freedom}
\newacronym{PNS}{PNS}{Photon Number Splitting}
\newacronym{S-DOF}{S-DOF}{Spatial Degree of Freedom}
\newacronym{QAV}{QAV}{Quantum Anonymous Veto}
\newacronym{QB}{QB}{Quantum Broadcasting}
\newacronym{QBER}{QBER}{Quantum Bit Error Rate}
\newacronym{QEC}{QEC}{Quantum Error Correction}
\newacronym{QIS}{QIS}{Quantum Information Splitting}
\newacronym{QKD}{QKD}{Quantum Key Distribution}
\newacronym{QSDC}{QSDC}{Quantum Secure Direct Communication}
\newacronym{QSS}{QSS}{Quantum Secret Sharing}
\newacronym{QST}{QST}{Quantum State Tomography}
\newacronym{QT}{QT}{Quantum Teleportation}
\newacronym{RIO}{RIO}{Remote Implementation of Operators}
\newacronym{RIHO}{RIHO}{Remote Implementation of Hidden Operators}
\newacronym{RIPUO}{RIPUO}{Remote Implementation of Partially Unknown Operators}
\newacronym{RSP}{RSP}{Remote State Preparation}
\newacronym{SDC}{SDC}{Super Dense Coding}
\newacronym{SMQC}{SMQC}{Secure Multi-party Quantum Computation}
\newacronym{SNSPD}{SNSPD}{Superconducting Nanowire Single Photon Detector}
\newacronym{SPD}{SPD}{Single Photon Detector}
\newacronym{TF}{TF}{Twin Field}
\newacronym{ULL}{ULL}{Ultra-Low-Loss}
\newacronym{VA}{VA}{Voting Authority}
\newacronym{WCP}{WCP}{Weak Coherent Pulses}

\newcommand\blankpage{%
	\null
	\thispagestyle{empty}%
	\addtocounter{page}{-1}%
	\newpage}
\titleformat{\section}
{\normalfont\fontsize{14}{16}\bfseries}
{\thesection}{1em}
{\MakeUppercase}

\titleformat{\subsection}
{\normalfont\fontsize{12}{16}\bfseries}
{\thesubsection}{1em}
{\MakeUppercase}

\titleformat{\subsubsection}
{\normalfont\fontsize{12}{16}\bfseries}
{\thesubsubsection}{1em}
{\MakeUppercase}

\makeatletter
\renewcommand*\l@figure{\@dottedtocline{1}{1.5em}{2.3em}}
\addtocontents{lof}{\def\string\@dotsep{1000}}
\addtocontents{lot}{\def\string\@dotsep{1000}}
\let\l@table\l@figure
\makeatother

\Title{Design and Experimental Realization of Various Protocols for Secure Quantum Computation and Communication}

\Author{Satish Kumar}
\Roll{ENROLLMENT NO. 20410007}

\SupervisorA{Prof. (Dr.) Anirban Pathak}

\Department{Department of Physics and Materials Science and Engineering }

\Institute{JAYPEE INSTITUTE OF INFORMATION TECHNOLOGY}
\InstituteTag{(Declared Deemed to be University U/S 3 of UGC Act)}
\InstituteAdd{A-10, SECTOR-62, NOIDA, INDIA}

\InstituteLogo{\includegraphics[scale=0.10]{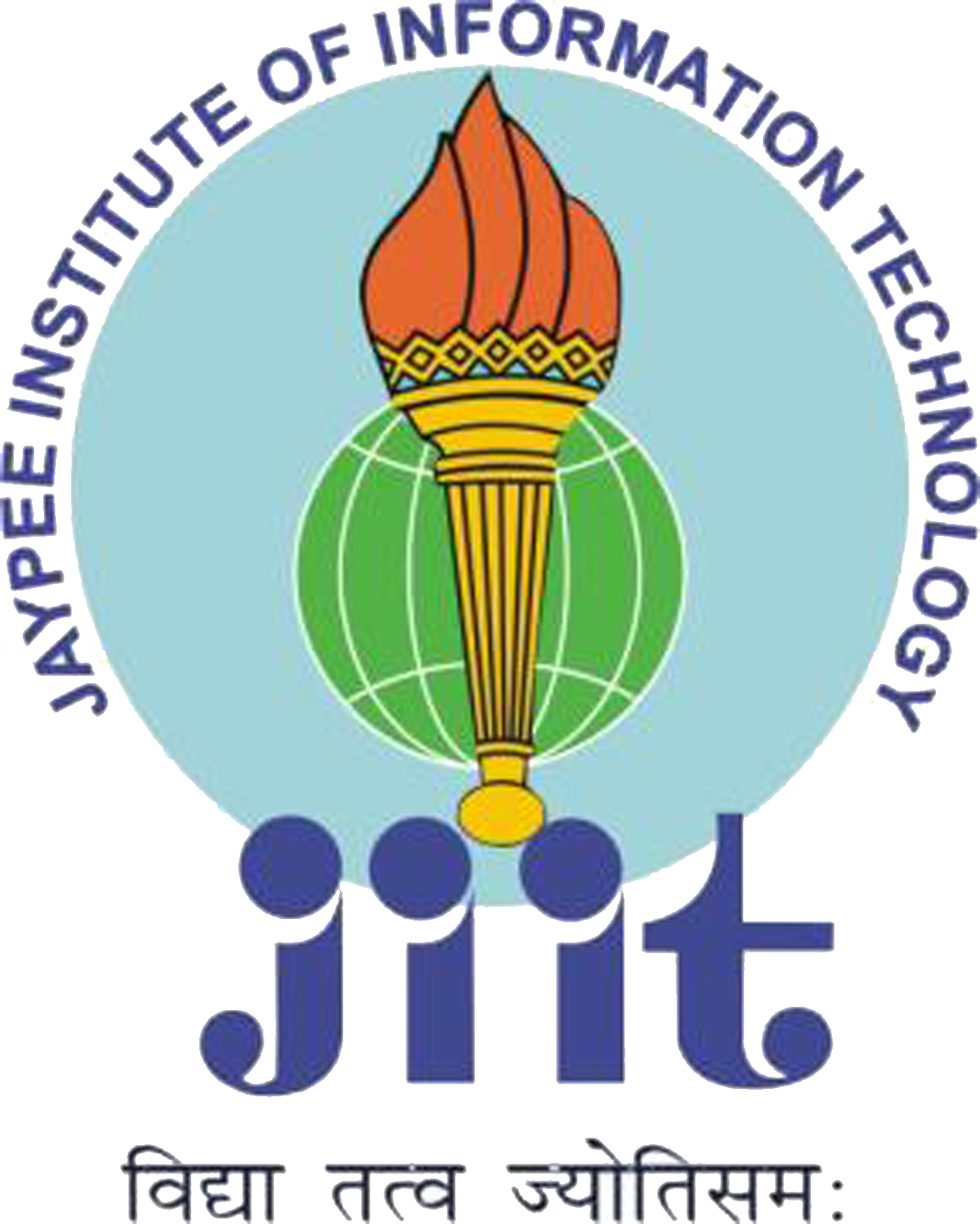}}

\Degree{DOCTOR OF PHILOSOPHY}

\StartDegreeDate{(August), (2020)}

\EndDegreeDate{\textnormal{February, 2025}}

\begin{document}

\frontmatter

\begin{titlepage}
	\maketitle
\end{titlepage}
\cleardoublepage
\pagenumbering{roman}
\addcontentsline{toc}{chapter}{\fontsize{12.5}{10}\selectfont{INNER FIRST PAGE}}
\newpage
\textbf{}\\
\textbf{}\\
\textbf{}\\
\textbf{}\\
\textbf{}\\
\textbf{}\\
\textbf{}\\
\textbf{}\\
\textbf{}\\
\textbf{}\\
\textbf{}\\
\textbf{}\\
\textbf{}\\
\textbf{}\\
\textbf{}\\
\textbf{}\\
\textbf{}\\
\textbf{}\\
\textbf{}\\
\textbf{}\\
\textbf{}\\
\textbf{}\\
\begin{Large} 
\centering
\singlespacing
\fontsize{10}{12}\selectfont{
{\copyright \ Copyright  JAYPEE INSTITUTE OF INFORMATION TECHNOLOGY, NOIDA}\\
{(Declared Deemed to be University U/S 3 of UGC Act)}\\
February, 2025\\
ALL RIGHTS RESERVED\\
}
\end{Large}
\newpage
\textbf{}\\
\textbf{}\\
\textbf{}\\
\textbf{}\\
\textbf{}\\
\textbf{}\\
\textsl{\centering This thesis is dedicated to my\\
		\centering late grandfather (Shri Gauri Shankar Seth)\\	
		\centering $\&$\\
		\centering my parents (Smt. Sunita Devi and Mr. C. L. Sahu)\\
		\centering for their unwavering support and belief in my abilities.\\
	}
\afterpage{\blankpage}
\newpage

\addcontentsline{toc}{chapter}{\fontsize{12}{10}\selectfont{TABLE OF CONTENTS}}
\renewcommand{\contentsname}{}
\vspace*{-2em} 
\begin{center}
	\fontsize{16}{10}\bfseries\MakeUppercase{TABLE OF CONTENTS}
\end{center}
\vspace{-4em} 
\tableofcontents
\cleardoublepage
\afterpage{\blankpage}
\newpage
\cleardoublepage
\addcontentsline{toc}{chapter}{\fontsize{12}{10}\selectfont{DECLARATION BY THE SCHOLAR}}

\begin{declaration}

\doublespacing
I hereby declare that the work reported in the Ph.D. thesis entitled {\fontsize{14}{10}{\bfseries{{\enquote{\TITLE}}}}} submitted at  \ \fontsize{14}{10}{\bfseries{Jaypee Institute of Information Technology, Noida, India,}} is an authentic record of my work carried out under the supervision of \fontsize{14}{10}{\bfseries{{\SUPERVISORA}}}. I have not submitted this work elsewhere for any other degree or diploma. I am fully responsible for the contents of my Ph.D. thesis. 
\\
\\
\\
\fontsize{14}{16}{{{(Signature of the Scholar)}}}\\
\fontsize{14}{16}{{{(\AUTHOR)}}}\\
\fontsize{14}{10}{{{Department of Physics and Materials Science and Engineering}}}\\
\fontsize{14}{16}{{{Jaypee Institute of Information Technology, Noida, India}}}\\
\fontsize{14}{16}{{{Date: 25 April, 2025}}}
\end{declaration} 
\afterpage{\blankpage}
\newpage
\cleardoublepage
\addcontentsline{toc}{chapter}{\fontsize{12}{10}\selectfont{SUPERVISOR'S CERTIFICATE}}


\begin{Supdeclaration}
\doublespacing
This is to certify that the work reported in the Ph.D. thesis entitled {\fontsize{14}{10}{\bfseries{{\enquote{\TITLE}}}}}, submitted by \fontsize{14}{10}{\bfseries{{\AUTHOR}}} at \fontsize{14}{10}{\bfseries{{Jaypee Institute of Information Technology, Noida, India}}}, is a bonafide record of his original work carried out under my supervision. This work has not been submitted elsewhere for any other degree or diploma.
\newline
\\
\\
\\
\\
\\
\fontsize{14}{16}{{{(Signature of Supervisor)}}}\\
\fontsize{14}{16}{{{(\SUPERVISORA})}}\\
\fontsize{14}{16}{{{Professor and HOD}}}\\
\fontsize{14}{16}{{{Department of Physics and Materials Science and Engineering}}}\\
\fontsize{14}{16}{{{Jaypee Institute of Information Technology, Noida, India}}}\\
\fontsize{14}{16}{{{Date: 25 April, 2025}}}
\end{Supdeclaration} 

\newpage

\afterpage{\blankpage}
\newpage
\cleardoublepage
\addcontentsline{toc}{chapter}{\fontsize{12}{10}\selectfont{ACKNOWLEDGEMENT}}


\begin{acknowledgements}

The journey from Mr. Satish to Dr. Satish would not have been possible without support from numerous people in my life. I would like to thank everyone who supported me throughout this entire journey in this acknowledgment.

First, I would like to thank my mentor, supervisor and a great teacher Prof. (Dr.) Anirban Pathak. I was inspired to pursue a PhD because of his insightful and informative lecture during the QIQT-2019 summer school. Throughout my PhD journey, I believe that every decision he made for me was driven by a good reason. I have always felt that he has been my pillar of support, much like how Shri Krishna mentored Arjuna during the battle of Mahabharata. I will always be grateful for his guidance and mentorship throughout my research journey.

Secondly, I would like to thank my seniors Dr. Kishore Thapliyal, Dr. Abhishek Shukla, Dr. Chitra Shukla, Dr. Amit Verma and Dr. Anindita Banerjee for their support and guidance wherever required. Sometimes, you find friends in form of your seniors who not only support in your research but also help you become aware of the hurdles during PhD journey and sharing their experiences to resolve them. I would like to thank such friendly seniors: Dr. Sandeep Mishra, Dr. Priya Malpani, Dr. Ashwin Saxena, Dr. Kathakali Mandal, Dr. Rishi Dutt Sharma and Dr. Mitali Sisodia.

Colleague and friends are ones who are on the same height of a mountain on which you are climbing. Their support is much more needed because they can understand situations from your perspectives. I would like to thank such friends of mine Mr. Narendra Hegade, Mr. Arindam Dutta, Mr. Vaisakh Mannalath, Mrs. Nikhitha Nunavath, Mr. Kuldeep Gangwar and Mr. Britant.

Juniors are always curious about new things. Sometimes their curiosity leads to have a new information. I would also like to thank my juniors too Mr. Ajay Kumar and Mr. Kamran Dar.

Finally, I would like to thank my DPMAC members Dr. Anuraj Panwar and Prof. (Dr.) Bhagwati Prasad Chamola for their encouraging words and continuous monitoring of my PhD progress. Also, thank to technical lab staff of our department (PMSE)-Mr. Munish Kumar, Mr. Tarkeshwar Mishra and Mr. Kailash Chandra-for their supports in the process of institute documentations whenever required. This thanks list is incomplete without thanking JIIT administration who successfully managed to provide all research scholars with many of facilities like, pantry for tea services, supportive guards, a sitting place and housekeeping staffs for cleaning.

\begin{flushright}
    (Satish Kumar)
\end{flushright}

\end{acknowledgements}

\newpage 
\cleardoublepage
\addcontentsline{toc}{chapter}{\fontsize{12}{10}\selectfont ABSTRACT}


\begin{abstract}
A set of new schemes for quantum computation and communication have been either designed or realized using optimal quantum resources. A scheme of multi-output quantum teleportation, where a sender (Alice) sends a $m$ and $m+1$-qubit GHZ like unknown quantum state to a receiver (Bob), has been realized using two-copies of the Bell state instead of a five-qubit cluster state. The scheme is also experimentally realized on IBM quantum computer for $m=1$ case. Another scheme where a known state is sent to two party (say, Bob and Charlie) which are geographically far apart, is popularly known as quantum broadcasting which uses higher qubit entangled state to accomplish the task. We observed that the quantum broadcasting schemes reported earlier can be reduced to the task of multiparty remote state preparation. Using this fact, we realized the quantum broadcasting scheme on IBM quantum computer and checked the fidelity in presence of various noisy environments. Once unknown and known state is sent from one place to another, one can think of sending an operator. Such scheme where an operator is operated remotely to an unknown qubit is known as remote implementation of operators (RIO). A scheme of controlled joint-RIO (CJRIO) has been proposed using a four-qubit hyper-entangled state which is entangled in two-DOF, spatial and polarization degree of photons. The proposed scheme has a direct application in distributed photonic quantum computing. We use the cross-Kerr interaction technique to allow interactions between photons. In the same line, two more schemes of variants of RIO have been proposed which are known as remote implementation of hidden or partially unknown operators (RIHO $\&$ RIPUO). The RIHO scheme has a direct application in blind quantum computing as the operator operated remotely is blind to operand. The success probability is also estimated in terms of dissipation of an auxiliary coherent state with interaction to the environment. There are certain tasks where we require multiple party in the secure computation. Such scenario is known as secure multi-party quantum computation (SMQC), for example, quantum voting, auction, etc. Such SMQC scheme quantum anonymous voting (QAV) is experimentally realized on quantum computer available at IBM cloud service. Quantum key distribution (QKD) is a process to generate a secure key between two authentic users which is used further for secure communication. Two such QKD schemes, coherent one way (COW) and differential phase shift (DPS) is experimentally demonstrated in the quantum cryptography lab. We have analyzed the effect of secure key rate as a function of different post-processing parameters and dead time of detector at various distances.\\


\end{abstract}

\newpage

\cleardoublepage
\addcontentsline{toc}{chapter}{\fontsize{12}{10}\selectfont LIST OF ACRONYMS \& ABBREVIATIONS}

\section*{\hfill LIST OF ACRONYMS \& ABBREVIATIONS\hfill}
\begin{tabular}{lll}
	BM      && Bell Measurements \\
	CJRIO   && Controlled Joint Remote Implementation of Operators \\
	COW     && Coherent One Way \\
	CR      && Compression Ratio \\
	CRIO    && Controlled Remote Implementation of Operators \\
	DOF     && Degree of Freedom \\
	DPR     && Distributed Phase Reference \\
	DPS     && Differential Phase Shift \\
	DPTS    && Differential Phase Time Shifting \\
	DR      && Disclose Rate \\
	DT      && Dead Time \\
	JRIO    && Joint Remote Implementation of Operators \\
	KR      && Key Rate \\
	LOCC    && Local Operation and Classical Communication \\
	MDI     && Measurement Device Independent \\
	MQT     && Multi-output Quantum Teleportation \\
	MZI     && Mach Zehnder Interferometer \\
	NISQ    && Noisy Intermediate-Scale Quantum \\	
	P-DOF   && Polarization Degree of Freedom \\
	PNS     && Photon Number Splitting \\
	S-DOF   && Spatial Degree of Freedom \\
	QAV     && Quantum Anonymous Veto \\
	QB      && Quantum Broadcasting \\
	QBER    && Quantum Bit Error Rate \\
	QEC     && Quantum Error Correction \\
	QIS     && Quantum Information Splitting \\
	QKD     && Quantum Key Distribution \\
	QSDC    && Quantum Secure Direct Communication \\
	QSS     && Quantum Secret Sharing \\
	QST     && Quantum State Tomography \\
\end{tabular}
\newpage
\begin{tabular}{lll}
	QT      && Quantum Teleportation \\
	RIO     && Remote Implementation of Operators \\
	RIHO    && Remote Implementation of Hidden Operators \\
	RIPUO   && Remote Implementation of Partially Unknown Operators \\
	RSP     && Remote State Preparation \\
	SDC     && Super Dense Coding \\
	SMQC    && Secure Multi-party Quantum Computation \\
	SNSPD   && Superconducting Nanowire Single Photon Detector \\
	SPD     && Single Photon Detector \\
	TF      && Twin Field \\
	ULL     && Ultra-Low-Loss \\
	VA      && Voting Authority \\
	WCP     && Weak Coherent Pulses \\	
\end{tabular}
\newpage
\cleardoublepage
\addcontentsline{toc}{chapter}{\fontsize{12}{10}\selectfont LIST OF FIGURES}
\section*{\hfill LIST OF FIGURES\hfill}
\begin{tabular}{p{1.5cm}p{12cm}p{1.5cm}}
	\textbf{Figure Number} &\centering \textbf{Caption} & \textbf{Page Number}\\
	1.1 & \centering A brief structure of the thesis. & 3\\
	1.2 & \centering An example of quantum communication schemes that does not require security. & 4\\
	1.3 & \centering One-dimensional system which has six sites. Particle can only exist in one of the six sites. & 6\\
	1.4 & \centering Representation of a qubit on Bloch sphere. & 8\\
	1.5 & \centering Single qubit quantum gates with their representation and corresponding matrices. & 11\\
	1.6 & \centering Symbolic representation of CNOT and SWAP gates. & 12\\
	1.7 & \centering Snap of few platforms where one can access quantum computers on cloud. & 13\\
	1.8 & \centering Quantum teleportation circuit. & 18\\
	1.9 & \centering Circuit illustrating the RSP process. & 19\\
	1.10 & \centering Pictorial representation for modeling of (a) close and (b) open quantum system. & 21\\
	2.1 & \centering A sketch to visualize the MQT scheme of Yan Yu et al. & 35\\
	2.2 & \centering A quantum circuit illustrating the generalized MQT scheme of Yan Yu et al. & 37\\
	2.3 & \centering A quantum circuit illustrating the MQT scheme using optimal resources. & 39\\
	2.4 & \centering A quantum circuit illustrating the MQT scheme for m = 1 case using optimal
	resources. & 40\\ 
	2.5 & \centering (a) A quantum circuit to teleport $|+\rangle=\frac{1}{\sqrt{2}}(|0\rangle+|1\rangle)$ to two distinct receivers at qubit $Q_2$ and $Q_6$ using two copies of a Bell states $|\phi^{+}\rangle^{\otimes2}$ (b) Topology of the quantum computer used
	(ibmq\_casablanca). & 41\\
	2.6 & \centering Experimentally obtained result after executing the quantum circuit depicted in Figure 2.5 on IBM quantum computer (ibmq\_casablanca). & 42\\
\end{tabular}
\newpage
\begin{tabular}{p{1.5cm}p{12cm}p{1.5cm}}
	2.7 & \centering Experimental quantum state tomography result with (a) real and (b) imaginary parts for the circuit shown in Figure 2.5. & 43\\
	2.8 & \centering Plot for success probability of the MQT scheme under various noisy environment having variable noise percentage. & 45\\
	3.1 & \centering A sketch to visualize the QB scheme of Yan Yu et al. & 50\\
	3.2 & \centering A quantum circuit for the generation of (a) Bell state and (b) cluster state. & 54\\
	3.3 & \centering Impact of (a) amplitude damping, (b) phase damping, (c) bit-flip and (d)
	depolarizing noise on 5a5pair of Bell states and cluster states. & 55\\
	3.4 & \centering A quantum circuit for broadcasting the state $\alpha|0\rangle + \beta|1\rangle$ with $|\alpha|^{2}=|\beta|^{2}=\frac{1}{2}$ to two distinct receivers using (a) the four-qubit cluster state and (b) two copies of the Bell state. Here, c.c. stands for classical communication. & 56 \\
	3.5 & \centering The obtained results after executing the quantum circuit depicted in (a) Figure 3.4 (a) and (b) Figure 3.4 (b) on ibmq\_manila. & 59\\
	4.1 & \centering A sketch to visualize (a) the CJRIO task and (b) the quantum resources used. & 63\\
	4.2 & \centering A schematic illustrating the first two steps of the CJRIO protocol is provided. A circle labeled $V, H$ represents a photon simultaneously in vertical and horizontal polarization, while a circle labeled $V$ only represents a photon in vertical polarization. Circles with two attached lines depict photons existing in two spatial paths simultaneously, whereas those with one line indicate photons with a single spatial path. The interaction between a photon path with a coherent state (CS) via cross-Kerr nonlinearity is represented by a line terminating in a bold dot on the photon path. Double arrows originating from the CS denote measurement outcomes. Vertical solid (dashed) lines represent entanglement in P-DOF (S-DOF). The BBS denotes a balanced beam splitter. & 69\\
\end{tabular}
\newpage
\begin{tabular}{p{1.5cm}p{12cm}p{1.5cm}}
	4.3 & \centering A schematic illustrating Step~3 and Step~4 of the CJRIO protocol. Here, Charlie as a controller, first blends paths of her photon using BBS and enables the interaction $K_{c_k}(\theta)|z\rangle|c_k\rangle$ and perform X-quadrature measures the CS, yielding the outcome $s$, which destroys the entanglement of photon $\text{C}$ from remaining photons in S-DOF. After that Bob$^1$ performs a similar operation by mixing his photon paths on a BBS and enabling the interaction $K_{b^1_k}(\theta)|z\rangle|b^1_k\rangle$ and measures the CS, yielding the outcome $l$, which allows Bob$^2$ to apply appropriate unitary operations to recover the state $\alpha_{B^2}|b^2_0\rangle+\beta_{B^2}|b^2_1\rangle$. & 71\\
	4.4 & \centering A schematic illustrating Step~5 and Step~6 of the CJRIO protocol is provided. Here, Bob$^1$ first initiates a new photon path using a BBS and enables the non-linear interaction $K_{b^1_{k\oplus l\oplus1}}(\theta)|z\rangle|b^1_{k\oplus l\oplus1}\rangle$ and forwards it to Bob$^2$ which enables the interaction $K_{b^2_0}(-\theta)|z\rangle|b^2_0\rangle$ and measures the CS, yielding the outcome $r$. Bob$^2$ then blends the two spatial paths of his photon and enables the cross-kerr interaction between one of his photon path and a CS followed by measuring the CS with outcome $g$ as depicted in the figure, that allows Bob$^1$ to implement a suitable unitary operation to achieve the state $\alpha_{B^1B^2}|b^1_0\rangle+\beta_{B^1B^2}|b^1_1\rangle$. & 73\\
	4.5 & \centering A schematic illustrating Step~7 to Step~9 of the CJRIO protocol is presented. Here, the joint parties Bob$^1$ and Bob$^2$ and the controller Charlie measure their respective photons in suitable basis. The measurement leads the collapse of photons $\text{B}^1$, $\text{B}^2$, and $\text{C}$ and we are only left with photon $\text{A}$. Alice then applies the suitable unitary operation followed by PBS and HWP along one of the paths to obtain the desired state $\alpha_{B^1B^2}|a_{0}\rangle+\beta_{B^1B^2}|a_{1}\rangle$. & 76\\
	4.6 & \centering This figure illustrates the steps involved in the protocol for remote implementation of hidden operators. An unpolarized photon is represented by a circle. & 86\\
	4.7 & \centering This figure illustrates the steps involved in the protocol for remote implementation of a partially unknown operator. & 88\\
\end{tabular}
\newpage
\begin{tabular}{p{1.5cm}p{12cm}p{1.5cm}}
	4.8 & \centering The modified success probabilities is shown in (a) and (b) as functions of dissipative parameter $D$  and initial amplitude $z$ of coherent state, with $\theta=\pi$ radians phase shift for both the RIHO ($P_{\rm{1Suc}}$) and the RIPUO ($P_{\rm{2Suc}}$) protocols. In (c), the behavior of the same probabilities $P_{\rm{1Suc}}$ and $P_{\rm{2Suc}}$ as a function of $D$ is shown, with phase shift $\theta=\pi$ radians and an initial amplitude of the coherent state $z=1$. & 90\\
	5.1 & \centering Topology of the IBMQ Manila. & 103\\
	5.2 & \centering A quantum circuit designed for the experimental realization of Protocol A. & 104\\
	5.3 & \centering Results obtained from the real device and simulator for Protocol A using Bell state for the (a) conclusive and (b) inconclusive outcomes. & 107\\
	5.4 & \centering A quantum circuit for implementing Protocol B using (a) the cluster state and (b) the GHZ state. & 108\\
	5.5 & \centering Result obtained from the real device and simulator for Protocol B using cluster state and GHZ state, showing the (a,c) inconclusive and (b,d) conclusive outcomes. & 112\\
	5.6 & \centering Impact of phase damping, amplitude damping, depolarization, and bit flip noise on: (a) Protocol A with the Bell state (b) Protocol B with the GHZ state (c) Protocol B with the cluster state. & 114\\
	5.7 & \centering Impact of (a) phase damping (b) amplitude damping (c) bit flip and (d)
	depolarizing noise on Protocol A and Protocol B. & 115\\
	6.1 & \centering Classification of QKD where P \& M refers to prepare-and-measure-based and EB refers to entanglement based QKD protocol. & 119\\
	6.2 & \centering A block diagram illustrating the DPS protocol. PM: phase modulator, ${\rm BS_{1}}$, ${\rm BS_{2}}$ are beam splitters and $M_{1}$, $M_{2}$ are mirrors. & 122\\
	6.3 & \centering A block diagram illustrating the COW protocol. IM: intensity modulator, BS1: beamsplitter. & 123\\
	6.4 & \centering A snapshot of the experimental set-up used for the DPR QKD implementation. & 125\\
\end{tabular}
\newpage
\begin{tabular}{p{1.5cm}p{12cm}p{1.5cm}}
	6.5 & \centering Plots for the obtained KR of the DPS protocol as a function of CR for various DR (a) at $80$ km (b) at $100$ km, (c) at $120$, with detector's DT $=50\,\mu$s. The KR decreases with increasing distance and increases with higher DR. & 130\\
	6.6 & \centering Plots for the obtained KR of the DPS protocol as a function of DR (CR) for various CR (DR) at detector's DT $=40\,\mu$s, $=30\,\mu$s and $=20\,\mu$s for $80$ km communication distance. & 132\\
	6.7 & \centering Plots for the obtained KR of the COW QKD protocol as a function of (a) DR for various CR and (b) CR for various DR, for $100$ km distance with detector's DT $=50\,\mu$s. & 133\\
	6.8 & \centering Plots for obtained KR as a function of DR for various CR for both DPS (Purple dashed and Red solid line) and COW (Blue and Green line with plot markers) QKD protocol. It is obvious that KR for DPS is higher than COW. & 134\\
	6.9 & \centering Plots for obtained KR as function of channel losses for both COW and DPS QKD protocol. & 134\\
	6.10 & \centering The stability plot for (a) KR and (b) QBER for both COW and DPS QKD protocol with fixed value of DS to $3.125$ \%, CR to $90$ \% and DT to $=50\,\mu$s for $80$ km communication distance. & 135\\
\end{tabular}
\afterpage{\blankpage}
\addcontentsline{toc}{chapter}{\fontsize{12}{10}\selectfont LIST OF TABLES}
\section*{\hfill LIST OF TABLES\hfill}
\begin{tabular}{p{1.5cm}p{12cm}p{1.5cm}}
	\textbf{Table Number} &\centering \textbf{Caption} & \textbf{Page Number}\\
	1.1 & \centering An example of generating a secure key using the BB84 protocol. & 20\\
	2.1 & \centering The calibration data of all active qubits of ibmq\_casablanca on December 01, 2021. cxi\_j, representing a $\rm{CNOT}$ gate with qubit i as the control and qubit j as the target. & 44\\
	3.1 & \centering The reduced resources to accomplish various QB schemes reported previously. Here, $|\phi^{+}\rangle=\frac{1}{\sqrt{2}}|00\rangle+|11\rangle$. & 50\\
	3.2 & \centering The calibration data of ibmq\_casablanca on Nov 28, 2022. cxi\_j, representing a $\rm{CNOT}$ gate with qubit i as the control and qubit j as the target. & 58\\
	4.1 & \centering Variations in operations at multiple stages when utilizing different quantum channels for the proposed RIHO protocol. Here, $|\Omega^{\pm}\rangle=|a_0\rangle|b_0\rangle\pm|a_1\rangle|b_1\rangle$, $|\Pi^{\pm}\rangle=|a_0\rangle|b_1\rangle\pm|a_1\rangle|b_0\rangle$. & 86\\
	4.2 & \centering Variations in operations at multiple stages when utilizing different quantum channels for the proposed RIPUO protocol. Here, $|\Omega^{\pm}\rangle=|a_0\rangle|b_0\rangle\pm|a_1\rangle|b_1\rangle$, $|\Pi^{\pm}\rangle=|a_0\rangle|b_1\rangle\pm|a_1\rangle|b_0\rangle$. & 88\\
	4.3 & \centering Optimal quantum resources for different variants of the proposed protocols for RIHO and RIPUO. Here, $|\Omega^{\pm}\rangle=|a_0\rangle|b_0\rangle\pm|a_1\rangle|b_1\rangle$, $|\Pi^{\pm}\rangle=|a_0\rangle|b_1\rangle\pm|a_1\rangle|b_0\rangle$. & 93\\
	5.1 & \centering An examples of quantum states and their corresponding quantum operations that can be utilized to implement Protocol B. & 102\\
	5.2 & \centering The calibration data of IBMQ Manila on March 26, 2022. Here, cxi\_j, representing a $\rm{CNOT}$ gate with qubit i as the control and qubit j as the target. & 103\\
	5.3 & \centering Comparison between the theoretically expected results for implementing the quantum veto protocol using Bell state (Protocol A) and the experimentally obtained results. Here, $|\phi^{\pm}\rangle=\frac{1}{\sqrt{2}}\left(|00\rangle+|11\rangle\right)$
	and simulator results corresponds to the measurement outcomes obtained by executing the circuit shown in Figure \ref{Fig5.2:Ckt-ProtocolA} on IBM
	Qasm simulator. & 105\\
\end{tabular}
\newpage
\begin{tabular}{p{1.5cm}p{12cm}p{1.5cm}}
	5.4 & \centering Results obtained from the implementation of quantum veto protocol using the cluster state (Protocol B). Here, $|\psi_{c}\rangle=\frac{1}{2}(|0000\rangle+|0011\rangle+|1100\rangle-|1111\rangle)$. & 109\\
	5.5 & \centering Results obtained from the implementation of quantum veto protocol using the GHZ state (Protocol B). Here, $|\psi_{GHZ}\rangle=\frac{1}{\sqrt{2}}(|000\rangle+|111\rangle)$. & 111\\
	6.1 & \centering A table to compare the COW and DPS QKD protocols. & 120\\
	6.2 & \centering The table outlines the
	advancements in experimental realization of the DPS QKD protocol. The symbol \# indicates that the information
	is not provided in the mentioned paper while {*} denotes dispersion shifted
	fiber. & 126\\
	6.3 & \centering The table outlines the
	advancements in experimental realization of the COW QKD protocol. The symbol \# indicates that the information is not provided clearly in the mentioned paper. & 127\\
	6.4 & \centering Details of the various components that are utilized in the experiment. & 127\\
	6.5 & \centering Parameters that are influencing the key rate of the COW and DPS protocols. & 129\\
\end{tabular}
\newpage
\afterpage{\blankpage}
\newpage
\setcounter{secnumdepth}{3}
\setcounter{tocdepth}{3}


\mainmatter
\pagenumbering{arabic}
\pagestyle{plain}

\chapter{INTRODUCTION}
\label{Ch1: Introduction}
\graphicspath{{Chapter1/Chapter1Figs/}{Chapter1/Chapter1Figs/}}

\section{Introduction to quantum information}
Everyday we exchange data with someone but every data exchange is not essentially the information. Information is a set of data that has been organized, analyzed and interpreted in a way that is useful for a specific analysis or decision-making process, providing insight beyond just raw data. Also, information is something that we don't know already \cite{Gershenfeld2022book, Pathak2013book}. Suppose, there is a timetable according to which lectures planned for your batch and you have a copy of that and you are familiar with the schedule. If one of your batch-mates says, hey let's go, there is a lecture now then this is not information for you, but when he says, hey this lecture is suspended today due to some specific reason then this is information. As you already know the lecture time so it is not an information but you are not aware of the reason behind suspension of the lecture so it is an information for you. The various modes to convey the information are words, numbers, images and symbols. Earlier, the idea of information was qualitative, but it is actually quantitative. Consider a lecture where a professor is teaching a quantum information course. Everybody in the lecture does not get the same amount of information despite being taught by the same professor. The amount of information depends on your focus and earlier knowledge. It is clear that information is quantitative. If information is quantitative, then there must be some measure of it. Claude Shannon first introduced such a measure of information in 1948 \cite{Shannon1948}. Information is not a purely abstract concept and it is associated with some physical entity. When you talk to your friend near you, the air molecules near you vibrate which also vibrates their nearest molecules and finally, it hits your friend's eardrum; nerves convert the mechanical energy into the electrical signals which are received by your friend's brain. If information is physical then all the physical laws must be applicable to it. Rolf Landauer was first to introduce this concept in 1991 \cite{Landauer1991info}. \\
The amount of information gained depends on the possible outcomes of an event. Consider two popular events (i) throwing a fair dice (ii) tossing a fair coin. The information obtained in event (i) is more than (ii) because there are six possible outcomes in (i) and only two possible outcomes in (ii). Suppose $X$ is an event and $x_i$ $(i=0,1,2,3,..,n)$ are possible outcomes with the probability $p_i$ then the amount of information obtained from the event $X$ is
\begin{equation}
	H(X)=-\sum_{i=1}^{n}p(x_i)\log_{2}p(x_i)
\end{equation}
This equation is popularly known as Shannon's entropy as it is introduced by Claude Shannon in 1948 \cite{Shannon1948}. The logarithmic base $2$ represents the unit of information in bits. It is to be noted that Shannon's entropy for an event is maximum if it has equal probable outcomes. Information that is processed in the form of bits is classical information while those processed in the form of quantum bits, called qubits, are known as quantum information. Qubits are explained in more detail in Section \ref{ssec:qubits}. There are various domains where information is carried in the form of qubits, two such domains, quantum computation and quantum communication will be briefly described. 
\begin{figure}
	\centering
	\includegraphics[width=\linewidth]{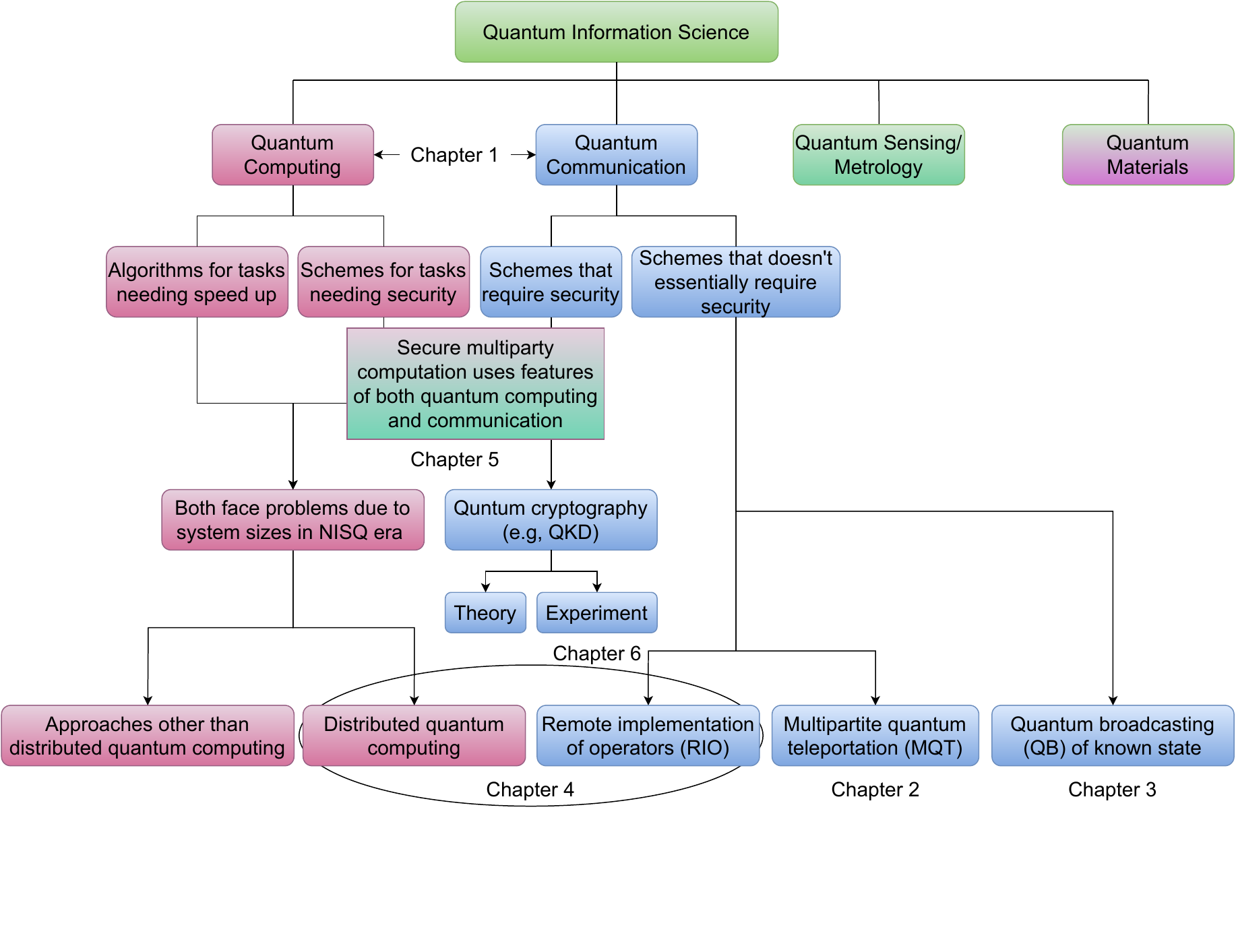}
	\caption{\centering A brief structure of the thesis}
	\label{fig:thesis_outline}
\end{figure}
\subsection{Quantum computation}
Computation may be viewed as a way of solving problems using computers. A classical computer uses bits to solve a problem while a quantum computer uses qubits to solve it. A detailed description of qubits is discussed in Section \ref{ssec:qubits}. The unique properties of qubit like superposition and entanglement, make quantum computers more powerful than classical computers. The advantage of using quantum computers overs classical computers is that quantum computers can solve certain problems much faster than classical computers.\\
Quantum processors approach mathematical computations fundamentally differently from classical computers. Instead of sequentially executing each step of a complex calculation, as classical systems do, quantum circuits leveraging logical qubits can process enormous amount of data simultaneously. This parallelism enables quantum processors to solve certain problems with higher efficiency, outperforming classical methods by orders of magnitude.\\
Due to this probabilistic operation of a quantum computers, their computational capability is more. However, traditional computers follow a deterministic approach, performing exhaustive calculations to arrive at a single definite outcome for any given inputs.\\
Traditional computers typically provide a definite answer, whereas quantum computers provides a range of possible solutions. This might seem less precision in quantum computers. However, the unique problem solving approach of a quantum computers could dramatically accelerate its capability to solve highly complex problems which potentially reducing computation times from hundreds of thousands of years to few hours or minutes.

\subsection{Quantum communication}
Communication is a process of conveying a message to an intended party. To make communication, we need to encode our messages. When encoding is done using bits, then it is referred to as classical communication and when encoding is done using qubits then it is called quantum communication. The advantage of performing quantum communication over classical communication is that quantum communication is more secure than classical communication. The security of quantum communication lies with the basic principles of quantum mechanics like uncertainty principle and no-cloning theorem. Several no-go theorems are discussed in Section \ref{sec:no-go theorems}. 
\begin{figure}
	\centering
	\includegraphics[width=\linewidth]{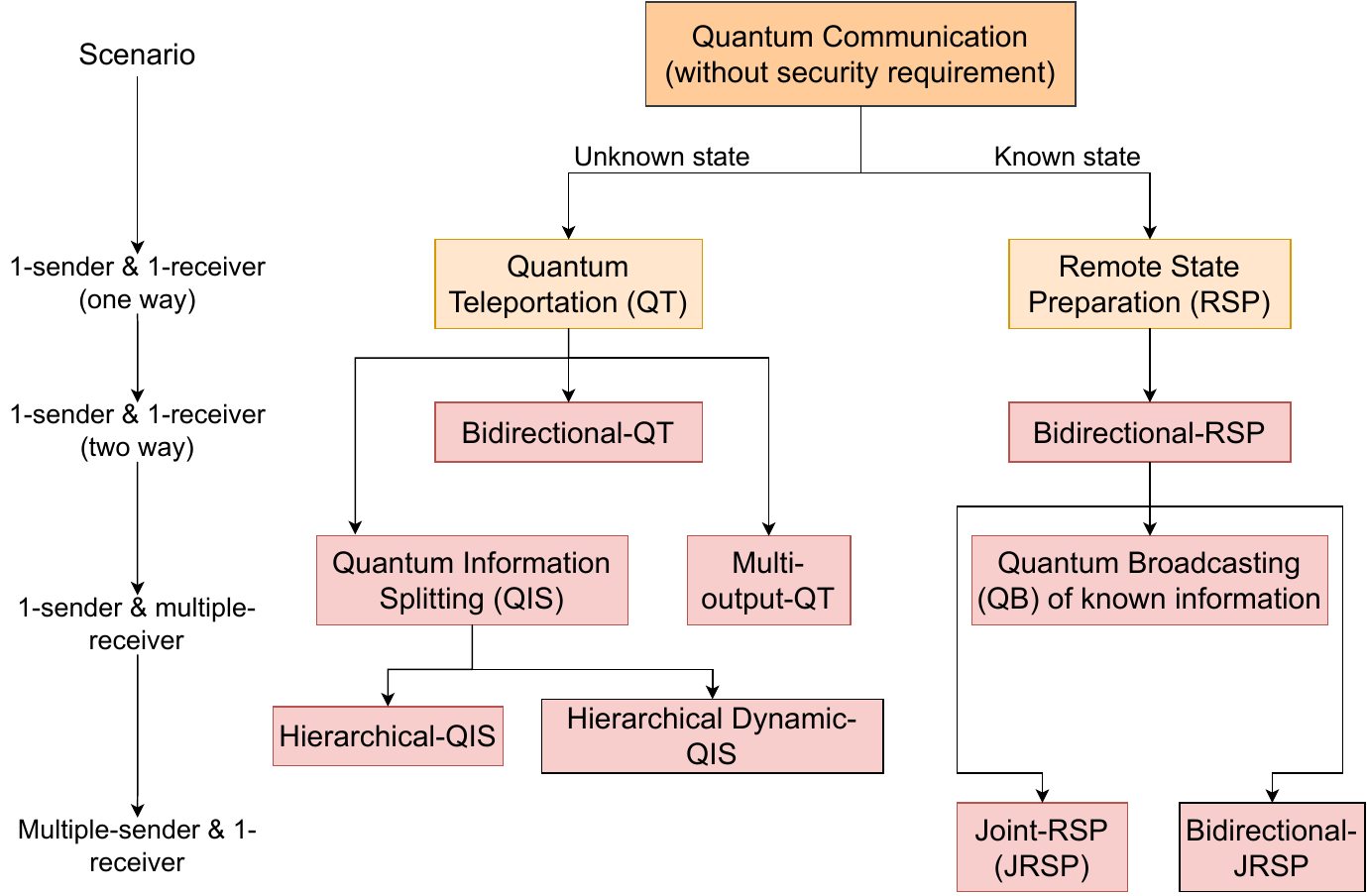}
	\caption{\centering An example of quantum communication schemes that does not require security.}
	\label{fig:QCscheme_classification}
\end{figure}
Quantum communication schemes are of two types: one which requires security (e.g., \acrfull{QKD} and one which does not essentially require security (e.g., \acrfull{QT}). First protocol for \acrshort{QT} was published by Bennett et al. in 1993 \cite{Bennett1993QT}. After that several variant of it have been proposed. Some of these schemes are generalization of the scheme of the standard teleportation scheme and some are special cases. For example, controlled-\acrshort{QT} is refer to a task that allow a sender (Alice) to teleport an unknown quantum state to a receiver (Bob) only when a controller (Charlie) allows Alice and Bob to do so. Thus this is a generalization of the \acrshort{QT} scheme and the capability of performing controlled-\acrshort{QT} would imply the capability of performing \acrshort{QT}. Similarly a protocol for bidirectional-\acrshort{QT} is a generalization of standard protocol for \acrshort{QT} as capability of teleporting in both direction (i.e., Alice to Bob and Bob to Alice) essentially implies capability of one way teleportation. However \acrfull{RSP} is a special case of \acrshort{QT} as in \acrshort{RSP} we teleport a known quantum state and capability of teleporting an unknown quantum state essentially ensures capability of teleporting a known quantum state. Interestingly, generalization of standard \acrshort{RSP} protocol (where, Alice teleport a known quantum state to Bob) to a protocol of joint-\acrshort{RSP} scheme where two senders sends a known quantum state to Bob. As for a known single qubit state $|\psi\rangle=\cos(\frac{\theta}{2})|0\rangle+\exp(i\phi)\sin(\frac{\theta}{2})|1\rangle$, one sender may have knowledge of $\theta$ and another sender may have knowledge of $\phi$ and they may jointly prepare this state at Bob's port. Such a sharing of knowledge is not possible when we have to teleport an unknown state. A relatively detailed discussion on variants of teleportation and \acrshort{RSP} can be found in Reference \cite{Pathak2013book,vishal2015cb_RSP_NC,Vikram2021CC-QT}. Without describing these variants in detail, different variants of teleportation and \acrshort{RSP} are summarized in Figure \ref{fig:QCscheme_classification}. Among different variants of teleportation and \acrshort{RSP} mentioned in Figure \ref{fig:QCscheme_classification}, this thesis has reported work on multi-output \acrshort{QT} and \acrfull{QB}.
\section{Preliminaries}
\subsection{Quantum states\label{ssec:qGates}}
\begin{figure}
	\centering
	\includegraphics[width=0.5\linewidth]{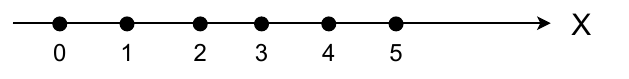}
	\caption{\centering One-dimensional system which has six sites. Particle can only exist in one of the six sites.\label{fig:qState}}
\end{figure}

A quantum state is used to describe the state of a quantum system. A quantum state is represented by a complex wave function $\Psi$, which depends on the coordinate $x$ and on time $t$, $\Psi(x,t)$. Consider a particle in a system that can only be in one of the six points along the $x$ axis labeled as $0,1,...5$ shown in Figure \ref{fig:qState}. Of course, such particles cannot be visualized in the classical world. The wave function $\Psi(x,t)$ encodes the entire information about the quantum system in a probabilistic sense. The probability of finding the particle in the interval $x\rightarrow x+dx$ is estimated by $|\Psi(x,t)|^2dx$. The probability of locating the particle along the real axis must be one, that is, 

\begin{equation}
||\psi||^{2}=\int{|\psi(x,t)|^{2}dx}=1.
\end{equation}
In any such discretized system, the state of the system is represented by a vector. The popular notations used to represent a quantum state is Dirac notation. The symbol $|\quad\rangle$ is called \textit{ket} used to represent the state of a quantum system. The complex conjugate of \textit{ket} is known as \textit{bra}, denoted as $\langle\quad|$. Below are few quantum states with their corresponding matrices:
    \begin{equation*}
	|0\rangle\quad\text{corresponds to}\quad\begin{pmatrix}
		1\\
		0
	\end{pmatrix},
\end{equation*}
\begin{equation*}
	|1\rangle\quad\text{corresponds to}\quad\begin{pmatrix}
		0\\
		1
	\end{pmatrix},
\end{equation*}
\begin{equation*}
	\alpha_{0}|0\rangle+\alpha_{1}|1\rangle\quad\text{corresponds to}\quad\alpha_{0}\begin{pmatrix}
		1\\
		0
	\end{pmatrix}
	+\alpha_{1}\begin{pmatrix}
		0\\
		1
	\end{pmatrix}
	=\begin{pmatrix}
		\alpha_{0}\\
		\alpha_{1}
	\end{pmatrix}.
\end{equation*}
In general, a quantum state can be written as:
\begin{equation}
	|\psi\rangle=\sum_{n}c_{n}|n\rangle
\end{equation}
where, $|n\rangle$ is the basis vectors. A detailed discussion about the measurement basis has done in Section \ref{sec:measurement-basis}.\\
Apart from vector representation of a quantum state, another popular way to represent a quantum state is the density matrix representation. The density matrix representation is mainly required when we describe an interaction of a quantum state. The density matrix state corresponding to the state $|\psi\rangle$ is given as
\begin{equation}
	\begin{split}
	    \rho=|\psi\rangle\langle\psi|=&\sum_{n}\sum_{m}c_nc_m^{*}|n\rangle\langle m|\\
	    & =\sum_{n,m}\rho_{nm}|n\rangle\langle m|
	\end{split}
\end{equation}
where, $\rho_{nm}$ are elements of the density operator matrix for state $|\psi\rangle$. A quantum state can be pure or mixed depending on the criterion given below:
\begin{equation}
	\text{Pure state:}\quad \rho=|\psi\rangle\langle\psi|;\quad\text{tr}(\rho^2)=1
\end{equation}
\begin{equation}
	\text{Mixed state:}\quad \rho=\sum_{i}p_{i}|\psi_{i}\rangle\langle\psi_{i}|;\quad\text{tr}(\rho^2)<1.
\end{equation}
where, tr stands for trace. Trace of an operator is estimated as sum of its diagonal elements. It is to be noted that $\sum_{i}p_{i}=1$.
\subsection{Quantum bits\label{ssec:qubits}}
\begin{figure}
	\centering
	\includegraphics[width=0.5\linewidth]{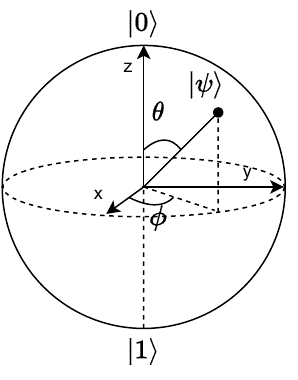}
	\caption{\centering Representation of a qubit on Bloch sphere\label{fig:bloch-sphere}}
\end{figure}

Quantum bits or qubits serve as the fundamental units of quantum information. Similar to classical bit $'0'$ and $'1'$, qubit also has two possible states $|0\rangle$ and $|1\rangle$. Unlike a bit which can have two values $0$ and $1$ at a time, a qubit can have many possible values in a linear combination of $|0\rangle$ and $|1\rangle$. Hence, qubits can be represented as the linear combination of $|0\rangle$ and $|1\rangle$ which is given as
\begin{equation}
	|\psi\rangle=\alpha|0\rangle+\beta|1\rangle
	\label{eq:qubit1}
\end{equation}
where, $\alpha$ and $\beta$ are arbitrary complex numbers which follows the normalization condition $|\alpha|^2+|\beta|^2=1$. A vector in two-dimensional hilbert space as $\begin{pmatrix} \alpha\\ \beta \end{pmatrix}$ can also be used to represent the qubit. When a qubit is measured in the computational basis $\{|0\rangle,|1\rangle\}$ (the measurement basis is discussed in more detail in Section \ref{sec:measurement-basis}) then the qubit collapses to either $|0\rangle$ or $|1\rangle$ with probability $|\alpha|^2$ or $|\beta|^2$ respectively. The qubit state representation in Equation \eqref{eq:qubit1} can also be rewritten as 
\begin{equation}
	|\psi\rangle=\cos(\frac{\theta}{2})|0\rangle+e^{i\phi}\sin(\frac{\theta}{2})|1\rangle.
	\label{eq:qubit2}
\end{equation}
which can be visualized in the Bloch sphere shown in Figure \ref{fig:bloch-sphere}.
Qubit can be physically realized in many different ways. The most popular ways to realize qubits are two distinct polarizations of a photon, two quantum states of an electron orbiting around a single atom, and orientation of a nuclear spin within a uniform magnetic field.
\subsection{Measurement basis\label{sec:measurement-basis}}
A measurement basis is required to measure a qubit. A qubit can be measured using different bases. Each basis gives different measurement outcomes. A set of vectors $\{v_1,v_2,v_3,...,v_n\}$ can form a measurement basis if it satisfies the following three conditions:
\begin{enumerate}
	\item Elements in the set are linearly independent i.e., $\sum_{i=1}^{n}b_iv_i=0$ iff all $b_i=0$,
	\item Elements must satisfies orthogonality condition $\langle v_i|v_j\rangle=\delta_{ij}$, and
	\item They also satisfies the completeness relation $\sum_{i=1}^{n}|v_i\rangle\langle v_i|=1$. 
\end{enumerate}
The basis sets used in this thesis are computational, diagonal, and Bell basis which are described below: 
\begin{description}
	\item[Computational basis] Single-qubit computational basis set is $\{|0\rangle,|1\rangle\}$. This can be extended to multi-qubit also as a tensor product of multiple single qubit basis. The two-qubit computational basis is $\{|00\rangle,|01\rangle,|10\rangle,|11\rangle\}$.
	\item[Diagonal basis] Single-qubit diagonal basis set is $\{|+\rangle,|-\rangle\}$ where, $|\pm\rangle=\frac{|0\rangle\pm|1\rangle}{2}$. This can also be extended to multi-qubit also. The two-qubit diagonal basis is $\{|++\rangle,|+-\rangle,|-+\rangle,|--\rangle\}$.
	\item[Bell basis] The Bell basis is given as $\{|\phi^{+}\rangle,|\phi^{-}\rangle,|\psi^{+}\rangle,|\psi^{-}\rangle\}$ where, $|\phi^{\pm}\rangle=\frac{1}{2}(|00\rangle\pm|11\rangle),|\psi^{\pm}\rangle=\frac{1}{2}(|01\rangle\pm|10\rangle)$.  
\end{description}
It is to be noted that there can be many such measurement bases to measure a qubit. 
\subsection{Tensor products}
Tensor product is a mathematical tool that acts between two small vector spaces to produce an enlarged vector space. The tensor product is represented by the symbol $\otimes$. The tensor product between $|0\rangle$ and $|1\rangle$ is given as
\begin{equation}
	|01\rangle=|0\rangle\otimes|1\rangle=\begin{pmatrix} 1 {\begin{pmatrix} 0 \\ 1 \end{pmatrix}} \\ 0 {\begin{pmatrix} 0 \\ 1 \end{pmatrix}} \end{pmatrix}=\begin{pmatrix} 0 \\ 1 \\ 0\\ 0\end{pmatrix}
\end{equation}
It is to be noted that the dimension of $|0\rangle$ or $|1\rangle$ is $2\times1$ but the dimension of $|01\rangle$ is $4\times1$. Let's take another example, consider a matrix $A$ and $B$ given as
\begin{equation}
	A=\begin{pmatrix} a_{00} & a_{01} \\ a_{10} & a_{11} \end{pmatrix} \quad B=\begin{pmatrix} b_{00} & b_{01} \\ b_{10} & b_{11} \end{pmatrix}
\end{equation}
then the tensor product between them is given as follows
\begin{equation}
	A\otimes B=\begin{pmatrix} a_{00}B & a_{01}B \\ a_{10}B & a_{11}B \end{pmatrix}=\begin{pmatrix} a_{00}b_{00} & a_{00}b_{01} & a_{01}b_{00} & a_{01}b_{01} \\ a_{00}b_{10} & a_{00}b_{11} & a_{01}b_{10} & a_{01}b_{11}\\
	a_{10}b_{00} & a_{10}b_{01} & a_{11}b_{00} & a_{11}b_{01}\\ 
a_{10}b_{10} & a_{10}b_{11} & a_{11}b_{10} & a_{11}b_{11}\end{pmatrix}
\end{equation}
One thing you can notice is that all the possible scalar multiplication between elements of the matrix $A$ and $B$ is there in the tensor product $A\otimes B$. 
\subsection{Quantum gates}
You may have played with kids in your home and encountered with the famous Lego blocks game. You can make various possible structures from the Lego blocks. Similarly, quantum gates are building blocks of quantum circuits so as classical gates. Then what makes quantum gates differ from the classical gates? All quantum gates must be reversible in nature while all classical gates are not necessarily reversible. Consider AND gate which gives $1$ only when both the inputs are set at $1$. It produces one output bit from two input bits which implies one bit has been erased. The eraser of one bit involves in the loss of energy $kT\ln(2)$ where, $K$ is the Boltzmann constant and $T$ is temperature.\\
An $N$-qubit quantum gate can be described by a $2^{N}\times2^{N}$ matrix. The most popular single-qubit quantum gates are the Pauli matrices listed below:
\begin{equation}\label{eq:Pauli-X}
	\text{Pauli-$X$ gate or NOT gate:}\quad
	X=\begin{bmatrix}
		0 & 1\\ 
		1 & 0
	\end{bmatrix}
=|0\rangle\langle1|+|1\rangle\langle0|,
\end{equation}

\begin{equation}\label{eq:Pauli-Y}
	\text{Pauli-$Y$ gate:}\quad
	Y=\begin{bmatrix}
		0 & -i\\ 
		i & 0
	\end{bmatrix}
=-i|0\rangle\langle1|+i|1\rangle\langle0|,
\end{equation}
\begin{equation}\label{eq:Pauli-Z}
	\text{Pauli-$Z$ gate:}\quad
	Z=\begin{bmatrix}
		1 & 0\\ 
		0 & -1
	\end{bmatrix}
=|0\rangle\langle0|-|1\rangle\langle1|.
\end{equation}

Apart from Pauli gates others single qubit gates are listed as follows:
\begin{equation}
	\text{Hadamard gate:}\quad
	H=\frac{1}{\sqrt{2}}|0\rangle \begin{bmatrix}
		1 & 1\\ 
		1 & -1
	\end{bmatrix}
=\frac{1}{\sqrt{2}}(|0\rangle+|1\rangle)\langle0|+\frac{1}{\sqrt{2}}(|0\rangle-|1\rangle)\langle1|,
\end{equation}
\begin{equation}
	\text{Phase gate:}\quad
	P(\phi)=\begin{bmatrix}
		1 & 0\\ 
		0 & e^{i\phi}
	\end{bmatrix}
	=|0\rangle\langle0|+e^{i\phi}|1\rangle\langle1|,
\end{equation}
\begin{equation}
	\text{Identity gate:}\quad
	I=\begin{bmatrix}
		1 & 0\\ 
		0 & 1
	\end{bmatrix}
	=|0\rangle\langle0|+|1\rangle\langle1|.
\end{equation}
Three different classes of unitary matrix are derived from the Pauli matrices, these are \textit{rotation operators} given as follows:
\begin{equation}
	R_x(\theta)\equiv e^{-i\frac{\theta}{2}X}
	=\cos\frac{\theta}{2}I-i\sin\frac{\theta}{2}X=
	\begin{bmatrix}
		\cos\frac{\theta}{2} & -i\sin\frac{\theta}{2}\\ 
		-i\sin\frac{\theta}{2} & \cos\frac{\theta}{2}
	\end{bmatrix},
\end{equation}
\begin{equation}
	R_y(\theta)\equiv e^{-i\frac{\theta}{2}Y}
	=\cos\frac{\theta}{2}I-i\sin\frac{\theta}{2}Y=
	\begin{bmatrix}
		\cos\frac{\theta}{2} & -\sin\frac{\theta}{2}\\ 
		\sin\frac{\theta}{2} & \cos\frac{\theta}{2}
	\end{bmatrix},
\end{equation}
\begin{equation}
	R_z(\theta)\equiv e^{-i\frac{\theta}{2}Z}
	=\cos\frac{\theta}{2}I-i\sin\frac{\theta}{2}Z=
	\begin{bmatrix}
		e^{-i\frac{\theta}{2}} & 0\\ 
		0 & e^{i\frac{\theta}{2}}
	\end{bmatrix}.
\end{equation}
These rotation operators $R_x$, $R_y$ and $R_z$ can be physically interpreted as the rotation of a qubit along $X$, $Y$ and $Z$ axes, respectively. A single qubit arbitrary unitary operator can be written as a combination of the three rotation operators with an additional global phase shift given as
\begin{equation}
	U=e^{i\phi}R_z(\gamma)R_y(\beta)R_x(\alpha).
\end{equation} 
\begin{figure}
	\centering
	\includegraphics[width=0.6\linewidth]{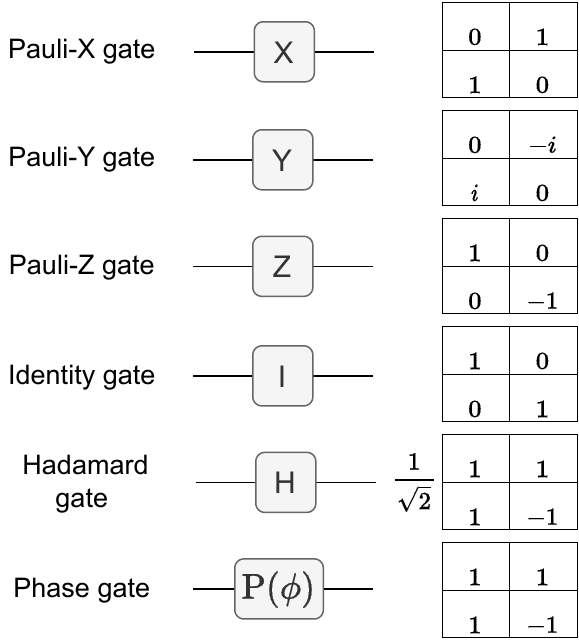}
	\caption{\centering Single qubit quantum gates with their representation and corresponding matrices.\label{fig:represent-single-qGate}}
\end{figure}
All single qubit gates have a way to represent. The representation of a few important single qubit gates discussed in this thesis are shown in Figure \ref{fig:represent-single-qGate}.\\

Like single qubit gates, there are multi-qubit quantum gates also. Here, this discussion is restricted to few two-qubit quantum gates which are used in this thesis. The two-qubit gates used are CNOT and SWAP gates. The CNOT gate flips the target qubit when the control qubit is set at $|1\rangle$. The matrix representation of the CNOT gate is given as follows:
\begin{equation}
 	\text{CNOT}=
 	\begin{bmatrix}
 		1 & 0 & 0 & 0\\ 
 		0 & 1 & 0 & 0\\
 		0 & 0 & 0 & 1\\
 		0 & 0 & 1 & 0\\
 	\end{bmatrix}
 	=|00\rangle\langle00|+|01\rangle\langle01|+|11\rangle\langle10|+|10\rangle\langle11|.
\end{equation}
The SWAP gate swaps the states of two qubits like $|ab\rangle\rightarrow|ba\rangle$. The matrix representation of the SWAP gate is given as follows:
\begin{equation}
	\text{SWAP}=
	\begin{bmatrix}
		1 & 0 & 0 & 0\\ 
		0 & 0 & 1 & 0\\
		0 & 1 & 0 & 0\\
		0 & 0 & 0 & 1\\
	\end{bmatrix}
	=|00\rangle\langle00|+|10\rangle\langle01|+|01\rangle\langle10|+|11\rangle\langle11|.
\end{equation}
\begin{figure}
	\centering
	\includegraphics[width=0.8\linewidth]{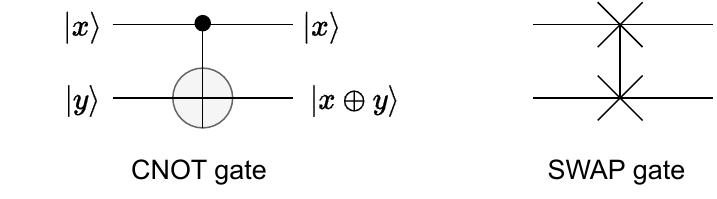}
	\caption{\centering Symbolic representation of CNOT and SWAP gates.\label{fig:represent-two-qGate}}
\end{figure}
The representation of CNOT and SWAP gates is shown in Figure \ref{fig:represent-two-qGate}.
\subsection{Available platforms to access quantum computers on cloud}
Recent developments in the domain of quantum computing have led to various platforms providing easy access to quantum computers by making them available on the cloud. There are various service providers (organizations) who provides cloud access to quantum computers. A few of them are listed below
\begin{figure}
	\centering
	\includegraphics[width=0.9\linewidth]{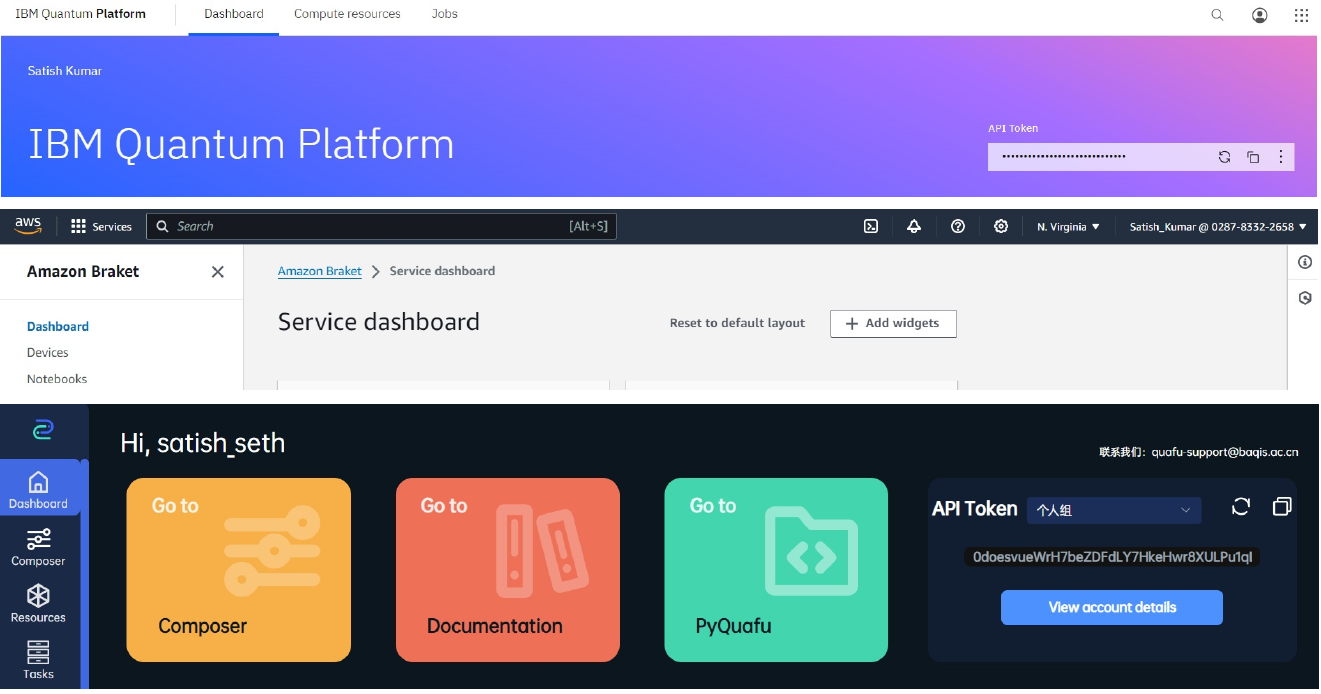}
	\caption{\centering Snap of few platforms where one can access quantum computers on cloud.}
	\label{fig:qc_platforms}
\end{figure}
\begin{description}
	\item[IBM Quantum] IBM provides an online platform called \textit{IBM Quantum Platform} where one can access quantum computers on the cloud. Qiskit is a python library that is used to give commands to quantum computers available at IBM.
	\item[Amazon Bracket] This platform is offered by \textit{Amazon Web Service} which is used to access quantum computers from various companies like, IonQ, Rigetti Computing and Oxford Quantum Circuits (OQC).
	\item[Quafu] This platform was recently launched by \textit{Beijing Academy of Quantum Information Sciences (baqis)} which also provides free access to quantum computers on the cloud.
\end{description}
A snap of the above-described platforms is shown in Figure \ref{fig:qc_platforms}. It is to be noted that even though various platforms are available to access quantum computers on the cloud, the results reported in this thesis are mainly obtained from IBM quantum computer.  
\subsection{Characteristics of qubits}
When a quantum computer is developed, it is needed to calibrate its qubits. Quantum computers can access through cloud service at various platforms like IBM, Quafu, AWS, etc. The quantum computers available at these platforms are based on superconducting qubits. Each quantum processor has their calibration data for their qubits. The several parameters used for calibrating qubits are listed below:
\begin{description}
	\item[Relaxation time $(T_1):$] It is a measure of how quickly a qubit in the excited $(|1\rangle)$ state spontaneously relaxes to the ground $(|0\rangle)$ state.
	\item[Dephasing time $(T_2):$] It is a measure of how fast a coherent superposition $(\cos\frac{\theta}{2}|0\rangle+e^{i\phi}\sin\frac{\theta}{2}|1\rangle)$ losses its relative phase. A completely dephased qubit is just a probabilistic classical bit.
	\item[] $T_1$ and $T_2$ together known as \textit{decoherence time}.
	\item[Frequency:] Superconducting qubits interact with each other via microwave resonators. Each qubit must have a different frequency to address them separately. The frequency is defined as the energy difference between the ground ($|0\rangle$) and excited states ($|1\rangle$).
	\item[Readout assignment error:] It measures the accuracy of performing the measurement on a qubit. It actually measures the probability of obtaining wrong measurement values for the measurement performed. 
	\item[Single-qubit Pauli-X-error:] It measures the error caused by applying the Pauli-X gate, on an average over all the different qubits.
	\item[CNOT error:] It is the measure of error induced by applying the two-qubit CNOT gate on the connecting qubits of the device. Average over all possible CNOT gates.
\end{description}
There are more characteristic parameters than listed above for qubit calibration. But here, a brief about the parameters used in the thesis is provided.
\subsection{Quantum state tomography}
As discussed before, a quantum state collapses into one possible state after measurement. So, it is difficult to obtain the entire picture of the quantum state which has many different possibilities. To obtain a quantum state having all possible superposition states, one needs to make several measurements in all possible bases. Such a technique which is used to determine an unknown quantum state of a system is known as \acrfull{QST}. A density matrix for a $n$-qubit quantum state is obtained as
\begin{equation}
	\rho=\sum_{\vec{v}}\frac{\text{tr}(\sigma_{v_1}\otimes\sigma_{v_2}\otimes...\sigma_{v_n}\rho)\sigma_{v_1}\otimes\sigma_{v_2}\otimes...\sigma_{v_n}}{2^n}
\end{equation}
where, the sum is over vector $\vec{v}=(v_1,...,v_n)$ and $\{\sigma_{v_i}\}=\{I,X,Y,Z\}$. For a single qubit ($n=1$), the density matrix is obtained as
\begin{equation}
	\rho=\frac{\text{tr}(\rho)I+\text{tr}(X\rho)X+\text{tr}(Y\rho)Y+\text{tr}(Z\rho)Z}{2} 
\end{equation}
where, $\text{tr}(A\rho)$ estimates the average value of observable $A$.  
\subsection{Quantum entanglement}
Consider two systems $A$ and $B$. The two systems are entangled when the physical property of one system is correlated with the physical property of the other system. Entanglement can occur between more than two systems also. When two or more systems are correlated with their quantum properties, then it is called as \textit{quantum entanglement}. In another way, entangled states are those quantum states which are not separable. Let $|\psi_{AB}\rangle$ is an entangled state with $|\psi_A\rangle$ and  $|\psi_B\rangle$ are quantum state of the system $A$ and $B$ respectively, then $|\psi_{AB}\rangle\neq|\psi_A\rangle\otimes\psi_B\rangle$. We often come across the singlet quantum state where the total spin of electron pair in a system is zero, given as
\begin{equation}
	|\psi\rangle=\frac{1}{\sqrt{2}}(|\uparrow\rangle|\downarrow\rangle-|\downarrow\rangle|\uparrow\rangle)
\end{equation}
When one electron in the system is found to be in the spin-up state then the other electron must be in the spin-down state. So, the singlet state is entangled in spin property of electrons. There can be many such entangled quantum states. These entangled states are used to accomplish various tasks of quantum computing and quantum communication. Here, some entangled quantum states discussed in this thesis are described which are listed as follows:
\begin{description}
	\item[Bell state:] These are maximally two-qubit entangled state, named after physicist John Bell, which are given as
	\begin{equation}
		|\phi^{\pm}\rangle=\frac{1}{\sqrt{2}}(|00\rangle\pm|11\rangle),\quad|\psi^{\pm}\rangle=\frac{1}{\sqrt{2}}(|01\rangle\pm|10\rangle).
	\end{equation}
    It is to be noted that these states form an orthogonal set which form the basis for any two-qubit system. These states are used as a resource in quantum tasks like, \acrshort{QT}, \acrshort{RSP}, \acrfull{SDC}, \acrfull{QEC}, etc.
    \item[GHZ state:] The GHZ state is the three or more qubit maximally entangled states, names after three scientist Daniel Greenberger, Michael Horne and Anton Zeilinger. The GHZ state stands for Greenberger–Horne–Zeilinger state. The simplest form of the GHZ state is given as
    \begin{equation}
    	|\psi_{GHZ}\rangle=\frac{1}{\sqrt{2}}(|000\rangle\pm|111\rangle).
    \end{equation}
    The GHZ state can be extended to multiple qubit also. The GHZ state for $n-$qubit is given as
    \begin{equation}
    	|\psi_{GHZ_n}\rangle=\frac{1}{\sqrt{2}}(|0\rangle^{\otimes n}\pm|1\rangle^{\otimes n}).
    \end{equation}
    This state has various applications in different versions of quantum computing and communication tasks.
    \item[Cluster state:] A cluster state is a multi-qubit entangled state. It is a type of \textit{graph state} where qubits are represented by vertexes of a graph and entanglement is created on the edges of a graph. In this thesis, a four-qubit cluster state will be used which is given as
    \begin{equation}
    	|\psi_{C}\rangle=\frac{1}{\sqrt{2}}(|0000\rangle+|0011\rangle+|1100\rangle-|1111\rangle)
    \end{equation}
    \item[Hyper-entangled state:] Entanglement is not restricted to only one quantum property or one \acrfull{DOF}. Two or more quantum states can be entangled in more than one \acrshort{DOF} simultaneously, quantum states with such properties are known as \textit{hyper-entangled states}. Usually hyper-entangled state is realized with photon as a qubit. Photons can entangle in various \acrshort{DOF} like, spatial, time-bin, orbital angular momentum (OAM) or frequency simultaneously. The two-qubit hyper-entangled state is given as
    \begin{equation}
    	|Q\rangle=\frac{1}{2}[(|HH\rangle+|VV\rangle)\otimes(|a_0b_0\rangle+|a_1b_1\rangle)]
    \end{equation}
    where, $|H\rangle$ ($|V\rangle$) represents a horizontal (vertical) polarization state of a photon and $a_0\,\&\,a_1$ and $b_0\,\&\,b_1$ represents two-paths of photon $A$ and $B$ respectively.
\end{description}
Till now, various maximally entangled and hyper-entangled states have been discussed but there are also non-maximally entangled states where correlation occurs in a probabilistic manner. The non-maximally entangled states are beyond the scope of this thesis.
\section{Few no-go theorems}\label{sec:no-go theorems}
A no-go theorem corresponds to an event (task or result) that seems to be physically possible but it is not possible in reality. No-go theorems are not only quantum, a classical example of a no-go theorem is Heisenberg's uncertainty principle. A few no-go theorems are described as follows: 
\begin{description}
	\item[No-cloning theorem:] Consider two quantum systems $A$ and $B$ belong to a common Hilbert space $H=H_A=H_B$. Given an unknown quantum state $|\phi\rangle_A\in H_A$ and an auxiliary state $|\acute{a}\rangle_B \in H_B$, there is no unitary which can map $|\phi\rangle_A\otimes|\acute{a}\rangle_A\longrightarrow|\phi\rangle_A\otimes|\phi\rangle_A$.
	\item[No-deletion theorem:] It states that given two copies of an unknown quantum state, one cannot be erased while preserving the other intact. In another way, there is no unitary $U$ that satisfies the following relation 
	\begin{equation}
		U(|\psi\rangle|\psi\rangle|\acute{a_i}\rangle)=|\psi\rangle|e\rangle|\acute{a_f}\rangle,
	\end{equation}
	where, $|\acute{a_i}\rangle$ ($|\acute{a_f}\rangle$) is initial (final) ancilla state and $|e\rangle$ is an empty state of dimension same as $|\psi\rangle$.	
	\item[No-broadcast theorem:] Quantum broadcasting is a weaker version of cloning. The theorem states that given an unknown state $\rho\in H$, there is no physical process which can create $\rho_{AB}$ in a Hilbert space $H_A\otimes H_B$ such that $\text{tr}_A(\rho_{AB})=\text{tr}_B(\rho_{AB})=\rho_i$.
	\item[No-hiding theorem:] It shows that a quantum information lost due to decoherence cannot be hidden or removed completely in a way that it is inaccessible. In another way, if a quantum information is lost in its surrounding environment then it is hidden somewhere in the correlation between the system and its environment.  
\end{description}
There are many such no-go theorems like no-teleportation, no-communication theorem, etc. Here, we restrict ourselves to a few important no-go theorems which are used in the thesis.   
\section{Various facets of quantum communication}
Quantum communication is a process to communicate an intended message using qubits. Basic principles of quantum mechanics ensure the security of quantum communication. There are various schemes where secure communication is required. A few of them are \acrshort{QT}, \acrshort{RSP}, \acrshort{SDC}, quantum cryptography, etc. Here, three such schemes will be studied. 
\subsection{Quantum teleportation}
\acrshort{QT} is a process to teleport an unknown quantum state (information) from one place to another. There is no classical analogue to the \acrshort{QT} scheme. The first scheme of \acrshort{QT} was introduced in the year 1993 by Charles Bennett et al. \cite{Bennett1993QT}. One can think of posting a letter as a classical analogue to the QT scheme but it is not because message travels through the channel during posting a letter. Another classical analogue one can think is Faxing but it is to be noted that original message remains with the sender during Faxing. In QT, an unknown message gets disappear at sender's side and appear at receiver's side without traveling through the channel. \\
\begin{figure}
	\includegraphics[width=\linewidth]{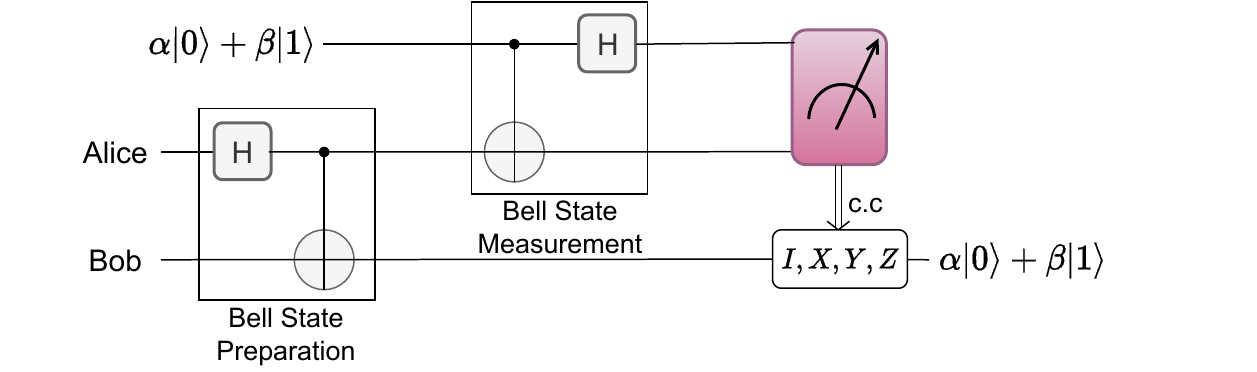}
	\caption{\centering Quantum teleportation circuit.}
	\label{fig:QT_ckt}
\end{figure}
Let Alice and Bob be two parties where Alice is a sender and Bob is a receiver. Alice wants to teleport an unknown state $|\psi\rangle=\alpha|0\rangle+\beta|1\rangle$ to Bob. To do so, Alice first prepares a Bell state (say $|\phi^{+}\rangle=\frac{1}{\sqrt{2}}(|00\rangle+|11\rangle)$), keeping the first qubit with herself and sends the second qubit to Bob. Now, Alice measures her both qubits (unknown and entangled qubit) using the Bell basis and announces her measurement outcomes. Then, Bob will apply unitary operation corresponding to the measurement outcomes to get the desired unknown quantum state. The process of QT is described in the Figure \ref{fig:QT_ckt}. It is to be noted that when Alice and Bob will use a non-maximally two-qubit entangled state for the QT scheme then it will reduce to the probabilistic-QT scheme. There are many variants of such schemes like controlled-QT, joint-QT, controlled-joint QT, and so on. 

\subsection{Remote state preparation}
\begin{figure}
	\includegraphics[width=\linewidth]{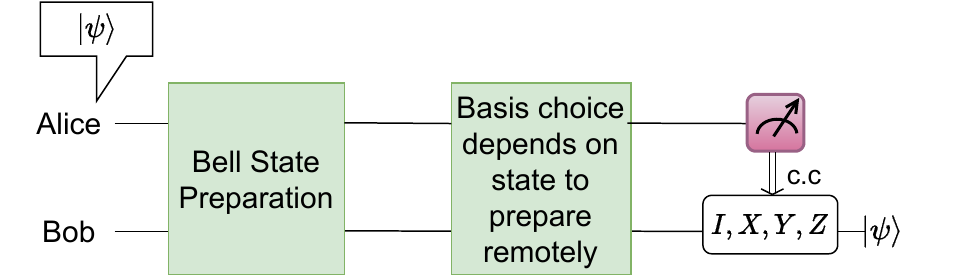}
	\caption{\centering Circuit illustrating the \acrshort{RSP} process.}
	\label{fig:RSP_ckt}
\end{figure}

Similar to the \acrfull{QT} scheme, where an unknown qubit is teleported, here in \acrfull{RSP}, a known state is remotely prepared. The concept of \acrshort{RSP} was first introduced by A. K. Pati in the year 2000 \cite{AkPati2000RSP}. The scheme of \acrshort{RSP} also has no classical analogue. In \acrshort{RSP}, Alice has a known state which she wants to remotely prepare at Bob's qubit. Let, Alice want to prepare $|\psi\rangle=\cos(\frac{\theta}{2})|0\rangle+e^{i\phi}\sin(\frac{\theta}{2})|1\rangle=|q_1\rangle$ (note, Alice knows $\theta$ and $\phi$) remotely at Bob's side. For that, Alice first prepares the Bell state (say, $|\phi^{-}\rangle=\frac{1}{\sqrt{2}}(|01\rangle-|10\rangle$) and then chooses a measurement basis based on the state $|\psi\rangle$. Here, Alice chooses $\{|q_1\rangle,|q_2\rangle\}$ as a measurement basis where $|q_2\rangle=-\sin(\frac{\theta}{2})e^{i\phi}|0\rangle+\cos(\frac{\theta}{2})|1\rangle=U_{RSP}|q_1\rangle$ ($U_{RSP}=XZP(-2\phi)$). The measurement basis $\{|q_1\rangle,|q_2\rangle\}$ transforms the Bell state to $|\phi^{-}\rangle=\frac{1}{\sqrt{2}}(|q_1q_2\rangle-|q_2q_1\rangle$). Alice then shares the second qubit of the Bell state to Bob and measures her first qubit. Bob then applies $I$ ($U^{-1}_{RSP}=-P(2\phi)ZX$) on his qubit when Alice's gets $|q_2\rangle$ ($|q_1\rangle=|\psi\rangle$). The quantum circuit illustrating the \acrshort{RSP} is shown in the Figure \ref{fig:RSP_ckt}.
\subsection{Quantum cryptography}
Cryptography is a technique to protect information from an unauthorized party. Quantum cryptography is a method to protect information using principles of quantum mechanics. \acrshort{QKD} is the best example of quantum cryptography. In \acrshort{QKD}, two parties generate a secure key at both ends with which they can make secure communication. The first \acrshort{QKD} protocol was introduced by Charles Bennett and Gilles Brassard, known as BB84 protocol in the year $1984$ \cite{Bennett1984QCrypt}. After that many \acrshort{QKD} protocols have been proposed like E91 \cite{Ekert1991QCrypt}, B92 \cite{Bennett1992QKD}, and so on. Here, the famous BB84 protocol is briefly explained with an example shown in Table \ref{tab:Example-BB84}. 
\begin{table}
	\caption{\centering An example of generating a secure key using the BB84 protocol.\label{tab:Example-BB84}}
	\centering
	\footnotesize
	\begin{tabular}{|c|c|c|c|c|c|c|c|c|}
		\hline
		Alice bits & 0 & 1 & 0 & 0 & 0 & 1 & 0 & 1 \\
		\hline
		Alice basis & $\{|0\rangle,|1\rangle\}$ & $\{|0\rangle,|1\rangle\}$ & $\{|+\rangle,|-\rangle\}$ & $\{|0\rangle,|1\rangle\}$ & $\{|+\rangle,|-\rangle\}$ & $\{|0\rangle,|1\rangle\}$ & $\{|+\rangle,|-\rangle\}$ & $\{|+\rangle,|-\rangle\}$\\
		\hline
		Bob basis & $\{|0\rangle,|1\rangle\}$ & $\{|0\rangle,|1\rangle\}$ & $\{|0\rangle,|1\rangle\}$ & $\{|+\rangle,|-\rangle\}$ & $\{|+\rangle,|-\rangle\}$ & $\{|+\rangle,|-\rangle\}$ & $\{|0\rangle,|1\rangle\}$ & $\{|+\rangle,|-\rangle\}$\\
		\hline
		Match & Y & Y & N & N & Y & N & N & Y\\
		\hline
		Keep & Y & Y & N & N & Y & N & N & Y\\
		\hline
	\end{tabular}
\end{table}
Alice encodes her random bits in the form of qubits. The encoding is such that $0$ ($1$) bit encoded as $\{|0\rangle,|+\rangle\}$ ($\{|1\rangle,|-\rangle\}$) randomly. Alice then sends the encoded states to Bob, who does measure it in computational $\{|0\rangle,|1\rangle\}$ and diagonal basis $\{|+\rangle,|-\rangle\}$ randomly. Alice and Bob then compare their measurement basis and keep the corresponding bits where bases match and discard others. In this way, Alice and Bob will have a common secret key.
\section{Modeling noisy quantum system}\label{sec1.5:Noise Model}
\begin{figure}
	\begin{centering}
		\begin{tabular}{cc}
			\includegraphics[scale=0.95]{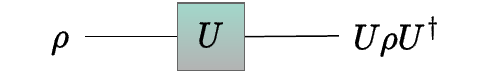} &
			\includegraphics[scale=0.95]{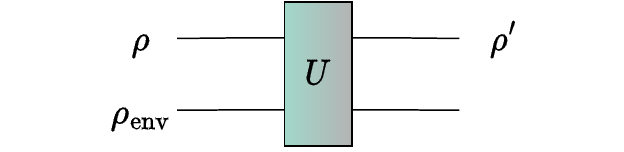}\tabularnewline
			(a) & (b)\tabularnewline
		\end{tabular}
	\end{centering}
	\caption{\centering Pictorial representation for modeling of (a) close and (b) open quantum system.\label{fig:modelSystem}}
\end{figure}

Any real system, whether classical or quantum, is not perfectly closed. A real system must have an environment surrounding it which causes noise in the system. Modeling noise in classical and quantum systems is different. Also, there are many ways to model noise in a system. Here, we restrict ourselves to model noise in quantum systems using \textit{operator-sum representation}.\\
A closed system can be modeled as a unitary evolution of an initial quantum state $\rho$. Consider a box that represents a unitary evolution $U$, which can either be a quantum circuit or Hamiltonian of some quantum system, and a solid line that represents a quantum state (see Figure \ref{fig:modelSystem}). A state $\rho$ of a closed system is evolved as $U\rho U^{\dagger}$ which is shown in Figure \ref{fig:modelSystem} (a). However, open system evolution behaves differently. For an open system, there must be some environmental state with which the principal system interacts. Let, $\rho$ and $\rho_{\text{env}}$ be a quantum state of a principal system and an environment, respectively. The initial joint state of the open system can be written as $\rho\otimes\rho_{\text{env}}$. The joint state interacts for a certain time period, after that the system is no longer interacts with the environment. The final state of an open system after the interaction with the environment is given as
\begin{equation}
	\rho^{\prime}=\text{tr}_{\text{env}}[U(\rho\otimes\rho_{\text{env}}) U^{\dagger}].
	\label{eq:noisyQsys}
\end{equation}

where, $\text{tr}_{\text{env}}$ stands for trace out environmental qubit. It is to be noted that in order to model a noisy system, the dimension of Hilbert space associated with the environment should be no more than $d^2$, if $d$ is the dimension of Hilbert space associated with the principal system.\\

Equation \eqref{eq:noisyQsys}, describing the noisy quantum system can also be written in the form of \textit{operator-sum representation}. Let $\{|e_k\rangle\}$ be an orthonormal basis of a finite-dimensional environment and $|e_0\rangle\langle e_0|$ be the initial state of the environment then Equation \eqref{eq:noisyQsys} can be rewritten as
\begin{equation}
	\begin{split}
		\rho^{\prime} &=\sum_{k}\langle e_k|U[\rho\otimes|e_0\rangle\langle e_0|]U^{\dagger}|e_k\rangle\\
		&=\sum_{k}E_k\rho E_{k}^{\dagger},
	\end{split}
\end{equation}
where, $E_k=\langle e_k|U|e_0\rangle$ is an operator that acts upon the principal system, popularly known as \textit{Kraus operator} which follows the completeness relation $\sum_{k}E_{k}^{\dagger}E_{k}=I$. The Kraus operator for various noisy channels are different which are described as follows:
\begin{description}
	\item[ Bit flip noise:] This noise in the quantum channel flips the state of a qubit from $|0\rangle$ to $|1\rangle$ with some probability $p$ and conversely. The Kraus operator for the same is given as
	\begin{equation}
		E_{0}^{BP}=\sqrt{p}
		\begin{pmatrix} 
			1 & 0 \\
			0 & 1
		\end{pmatrix},
		\qquad
		E_{1}^{BP}=\sqrt{1-p} 
		\begin{pmatrix} 
			0 & 1 \\
			1 & 0
		\end{pmatrix}.
	\end{equation}
  \item[Depolarizing noise:] This noise in the quantum channel converts a pure quantum state into a maximally mixed state. The Kraus operator for the same is given as
  \begin{equation}
  	E_{0}^{DP}=\sqrt{1-\frac{3p}{4}}
  	\begin{pmatrix} 
  		1 & 0 \\
  		0 & 1
  	\end{pmatrix},
  	\quad
  	E_{1}^{DP}=\frac{\sqrt{p}}{2}
  	\begin{pmatrix} 
  		0 & 1 \\
  		1 & 0
  	\end{pmatrix},
  	\quad
  	E_{2}^{DP}=\frac{\sqrt{p}}{2}
  	\begin{pmatrix} 
  		0 & -i \\
  		i & 0
  	\end{pmatrix},
  	\quad 
  	E_{3}^{DP}=\frac{\sqrt{p}}{2}
  	\begin{pmatrix} 
  		1 & 0 \\
  		0 & -1
  	\end{pmatrix},
  \end{equation}
  where, $p$ is the probability of depolarizing a pure quantum state.
  \item[Amplitude damping noise:] This noise in the quantum channel causes the spontaneous emission of photons. The Kraus operator for the same is given as
  \begin{equation}
  	E_{0}^{AD}=
  	\begin{pmatrix} 
  		1 & 0 \\
  		0 & \sqrt{1-p}
  	\end{pmatrix},
  	\qquad
  	E_{1}^{AD}= 
  	\begin{pmatrix} 
  		0 & \sqrt{p} \\
  		0 & 0
  	\end{pmatrix},
  \end{equation}
  where, $p$ represents the amplitude damping probability.
  \item[Phase damping noise:] This noise in the quantum channel causes losses of the relative phase information of the quantum state in the channel while conserving its energy. The corresponding Kraus operator for the same is given as
  \begin{equation}
  	E_{0}^{PD}=\sqrt{1-p}
  	\begin{pmatrix} 
  		1 & 0 \\
  		0 & 1
  	\end{pmatrix},
  	\qquad
  	E_{1}^{PD}=\sqrt{p} 
  	\begin{pmatrix} 
  		1 & 0 \\
  		0 & 0
  	\end{pmatrix},
  	\qquad
  	E_{2}^{PD}=\sqrt{p} 
  	\begin{pmatrix} 
  		0 & 0 \\
  		0 & 1
  	\end{pmatrix},
  \end{equation}
  where, $p$ represents the phase damping probability.
\end{description}

\section{A chronological history of quantum computation and quantum communication}
The chronological history given is basically covering the major developments in the domain of quantum computing and communication. Also, it covers the developments related to this thesis work.
\subsection*{Quantum Computation}
\begin{description}
	 \item[1970-79] Several developments have been made to control the single quantum system. Specially, atom traps and scanning tunneling microscopes methods were developed to control over single quantum systems \cite{Wiesner1983ConjuCode}. 
	 \item[1973] \textit{Reversible Turing machine} model was provided by Charles Bennett \cite{Bennett1973qTuringM}. This lays the foundation stone for the \textit{quantum Turing machine}.
	 \item[1980-82] The idea of quantum mechanical Hamiltonian model of the Turing machine was introduced by Paul Benioff \cite{Benioff1980qTuringM1}. This established that quantum machines can be used to efficiently simulate classical computers.
	 \item[1981] First conference on the \textit{Physics of Computation} was held at MIT in May, where Paul Benioff and Richard Feynman delivered talks on \textit{Quantum Computing}.
	 \item[1982] \begin{itemize}
	 	\item Richard Feynman claims that no classical computer can simulate a quantum mechanical system \cite{Feynman1982QC}.
	 	\item W. K. Wootters and W. H. Zurek introduced the \textit{no-cloning theorem} which states that an unknown quantum state cannot be cloned \cite{Wootters1982No-cloning}.
	    \end{itemize}
	 \item[1985] David Deutsch developed a simple algorithm, now known as \textit{Deutsch's algorithm}, demonstrating that quantum computers can solve certain computational tasks much faster than classical computers \cite{Deutsch1985algo}.
	 \item[1988] First ever quantum computer has been physically realized which uses photon as a qubit \cite{Igeta1988QC1st}.
	 \item[1989] Fredkin gate was optically realized by G. J. Milburn \cite{Milburn1989RealizeFredkinGate}.
	 \item[1994] Peter Shor came up with a quantum algorithm to find prime factors of an integer. Classical secure cryptography protocol RSA, which is based on the factorization problem comes under threat  \cite{Shor1994threatCcrypt}. \textit{A threat to classical cryptography}.
	 \item[1995] \begin{itemize}
	 	\item A theory for \textit{quantum entropy} was developed. The term \textit{'qubit'} was coined first, which is a unit to measure quantum entropy \cite{Schumacher1995qCode}.
	 	\item In the same year, Peter Shor came up with a $9$-qubit \textit{\acrshort{QEC}} code was for reducing decoherence in quantum computers \cite{Shor1995QEC}.
	    \end{itemize} 
	 \item[1996] \begin{itemize}
	 	\item Shor's original idea of \acrshort{QEC} was improved by Laflamme et al. and showed that $5$-qubit \acrshort{QEC} is sufficient to correct single-qubit error \cite{Laflamme1996QEC}.
	 	\item Peter Shor introduced the concept of \textit{fault-tolerant quantum computation} \cite{Shor1996ftQC}.
	    \end{itemize}
	 \item[1997] \begin{itemize}
	 	\item A quantum algorithm was developed by Lov Grover for searching an unstructured database. The algorithm is popularly known as \textit{Grover's search algorithm} \cite{Grover1997algoQC}.
	 	\item DiVincenzo from IBM Research Division provided a set of requirements for building a quantum computer which is popularly known as \textit{DiVincenzo's criteria} \cite{Divincenzo1997QC}.
	    \end{itemize}
	 \item[1998] \begin{itemize}
	 	\item A quantum computer using spin state of couples single-electron quantum dots was proposed \cite{Daniel1998SpinQC}. 
	 	\item The first experimental demonstration of a two-qubit quantum computer using \textit{nuclear magnetic resonance (NMR)} qubit was achieved.
	    \end{itemize}
	 \item[1999] \begin{itemize}
	 	\item A NMR based quantum computer was used to implement Grover's database search algorithm \cite{Chuang1999ImpQC}.
	 	\item Nakamura et al. was proposed that superconducting circuit can be used as a qubit \cite{Nakamura1999SCqubit}.
	    \end{itemize}
	 \item[2001] A $7$-qubit NMR based quantum computer was used to successfully demonstrate Shor's algorithm \cite{Chuang2001algoQC}.
	 \item[2003] Deutsch-Jozsa algorithm has been implemented on an ion-trap quantum computer \cite{Gulde2003DJalgoQC}.
	 \item[2005] The $8$-qubit maximally entangled W-type state with trapped ions was constructed by H\H{a}ffner et al. \cite{Haffner2005WstateQC}.
	 \item[2007] The CNOT quantum gate on superconducting qubit was experimentally realized by Plantenberg et al. \cite{Plantenberg2007CNOT-QC}.
	 \item[2008] National Science Foundation (NSF) news released about a breakthrough in storing information in the nucleus of an atom. The scientists found that information stored in the nucleus could last for about $3/4$ seconds.
	 \item[2009] HHL algorithm was introduced for solving a linear equation \cite{HarrowR2009HHLalgoQC}. Specially, finding $\vec{x}$ in the equation $A\vec{x}=\vec{a}$ given a matrix $A$ and a vector $\vec{a}$.
	 \item[2011] \textit{D-Wave}, a Canadian firm, claimed first commercially available $128-$qubit quantum computer called \textit{D-Wave One}.
	 \item[2012] Nitrogen vacancy (NV) center in diamond was claimed as a promising candidate for a qubit \cite{Grotz2012NVcenterQC}.
	 \item[2016] IBM launched an online platform \textit{IBM Quantum Experience} where one can cloud access their superconducting system.
	 \item[2017] IBM revealed a $17-$qubit quantum computer and then extend up to $50-$qubit which can maintain a quantum state for $90\mu s$.
	 \item[2018] \begin{itemize}
	 	\item \acrfull{NISQ} technology was introduced by John Preskill which help in studying many-body quantum physics \cite{Preskill2018NISQera}.
	 	\item IonQ launched the first commercial trapped-ion quantum computer..
	    \end{itemize}
	 \item[2019] \textit{Quantum supremacy} was achieved by Google which claims that their $53-$qubit quantum processor, Sycamore, is able to solve a problem in $3$ minutes and $20$ seconds while a supercomputer takes some $10,000$ year \cite{Arute2019qSupremacy}. This claim was challenged by IBM researchers who showed that the task can be solved in $2.5$ days on the Summit supercomputer \cite{Pednault2019commentQsupremacy}.
	 \item[2020] Honeywell introduced a quantum computer having quantum volume of $64$ which was the highest achieved at that time.
	 \item[2021] Microsoft started providing cloud access to quantum computing services at a platform called \textit{Azure Quantum}.
	 \item[2022] \textit{SpinQ Technology}, a Chinese quantum computing company, launched the world's first portable quantum computer which is designed for educational purpose only.
	 \item[2023] IBM introduced $1121-$qubit quantum processor called 'Condor'.
\end{description}
\subsection*{Quantum Communication}
\begin{description}
	\item[1970] The concept of \textit{conjugate coding} was introduced by Stephan Wiesner which ensures the security of quantum communication.
	\item[1984] Charles Bennett and Gilles Brassard was proposed the first quantum cryptography protocol, which is popularly known as the \textit{BB84 protocol} \cite{Bennett1984QCrypt}.
	\item[1991] Artur K. Ekert came up with another quantum cryptography protocol based on quantum entanglement, which is popularly known as \textit{E91 protocol} \cite{Ekert1991QCrypt}.
	\item[1992] \begin{itemize}
		\item The idea of \textit{\acrshort{SDC}} was introduced by Charles Bennett and Stephan Wiesner \cite{Bennett1992SDC-QT}.
		\item Another quantum cryptography protocol was proposed using two non-orthogonal states, which was popularly known as \textit{B92} protocol. 
	    \end{itemize}
	\item[1993] The concept of \textit{\acrshort{QT}} was introduced \cite{Bennett1993QT}.
	\item[1996] It is shown that broadcasting of a general mixed state cannot be possible into two separate quantum systems \cite{BarnumH1996theoremQB}.
	\item[1997] Experimental demonstration of \acrshort{QT} was reported \cite{bouwmeester1997ExpQT}. Polarization of photons had been used as a qubit.
	\item[2000] The concept of \textit{\acrshort{RSP}} was introduced in which an known qubit can remotely be simulated using one ebit of shared entanglement and classical communication of one cbit \cite{AkPati2000RSP}.
	\item[2001] Possibility of teleporting an arbitrary unitary had been checked which was viewed as \textit{quantum remote control} \cite{Huelga2001QRC}.
	\item[2002] \begin{itemize}
		\item It was shown that a restricted set of quantum operations can be remotely controlled \cite{Huelga2002QRC}.
		\item A scheme was proposed to perform any \acrfull{SMQC} task. Specially, an upper bound on dishonest parties was provided for \acrshort{SMQC} \cite{Crepeau2002bookQC}.
		\item Another scheme of quantum cryptography was proposed where relative phase of sequential pulses is utilized. The proposed scheme is popularly known as \textit{\acrfull{DPS} \acrshort{QKD}} \cite{Inoue2002DPS-QKD}.
	    \end{itemize} 
	\item[2003] First ever, standard telecom fibre-based \acrshort{QKD} over $101$ km was experimentally demonstrated \cite{Yuan2003QKD}. 
	\item[2004] \begin{itemize}
		\item The BB84 protocol was modified by changing the classical shifting procedure \cite{Scarani2004QCrypt}. The modified scheme is robust against \acrfull{PNS} attack.
		\item Deterministic \acrshort{QT} was demonstrated between a pair of trapped calcium ions \cite{Riebe2004ImplementQT}.
	    \end{itemize}
	\item[2005] \begin{itemize}
		\item The experimental demonstration of remote implementation of a rotation angle was conducted.\cite{XiangG2005ExpRIO}.
		\item Stucki et al. presented a new protocol for \acrshort{QKD} using weak coherent pulses and demonstrated experimentally. The protocol is known as \textit{\acrfull{COW} protocol} \cite{Stucki2005COW-QKD}.
	    \end{itemize}
	\item[2006] A scheme for quantum voting, an example of \acrshort{SMQC}, was proposed \cite{Hillery2006qVoting}. The proposed scheme ensures the privacy of voters and the security of votes.
	\item[2007] \begin{itemize}
		\item Nguyen Ba An introduced an idea of \textit{remote implementation of hidden operator} where, an operator hidden inside some operator is remotely implemented \cite{BaAn2007RIHO}.
		\item Another scheme for quantum voting was proposed \cite{VaccaroJ2007qVoting}. The anonymity of voters was ensured using an entangled state. The advantage of this protocol is that it reduces computational complexity.
		\item Michel Boyer and Tal Mor came up with an idea of \textit{semiquantum key distribution} where one party uses quantum features while the another party uses classical feature, to generate a secure key \cite{TalMor2007SemiQKD}.
		\item The \acrshort{DPS} \acrshort{QKD} system was implemented with $10$-GHz clock frequency and \acrfull{SNSPD}. The quantum key was distributed over a channel loss of $45$ dB. The key rate obtained was $17000$ bps and $12$ bps over a distance of $105$ km and $200$ km respectively \cite{Takesue2007DPS-QKD}.
	    \end{itemize} 
	\item[2009] \acrshort{COW} protocol based \acrshort{QKD} prototype was presented with $625$ MHz clock rate, \acrfull{ULL} fibers and low noise detectors. The key rate obtained was $6000$ bps and $15$ bps over a distance of $100$ km and $250$ km, respectively \cite{Stucki2009ExpCOW-QKD}.
	\item[2010] \begin{itemize}
		\item Quantum cryptography was employed for secure communication during the Soccer World Cup held in South Africa \cite{Mirza2010fieldQKD}.
		\item Free space \acrshort{QT} over a distance of $10$ miles ($16$ km) was successfully achieved \cite{Jin2010ExpQT}.
		\item Transmission distance through quantum channel has been enhanced using multiplexed network design \cite{Munro2010mpxQCom}.
	    \end{itemize}
	\item[2011] It was shown that a Bell state is sufficient to teleport a quantum state of the from $\alpha|x\rangle+\beta|\bar{x}\rangle:|\alpha|^2+|\beta^2|=1$ where, $\bar{x}=1^{\otimes n}\oplus x$ in modulo $2$ and $x$ varies from $0$ to $2^n-1$ \cite{Pathak2011QC_QT}.
	\item[2012] \begin{itemize}
		\item To overcome side-channel attack in quantum cryptography, a protocol was proposed called \acrfull{MDI}-\acrshort{QKD} \cite{Lo2012MDI-QKD}.
		\item Entanglement distribution and free space \acrshort{QT} was successfully achieved over a distance of $100$-kilometer \cite{Yin2012ExpQT}.
	    \end{itemize}
    \item[2014] Quantum information transferred via \acrshort{QT} over a distance of a $3$ meter \cite{Pfaff2014qT-QC}.
    \item[2017] \begin{itemize}
    	\item Protocols were proposed for two schemes of \acrshort{SMQC}, one was for \acrshort{QB} and the other was for \acrfull{MQT} \cite{YuY2017QBandMQT}.
    	\item It was shown that $\lceil\log_{2}m\rceil$ Bell states is sufficient to teleport a quantum state having $m$-unknown coefficients \cite{Sisodia2017bmIBM}.
    	\item \acrfull{QSDC} was experimentally demonstrated over fiber link up-to $10$ km distance \cite{Zhu2017ExpQSDC}.
    	\item Quantum teleportation between two remote nodes on Earth was experimentally demonstrated through satellite over distances up to $1400$ km \cite{Ren2017satelliteQT}.
        \end{itemize}
    \item[2018] Transmission channel loss is a major barrier to achieve longer distance \acrshort{QKD}. To overcome this, a new \acrshort{QKD} scheme was proposed based on phase randomized optical fields which is called as \textit{\acrfull{TF} \acrshort{QKD}} \cite{Lucamarini2018TF-QKD}.
    \item[2019] \begin{itemize}
    	\item The solution to the \acrfull{QAV} scheme was provided using GHZ-type entangled state \cite{RahamanR2015QV}.
    	\item Scheme for \acrfull{RIO} using two-photon hyperentangled state was proposed \cite{Jiao2019RIO}.
        \end{itemize}
    \item[2021] \begin{itemize}
    	\item Optical circuits for a set of quantum cryptographic schemes were designed for their experimental realization \cite{Sisodia2021OdesignQKD}.
    	\item IBM quantum computer was used to experimentally verified a proposed scheme for \acrshort{QAV} for four voters case \cite{WangQ2021QV}.
    	\item Chinese researches were able to distribute entangled photons between two flying drones \cite{Liu2021droneQKD}.
        \end{itemize}
    \item[2022] \begin{itemize}
    	\item A \acrshort{TF}-\acrshort{QKD} which has higher-speed and lower-noise was experimentally realized by optimizing several parameters associated with the light source, the quantum channel and the detector. The \acrshort{QKD} system was able to tolerate channel loss beyond $140$ dB and achieve a secure communication distance of $833.8$ km \cite{Wang2022TF-QKD}.
    	\item To enhance the security of \acrshort{RIO}, a protocol for controlled-\acrshort{RIO} had been proposed \cite{BaAn2022CRIO}.
    	\item A protocol for joint-\acrshort{RIO} had been proposed which can be used for distributed tasks throughout a quantum network \cite{BaAn2022JRIO}.
    	\item Mishra et al. were proposed a new set of schemes for \acrshort{QAV} \cite{SandeepM2021QV}.
    	\item Experimentally verified the Mishra et al. schemes for QAV on IBMQ. Also analyzed the impact of noise on the proposed schemes \cite{Satish2022QV}.
    	\item It was shown that if a qubit and qutrit are restricted upon some conditions then it is possible to broadcast the message \cite{HilleryM2022QB}.
    	\item \acrshort{QSDC} was experimentally demonstrated over fiber link that can tolerate transmission loss up-to 18.4 dB which is equivalent to $102.2$ km distance \cite{Long2022QSDC}.
        \end{itemize}
    \item[2023] \begin{itemize}
    	\item Scheme for \acrshort{MQT} was proposed using optimal resources. Effect of various noises have also been studied \cite{Satish2023MQT}.
    	\item It was shown that existing schemes for \acrshort{QB} could be done with multi-party \acrshort{RSP} \cite{Satish2024QB}.
    	\item India's first quantum secure communication link between Sanchar Bhawan and NIC, CGO complex became operational.
        \end{itemize}
    \item[2024] \begin{itemize}
    	\item Experimental realization of \acrshort{COW} \acrshort{QKD} was tested up-to $145$ km \cite{Malpani2024ExpCOW-QKD}.
    	\item Experimental realization of \acrshort{COW} and \acrshort{DPS} \acrshort{QKD} were tested with different classical shifting parameters \cite{Satish2024ExpDPR-QKD}.
    	\item A protocol for controlled-joint \acrshort{RIO} is proposed and generalized it to multi-party scenario \cite{Satish2024CJRIO}.
    	\item Two protocols were proposed for operating a hidden and a partially unknown operators on an arbitrary qubit remotely using the Bell state only \cite{Satish2024RIHO}.
        \end{itemize}
\end{description}

\section{A brief overview and structure of rest of the thesis}
Quantum mechanics is not deterministic and it contains various features with no classical analogue. Such nonclassical features (i.e., features having no classical analogue) have led to a set of no-go theorems (see Chapter $3$ of Reference \cite{Pathak2013book}). Initially, such no-go theorems (say, Heisenberg uncertainty principle) and the indeterministic nature of quantum mechanics manifested through the collapse on measurement postulate were viewed as the limitations of quantum mechanics. However, with the recent progress in quantum computation, quantum communication and quantum foundation, it has been realized that such limitations of quantum mechanics (as reflected through the early no-go theorems like Heisenberg's uncertainty principle and recently introduced variants including no-cloning theorem, no-broadcasting theorem, no-deletion theorem) can be exploited for various useful purposes. For example, the Heisenberg's uncertainty principle and the no-cloning theorem which states that physical properties corresponding to two non-commutative operators cannot be simultaneously measured with an arbitrary accuracy, can be used to generate an unconditionally secure key using any one of a large number of conjugate coding-based protocols for \acrshort{QKD} (e.g., BB84 and B92 protocols for QKD) \cite{Bennett1984QCrypt,Bennett1992QKD}. Similarly, other nonclassical features of the quantum world (like entanglement and nonlocality) have recently been used for performing tasks like quantum teleportation \cite{Bennett1993QT}, dense coding \cite{Bennett1992SDC-QT}, and \acrshort{MDI} \acrshort{QKD} \cite{fan2022robust,xie2022breaking} which do not have classical counterparts. Quantum features are also used to build quantum computers and  performing certain computational tasks in quantum computers at a speed that cannot be  matched by classical computers. In short, with the recent progress in the domain of quantum information sciences, we have realized that nonclassical features of quantum mechanics including the no-go theorems have many applications. These applications can be broadly divided into three areas: (i) quantum communication, (ii) quantum computing, and (iii) quantum sensing. This thesis will discuss some topics related to the first two with a specific discussion on the role of no-go theorems and nonclassical features in realizing the quantum advantages in the tasks performed as part of this thesis. To be specific, this thesis reports new works in the area of quantum computing and quantum communication. Interestingly, it also reports a reasonable amount of new works on the theory (where a set of new protocols for quantum communication and secure multiparty computation tasks have been designed and analyzed) and experiments (partly performed in our lab and partly performed using the quantum computers available over the cloud). The domain of quantum computing and quantum communication is extremely broad and naturally, it will not be possible to touch every aspect of quantum computing and communication in this thesis. The thesis primarily discusses a set of closely connected but independent problems with an aim to either experimentally realize an existing quantum protocol/scheme or to design and analyze a new one. In both cases, efforts have been made to add a conceptual clarity to the existing know-how of the related topics. Efforts have also been made to reduce the amount of quantum resources used for realizing a particular scheme and to analyze the impact of noise on the protocols studied here. For example, in Chapter \ref{Ch2:MQT} of the thesis, a new protocol for multi-output quantum teleportation that allows a sender to teleport two different quantum states to two receivers with an optimal amount of quantum resources is proposed. A proof-of-principle realization of the scheme is performed using a cloud-based IBM quantum computer, and the impact of different types of noise on the proposed protocol is also investigated. Interestingly, this scheme and any other scheme designed for a similar task cannot be used for sending an unknown (arbitrary) quantum state simultaneously to two or more receivers as that would violate no-broadcasting theorem \cite{BarnumH2007theoremNoQB} as well as no-cloning theorem \cite{Wootters1982No-cloning}. However, if the state to be sent is known the above restriction disappears and keeping that in mind, in Chapter \ref{Ch3:QB} of the thesis, a scheme for quantum broadcasting of a known quantum state to multiple receivers has been proposed and analyzed and as before a proof-of-principle realization of the scheme have been reported with the help of cloud-based IBM quantum computer and also investigate the impact of noise. Up to this point, the thesis discusses the teleportation of unknown  (in Chapter \ref{Ch2:MQT}) and known (in Chapter \ref{Ch3:QB}) quantum states to the receiver, but there may be situations where we need to teleport an "operator" instead of a quantum state. Such schemes of remote implementation of operators have many variants including but not restricted to remote implementation of an unknown operator, a hidden operator or a partially known operator. These tasks are analogous to quantum teleportation, but being teleportation of the operators these schemes have potential applications in distributed quantum computing. Especially, in the present era when the available quantum computers are small in size. Keeping these facts in mind, a new schemes for variants of the quantum remote implementation of unknown, hidden and partially known operators has been proposed in Chapter \ref{Ch4:RIO}. This chapter establish a kind of link between the schemes of quantum communication (teleportation and remote state preparation) and those of quantum computing as remote implementation of operators have potential applications in the distributed quantum computing. This link is not an exception. In fact, there are several computing tasks where security is needed and it is not difficult to visualize that in a modern society where cyber security plays a crucial role. For example, banking to voting, auction to e-commerce may be viewed as computing tasks involving multiple parties (as in our schemes of teleportation) needing security. Such computing tasks involving multiple parties and needing security are known as secure multiparty computation (SMC) tasks. We have already noted that schemes of quantum cryptography can provide unconditional security. Extending this feature of the schemes of quantum communication (specifically of quantum cryptography), it is possible to obtain quantum secure schemes for certain	SMC tasks. One such SMC task is quantum anonymous veto. In Chapter 5, the experimental realization of a scheme of quantum anonymous veto proposed by Mishra et al. \cite{SandeepM2021QV} using an IBM quantum computer is described. Up to this point, the thesis is primarily theoretical with flavor of experimental realization involving quantum computers available over the cloud. However, the proposed schemes are experimentally realizable in true sense. Limitations of the available resources have restricted us to experimentally realize the schemes discussed in Chapter \ref{Ch2:MQT}-\ref{Ch5:QV}, but for the sake of completeness and to establish the claim that unconditionally secure quantum key distribution can be realized in the lab, in Chapter 6, the experimental realization of two schemes for quantum key distribution (namely, \acrshort{COW} and \acrshort{DPS} protocols) is reported for various physical situations, including various distances. Finally, the thesis is concluded in Chapter \ref{Ch7:Conclusion}, where the scope of future works is discussed. In the remaining part of this thesis, the content of each chapter is elaborated. A brief structure of the thesis is shown in Figure \ref{fig:thesis_outline}.
\newpage


\chapter{MULTI-OUTPUT QUANTUM TELEPORTATION}\label{Ch2:MQT}
\graphicspath{{Chapter2/Chapter2Figs/}{Chapter2/Chapter2Figs/}}
\section{Introduction}
A scheme for quantum communication referred to as quantum teleportation or simply teleportation was first proposed by Bennett et al. in 1993 \cite{Bennett1993QT}. This scheme allows a sender (Alice) to send an unknown quantum state to a receiver (Bob) in an interesting manner such that the state (equivalently all the information about the state) disappears at Alice's end and appears at Bob's end, but the state is never found to exist in the channel connecting Alice \& Bob. The theoretical proposal of Bennet et al. uses a maximally entangled Bell state to teleport a single qubit unknown quantum state. Since then, a large number of other schemes for quantum teleportation and various variants of it such as, bidirectional teleportation, \acrfull{QIS}, \acrfull{RSP}, and \acrfull{QSS} (see \cite{Pathak2013book,vishal2015cb_RSP_NC} and references therein) have been reported. Recently Yan Yu et al. \cite{YuY2017QBandMQT} modified the quantum teleportation scheme and reported it as a scheme of \acrfull{MQT}. In the MQT scheme, a sender (Alice) teleports two distinct quantum states to two distinct receivers ($\rm{Bob}_1$ and $\rm{Bob}_2$) at the same time. In this scheme, two distinct unknown quantum states to be teleported are $m$ and $m+1$-qubit GHZ like states. The MQT scheme is a new and interesting variant of the quantum teleportation. In comparison to the previously reported variants of quantum teleportation, the MQT scheme is less studied. The MQT scheme proposed by Yan Yu et al. utilizes $5$-qubit cluster state whereas a conventional teleportation scheme requires a Bell state which is a two-qubit entangled state. This observation and the fact that preparation and maintenance of a $5$-qubit cluster state are much more demanding compared to the preparation and maintenance of Bell state(s), motivates us to design a new scheme for the same MQT task using optimal resources. Here, a protocol for the MQT scheme will be proposed where a sender sends two distinct unknown quantum states ($m$ and $m+1$-qubit GHZ like state) to two distinct receivers ($\rm{Bob}_1$ and $\rm{Bob}_2$) which utilize two copies of the Bell state. In what follows, the MQT task proposed by Yan Yu et al. will be briefly explained and then present a new approach (proposal) to accomplish the same task using an optimal amount of quantum resources. The MQT task of Yan Yu et al. has three parties, one sender (Alice) and two receivers ($\rm{Bob}_1$ and $\rm{Bob}_2$). Alice wants to send the following two unknown quantum states: 
\begin{eqnarray}
	|\chi_{a}\rangle & = & \alpha_{1}|0\rangle^{\otimes m}+\beta_{1}|1\rangle^{\otimes m},\label{eq2.1}
\end{eqnarray}
and another one is 
\begin{eqnarray}
	|\chi_{b}\rangle & = & \alpha_{2}|0\rangle^{\otimes(m+1)}+\beta_{2}|1\rangle^{\otimes(m+1)}\label{eq2.2}
\end{eqnarray}
to $\rm{Bob}_1$ and $\rm{Bob}_2$, respectively. To do so, they use a five-qubit maximally entangled cluster state
\begin{eqnarray}
	|\psi\rangle_{A_1A_2B_1B_2B_2^{'}} & = & \frac{1}{2}(|00000\rangle+|01011\rangle+|10100\rangle+|11111\rangle)_{A_1A_2B_1B_2B_2^{'}}\label{eq2.3}
\end{eqnarray}
\begin{center}
	\begin{figure}
		\centering
		\includegraphics[width=0.6\paperwidth]{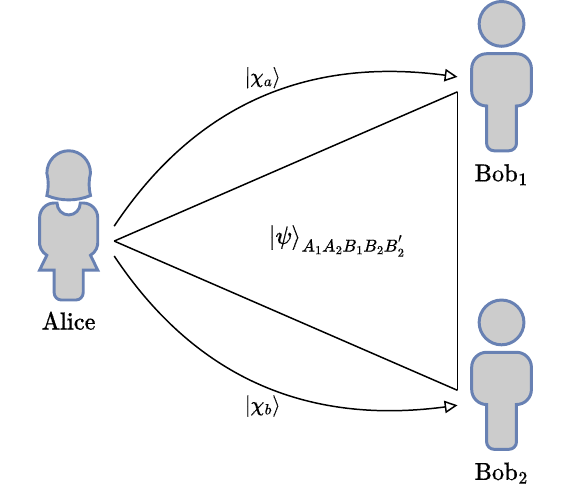}
		\caption{\centering A sketch to visualize the MQT scheme of Yan Yu et al.}
		\label{fig2:MQT_Task_YanYu}
	\end{figure}
\end{center}
which is shared among all the three parties such that qubit(s) labeled as $A_1$ $\&$ $A_2$ are with Alice, $B_1$ is with $\rm{Bob}_1$ and $B_2$ $\&$ $B_2^{'}$ are with $\rm{Bob}_2$. Alice then performs several \acrfull{BM} in a manner depicted in Figure \ref{fig2.1:MQT.ckt-Yan}. Depending on the outcomes of the \acrshort{BM}, $\rm{Bob}_1$ and $\rm{Bob}_2$ apply corresponding unitary operations to get the desired states $|\chi_a\rangle$ and $|\chi_b\rangle$ which are $m$ and $(m+1)$-qubit GHZ class states, respectively. Some reservations are held regarding calling the desired state a GHZ-class state, and these states would like to be referred to as generalized Bell-type states, as was rationally done in a number of previous papers (\cite{Pathak2011QC_QT,Panigrahi2006QC_QM} and references therein). It was shown in Reference \cite{Pathak2011QC_QT} that a Bell state is sufficient to teleport a quantum state of the form $\alpha|x\rangle+\beta|\bar{x}\rangle:|\alpha|^{2}+|\beta|^{2}=1$ where, $\bar{x}=1^{\otimes n}\oplus x$ in modulo 2 arithmetic and $x$ changes from $0$ to $2^{n}-1$. As both the states to be teleported $|\chi_{a}\rangle$ and $|\chi_{b}\rangle$ are of the form $\alpha|x\rangle+\beta|\bar{x}\rangle$, each of them can be sent with the help of a Bell state. To clearly justify this point, we can note that the quantum state of the form $\alpha|x\rangle+\beta|\bar{x}\rangle$ to be sent can dissolve into a single qubit quantum state with the help of $n-1$ $\rm{CNOT}$ operations with first qubit acts as the control qubit and the remaining $(n-1)$-qubits acts as target qubits (see left part of Figure \ref{fig2.2:Mqt.ckt2_SK}). We can understand this point with an example, suppose sender wants to teleport an unknown quantum state $|\psi\rangle=\alpha|0\rangle^{\otimes n}+\beta|1\rangle^{\otimes n}:|\alpha|^2+|\beta|^2$ then the application of the $\rm{CNOT}$ gates in a manner $\otimes_{i=2}^{n}\rm{CNOT}_{1\rightarrow i}|\psi\rangle$ will transform the initial state to a product state  $(\alpha|0\rangle+\beta|1\rangle)|0\rangle^{\otimes n-1}$ where all the necessary information to be teleported is contained in the single qubit state $(\alpha|0\rangle+\beta|1\rangle)$; here $\rm{CNOT}_{i\rightarrow j}$ is operated such that $i^{th}$ is the control qubit and $j^{th}$ are the target qubits. A single Bell state is sufficient to teleport the reduced quantum state $(\alpha|0\rangle+\beta|1\rangle)$ as demonstrated in the central block of Figure \ref{fig2.2:Mqt.ckt2_SK}. To reconstruct the desired state at receiver's port, $n-1$ $\rm{CNOT}$ gates are to be applied in a sequence $\rm{CNOT}_{1\rightarrow2}\otimes\rm{CNOT}_{1\rightarrow3}\otimes\rm{CNOT}_{1\rightarrow4}\otimes\cdots\otimes\rm{CNOT}_{1\rightarrow n}(\alpha|0\rangle+\beta|1\rangle)|0\rangle ^{\otimes n-1}=\alpha|000\cdots0\rangle+\beta|111\cdots1\rangle$ as shown in the right block of Figure \ref{fig2.2:Mqt.ckt2_SK}. Thus, the states considered by Yan Yu et al. ($|\chi_{a}\rangle$ and $|\chi_{b}\rangle$), which take the form $\alpha|x\rangle+\beta|\bar{x}\rangle$, can independently be teleported to two separate receivers using two Bell states, as illustrated in Figure \ref{fig2.2:Mqt.ckt2_SK}. Specially, the unknown $m$ and $m+1$-qubit GHZ like states can be dissolved to two distinct single qubit states by locally operating $m-1$ and $m$ $\rm{CNOT}$ gates with the first qubit serving as control qubit and subsequent qubits as the target qubit respectively as illustrated in Figure \ref{fig2.2:Mqt.ckt2_SK}. Now, the standard teleportation protocol is used to teleport two single qubit states. To get the original information, $\rm{Bob}_1$ and $\rm{Bob}_2$ locally apply unitary operation which Alice applied for dissolving the quantum states. This implies that the utilization of multi-qubit entangled state or the five-qubit cluster state used by Yan Yu et al. is not required to achieve the MQT task. Building on the aforementioned observations, it is appropriate to add that a generalized teleportation strategy was reported in Reference \cite{Sisodia2017OpticalDesign_QT} which claims that $\lceil\log_{2}m\rceil$ Bell state is sufficient to teleport a quantum state with $m$-unknown coefficients. The proposed MQT scheme of Yan Yu et al. has four unknown coefficients $\alpha_{1}$, $\beta_{1}$, $\alpha_{2}$ and $\beta_{2}$ of the state to be teleported, hence only $\lceil\log_{2}4\rceil=2$ Bell states are needed for teleportation. It is to be noted that the scheme reported in Reference \cite{Sisodia2018optimalQT} gives an insight on how to teleport a more general arbitrary two qubit quantum state with the help of two Bell states. Despite the availability of this approach of using optimal resources for teleportation, several researchers have recently documented other teleportation strategies that use a substantially higher amount of quantum resources. For instance, in Reference \cite{Bikash2020IBM}, two qubit unknown quantum state is teleported using a four-qubit cluster state as a quantum resource. The teleported two qubit state is 
\begin{eqnarray}\label{eq2.4}
	|\lambda\rangle_{ab} & = & \alpha|00\rangle+\beta|01\rangle+\gamma|10\rangle+\delta|11\rangle
\end{eqnarray}
which has four unknown coefficients, hence require $2$ Bell state ($\lceil\log_{2}4\rceil=2$) is enough instead of the four-qubit cluster state. The aforementioned explanation so far has constructed a base allowing us to state that the resource used by Yan Yu et al. for the MQT scheme is not optimal. Also, using the fact that their scheme is realized for the case $m=1$ on IBM quantum computer motivates us to realize the MQT scheme using the optimal resource (two Bell state) which is discussed in the next section. To understand how well different quantum schemes will work in the actual physical scenario, the experimental realization of these schemes is crucial. An IBM quantum computer has already been used to analyze the performance of different schemes \cite{Satish2022QV,LiuX2022controlledQT,Pranav2020gameNC}. 
\begin{figure}
	\centering	
	\includegraphics[width=0.8\paperwidth]{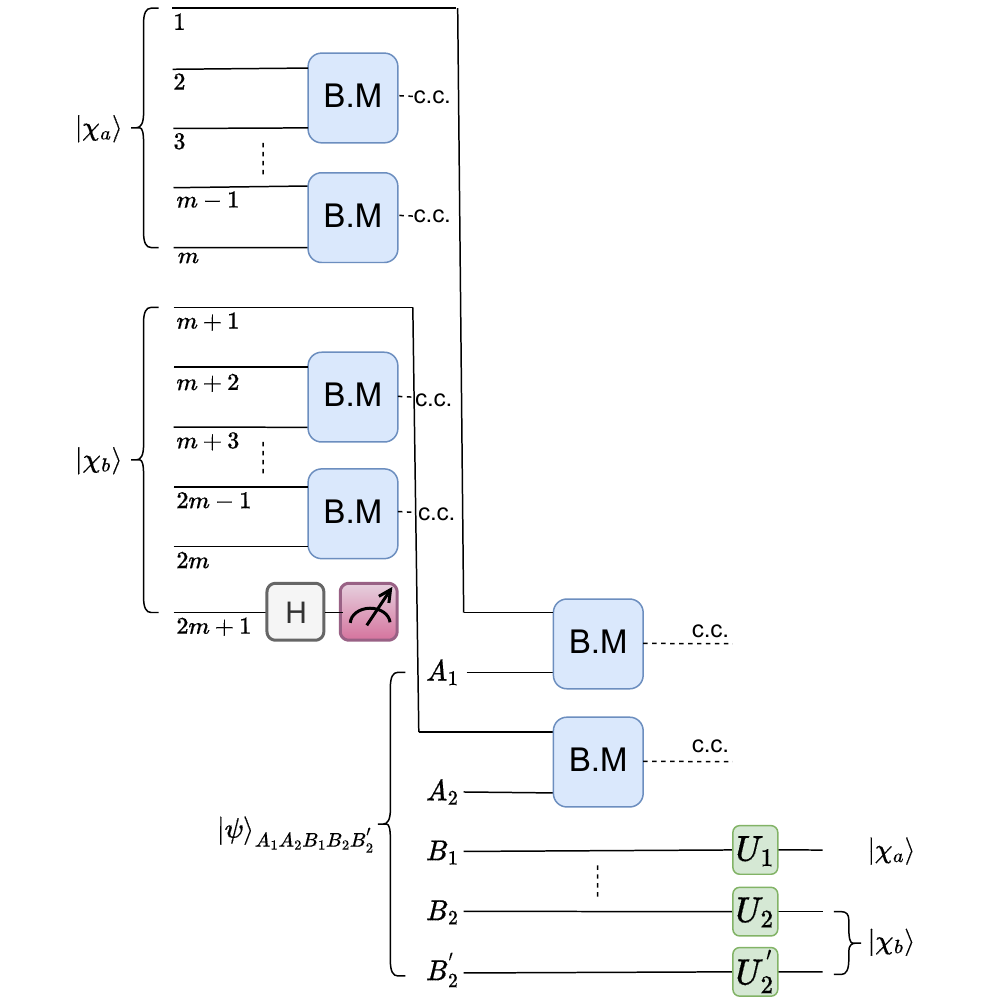}
	\caption{\centering A quantum circuit illustrating the generalized MQT scheme of Yan Yu et al. \label{fig2.1:MQT.ckt-Yan}}
\end{figure}
\section{Multi-output quantum teleportation using optimal quantum resource}\label{sec2.2:MQToptimal}
As previously stated, Yan Yu et al. first used the term \acrshort{MQT} in 2017 \cite{YuY2017QBandMQT} to describe a quantum communication protocol that enables a sender (Alice) to teleport two distinct single qubit states $|\chi_{1}\rangle=\alpha_1|0\rangle+\beta_1|1\rangle$ and
$|\chi_{2}\rangle=\alpha_2|0\rangle+\beta_2|1\rangle$ to two distinct receivers ($\rm{Bob}_1$ and $\rm{Bob}_2$) with the help of a four-qubit
cluster state 
\begin{eqnarray}
	|\psi\rangle_{A_{1}A_{2}B_{1}B_{2}} & = & \frac{1}{2}(|0000\rangle+|0101\rangle+|1010\rangle-|1111\rangle).
\end{eqnarray}
In the proposed \acrshort{MQT} scheme, Alice holds the first two qubits (labeled by subscripts $A_{1}$and $A_{2}$) of the cluster state and sends the remaining qubits to the two distinct receivers (qubits shared to $\rm{Bob}_{i}$ is labeled by $B_{i}$). Alice now perform measurement on the first four qubit $|\psi_{i}\rangle_{12A_{1}A_{2}}$ in cluster basis. It is to be noted that two qubits labeled as $1$ $\&$ $2$ of $|\psi_{i}\rangle_{12A_{1}A_{2}}$ are those which contains the information while the remaining qubits are those which are held by Alice. The desired state $|\chi_{1}\rangle$ and $|\chi_{2}\rangle$ appear at two receiver sites, respectively, after the application of the corresponding unitary operators by the receivers. Along the same line, Yan Yu et al. came up with a different \acrshort{MQT} scheme in $2021$ where Alice wanted to teleport $m$ and $(m+1)$-qubit GHZ like state (cf. Equation \eqref{eq2.1} and \eqref{eq2.2}) to the two distinct receivers using a five-qubit cluster state (see Equation \eqref{eq2.3}). It has been previously stated that the two copies of the Bell states are sufficient to accomplish the \acrshort{MQT} task. As the experimental realization of the \acrshort{MQT} scheme proposed by
\begin{center}
	\begin{figure}[h!]
		\centering
		\includegraphics[width=0.8\paperwidth]{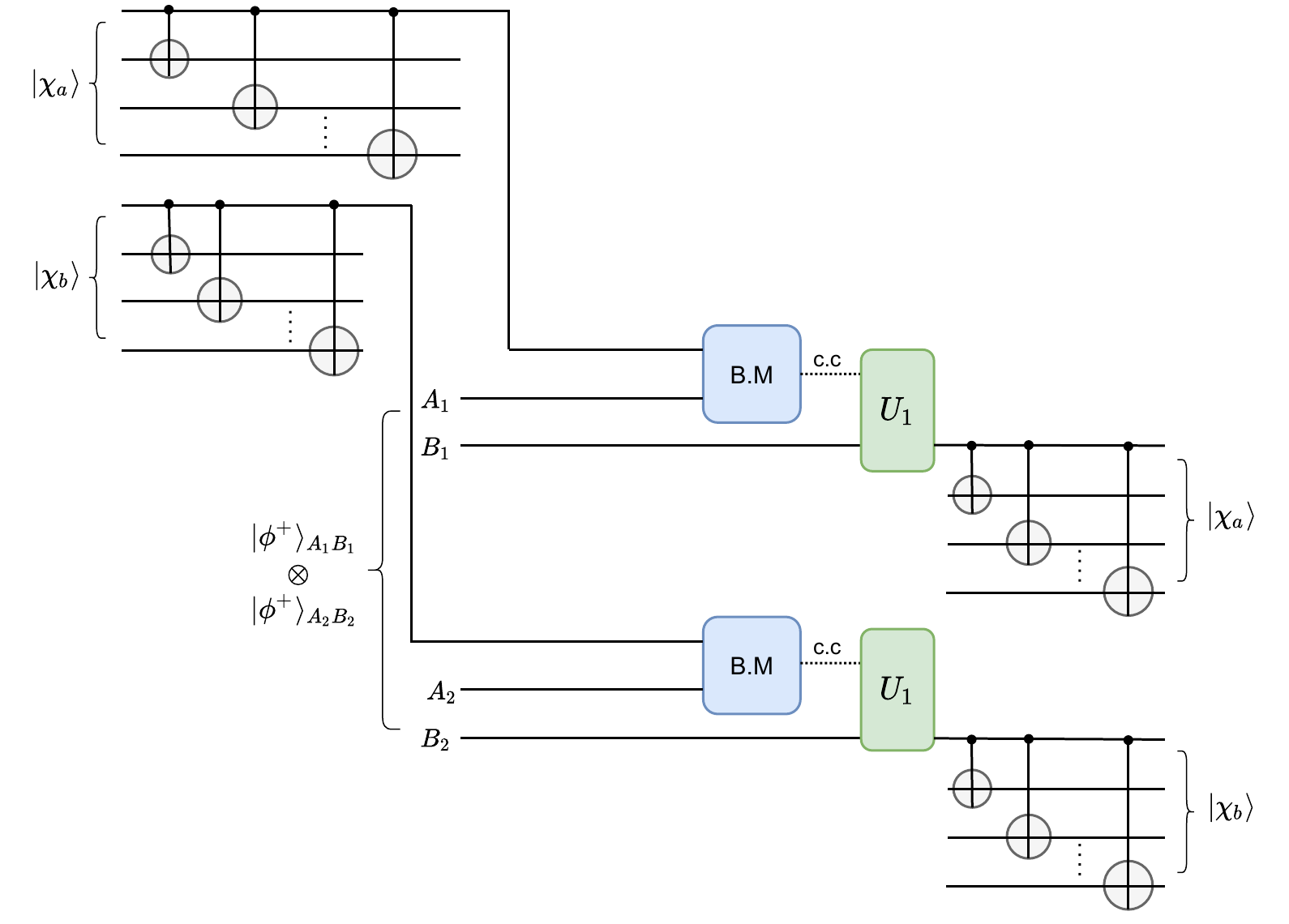}
		\caption{\centering A quantum circuit illustrating the MQT scheme using optimal resources.}
		\label{fig2.2:Mqt.ckt2_SK}
	\end{figure}
\end{center}

\begin{center}
	\begin{figure}
		\centering
		\includegraphics[width=0.8\paperwidth]{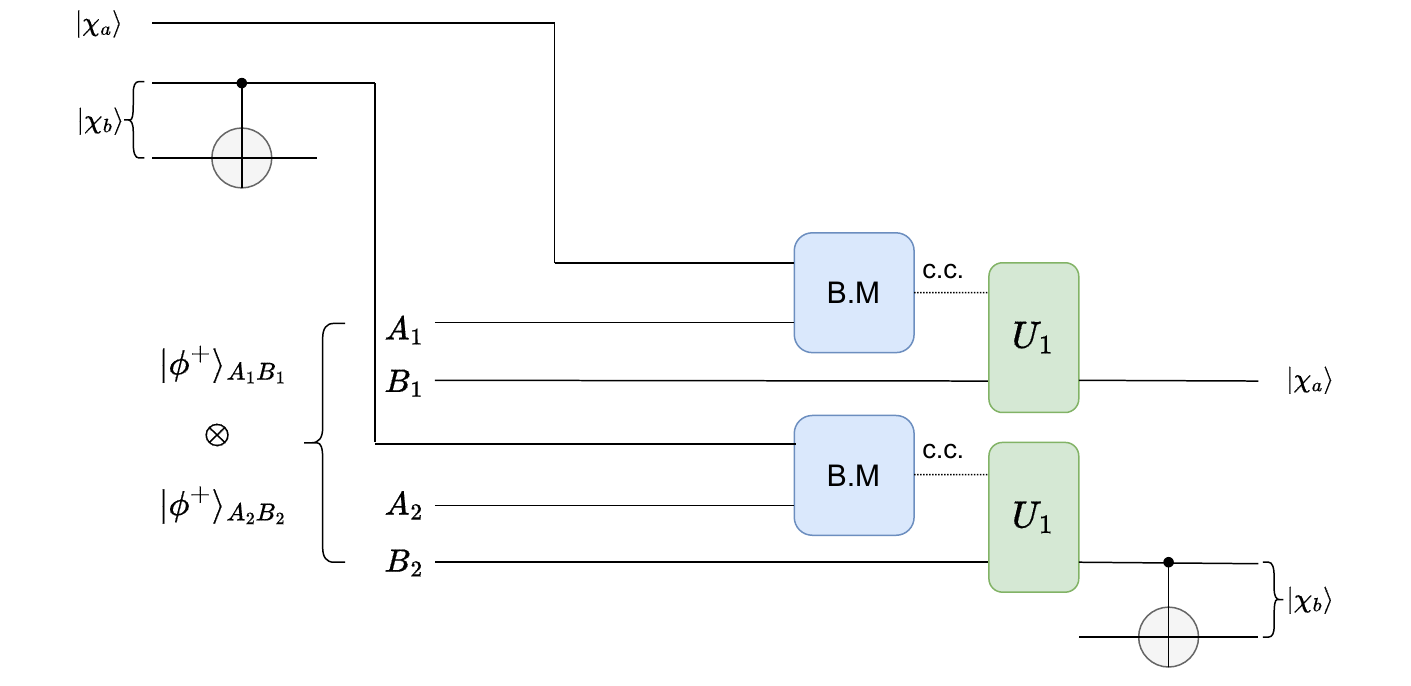}
		\caption{\centering A quantum circuit illustrating the MQT scheme for $m=1$ case using optimal resources.\label{fig2.3:Mqt.ckt1_SK}}
	\end{figure}
\end{center}
Yan Yu et al. is limited to the case $m=1$, the quantum circuit design is specially employed for $m=1$ scenario which is necessary to complete the \acrshort{MQT} task utilizing two copies of a Bell state, shown in Figure \ref{fig2.3:Mqt.ckt1_SK}. Let us assume $|\chi_{a}\rangle$
and $|\chi_{b}\rangle$ are the two distinct quantum states to be sent (Equation \eqref{eq2.1} and \eqref{eq2.2} for $m=1)$. The $\rm{CNOT}$ operation on the quantum state $|\chi_{b}\rangle$ with its first qubit acting as the control while its second qubit as the target, can be
simplified to a state $|\chi_{b}^{\prime}\rangle$ as follows:
\begin{eqnarray}
	\rm{CNOT}|\chi_{b}\rangle & \longrightarrow & |\chi'_{b}\rangle|0\rangle=(\alpha_{2}|0\rangle+\beta_{2}|1\rangle)\otimes|0\rangle.\label{eq2.5}
\end{eqnarray}
The $\rm{CNOT}$ operation reduces the problem of teleportation of the
product state of $|\chi_{a}\rangle$ and |$\chi_{b}^{\prime}\rangle=\alpha_{2}|0\rangle+\beta_{2}|1\rangle$.
\section{Experimental realization of the MQT scheme on IBM quantum computer}\label{sec:Experimental-realization-usingIBM}
\begin{figure}
	\begin{centering}
		\begin{tabular}{c}
		\includegraphics[scale=0.9]{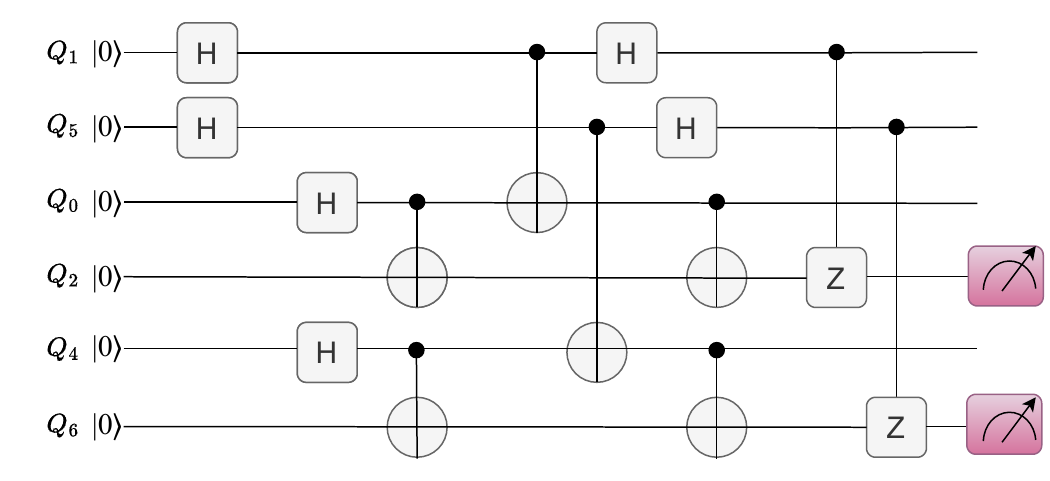}\\
		(a)\\[2ex]
		\includegraphics[scale=0.9]{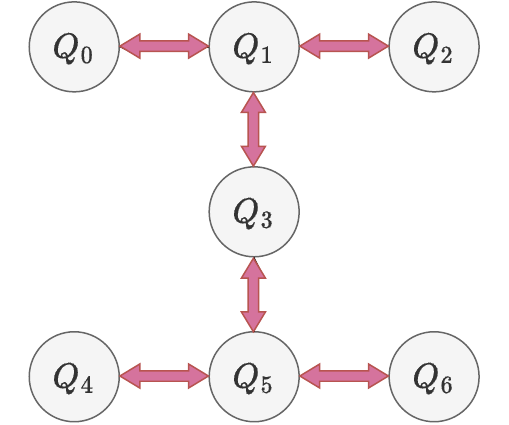}\\
		(b)\\[1ex]
	\end{tabular}
\par\end{centering}
	\caption{\centering (a) A quantum circuit to teleport $|+\rangle=\frac{1}{\sqrt{2}}(|0\rangle+|1\rangle)$ to two distinct receivers at qubit $Q_2$ and $Q_6$ using two copies of a Bell states $|\phi^{+}\rangle^{\otimes2}$ (b) Topology of the quantum computer used
		(ibmq\_casablanca). \label{fig2.4:MQTCkt}}
\end{figure}

Here, a simple quantum circuit has been built, as shown in Figure \ref{fig2.4:MQTCkt} (a), to experimentally verify the \acrshort{MQT} scheme on a quantum computer based on superconducting qubits, available on the cloud at IBM quantum experience. The designed quantum circuit is almost similar to the circuit depicted in Figure \ref{fig2.3:Mqt.ckt1_SK} except the locally applied $\rm{CNOT}$ gates to simplify the teleported state. IBM quantum composer is used to run this circuit and the obtained results are reported in Section \ref{sec2.3.1:Results-MQT}. To implement the quantum circuit depicted in Figure \ref{fig2.4:MQTCkt} on IBM quantum computer, ibmq\_casablanca which has Falcon processor and seven qubit active chip, has been used\cite{Ibm2021} that is not enough to implement the quantum circuit depicted in Figure \ref{fig2.3:Mqt.ckt1_SK}, but sufficient to verify the circuit depicted in Figure \ref{fig2.4:MQTCkt}. Qubits labeled as $Q_1$ and $Q_5$ in Figure \ref{fig2.4:MQTCkt} are assigned to the quantum states $\frac{1}{\sqrt{2}}(|0\rangle+|1\rangle)\otimes\frac{1}{\sqrt{2}}(|0\rangle+|1\rangle)=|+\rangle\otimes|+\rangle$ which have to be teleported and the two copies of the Bell state $|\phi^{+}\rangle=\frac{1}{\sqrt{2}}(|00\rangle+|11\rangle)$ which is used as the resources for the \acrshort{MQT} scheme, are prepared at qubit $Q_0$ $\&$ $Q_2$ and $Q_4$ $\&$ $Q_6$. After the preparation of Bell states, the qubits are distributed such that $Q_1$, $Q_5$, $Q_0$ and $Q_4$ are with Alice and $Q_2$ \& $Q_6$ are with $\rm{Bob}_1$ \& $\rm{Bob}_2$, respectively. Now, Alice does Bell measurement by applying $\rm{CNOT}$ followed by Hadamard operation on qubit $Q_1$ $\&$ $Q_0$ and $Q_5$ $\&$ $Q_4$. Depending on the measurement outcomes, $\rm{Bob}_1$ and $\rm{Bob}_2$ apply unitary (here, Pauli-Z operation) to obtain the desired state $|+\rangle$ at $Q_2$ and $Q_6$, respectively. Here, it is to be noted that the topology of the quantum computer depicted in Figure \ref{fig2.4:MQTCkt} (b) decides the indexing of qubits in the designed quantum circuit depicted in Figure \ref{fig2.4:MQTCkt} so that its circuit cost reduces. It is important to minimize the quantum costs of a circuit while its implementation \cite{Dueck2018qc_Opt.}. Clearly, realization of the \acrshort{MQT} scheme using the quantum circuit designed as shown in Figure \ref{fig2.4:MQTCkt} are the two separate teleportation circuits, and they are sufficient to accomplish the goals of the prior studies that required expensive quantum resources like multi-qubit cluster state.

\subsection{Results}\label{sec2.3.1:Results-MQT}
\begin{figure}[h!]
	\begin{centering}
		\includegraphics[scale=0.8]{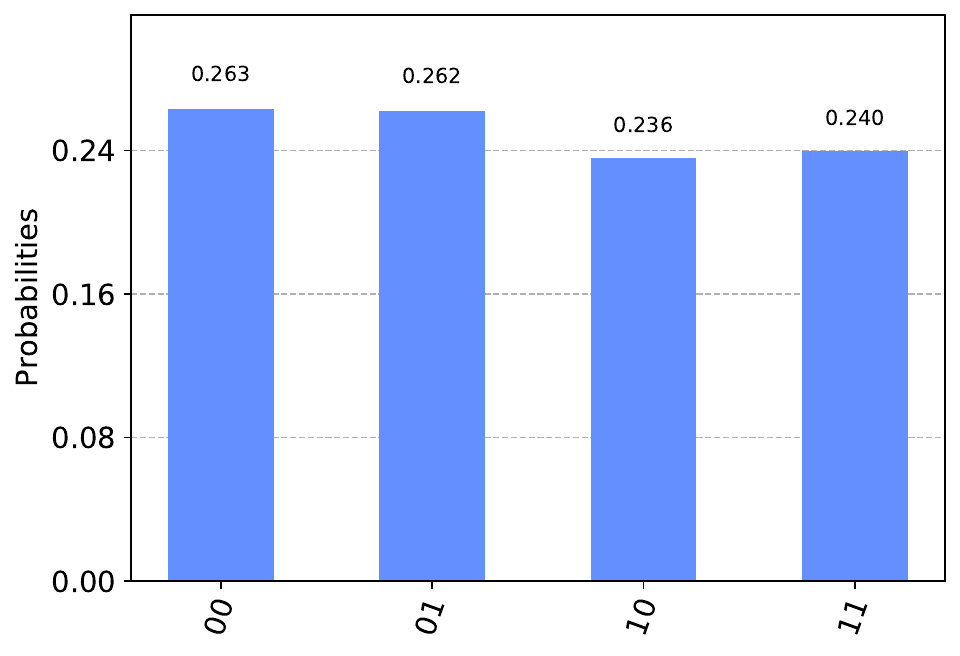}
		\par\end{centering}
	\caption{\centering Experimentally obtained result after executing the quantum circuit depicted in Figure \ref{fig2.4:MQTCkt} on IBM quantum computer (ibmq\_casablanca). \label{fig2.5:MQTresult}}
\end{figure}

\begin{figure}[h!]
	\begin{centering}
		\begin{tabular}{c}
		\includegraphics[scale=1]{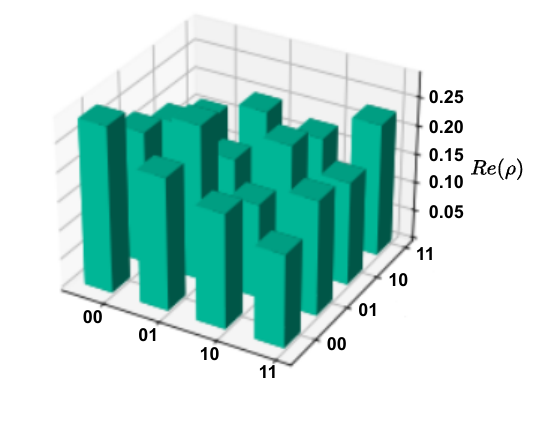}\\
		(a)\\
		\includegraphics[scale=1]{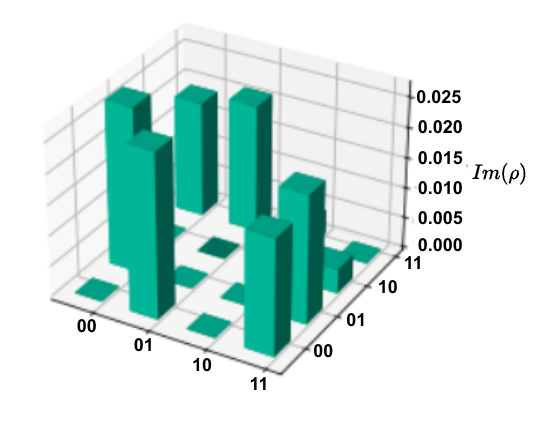}\\
		(b)
	\end{tabular}
\par\end{centering}
\caption{Experimental quantum state tomography result with (a) real and (b) imaginary parts for the circuit shown in Figure \ref{fig2.4:MQTCkt}.\label{fig2.6:MQT_DMresult}}
\end{figure}

A seven qubit quantum computer based on superconducting qubit named as ibmq\_casablanca is used to execute the designed quantum circuit as shown in Figure \ref{fig2.4:MQTCkt}. The desired result is $|++\rangle=\frac{1}{2}(|00\rangle+|01\rangle+|10\rangle+|11\rangle)$ at qubit $Q_2$ and $Q_6$ and the result obtained from the ibmq\_casablanca is illustrated in Figure \ref{fig2.5:MQTresult}, which is quite close to the desired result. The density matrix corresponding to the obtained result is shown in Figure \ref{fig2.6:MQT_DMresult}, is derived using a technique called \acrfull{QST}. The \acrshort{QST} technique has already been discussed in Chapter \ref{Ch1: Introduction}. The available quantum computers are calibrated twice or thrice a day to check its performance. The calibration data of each qubit of ibmq\_casablanca are provided in Table \ref{tab2.2:CalibData}.\\
Fidelity is an important measure to check closeness of any two quantum states, estimated using the formula $F(\sigma,\rho)=Tr\left[\sqrt{\sqrt{\sigma}\rho\sqrt{\sigma}}\right]^{2}$, where $\sigma$ is the density matrix of the final state obtained theoretically and $\rho$ is the density matrix of the final state obtained experimentally. Here, the fidelity between the desired and obtained result is obtained as $84.64$ $\%$ for a specific set of experiments with each experiment having total of $8192$ shots. To verify the consistency of the obtained results, the same experiment is performed ten times whose fidelities are (in \%) $79.31$, $77.51$, $78.98$, $84.64$, $76.17$, $83.64$, $81.33$, $80.21$, $79.92$, $74.65$. These ten data sets have a standard deviation of $3.096$ which indicates a fairly precise result. This proves that the resources used in prior works were not minimal. A comparison between the fidelity obtained using our optimal resources and previously used resources is not possible as Yan Yu et al. have not reported fidelity. It is obvious that the effect of noise on a scheme that uses a simpler entangled state as a resource will be less compared to a scheme for the same task that uses a relatively complex entangled state. The validity of this expected result is systematically investigated  in the next section. 

\begin{table}
	\caption{The calibration data of all active qubits of ibmq\_casablanca on December 01, 2021. cxi\_j, representing a $\rm{CNOT}$ gate with qubit i as the control and qubit j as the target.\label{tab2.2:CalibData}}
	\begin{centering}
		\begin{tabular}{|c|c|c|>{\centering}p{1.6cm}|>{\centering}p{2cm}|>{\centering}p{2cm}|>{\centering}p{4.4cm}|}
			\hline 
			Qubit & T1 ($\mu s$) & T2 ($\mu s$) & \centering{}Frequency (GHz) & \centering{}Readout assignment error & \centering{}Single-qubit Pauli-X-error & \centering{}$\rm{CNOT}$ error\tabularnewline
			\hline 
			$Q_{0}$ & 97.07 & 41.56 & \centering{}4.822 & \centering{}$3.52\times10^{-2}$ & \centering{}$2.73\times10^{-4}$ & \centering{}cx0\_1: $1.105\times10^{-2}$\tabularnewline
			\hline 
			$Q_{1}$ & 179.27 & 106.63 & \centering{}4.76 & \centering{}$1.56\times10^{-2}$ & \centering{}$1.56\times10^{-4}$ & \centering{}cx1\_3: $6.796\times10^{-3}$, cx1\_2: $1.013\times10^{-2}$,
			cx1\_0: $1.105\times10^{-2}$\tabularnewline
			\hline 
			$Q_{2}$ & 164.86 & 96.43 & \centering{}4.906 & \centering{}$8.50\times10^{-3}$ & \centering{}$3.54\times10^{-4}$ & \centering{}cx2\_1: $1.013\times10^{-2}$\tabularnewline
			\hline 
			$Q_{3}$ & 123.23 & 151.27 & \centering{}4.879 & \centering{}$1.70\times10^{-2}$ & \centering{}$3.40\times10^{-4}$ & \centering{}cx3\_1: $6.796\times10^{-3}$, cx3\_5: $1.139\times10^{-2}$\tabularnewline
			\hline 
			$Q_{4}$ & 128.4 & 54.14 & 4.871 & \centering{}$3.06\times10^{-2}$ & \centering{}$2.88\times10^{-4}$ & cx4\_5: $1.148\times10^{-2}$\tabularnewline
			\hline 
			$Q_{5}$ & 133.5 & 91.77 & 4.964 & \centering{}$9.60\times10^{-3}$ & \centering{}$3.17\times10^{-4}$ & cx5\_3: $1.139\times10^{-2}$, cx5\_4:$1.148\times10^{-2}$, cx5\_6:
			$1.156\times10^{-2}$,\tabularnewline
			\hline 
			$Q_{6}$ & 112.08 & 166.07 & 5.177 & \centering{}$2.18\times10^{-2}$ & \centering{}$4.70\times10^{-4}$ & cx6\_5:$1.156\times10^{-2}$\tabularnewline
			\hline 
		\end{tabular}
		\par\end{centering}
\end{table}

\subsection{Impact of noise}\label{sec2.3.2:Effect-of-noise-MQT}
\begin{figure}[h!]
	\begin{centering}
		\includegraphics[scale=0.9]{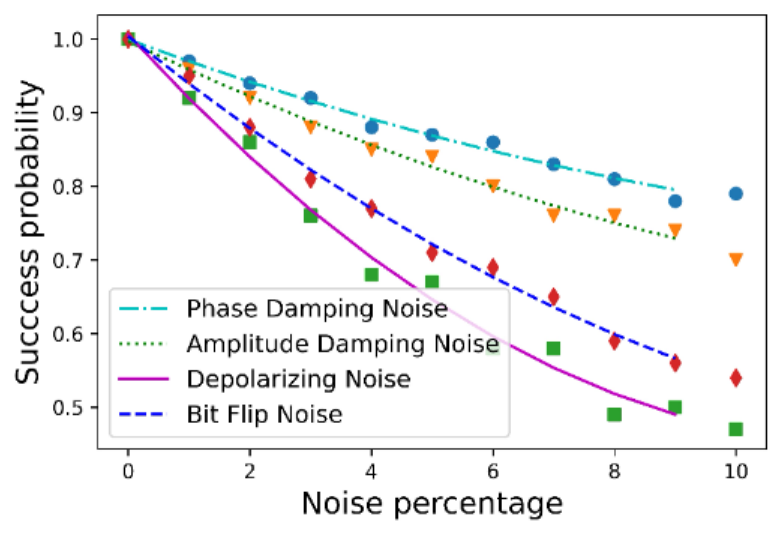}
		\par\end{centering}
	\caption{\centering Plot for success probability of the MQT scheme under various noisy environment having variable noise percentage. \label{fig2.7:Plot-noise-effect}}
\end{figure}

A physical system cannot be imagined without an environment. The surrounding environment has a considerable impact on the system. Environment near a quantum system causes diminish of its quantum mechanical behaviors. Hence, the performance of any scheme of quantum communication, which involves the traveling of qubits over the quantum channel gets impacted by the undesired noises. Therefore, the practical usefulness of any quantum communication scheme depends on its ability to withstand noise (for details see \cite{Anindita2017qDialogueNC,Anindita2018qConferenceNC,RishiD2016verifyQubitNC,Kishore2018qPrivateComparisionNC}). A complete positive trace preserving (CPTP) map is used to model noisy quantum system which is discussed in Section \ref{sec1.5:Noise Model}. There is a distinct Kraus operator corresponding to each noisy channels. Qiskit, a python library provides the noise model building technique has been used to investigate the impact of noise. The impact of different noisy environments such as depolarizing, bit flip, amplitude damping and phase damping noise are studied for the \acrshort{MQT} circuit illustrated in Figure \ref{fig2.4:MQTCkt}. It is found from the plot shown in Figure \ref{fig2.7:Plot-noise-effect} that the \acrshort{MQT} scheme using optimal resources is more resilient to the phase damping noise.
\section{Conclusion}
In this chapter, it is shown that the quantum resources used in the previously proposed schemes for MQT \cite{YuY2021MQT} are higher than the required (optimal) amount of quantum resources.  A modified \acrshort{MQT} scheme which is technically equivalent to the scheme for MQT proposed by Yan Yu et al. as it performs the same task but uses a lesser amount of quantum resources, is experimentally verified on a quantum computer (ibmq\_casablanca), available at IBM quantum experience on cloud. The purpose of the work reported in this chapter is to show that the multi-qubit entangled state is not necessarily required as a quantum resource for the \acrshort{MQT} scheme. Additionally, the robustness of the modified \acrshort{MQT} scheme is established and the impact of various types of noise on the \acrshort{MQT} scheme is examined. Also, the study reported in this chapter has shown that the modified \acrshort{MQT} scheme is resilient to several noises in an increasing order as phase damping, amplitude damping, depolarizing and bit-flip noise. It is hoped that the \acrshort{MQT} scheme that uses optimal quantum resources will soon be implemented in practical applications. The \acrshort{MQT} scheme discussed here teleports an unknown quantum state to multiple parties. In the next chapter, the \acrfull{QB} scheme which allows a sender to broadcast a known information to multiple parties will be discussed.
\newpage

\chapter{QUANTUM BROADCASTING OF KNOWN INFORMATION}\label{Ch3:QB}
\graphicspath{{Chapter3/Chapter3Figs/}{Chapter3/Chapter3Figs/}}

\section{Introduction}\label{sec3.1:intro-QB}
A groundbreaking study by Barnum et al. \cite{BarnumH1996theoremQB} in 1996 proved that a set of quantum states is broadcastable if they mutually commute, and therefore universal quantum broadcasting is not feasible. In other words, a noncommuting and hence nonorthogonal states cannot be broadcasted. Popularly, this theorem is recognized as the no-broadcasting theorem which is a class of no-go theorem \cite{LuoM2013QB}. Research community working in this domain has been attracted towards the no-broadcasting theorem since the introduction of the theorem and that led to further investigations leading to the generalization of the theorem for every finite-dimensional non-classical probabilistic theory that meets the no-signaling requirement \cite{BarnumH2007theoremNoQB}. One could argue that no-broadcasting is an obvious consequence of the no-cloning theorem because impossibility of cloning an unknown quantum state prohibits broadcasting of the state. However, a closer look into both the no-broadcasting and the no-cloning theorem would show that no-broadcasting is merely a restricted form of the no-cloning theorem. In particular, the no-cloning theorem states that existence of any CPTP map $\cal E$ is impossible that can generate $\cal E(\rho_{i})=\rho_{i}\otimes \rho_{i}$ $\forall i$ for any pure states pairs $\rho_{i}, i\in\{1,2\}$ that are non-orthogonal, while in the quantum broadcasting scenario, the final state not needed to be separable which implies a state $\rho$ is broadcastable on $\cal H$ to $\cal H_{A}\otimes H_{B}$ if there exist a CPTP map $\cal E^{\prime}$ such that $Tr_{A}$($\cal E^{\prime}(\rho))$=$Tr_{B}$(${\cal E^{\prime}}(\rho))=\rho$. Such CPTP map $\cal E^{\prime}$ is not available for any non-orthogonal unknown quantum state and referred as the well known no-go theorem called as no-broadcasting theorem \cite{BarnumH2007theoremNoQB}.\\
Even though the quantum no-broadcasting theorem is widely accepted, a number of recent publications assert that they have achieved quantum broadcasting \cite{LuoM2013QB,Giovannetti2008QB,YuY2017QBandMQT,ZhouY2018probableQB,YuY2018generalQB}. In particular, Yu et al. presented a method in Reference \cite{YuY2017QBandMQT} that allowed a sender (Alice) to use a four-qubit cluster state to broadcast a single-qubit arbitrary state to two geographically separated receivers (Bob and Charlie). Interestingly, Alice was fully aware of the state that was going to be broadcasted according to their protocol. Therefore, the job was basically to prepare many replicas of a known quantum state remotely. Neither the no-cloning theorem nor the no-broadcasting theorem forbids such a task. However, without explicitly mentioning the fact that the existing quantum broadcasting schemes actually broadcasting a known quantum state, these schemes were misleading or incorrectly called as quantum broadcasting schemes, especially in light of the well known no-broadcasting theorem. In References \cite{LuoM2013QB,BarnumH2007theoremNoQB,ZhaoN2020QB,WenX2008signatureQB,XiaoM2016blindSignatureQB,Zhang2016blindSignatureQB,YuY2018generalQB}, they also employed similar deceptive nomenclature. Later, a four-qubit cluster state which is non-maximally entangled, was used as a quantum resource in the Reference \cite{ZhouY2018probableQB} to modify the Yu et al. scheme and provides a slightly changed version of the original protocol with probabilistic flavour. In light of this, in the remaining chapter, broadcasting of a quantum information would actually mean broadcasting of a known quantum information. In this chapter, it is intended to make it very clear that broadcasting is only possible for known quantum state, and the best possible way to do so will be described. In what follows, it will be shown that the task at hand (i.e., broadcasting of a known quantum state) is equivalent to multiparty joint-\acrshort{RSP}, and an optimal strategy for the same will be specifically proposed here. A quantum computer from IBM will be used to implement the designed scheme for quantum broadcasting. Before moving ahead, it would be appropriate to clarify that the optimality discussed here pertains to broadcast a quantum state of general class, such as a qubit, that is achieved using entangled quantum states comprising of qubits only (implies that an entangled state of higher dimensional are not used). Such aspect requires particular attention because the recently proposed scheme for quantum broadcasting introduced by Hillery et al. uses lesser ebits than the scheme reported here. However, a quantum states belongs to a special class was broadcasted in their scheme. For example, a qubit which has the form $|\psi(\theta)\rangle =\frac{\exp(\iota \theta)|0\rangle+\exp(-\iota \theta)|1\rangle}{\sqrt{2}}$) is broadcasted to two receivers using an entangled state which is shared in a way that its two qubits are with the receivers and its qutrit is with the sender. The worthy point to mention here is that the study conducted by Hillery et al. was explicit in line with the statement discussed above that the implementation of a broadcasting protocol is basically a \acrshort{RSP}. It would be appropriate to mention here that the study conducted by Hillery et al. was explicit in stating that the implementation of a broadcasting protocol is basically a multi-party \acrshort{RSP} as previously emphasized. Compared to teleportation, a \acrshort{RSP} requires lesser number of classical communication. The capability of teleportation guarantees the capability of \acrshort{RSP} as the possibility of teleporting an unknown quantum state rises the possibility of teleporting a known state. It is now possible to teleport a quantum states of certain class using fewer classical resources. The study conducted by Hillery et al. took advantage of this fact. This method combined with the utilization of an entangled states of higher dimension can help in reducing resources requirements. However, it would neither be universally applicable nor easily achievable with current technologies.\\
The remaining part of this chapter is structured as follows. The proposed quantum broadcasting strategies utilizing the best available quantum resources are described in Section \ref{sec3.2:Broadcasting-of-known} and its potential generalizations for several aspects of quantum broadcasting is discussed in Section \ref{sec3.3:Possible-generalizations-QB}. In Section \ref{sec3.4:Effect-of-noise-QB}, the impact of various types of noise on the cluster state based \acrfull{QB} scheme and two Bell states based \acrshort{QB} scheme is examined and compared. Additionally, in Section \ref{sec3.5:Realization-IBM-QB}, the proposed methodology is experimentally verified on IBM quantum computer. Finally, Section \ref{sec3.6:Conclusion-QB} concludes this chapter.

\section{Optimal resources to broadcast a known quantum information\label{sec3.2:Broadcasting-of-known}}
Let us talk about the \acrshort{QB} scheme of Yan et al. \cite{YuY2017QBandMQT} involving three parties, one sender (Alice) and two receivers (Bob and Charlie). Alice wishes a message $\alpha|0\rangle+\beta|1\rangle$ where $|\alpha|^{2}+|\beta|^{2}=1$ to be broadcasted to Bob and Charlie. To do that, a cluster state having four-qubits is prepared by Alice which is given as
\begin{equation}
	|\phi\rangle_{1234}  = \frac{1}{2}(|0000\rangle+|0101\rangle+|1010\rangle-|1111\rangle.
\end{equation}
Alice transfers qubit $3$ and $4$ to Bob and Charlie, respectively, while keeping the qubit $1$ and $2$ for herself. Alice then selects a measurement basis set $\{|\varphi_{1}\rangle,|\varphi_{2}\rangle,|\varphi_{3}\rangle,|\varphi_{4}\rangle\}$ where,
\begin{equation}
	\begin{split}
		|\varphi_{1}\rangle =  \alpha^{2}|00\rangle+\alpha\beta|01\rangle+\alpha\beta|10\rangle-\beta^{2}|11\rangle,\\
		|\varphi_{2}\rangle = \alpha\beta|00\rangle-\alpha^{2}|01\rangle+\beta^{2}|10\rangle+\alpha\beta|11\rangle,\\
		|\varphi_{3}\rangle = \alpha\beta|00\rangle+\beta^{2}|01\rangle-\alpha^{2}|10\rangle+\alpha\beta|11\rangle,\\
		|\varphi_{4}\rangle = \beta^{2}|00\rangle-\alpha\beta|01\rangle-\alpha\beta|10\rangle-\alpha^{2}|11\rangle,
	\end{split}
\end{equation}
and uses this to measure her qubits $1$ and $2$. Ultimately, Bob and Charlie can recover the message by applying the proper unitary operation to their respective particles based on Alice's measurement results. Alice successfully broadcast her intended information to both Bob and Charlie in this manner. An identical work of Yun-Jing Zhou et al. \cite{ZhouY2018probableQB} demonstrated the probabilistic version of quantum broadcasting using a non-maximally entangled state of four-qubit. A scheme where distinct messages are transmitted to several recipients is known as \acrfull{MQT}. The \acrshort{MQT} scheme has already been discussed in Chapter \ref{Ch2:MQT}. Some recent articles have been generalized the \acrshort{QB} scheme \cite{LuoM2013QB,YuY2018generalQB}. As Alice is aware of the message she sent, therefore the plan might be seen as a multi-party \acrshort{RSP}. As discussed in Chapter \ref{Ch2:MQT}, $\lceil\log_{2}m\rceil$ Bell states are sufficient to teleport a quantum state with $m$ unknown coefficients \cite{Sisodia2017OpticalDesign_QT}. Similarly, $\lceil\log_{2}m^{n}\rceil$ Bell states would be needed to broadcast known quantum state with $m$-coefficient to $n$-receivers. Thus, the broadcasting scheme stated above (see Reference \cite{YuY2017QBandMQT}) needs $\lceil\log_{2}4=2\rceil$ Bell state instead of a maximally entangled four-qubit state. In this context, a generalized technique will be presented that uses optimal quantum resources to broadcast known quantum states to numerous receivers. However, before we proceed, it should be pointed out that the above-described scheme of Yan et al. \cite{YuY2017QBandMQT} and the majority of the subsequent schemes for the so-called quantum broadcasting, were actually the schemes for multi-party \acrshort{RSP}, also referred as the schemes for broadcasting of known quantum states, in order to avoid conflict with the quantum no-broadcasting theorem statement. Furthermore, it should be noted that the quantum channels employed in these schemes are not ideal. In the remaining part of this section, it will be explained how to carry out an identical task using a minimal amount of quantum resources. The most straightforward scenario might be discussed before expanding on it. Consider two Bell states $|\psi^{+}\rangle_{13}$ and $|\psi^{+}\rangle_{24}$ where $|\psi^{+}\rangle=\frac{1}{\sqrt{2}}(|00\rangle+|11\rangle)$. Alice keep particle 1 \& 2 to herself and send particle 3 \& 4 to Bob and Charlie, respectively. Alice uses a well known technique called \acrshort{RSP} \cite{Bennett2001RSP,Pathak2013book} to communicate with Bob and Charlie at their ends. The simultaneous delivery of the message to each recipient may cause confusion to readers. However, it should be emphasized that for the recipient to apply unitary in the \acrshort{RSP} process, the sender must send one classical bit of information to obtain the desired message. Each recipient receives these $1$-bit classical messages simultaneously over the conference call, allowing them to apply unitary operations and to receive the desired message simultaneously. Table \ref{tab3.1:Optimize-QB} reports the optimal state corresponding to resources employed for various reasons. In Reference \cite{Kishore2015generalControlledBQT}, a generalized technique for choosing a quantum channels for different controlled quantum communication schemes have been reported. Similarly, the next section aims to describe the potential generalizations of the scheme of quantum broadcasting of known quantum states leading to a large number of variants of it.
\begin{table}
	\caption{The reduced resources to accomplish various QB schemes reported previously. Here, $|\phi^{+}\rangle=\frac{1}{\sqrt{2}}|00\rangle+|11\rangle$.}
	\label{tab3.1:Optimize-QB}
	\centering
	\begin{tabular}{|>{\centering}m{0.8cm}|>{\centering}m{5.4cm}|>{\centering}m{3.9cm}|c|c|}
		\hline 
		S.No & \centering{Quantum channels used} & Scheme & Party & Optimal State \\
		\hline 
		1 & $\frac{1}{2}(|0000\rangle+|0101\rangle+|1010\rangle-|1111\rangle)_{A_{1}A_{2}B_{1}B_{2}}$\cite{YuY2017QBandMQT} & Quantum broadcasting & 3 & $|\phi^{+}\rangle_{A_{1}B_{1}}\otimes|\phi^{+}\rangle_{A_{2}B_{2}}$\\
		\hline
		2 & $\frac{1}{2}(|0001\rangle+|0110\rangle+|1011\rangle+|1100\rangle)_{A_{1}A_{2}B_{1}B_{2}}$\cite{LuoM2013QB} & Quantum broadcasting & 3 & $|\phi^{+}\rangle_{A_{1}B_{1}}\otimes|\phi^{+}\rangle_{A_{2}B_{2}}$\\
		\hline
		3 & $\frac{1}{4\sqrt{2}}(|00000000000000000000\rangle+|00000000010000000001\rangle+...+|11111111101111111110\rangle+|11111111111111111111\rangle)$\cite{ZarmehiF2021circularControlledQT_QB_NC} & Circular quantum broadcasting & 5 & $|\phi^{+}\rangle_{A_{i}B_{i}}^{\otimes10}$\\
		\hline 
	\end{tabular}
\end{table}
\begin{figure}
	\centering
	\includegraphics[scale=0.9]{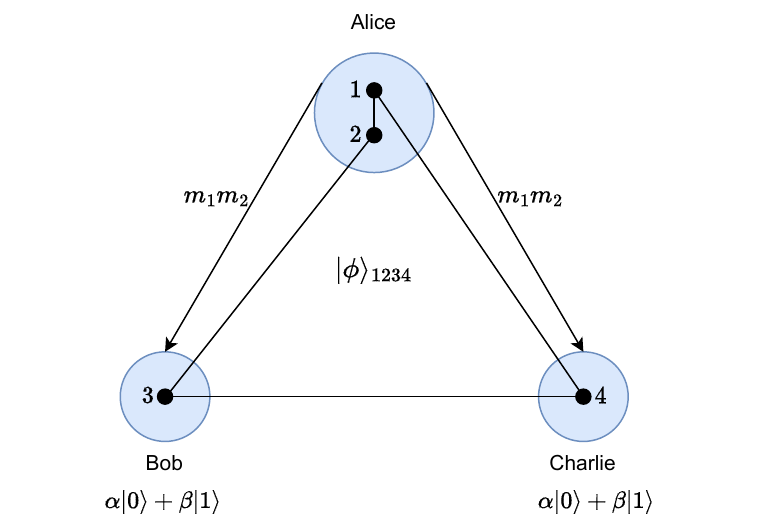}
	\caption{A sketch to visualize the \acrshort{QB} scheme of Yan Yu et al.}
	\label{Fig3:QB_task_YanYu}
\end{figure}
\section{Possible generalizations and potential applications \label{sec3.3:Possible-generalizations-QB}}
Recent work has proposed a potential generalization to teleport a $n-$qubit quantum information having $m-$unknown coefficients \cite{Sisodia2017OpticalDesign_QT}. Also, a general technique for choosing a quantum channel for schemes related to bidirectional teleportation has been established \cite{Kishore2015generalControlledBQT}. As mentioned in the previous section, \acrshort{QB} of known information can be thought of as a multi-party \acrshort{RSP}. The acquisition of \acrshort{RSP} requires a pre-shared entangled state. The scenario might be limited to one where the multi-party \acrshort{RSP} task is to be carried out using just bi-partite entangled states, as generating and maintaining multipartite entangled states is challenging. This visualization requires $m$ pre-shared bipartite entangled state to complete \acrshort{RSP} process having $m$ receivers. Consequently, the \acrshort{QB} task with $m$-receivers  can be accomplished using the following channel
\begin{eqnarray}
	|\chi\rangle & = & \otimes_{i=1}^{m}|\phi_{2i-1,2i}^{i}\rangle,
	\label{Eq3.3:General broadcast}
\end{eqnarray}
where, $|\phi^{i}\rangle$ is a two-qubit entangled state such that $|\phi^{i}\rangle\neq|\phi^{j}\rangle$. The entangled state is distributed such that all the first qubits $\{2i-1\}$ are with Alice and qubit $\{2i\}$ is with $i^{th}$ receiver. The nature of entanglement of $|\phi^{i}\rangle$ determines the variant of the \acrshort{QB} task, if the state $|\phi^{i}\rangle$ is maximally entangled then the scheme is perfect-\acrshort{QB} and if it is non-maximally entangled then the scheme is probabilistic-\acrshort{QB}. One may verify that scheme reported in Reference \cite{ZhouY2018probableQB} is a particular case of the suggested universal channel. For different broadcasting schemes, the generalized optimum channel suggested above can be used in place of previously published quantum channels (see Table \ref{tab3.1:Optimize-QB}). It is possible to achieve useful \acrshort{QB} scheme with lesser e-bits if we take into account higher dimensional entangled states (qudits). In Reference \cite{HilleryM2023QB}, a broadcasting scheme has been proposed which broadcasts a states belongs to restricted class to several receivers using higher dimensional entangled state. Here, the focus will be restricted to the \acrshort{QB} schemes that use qubits as a quantum channel, as they are simpler to verify experimentally on the IBM quantum computer.\\
Several variations of \acrshort{QB} of known quantum information are possible, just as there are several variations of \acrshort{RSP} (for further information, see the references in \cite{vishal2015cb_RSP_NC}). Variations of the \acrshort{QB} scheme can be probabilistic and deterministic joint-\acrshort{QB}, controlled-\acrshort{QB}, bidirectional-\acrshort{QB} and many more. Although these broadcasting variations have not yet been thoroughly investigated, they are clear expansions of the general methods for \acrshort{RSP} that are covered in Reference \cite{vishal2015cb_RSP_NC}. It is crucial to mention that two senders are often involved in joint-\acrshort{RSP}. Alice $1$ is the sender who is aware of the amplitude information, while Alice $2$ is the another sender who is aware of the phase information. To be more precise, if the qubit $\cos(\theta)|0\rangle+\exp(\iota \phi)|1\rangle$ is considered to be broadcasted such that Alice $1$ is aware of $\theta$ while Alice $2$ is aware of $\phi$, then they can jointly prepare the qubit state remotely. A similar scheme for joint-\acrshort{QB} will have two sender, each with insufficient knowledge about the broadcasted state. For such a scheme, the simplest quantum channel i.e., m copies of 3-qubit entangled states used as a quantum resource, where $m$ receivers are believed to receive a single qubit. A generalized channel for joint-\acrshort{QB} with two senders and $m$ receivers is as follows:
	\begin{eqnarray}
		|\varphi\rangle & = & \otimes_{i=1}^{m}|\phi_{3i-2,3i-1,3i}^{i}\rangle.
		\label{Eq3.4:Joint-QB}
	\end{eqnarray}
Equation \eqref{Eq3.4:Joint-QB} represents the generalized form of the quantum channel for joint-\acrshort{QB} (probabilistic joint-\acrshort{QB}) if $|\phi^{i}\rangle$ is considered to be a maximally (non-maximally) three-qubit entangled state. To generalize this to the $n$ sender case, the corresponding requirement would be of $(n+1)$-qubit entangled state, which is quite harder to generate and preserve as $n$ grows. In light of this, let's examine the potential for $n$ senders to transmit known information to $m$ receivers via quantum broadcasting. In such scenario, the known quantum information is distributed among all $n$ senders which are labeled as Alice 1, Alice 2,$\cdots$, Alice $n$. In order to broadcast, $\cos(\theta)|0\rangle+\exp(\iota \phi)\sin(\theta)|1\rangle$, the amplitude ($\theta$) and phase ($\phi$) information must be shared among all the senders. For simplicity, we can suppose that Alice 1 has knowledge of $\theta$, while each Alice $j$ (where, $j\in{2,3,...,n}$) is aware of $\phi_{j-1}:\phi_{1}+\phi_{2}+\phi_{3}+...\phi_{n}'$. In this case, senders will use the proper rotation to progressively encode the information at their ends. There are several choices for the implementations. A straightforward method would be to create $m$ replicas of the $(n+1)$-partite entangled state and another method is as follows. Among the senders and receivers, an entangled state(s) is shared with Alice $n$ only and the receivers. First, a state $\cos(\theta)|0\rangle+\sin(\theta)|1\rangle$ is prepared by Alice 1 and forwarded it to Alice 2 who operates the $P(\phi_{1})$ gate on it which transforms into a state $\cos(\theta)|0\rangle+\exp(\iota\phi_{1})\sin(\theta)|1\rangle$ and move forward towards Alice 3, who encodes her information using the $P(\phi_{2})$ gate and move towards the next sender. This procedure keeps going until Alice $n$ information is encoded, resulting in the quantum state $\cos(\theta)|0\rangle+\exp(\iota\phi)\sin(\theta)|1\rangle$, which she may then broadcast to $m$ recipients using $m$ copies of a shared two-qubit entangled states as previously mentioned. Reference \cite{HilleryM2023QB} has addressed such a generic system for $n$ senders and $m$ receivers, however, the quantum resources needed there are far more complex and involve qudits. Controlled-\acrshort{QB} schemes which has controller to control which recipients will obtain the message transmitted by Alice. Recipients will unable to extract information until the controller discloses the measurement outcomes. Similar to the channel used in controlled-\acrshort{RSP}, the channel used in controlled-\acrshort{QB} could be generalized as follows\\
\begin{eqnarray}
	|\tau\rangle & = & \sum_{k=1}^{m}|\chi\rangle_{k}|a_{k}\rangle\label{Eq3.5:controlled-QB},
\end{eqnarray}
where, $|\chi\rangle_{k}\neq|\chi\rangle_{k'}$ ($|\chi\rangle$ is already stated in Equation \eqref{Eq3.3:General broadcast}) and {$|a_{k}\rangle$} represents a set of $m$ mutually orthogonal single qubit states. Interestingly, this can readily be extended to the case of $n$ senders and $m$ receivers using the previously described strategy. Furthermore, to execute a controlled-\acrshort{QB} scheme using two-qubit entangled states only, the controller must randomly generate $m$ entangled states of two-qubit (for instance $m$ Bell states) and distribute them such that the first qubit of the $i^{th}$ entangled state is assigned to Alice $n$ while the second qubit is assigned to Bob $i$. However, the controller would have confidentially keep the details of the shared entangled state between Alice $n$ and Bob $i$. The controlled \acrshort{QB} will not be successful unless this information is made public. This means that, in theory, two-qubit entangled states can be used to perform any of the different variations of \acrshort{QB} scheme that can be designed. Generally, the state presented in Equation \eqref{Eq3.3:General broadcast} can be utilized to determine the optimal channel for different schemes of \acrshort{QB}. We don't need the complex multi-qubit states seen in \cite{LuoM2013QB,YuY2017QBandMQT,ZhouY2018probableQB,ZarmehiF2021circularControlledQT_QB_NC}, nor do we need the higher dimensional states found in \cite{HilleryM2022QB, HilleryM2023QB}.\\
As discussed above, bidirectional-\acrshort{RSP} and its related methods enables both Alice and Bob to share an information. In the communication scenario which involves only two-party, it is effectively bidirectional because there are two possible ways of communication. On the other hand, if we consider in general $n$ senders and $m$ receivers and permit all possible ways of broadcasting (without taking into account that the broadcasted state is jointly prepared), we will have multi-directional \acrshort{QB} and total $2(m\times n)$ possible direction of communication. We may consider a more general scenario where there are $n$ parties involved and each one broadcasts a known quantum state to the others. As a result, the notion of senders and receivers is hazy and $n(n-1)$ communication directions become apparent. A generalized quantum channels for the multi-directional-\acrshort{QB} scheme can be described as follows:
\begin{eqnarray}
	|\kappa\rangle & = & \otimes_{i=1}^{(n\times(n-1))}|\phi_{2i-1,2i}^{i}\rangle. 
	\label{Eq3.6:Multi-direction-QB}.   
\end{eqnarray}
This state is obviously experimentally achievable, and it is easy to build schemes for very generic scenarios, such as probabilistic and deterministic controlled multi-directional \acrshort{QB}, and then to build special cases of those scenarios.\\
Until now, the possible variations of the \acrshort{QB} scheme and the methods to execute them with the best available resources, or just the necessary copies of Bell states or Bell-type states, have mostly been discussed. However, the question remains, why should we perform broadcasting, or why might someone be interested in it? We are familiar with various applications of classical broadcasting in our everyday lives. Furthermore, it is explicitly demonstrated in a series of recent papers \cite{HilleryM2022QB, HilleryM2023QB} that quantum broadcasting may be used to realize systems for quantum cryptography, quantum voting, and quantum identity authentication.
\section{Effect of noise \label{sec3.4:Effect-of-noise-QB}}
\begin{figure}
	\centering
	\includegraphics[scale=0.8]{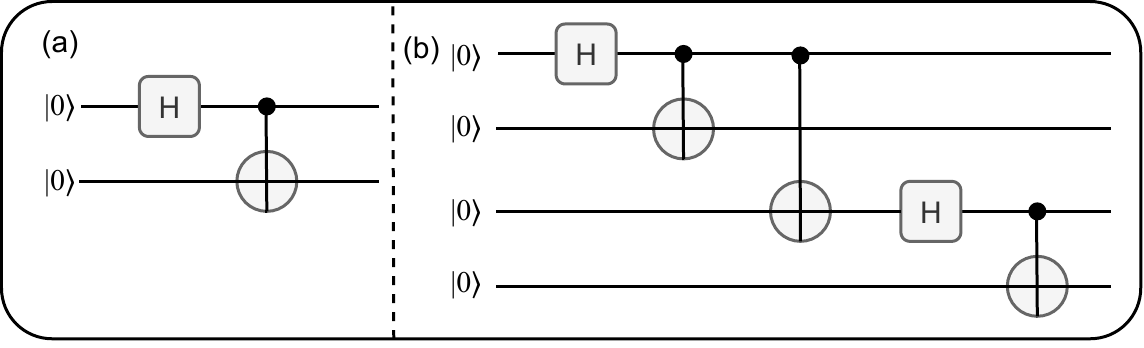}
	\caption{A quantum circuit for the generation of (a) Bell state and (b) cluster state.}
	\label{Fig3.1:Bell-cluster-circuit}
\end{figure}
\begin{figure}
	\begin{centering}
		\begin{tabular}{c}
			\includegraphics[scale=0.8]{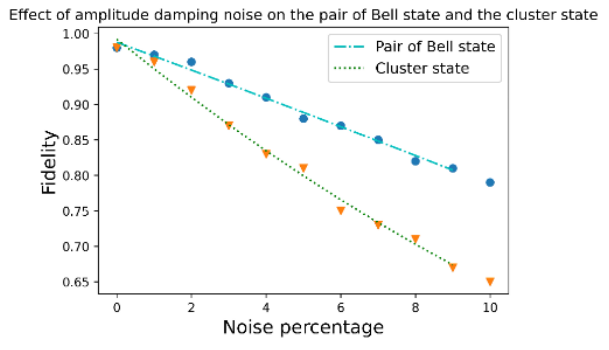}\\
			(a)\\
			\includegraphics[scale=0.8]{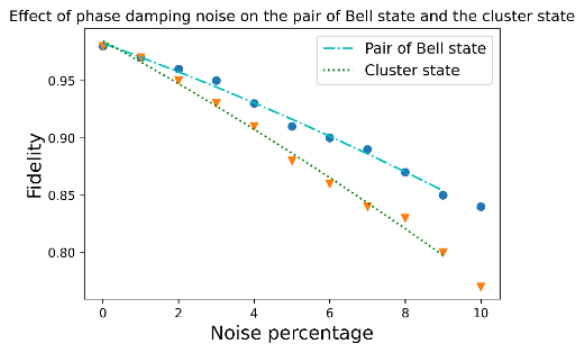}\\
			(b)\\
			\includegraphics[scale=0.8]{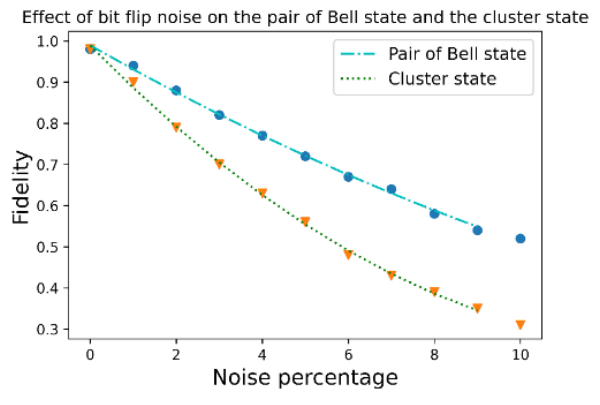}\\
			(c)\\ 
			\includegraphics[scale=0.8]{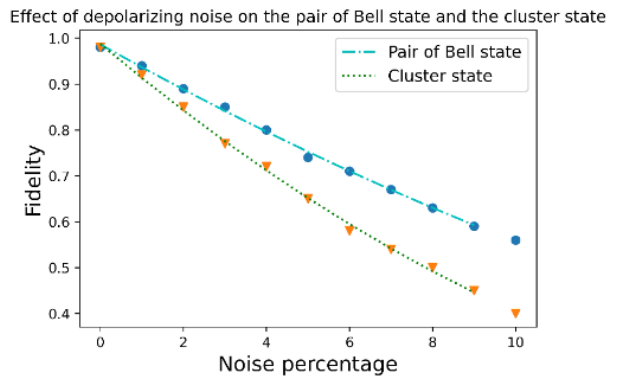}\\
			(d)
		\end{tabular}
		\par\end{centering}
	\caption{Impact of (a) amplitude damping, (b) phase damping, (c) bit-flip and (d) depolarizing noise on a pair of Bell states and cluster states.}
	\label{Fig3.2:Noise-plot-QB}
\end{figure}

Modeling of noise is already discussed in detail in Section \ref{sec1.5:Noise Model} and further described in Section \ref{sec2.3.2:Effect-of-noise-MQT}. We are familiar in the previous Chapter that technique to model noise on quantum circuits are available on qiskit \cite{NoiseQiskit}, an open source python library. A quantum circuit used to prepare Bell state and four-qubit cluster state is depicted in Figure \ref{Fig3.1:Bell-cluster-circuit}. These circuits are exposed to various kinds of noise, and the fidelity is examined by comparing the outputs of these circuits before and after the noise was added. A comparison has been made between the output of two copies of Bell state preparation circuit (see Figure \ref{Fig3.1:Bell-cluster-circuit} (a)) and the output of a cluster state preparation circuit (see Figure \ref{Fig3.1:Bell-cluster-circuit} (b)). Additionally, the task accomplished in Reference \cite{YuY2017QBandMQT} using four-qubit cluster state, can be completed with two copies of Bell states only. Also, it is found that the two copies of a Bell state outperforms the four-qubit cluster state in the presence of various noisy environments. Plots of fidelity measure as a function of noise percentage is displayed in Figure \ref{Fig3.2:Noise-plot-QB} which represents the performance of states or quantum channels.

\section{Implementation of the proposed scheme for QB on IBM quantum computer \label{sec3.5:Realization-IBM-QB}}
\begin{figure}
	\begin{centering}
		\begin{tabular}{c}
	\includegraphics[scale=0.9]{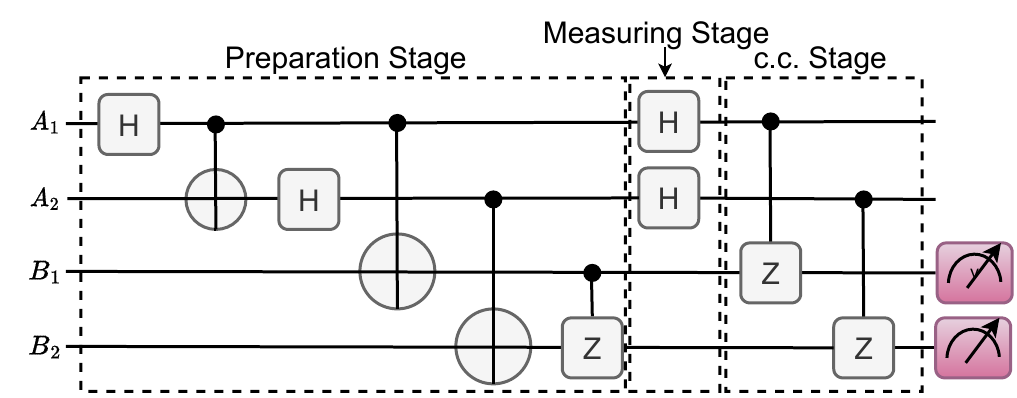}\\
	(a)\\
	\includegraphics[scale=1]{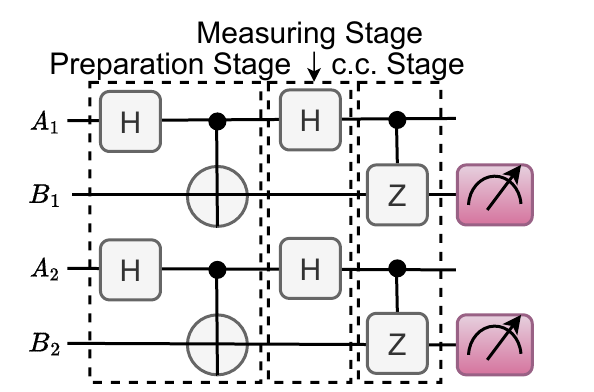}\\
	(b)
   \end{tabular}
   \par\end{centering}
	\caption{A quantum circuit for broadcasting the state $\alpha|0\rangle + \beta|1\rangle$ with $|\alpha|^{2}=|\beta|^{2}=\frac{1}{2}$ to two distinct receivers using (a) the four-qubit cluster state and (b) two copies of the Bell state. Here, c.c. stands for classical communication.}
	\label{Fig3.3:Comparative IBM circuits-QB}
\end{figure}

It has been observed that either a cluster state or two Bell states can be used to perform \acrshort{QB} task with one sender and two receivers. As seen earlier, two Bell states are the better resource for \acrshort{QB} and are more resilient to noise than a cluster state (see Figure \ref{Fig3.2:Noise-plot-QB}). This section describes a quantum circuit design to broadcast a known information $\alpha|0\rangle + \beta|1\rangle$ satisfies the condition $|\alpha|^{2}=|\beta|^{2}=\frac{1}{2}$.\\
The \acrshort{QB} scheme uses a four-qubit cluster state can be visualized as Alice first build a quantum circuit for the cluster state $\frac{1}{2}(|0000\rangle +|0101\rangle +|1010\rangle +|1111\rangle)$ preparation (see the block on left side of Figure \ref{Fig3.3:Comparative IBM circuits-QB} (a)) and measures the first two qubits in the Hadamard basis, that is, $\{|++\rangle + |+-\rangle +|-+\rangle +|--\rangle\}$ where, $|\pm \rangle=\frac{1}{\sqrt{2}}(|0\rangle \pm |1\rangle)$ (see the block in middle of Figure \ref{Fig3.3:Comparative IBM circuits-QB} (a)) and sends the remaining two qubits to the corresponding receivers. At last, based on the measurement results of Alice, the two receivers applies respective unitary. In this particular case, the receivers operates the pauli-$Z$ gate to their qubits if Alice obtains $11$ as measurement result (see the block on right side of Figure \ref{Fig3.3:Comparative IBM circuits-QB} (a)). This will end up in producing the desired state $\alpha|0\rangle + \beta|1\rangle$, for the two receivers.\\
The \acrshort{QB} scheme uses two copies of Bell state can be visualized as Alice first build a quantum circuit for two Bell state $\frac{1}{\sqrt{2}}(|00\rangle +|11\rangle)^{\otimes{2}}$ (see the block on left side of Figure \ref{Fig3.3:Comparative IBM circuits-QB} (b)) and measures the first two qubit of each Bell state in the Hadamard basis, that is, $\{|+\rangle, |-\rangle\}$ (cf. middle block of Figure \ref{Fig3.3:Comparative IBM circuits-QB} (b)) and sends the remaining qubits of each Bell state to the corresponding receivers. At last, based on the measurement results of Alice, the two receivers applies respective unitary operations. In this particular case, the receivers operate the pauli-$Z$ gate to their qubits if Alice obtains $1$ as measurement result (see the block on right side of Figure \ref{Fig3.3:Comparative IBM circuits-QB} (b)). This will end up in producing the desired state $\alpha|0\rangle + \beta|1\rangle,$ for the two receivers.\\
A superconducting based quantum computer named as ibmq\_manila accessible on cloud at IBM Quantum platform is used to simulate the quantum circuit shown in Figure \ref{Fig3.3:Comparative IBM circuits-QB}. The ibmq\_manila has total five superconducting qubits with the Falcon processor and having quantum volume (QV) 32 \cite{CrossA2019qComputer}. The QV is a benchmark which identify the performance of a quantum computer. It is based on a higher random circuit having same width and depth which can be implemented on a quantum computer. Obviously, a quantum computer having highest QV is the better choice. The quantum computers can be accessed on cloud at IBM quantum platform \cite{Ibm2021}. A ibmq\_manila has been chosen to execute the designed quantum circuit for the \acrshort{QB} scheme. The calibration data for each qubit of ibmq\_manila is provided in Table \ref{tab3.2:Calibration-data-QB}.\\
\begin{table}
	\caption{The calibration data of ibmq\_casablanca on Nov 28, 2022. cxi\_j, representing a $\rm{CNOT}$ gate with qubit i as the control and qubit j as the target.\label{tab3.2:Calibration-data-QB}}
	\begin{centering}
		\begin{tabular}{|c|c|c|>{\centering}p{1.8cm}|>{\centering}p{2.2cm}|>{\centering}p{2.2cm}|>{\centering}p{4.2cm}|}
			\hline
			Qubit & T1 ($\mu s$) & T2 ($\mu s$) & \centering{}Frequency (GHz) & \centering{}Readout assignment error & \centering{}Single-qubit Pauli-X-error & \centering{}CNOT error\tabularnewline
			\hline 
			$Q_{0}$ & 110.78 & 115.99 & \centering{}4.96 & \centering{}$2.61\times10^{-4}$ & \centering{}$2.73\times10^{-4}$ & \centering{}cx0\_1: $7.84\times10^{-3}$\tabularnewline
			\hline 
			$Q_{1}$ & 197.24 & 70.83 & \centering{}4.84 & \centering{}$4.63\times10^{-2}$ & \centering{}$2.12\times10^{-4}$ & cx1\_0: $7.84\times10^{-3}$, cx1\_2: $1.30\times10^{-2}$
			\tabularnewline
			\hline 
			$Q_{2}$ & 153.43 & 25.39 & \centering{}5.04 & \centering{}$1.99\times10^{-2}$ & \centering{}$4.85\times10^{-4}$ & \centering{}cx2\_1: $1.30\times10^{-2}$, cx2\_3:$1.65\times10^{-2}$\tabularnewline
			\hline 
			$Q_{3}$ & 124.01 & 62.45 & \centering{}4.95 & \centering{}$3.23\times10^{-2}$ & \centering{}$5.69\times10^{-4}$ & \centering{}cx3\_2: $1.65\times10^{-2}$, cx3\_4:$1.06\times10^{-2}$\tabularnewline
			\hline 
		\end{tabular}
	\end{centering}
\end{table}

\begin{figure}[h!]
	\begin{centering}
		\begin{tabular}{c}
	\includegraphics[scale=0.9]{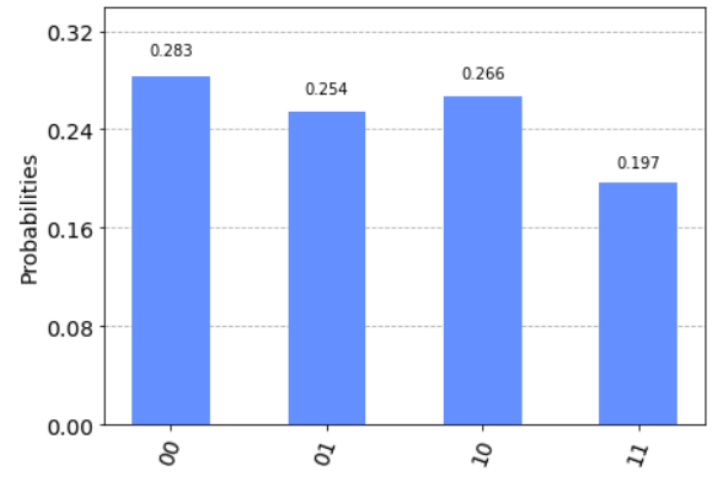}\\
	(a)\\
	\includegraphics[scale=0.9]{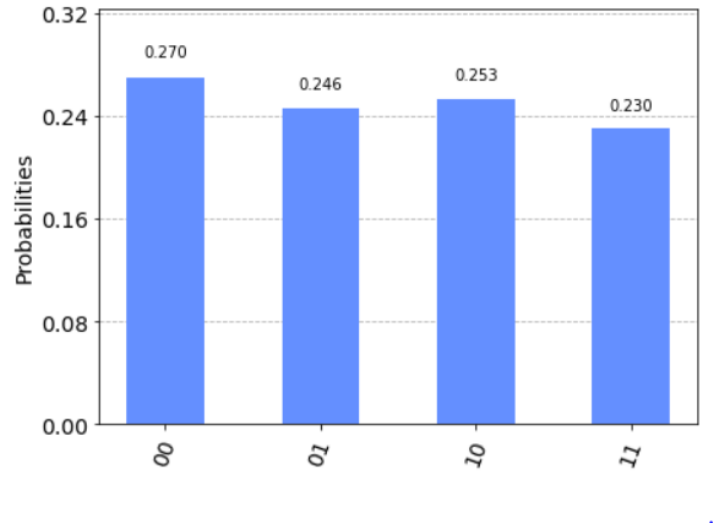}\\
	(b)
   \end{tabular}
   \par\end{centering}
	\caption{The obtained results after executing the quantum circuit depicted in (a) Figure \ref{Fig3.3:Comparative IBM circuits-QB} (a) and (b) Figure \ref{Fig3.3:Comparative IBM circuits-QB} (b) on ibmq\_manila.}
	\label{Fig3.4:Bar plot-QB}
\end{figure}

The obtained results after implementing the designed quantum circuit is shown in Figure \ref{Fig3.4:Bar plot-QB}. Theoretically, the expectation is to obtain all possible two bits 00, 01, 10 and 11 with equal probabilities, but the actually obtained values are slightly different due to the intrinsic errors in the quantum computer used. The designed circuit is run $8192$ times on ibmq\_manila to get the accurate result. Fidelity has already been discussed in Section \ref{sec2.3.1:Results-MQT} of the previous Chapter. The circuit depicted in Figure \ref{Fig3.3:Comparative IBM circuits-QB}(a) and Figure \ref{Fig3.3:Comparative IBM circuits-QB}(b) has an average fidelity of 86.47\% and 58.27\%, respectively. The fidelity is determined $10$ times and then averaging them to obtain the average fidelity value. The values of fidelity achieved indicates that the technique which is used for broadcasting known quantum information using two Bell states outperforms that using the cluster state.

\section{Conclusion\label{sec3.6:Conclusion-QB}}
Despite the existence of the no-broadcasting theorem, several research papers reported protocols for \acrshort{QB}. It is realized that all the existing schemes for \acrshort{QB} actually broadcast a known quantum state, which can easily be reduced to multi-party \acrshort{RSP}. It has been shown here how existing schemes of \acrshort{QB} can be realized with multi-party \acrshort{RSP}. It is clearly shown that the use of an adequate number of copies of the two-qubit entangled state (Bell state) is sufficient to accomplish the \acrshort{QB} task for the known quantum states. A Table \ref{tab3.1:Optimize-QB} is provided to summarize the optimal amount of quantum resources required for realizing the existing \acrshort{QB} scheme. A proof-of-principle experiment has also been performed, which shows that the \acrshort{QB} scheme works better with the technique provided than with the earlier reported techniques on the IBM quantum computer, accessible on cloud at IBM quantum platform. Performance of earlier reported resources and resources used by us is checked in the presence of various noisy environments. It is found that a pair of Bell state outperforms the cluster state. This chapter is concluded with the expectation that the current study will be useful for future research on several aspects of both strong and weaker forms of quantum broadcasting. The \acrshort{QB} scheme discussed here broadcasts a known quantum state. In the next chapter, the \acrfull{RIO} scheme where an unknown operator is teleported or operated remotely will be discussed.
\newpage

\chapter{REMOTE IMPLEMENTATION OF OPERATORS AND ITS VARIANTS}\label{Ch4:RIO}
\graphicspath{{Chapter4/Chapter4Figs/}{Chapter4/Chapter4Figs/}}

\section{Introduction}\label{sec:intro-CJRIO}
It is well known that entanglement is a crucial component of quantum communication and computing. The significance of the entangled state or a nonlocal state is due to its ability to accomplish several tasks that are impossible in the classical world. Limiting ourselves to the content of this chapter, it may be pointed out that entanglement can be used as a resource to achieve the so-called \acrfull{QT}, a concept first proposed by Bennett et al. in 1993 \cite{Bennett1993QT}. A traditional teleportation strategy uses a shared two-qubit entangled state, some local operations, and two bits of classical communication to transfer an unknown single-qubit quantum state from one location to another, without physically transmitting the qubit itself. Subsequently, Pati et al. \cite{AkPati2000RSP} demonstrated that a shared bipartite entangled state and one classical bit are sufficient to teleport a known quantum state. This method of teleportation of a known quantum state is popularly known as quantum \acrshort{RSP}. The introduction of this new concept called \acrshort{RSP} led to two questions for researchers to solve. The first one was: What are the possible variants of the \acrshort{RSP} scheme?. This question was addressed by introducing various variants of it such as, controlled-\acrshort{RSP}, joint-\acrshort{RSP}, controlled-joint-\acrshort{RSP}, and bidirectional-\acrshort{RSP} (see \cite{vishal2015cb_RSP_NC} and references therein). The other question was: Similar to \acrshort{RSP}, can we remotely apply operators on qubit(s)? The query was addressed first by Huelga et al. in 2001 \cite{Huelga2001QRC}, who demonstrated that shared entanglement, in conjunction with \acrfull{LOCC}, can be used to remotely implement a quantum operation. This technique of implementing a quantum operation remotely is referred to as quantum remote control, also referred to as \acrfull{RIO}. Before further proceeding, it would be appropriate to point out that any protocol for bidirectional quantum state teleportation \cite{Kishore2015generalControlledBQT,Sisodia2017OpticalDesign_QT} can be easily modified to implement \acrshort{RIO}. This is easily visualized if we assume that Bob wants to remotely implement an arbitrary operator $U_B$ on a quantum state $|\psi\rangle$ which is in the possession of Alice located in a spatially separated place. Alice may now teleport the state $|\psi\rangle$ to Bob, who may then teleport the state $|\psi^{\prime}\rangle$ to Alice after applying his operation on the received state to obtain $|\psi^{\prime}\rangle=U_B|\psi\rangle$. This simple approach would require a minimum of four bits of classical communication and a pair of Bell states. This establishes a kind of upper limit on the quantity of resources needed because it would be illogical to utilize more. An efficient \acrshort{RIO} scheme is expected to require a lesser amount of resources, and in Reference \cite{Huelga2002QRC}, it is shown that implementation of the \acrshort{RIO} scheme requires at least four bits of classical communication and a pair of Bell state, which is the minimal necessary requirements. Also, it was mentioned that if $U_B$ belongs to a specific unitary class then the \acrshort{RIO} task requires two bits of classical communication and one Bell state only. The \acrshort{RIO} concept is also widely generalized and experimentally realized for a particular version of \acrshort{RIO} \cite{XiangG2005ExpRIO}. The concept of \acrshort{RIO} is extended to different variants such as, controlled \acrshort{RIO} \cite{QiuX2023CRIO,BaAn2022CRIO}, joint \acrshort{RIO} \cite{BaAn2022JRIO}. The resources utilized to accomplish such tasks of \acrshort{RIO} are different. For instance, Reference \cite{BaAn2022CRIO,BaAn2022JRIO,WangM2024QRC} uses hyperentangled state as a quantum resource for the \acrshort{RIO}, \acrfull{CRIO} and \acrfull{JRIO} schemes, while Reference \cite{QiuX2023CRIO} uses graph state for the \acrshort{CRIO} scheme. There are existing schemes for variants of \acrshort{RIO} such as \acrshort{CRIO}, \acrshort{JRIO} but another variant \acrfull{CJRIO} is unexplored. This motivates to propose a scheme for \acrshort{CJRIO}. Interestingly, the \acrshort{CJRIO} scheme can easily be simplified to existing schemes of \acrshort{CRIO}, \acrshort{JRIO} and \acrshort{RIO}. The scheme of \acrshort{RIO} and its variants has a significant application in distributed quantum computing which allows nonlocal operation (see \cite{Pathak2007dQSearch} and references therein and \cite{CAF+24}).\\
Another related scheme to \acrshort{RIO} called \acrfull{RIHO} was introduced in 2006 where the required operator to be implemented remotely is hidden inside a lump operator \cite{BaAn2007RIHO}. Operating a lump operator at one end will reflect the required operator at another end. To understand it explicitly, consider an operator $U_B=\frac{1}{\sqrt{2}}(U_0+U_1)$, where $U_0$ and $U_1$ are suboperators of a particular form described in Section \ref{Sec4.6:Task-RIHO_RIPUO}. The \acrshort{RIHO} scheme works as an operation of the operator $U_B$ at one ends with operation of $U_0$ at the other end. To accomplish this \acrshort{RIHO} task, the three-qubit entangled GHZ state is used \cite{BaAn2007RIHO}. In this chapter, a protocol for the same task using a two-qubit maximally entangled state is proposed. This has a wide application in blind quantum computing.\\
A relatable scheme to \acrshort{RIO} was introduced by Wang in $2006$ called as \acrfull{RIPUO} where the nature of operator to operate at one end is partially known or unknown. The operator is partially unknown in the sense that structure of its matrix is known but not value of its matrix elements. An interesting aspect of the \acrshort{RIPUO} scheme is that it requires half of the resources as that of bidirectional-\acrshort{QT}. The concept of \acrshort{RIPUO} is expanded to the situation of numerous sender and numerous receiver by Wang himself in $2007$ \cite{WangM2007RIPUO}. Several variants since then have been emerged such as, controlled \acrshort{RIPUO} (CRIPUO) \cite{FanQ2008CRIPUO}, cyclic CRIPUO (CCRIPUO) \cite{PengJ2019CCRIPUO}, bidirectional CCRIPUO (BCCRIPUO) \cite{PengJ2022BCCRIPUO} and many-party CRIPUO (MCRIPUO) \cite{Peng22MPCRIPUO}. The resources used to realize these schemes are different. For instance, \acrshort{RIPUO} was realized using the Bell state, \cite{Wang06RIPUO} and different variants of \acrshort{RIHO} (CRIHO \cite{BaAn2007RIHO}) and \acrshort{RIPUO} (CRIPUO \cite{FanQ2008CRIPUO}, CCRIPUO \cite{PengJ2019CCRIPUO}, BCCRIPUO \cite{PengJ2022BCCRIPUO}, MCRIPUO \cite{Peng22MPCRIPUO}) were realized using $3+n$ and $2+n$ party entangled states, respectively, where $n$ are additional parties involved. In the following, it will be demonstrated that a bipartite entangled state can be used to realize not only \acrshort{RIHO} but also controlled \acrshort{RIHO} (CRIHO), \acrshort{RIPUO}, and CRIPUO. We are familiar with the problem in maintaining the higher qubit entangled state. Hence, an attempt has been made to design a novel protocol for a family of \acrshort{RIO}, such as \acrshort{RIHO}, CRIHO, \acrshort{RIPUO}, CRIPUO, etc. using optimal required resources. Additionally, the \acrshort{RIPUO} and \acrshort{RIHO} schemes have been proposed separately until now, but in what follows, it will be demonstrated that there is a common foundation and that a few minor adjustments to the \acrshort{RIHO} scheme can convert it into a \acrshort{RIPUO} scheme.\\
The remaining part of this chapter is structured as follows. In Section \ref{Sec4.2:Task-CJRIO}, the \acrshort{CJRIO} task in details will be explained. The designing of protocol for the \acrshort{CJRIO} scheme using a four qubit hyperentangled state, entangled in both spatial and polarization \acrfull{DOF} simultaneously, is discussed in Section \ref{Sec4.3:Protocol-CJRIO}. The \acrshort{CJRIO} scheme has been generalized to many joint parties and many controller in Section \ref{Sec4.4:Generalization-CJRIO}. In Section \ref{Sec4.5:Variants-CJRIO}, it is discussed that the generalized \acrshort{CJRIO} scheme can be reduced to the existing variants of the \acrshort{RIO} schemes and a specific focus is given to estimating the efficiency of the \acrshort{CJRIO}  and its related schemes. In Section \ref{Sec4.6:Task-RIHO_RIPUO}, the task for \acrshort{RIHO} and \acrshort{RIPUO} scheme is discussed in details. Protocols for both the schemes \acrshort{RIHO} and \acrshort{RIPUO} are described in Section \ref{Sec4.7:Protocols-RIHO_RIPUO}. Photon loss of the auxiliary coherent state has also been examined for both protocols as a result of interaction with the environment. Generalization of both the protocols to numerous joint parties and controllers is discussed in Section \ref{Sec4.8:General-RIHO_RIPUO}. Also it is shown that the controlled version of the proposed schemes does not actually required an additional quantum resource. Finally, Section \ref{Sec4.9:Conclusion-RIHO_RIPUO} concludes this chapter. 
\begin{figure}
	\begin{centering}
		\begin{tabular}{c}
			\includegraphics[scale=0.8]{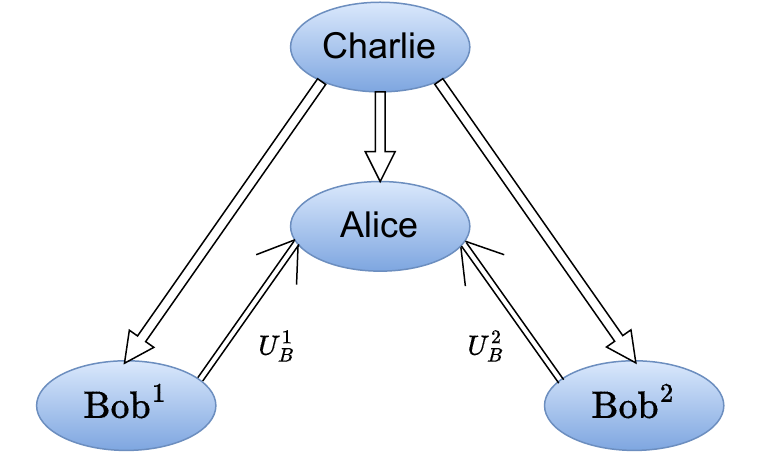}\\
			(a)\\[2ex]
			\includegraphics[scale=0.8]{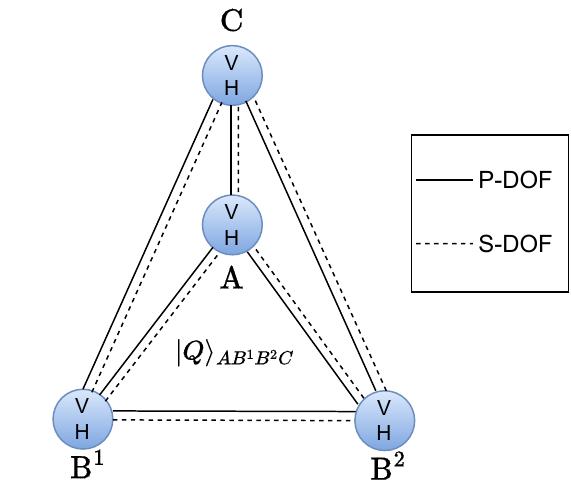}\\
			(b)\\[1ex]
		\end{tabular}
		\par\end{centering}
	\centering{}\caption{A sketch to visualize (a) the CJRIO task and (b) the quantum resources used. \label{Fig3:CJRIO_Task}}
\end{figure}
\section{The CJRIO task\label{Sec4.2:Task-CJRIO}}
The goal of \acrshort{CJRIO} is to controllably and cooperatively apply an unknown quantum operation to an arbitrary unknown quantum state at various nodes. There are four spatially separated parties involved in the \acrshort{CJRIO} task, Alice, Bob$^1$, Bob$^2$ and Charlie. Here, we assume Bob$^1$ and Bob$^2$ as the joint parties who jointly implement an arbitrary unitary $U$ that can be decomposed as $U=U^1_B.U^2_B$ on an unknown quantum state $ |\psi\rangle_X$ that is available with Alice under the supervision Charlie as a controller. For a broader perspective, Bob$^1$ and Bob$^2$ want to operate the operators $U^1_B$ and $U^2_B$, respectively, whose forms are as follows:
\begin{equation}\label{eq:Ub1}
	U^1_B=\begin{pmatrix}
		u^1_B & v^1_B\\ 
		-v^{1*}_B & u^{1*}_B
	\end{pmatrix},
\end{equation}
\begin{equation}\label{eq:Ub2}
	U^2_B=\begin{pmatrix}
		u^2_B & v^2_B\\ 
		-v^{2*}_B & u^{2*}_B
	\end{pmatrix}.
\end{equation}
It is to be noted that the operators listed above are the most general since they cover every possible qubit rotation (the state that Alice possesses, on which Bob's operator is to be applied remotely). This is true because a unimodular matrix of the following type can be used to describe the arbitrary rotation of a qubit:
\begin{equation}
U=\begin{pmatrix}
	u & v\\ 
	-v^{*} & u^{*}
\end{pmatrix}
\end{equation}
so that $|u|^2+|v|^2=1, u,v\in\mathbb{C}.$ The SU(2) group is made up of all such unitaries set, and unitary operations of the same form are employed in the current work. As previously mentioned, Alice, who is geographically isolated from Bob$^1$ and Bob$^2$, has an unknown quantum state $ |\psi\rangle_X$ of the following kind 
\begin{equation}\label{eq:Qs}
	|\psi\rangle_X=(\alpha|x_0\rangle+\beta|x_1\rangle)_X|\text{V}\rangle_X
\end{equation}
where, $|V\rangle_X$ describes the photon's polarization state and $\alpha$ and $\beta$ are unknown coefficients that satisfy the normalization requirement $|\alpha|^2+|\beta|^2=1$. Physically, it appears as Alice has a vertically polarized photon indexed by $X$ that is in the spatial superposition states of $|x_0\rangle$ and $|x_1\rangle$.\\
The unitary operator $U^1_B$'s operated on $|\psi\rangle_X$ can be explained as $|\psi_{B^1}\rangle=U^{1}_B|\psi\rangle_X=\alpha_{B^1}|x_0\rangle+\beta_{B^1}|x_1\rangle$ with $\alpha_{B^1}=\alpha{u^1_B}+\beta{v^1_B}$ and $\beta_{B^1}=-\alpha {v^{*1}_B}+\beta{u^{*1}_B}$. Additionally, the operator $U^2_B$ action on $|\psi\rangle_X$ can be explained as $|\psi_{B^2}\rangle=U^{2}_B|\psi\rangle_X=\alpha_{B^2}|x_0\rangle+\beta_{B^2}|x_1\rangle$ with $\alpha_{B^2}=\alpha{u^2_B}+\beta{v^2_B}$ and $\beta_{B^2}=-\alpha {v^{*2}_B}+\beta{u^{*2}_B}$.\\
This assignment involves Bob$^1$ and Bob$^2$ applying their operators remotely on Alice's state, which may be expressed mathematically as
\begin{equation}\label{eq:task}
	\begin{split}
		|\psi_{B^1B^2}\rangle & ={U^1_B}{U^2_B}|\psi\rangle_X\\
		& = (\alpha_{B^1 B^2}|x_0\rangle+\beta_{B^1 B^2}|x_1\rangle)|\text{V}\rangle_{X}
	\end{split}
\end{equation}
where, $\alpha_{B^1B^2}=\alpha_{B^2}{u^1_B}+\beta_{B^2}{v^1_B}$ and $\beta_{B^1B^2}=-\alpha_{B^2}{v^{*1}_B}+\beta_{B^2}{u^{*1}_B}$. \\
This particular task uses a four-qubit hyperentangled state as a quantum channel, which is described as
\begin{equation}\label{Eq4.6:QCused}
	|Q\rangle_{AB^1B^2C}=|Q^S\rangle_{AB^1B^2C} |Q^P\rangle_{AB^1B^2C}
\end{equation}
where ,
\begin{equation}\label{Eq4.7:HySEntangle}
	|Q^S\rangle_{AB^1B^2C}=|a_0\rangle_A |b^1_0\rangle_{B^1}
	|b^2_0\rangle_{B^2} |c_0\rangle_C + |a_1\rangle_A |b^1_1\rangle_{B^1}
	|b^2_1\rangle_{B^2} |c_1\rangle_C,
\end{equation}
\begin{equation}\label{Eq4.8:HyPEntangle}
	|Q^P\rangle_{AB^1B^2C}=|\text{H}\rangle_A |\text{H}\rangle_{B^1}
	|\text{H}\rangle_{B^2} |\text{H}\rangle_C + |\text{V}\rangle_A |\text{V}\rangle_{B^1}
	|\text{V}\rangle_{B^2} |\text{V}\rangle_C.
\end{equation}
Here, $\text{H}$ and $\text{V}$ represent the horizontal and vertical polarization, and $a_j$, $b^1_j$, $b^2_j$, and $c_j$ $(j = 0,1)$ indicate the spatial paths. The \acrfull{P-DOF} and \acrfull{S-DOF} are indicated by the superscripts $P$ and $S$, respectively. It should be observed that the normalization factor $1/\sqrt{2}$ is not included in Equation \eqref{Eq4.7:HySEntangle} and \eqref{Eq4.8:HyPEntangle}. In the subscript, the labels $A$, $B^1$, $B^2$, and $C$ indicate the photon states of Alice, Bob$^1$, Bob$^2$, and Charlie, respectively. Bob$^1$ and Bob$^2$ attempt to apply a unitary operation on an arbitrary unknown quantum state at Alice's node with all parties geographically separated from one another.  

\section{Protocol for CJRIO\label{Sec4.3:Protocol-CJRIO}}
The unknown quantum state $|\psi\rangle_X$ and the quantum channel used in the \acrshort{CJRIO} task can be expressed together as 
\begin{equation}\label{eq:totalS}
	|\psi\rangle_X|Q^{SP}\rangle_{AB^1B^2C}=|\phi^S\rangle_{XAB^1B^2C}|\text{V}\rangle_{X}|Q^{P}\rangle_{AB^1B^2C},
\end{equation}
where,
\begin{equation}\label{eq:9}
	|\phi^{S}\rangle=(\alpha|x_0\rangle+\beta|x_1\rangle)_X 
	\otimes(|a_0\rangle_A |b^1_0\rangle_{B^1}
	|b^2_0\rangle_{B^2} |c_0\rangle_C + |a_1\rangle_A |b^1_1\rangle_{B^1}
	|b^2_1\rangle_{B^2} |c_1\rangle_C).   
\end{equation}
The quantum channel that will be used for the \acrshort{CJRIO} protocol can be realized using photonic qubits, and due to bosonic nature of photons, their interaction is not possible. In other words, we can say photons are weakly interacting particles. The ability of photons to have the least interaction with the environment makes it a promising candidate for developing quantum computers based on photonic qubits \cite{TF19,SP19,KMN+07,JM24}. However, photon-photon interaction is required to communicate between qubits. A significant number of theoretical and practical attempts have been made in recent years to establish strong interactions between two optical beams \cite{FDS16,FHM+15,HWR+16,sun2018single,LCC+16,TSR+16,SSJ+20}. Remarkably, obtaining such interactions at the single-photon level is severely hampered by weak photon-photon scattering.\\
Despite the existence of the above-mentioned difficulty, there has been a lot of recent progress made in the development of photonic quantum technologies \cite{OFV09,WSL+20, TGC+21,S24}, such as photonic quantum computers (see Reference \cite{JM24} and references therein), which depend on the appropriate utilization of photon-photon interaction. For example, a strong cross-Kerr non-linearity employing electromagnetically induced transparency (EIT) \cite{FIM05,H97} has been used to obtain cross-phase shifts of $13$ $\mu$rad per photon \cite{FHM+15}. Moreover, rubidium atoms restrained to a hollow-core photonic bandgap fiber have been used to record cross-phase shifts of $0.3$ mrad per photon \cite{VSG+13}, and it has been discovered that exciton polaritons in micropillars with embedded quantum wells undergo shifts of $3$ mrad per polariton. In addition to cross-Kerr interactions, atom-cavity systems interaction \cite{HWR+16,sun2018single} and EIT \cite{TSR+16,LCC+16,SSJ+20} that causes a $\pi$ phase shift between photons have been reported.\\
With the foregoing background knowledge in mind, a brief overview will be given on how the auxiliary coherent state (CS) $|z\rangle$ obtains phase shift $\theta$ through the cross-Kerr interaction. To do this, the cross-Kerr interaction between the signal and probe modes is employed. Hamiltonian for the two interacting modes $a$ and $b$ is given as $H=\hbar\chi a^{\dagger} a b^{\dagger} b$, where $\chi$ represents the non-linear coupling constant and $a^{\dagger}(b^{\dagger})$ and $a(b)$ are creation and annihilation operators of the two interacting modes. Assume, Fock state $|n\rangle$ as a signal mode and coherent state $|z\rangle$ as a probe mode. The two modes would jointly evolve as $e^{-i \chi a^{\dagger} a b^\dagger b t}|n\rangle |z\rangle=|n\rangle|ze^{-i n \theta}\rangle$, where $\theta=\chi t$, and $t$ is the time needed for interaction to gain the desired phase. It is noticeable that a phase shift of $\theta$ is acquired by the probe mode while the phase of signal mode is unaffected. Whatever discussion have done till now to give a physical insight of the photon-photon interaction picture, will be used to explain the proposed protocols in a step-wise manner in this Chapter.\\
The \acrshort{CJRIO} protocol is described in two parts, one is when the \acrshort{S-DOF} of the quantum channel is used and other when \acrshort{P-DOF} is used. In Section \ref{Sec4.3.1:S-DOF_CJRIO} (\ref{Sec4.3.2:P-DOF_CJRIO}), the steps involved in the \acrshort{CJRIO} protocol will be discussed which uses \acrshort{S-DOF} (\acrshort{P-DOF}) of the quantum channel.

\subsection{Utilization of S-DOF \label{Sec4.3.1:S-DOF_CJRIO}}
\begin{figure}
	\centering
	\includegraphics[scale=1.2]{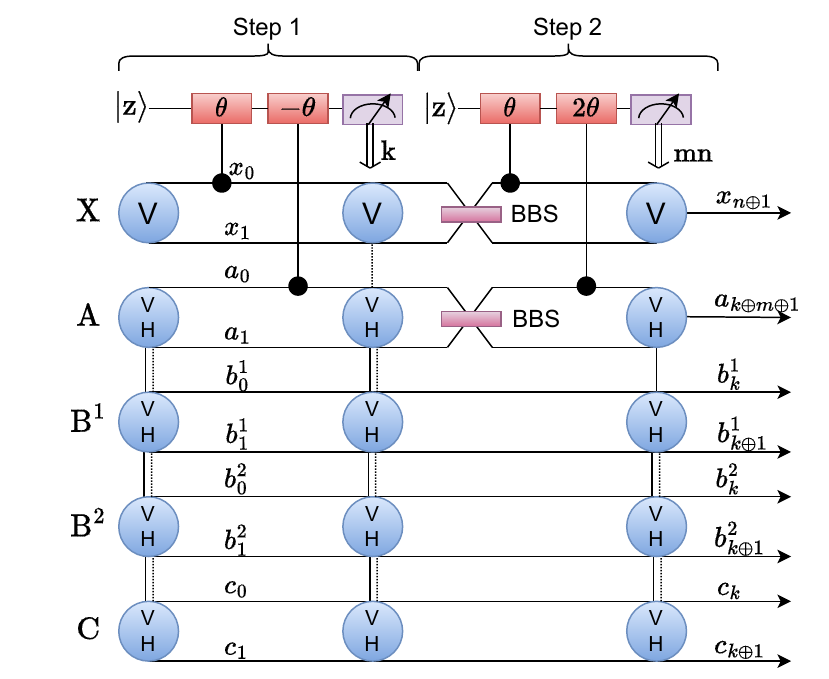}
	\caption{A schematic illustrating the first two steps of the CJRIO protocol is provided. A circle labeled $V, H$ represents a photon simultaneously in vertical and horizontal polarization, while a circle labeled $V$ only represents a photon in vertical polarization. Circles with two attached lines depict photons existing in two spatial paths simultaneously, whereas those with one line indicate photons with a single spatial path. The interaction between a photon path with a coherent state (CS) via cross-Kerr nonlinearity is represented by a line terminating in a bold dot on the photon path. Double arrows originating from the CS denote measurement outcomes. Vertical solid (dashed) lines represent entanglement in P-DOF (S-DOF). The BBS denotes a balanced beam splitter.}
	\label{Fig4.1:Step1-2_CJRIO}
\end{figure}

\begin{description}
	\item[Step~1] The protocol starts by entangling the photon $X$ with the quantum channel. This can be achieved when Alice takes an auxiliary CS $|z\rangle$ and allows it to interact with one of the two paths of photon $X$ and $A$ (here, path $|x_0\rangle$ and $|a_0\rangle$ are chosen corresponding to photons $X$ and $A$, respectively) to interact through a non-linear cross-Kerr interaction with a phase shift of $\theta$ and $-\theta$ respectively. The joint state after this interaction can be written as follows:
	\begin{equation}\label{Eq4.11:Interaction-Step1}
		\begin{split}
			|\xi\rangle & =(\alpha|x_0\rangle_X |a_0\rangle_A |b^1_0\rangle_{B^1}
			|b^2_0\rangle_{B^2} |c_0\rangle_C + \beta|x_1\rangle_X |a_1\rangle_A |b^1_1\rangle_{B^1}|b^2_1\rangle_{B^2} |c_1\rangle_C)|z\rangle\\
			& + (\alpha|x_0\rangle_X |a_1\rangle_A |b^1_1\rangle_{B^1}
			|b^2_1\rangle_{B^2} |c_1\rangle_C|ze^{i\theta}\rangle + \beta|x_1\rangle_X |a_0\rangle_A |b^1_0\rangle_{B^1}|b^2_0\rangle_{B^2} |c_0\rangle_C)|ze^{-i\theta}\rangle
		\end{split}
	\end{equation}
	The coherent state is then measured to give two potential results $k=0$ $(1)$ corresponding to the states $|z\rangle$  $(|ze^{\pm i\theta}\rangle)$ respectively. Depending on the two potential outcomes, the state in Equation \eqref{Eq4.11:Interaction-Step1} can be written as follows:
	\begin{equation}\label{Eq4.12:Measurement-Step1}
		|\xi_k\rangle=\alpha|x_0\rangle_X |a_k\rangle_A |b^1_k\rangle_{B^1}
		|b^2_k\rangle_{B^2} |c_k\rangle_C + \beta|x_1\rangle_X |a_{k\oplus1}\rangle_A |b^1_{k\oplus1}\rangle_{B^1}
		|b^2_{k\oplus1}\rangle_{B^2} |c_{k\oplus1}\rangle_C
	\end{equation}
	where, the symbol $\oplus$ denotes the addition mod $2$. It can be clearly seen that the unknown qubit with Alice is now entangled with the quantum channel used. Optical quantum circuit illustrating Step $1$ of the \acrshort{CJRIO} protocol is shown in Figure \ref{Fig4.1:Step1-2_CJRIO}. In the Figure \ref{Fig4.1:Step1-2_CJRIO}, circle labelled as $\text{X}$ represents the unknown qubit which is with the Alice and circles labelled as $\text{A}$, $\text{B}^1$, $\text{B}^2$ and $\text{C}$ are photonic qubits which are with Alice, Bob$^1$, Bob$^2$ and Charlie, respectively.
	
	\item[Step~2] The entanglement developed in the previous step is destroyed by Alice when she mixes her photon paths $|x_0\rangle$ and $|x_1\rangle$ (|$a_k\rangle$ and $|a_{k\oplus1}\rangle$) of photon $X$ ($A$) on a balanced beam splitter (BBS) (for more details see Section 2 of Reference \cite{Kishore2019BS} or that of Reference \cite{BaAn2022CRIO}). The BBS transforms the state as $|\sigma_j\rangle\rightarrow|\sigma_j\rangle+(-1)^j|\sigma_{j\oplus1}\rangle$ (ignoring the normalization factor). The mixing of photon's path changes the state described in Equation \eqref{Eq4.12:Measurement-Step1} to
	\begin{equation}\label{Eq4.13:Mixe-path-Step2}
		\begin{split}
			|\xi^{'}_k\rangle & =(|x_0\rangle|a_k\rangle+(-1)^k|x_1\rangle|a_{k+1}\rangle)\otimes(\alpha|b^1_k\rangle||b^2_k\rangle|c_k\rangle+(-1)^k\beta|b^1_{k+1}\rangle|b^2_{k+1}\rangle|c_{k+1}\rangle)\\
			&+(|x_0\rangle|a_{k+1}\rangle+(-1)^k|x_1\rangle|a_{k}\rangle)\otimes(\alpha|b^1_k\rangle||b^2_k\rangle|c_k\rangle-(-1)^k\beta|b^1_{k+1}\rangle|b^2_{k+1}\rangle|c_{k+1}\rangle).
		\end{split}
	\end{equation}
	It can be seen that the state of the photons $X$ and $A$ are still not separable from that of the channel. This separation  can be achieved by allowing a cross-Kerr interaction between an auxiliary CS $|z\rangle$ and one of the paths (say, $|x_0\rangle$ and $|a_k\rangle$) of photon $X$ and $A$ with the phase shift parameter of $\theta$ and $2\theta$ respectively. The X-quadrature measurement on the CS gives four possible outcomes $mn=00$, $01$, $10$ and $11$ corresponding to $|z\rangle$, $|ze^{i\theta}\rangle$, $|ze^{i2\theta}\rangle$ and $|ze^{i3\theta}\rangle$ respectively. This measurement leads to a collapse of the previous state to 
	\begin{equation}\label{Eq4.14:Measurement-Step2}
		|\xi_{kmn}\rangle=|x_{n\oplus1}\rangle|a_{k\oplus m\oplus1}\rangle(\alpha|b^1_k\rangle|b^2_k\rangle|c_k
		\rangle+(-1)^{k\oplus m\oplus n}\beta|b^1_{k\oplus1}\rangle|b^2_{k\oplus1}\rangle|c_{k\oplus1}\rangle)
	\end{equation}
	The spatial states of photon $X$ and $A$ are now isolated from the other photons in the channel which can now be observed in Equation \eqref{Eq4.14:Measurement-Step2}. For the forthcoming steps, we may drop the spatial state of photon $X$. The modified state can be written as  
	\begin{equation}\label{Eq4.15:Drop-Step2}
		|\Xi_{kmn}\rangle=|a_{k\oplus m\oplus1}\rangle(\alpha|b^1_k\rangle|b^2_k\rangle|c_k
		\rangle+(-1)^{k\oplus m\oplus n}\beta|b^1_{k\oplus1}\rangle|b^2_{k\oplus1}\rangle|c_{k\oplus1}\rangle).
	\end{equation}
	Optical quantum circuit illustrating Step~2 of the \acrshort{CJRIO} protocol is shown in Figure \ref{Fig4.1:Step1-2_CJRIO}.
\end{description}
   
\begin{figure}
	\centering
	\includegraphics[width=\textwidth]{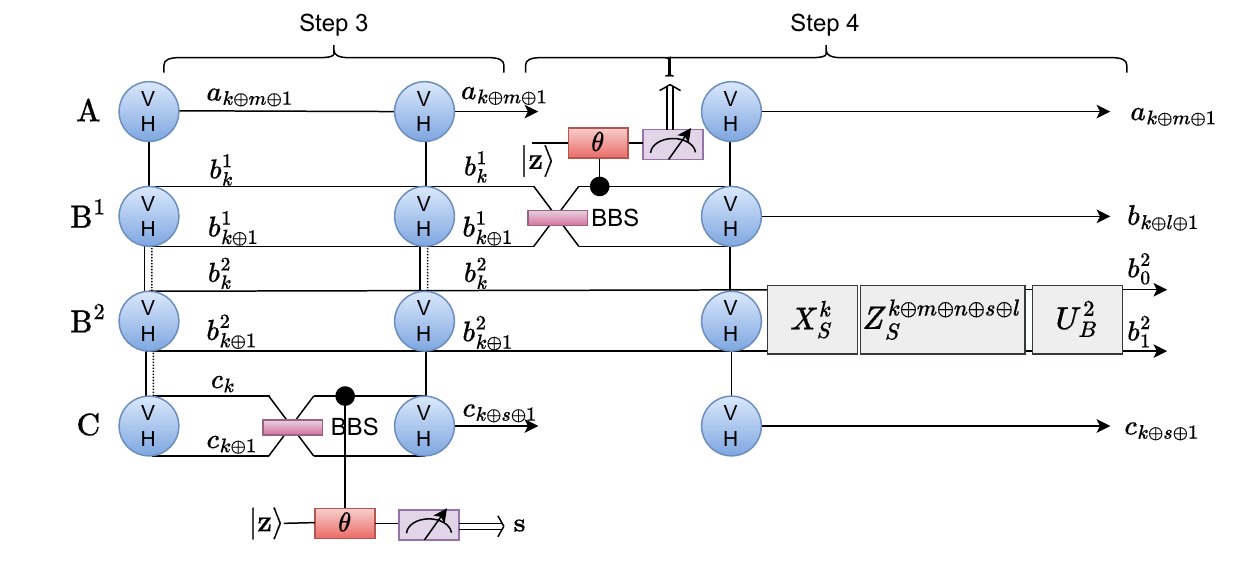}
	\caption{A schematic illustrating Step~3 and Step~4 of the CJRIO protocol. Here, Charlie as a controller, first blends paths of her photon using BBS and enables the interaction $K_{c_k}(\theta)|z\rangle|c_k\rangle$ and perform X-quadrature measures the CS, yielding the outcome $s$, which destroys the entanglement of photon $\text{C}$ from remaining photons in S-DOF. After that Bob$^1$ performs a similar operation by mixing his photon paths on a BBS and enabling the interaction $K_{b^1_k}(\theta)|z\rangle|b^1_k\rangle$ and measures the CS, yielding the outcome $l$, which allows Bob$^2$ to apply appropriate unitary operations to recover the state $\alpha_{B^2}|b^2_0\rangle+\beta_{B^2}|b^2_1\rangle$.}
	\label{Fig4.2:Step3-4_CJRIO}
\end{figure}

\begin{description}
	\item[Step~3] Charlie here will now act as a controller, if she allows the joint parties for the joint operation then she combines the  paths ($|c_k\rangle$ and $|c_{k\oplus1}\rangle$) of her photon $C$ using BBS otherwise she does nothing. Without the blending of paths, it would not be possible for the joint parties to operate their operations correctly. The combining of the photon paths transforms the state in Equation \eqref{Eq4.15:Drop-Step2} to
	\begin{equation}\label{Eq4.16:Mix-Step3}
		\begin{split}
			|\Xi^{'}_{kmn}\rangle &= |a_{k\oplus m\oplus1}\rangle[(\alpha|b^1_k\rangle|b^2_k\rangle+(-1)^{m\oplus n}\beta|b^1_{k\oplus1}\rangle|b^2_{k\oplus1}\rangle)|c_k\rangle\\
			&+(-1)^k(\alpha|b^1_k\rangle|b^2_k\rangle-(-1)^{m\oplus n}\beta|b^1_{k\oplus1}\rangle|b^2_{k\oplus1}\rangle)|c_{k\oplus1}\rangle]  
		\end{split}
	\end{equation}
	She then chooses one of the two paths of her photon $C$ (say, $|c_k\rangle$) and allow the cross-Kerr interaction with an auxiliary CS having phase shift of $\theta$. The X-quadrature measurement on the CS gives outcome as $s=0$ $(1)$ corresponding to $|z\rangle$ $(|ze^{i\theta}\rangle)$. The transformation of state described in Equation \eqref{Eq4.16:Mix-Step3} after this step can be written as follows:  
	\begin{equation}\label{Eq4.17:Measurement-Step3}
		|\Xi_{kmns}\rangle=|a_{k\oplus m\oplus1}\rangle(\alpha|b^1_{k}\rangle|b^2_k\rangle-(-1)^{m\oplus n\oplus s}\beta|b^1_{k\oplus1}\rangle|b^2_{k\oplus1}\rangle)|c_{k\oplus s\oplus1}\rangle.
	\end{equation}
	This interaction leads to separation of the photon $C$ in \acrshort{S-DOF} with the remaining photons in the channel. This gives power to Charlie, who can now allow the joint parties Bob$^1$ and Bob$^2$ to perform their operations jointly. Optical quantum circuit illustrating Step~3 of the \acrshort{CJRIO} protocol is shown in Figure \ref{Fig4.2:Step3-4_CJRIO}.
	\item[Step~4] The joint parties can decide among themselves about who will first apply the operation. Here, let us consider the situation where Bob$^2$ first executes his operation. Bob$^2$ is allowed to operate $U^{2}_B$ only if his collaborator Bob$^1$ combines his photon paths $|b^1_{k}\rangle$ and $|b^1_{k\oplus1}\rangle$ and turn on the cross-Kerr interaction between one of his photon path and an auxiliary CS $|z\rangle$ with phase shift of $\theta$. The measurement on the CS gives $l=0$ $(1)$ as an outcome corresponding to $|z\rangle$ $(|ze^{i\theta}\rangle)$. This transforms the previous state in Equation \eqref{Eq4.17:Measurement-Step3} to
	\begin{equation}\label{Eq4.18:Measurement-Step3} 
		|\Xi_{kmnsl}\rangle=|a_{k\oplus m\oplus1}\rangle|b^1_{k\oplus l\oplus1}\rangle(\alpha|b^2_k\rangle+(-1)^{k\oplus m\oplus n\oplus s\oplus l}\beta|b^2_{k\oplus1}\rangle)|c_{k\oplus s\oplus1}\rangle.
	\end{equation}
	Equation \eqref{Eq4.18:Measurement-Step3} clearly depicts that the coefficient $\alpha$ and $\beta$ reflects at the Bob$^2$ end which was previously at Alice's end. An appropriate unitary operation $Z^{k\oplus m\oplus n\oplus s\oplus l}_{S}X^{k}_{S}$, where $X_S=|b^{2}_{0}\rangle\langle b^{2}_{1}|+|b^{2}_{1}\rangle\langle b^{2}_{0}|$ and $Z_S=|b^{2}_{0}\rangle\langle b^{2}_{0}|-|b^{2}_{1}\rangle\langle b^{2}_{1}|$, applied by Bob$^2$ produces the state $\alpha|b^2_0\rangle+\beta|b^2_1\rangle$. The operation of $U_B^2$ by Bob$^2$ transforms into the following state:
	\begin{equation}\label{Eq4.19:Recover-Step4}
		|\Lambda_{kmsl}\rangle=|a_{k\oplus m\oplus1}\rangle|b^1_{k\oplus l\oplus1}\rangle(\alpha_{B^2}|b^2_0\rangle+\beta_{B^2}|b^2_1)|c_{k\oplus s\oplus1}\rangle.
	\end{equation}
	Optical quantum circuit illustrating Step $4$ of the \acrshort{CJRIO} protocol is shown in Figure \ref{Fig4.2:Step3-4_CJRIO}.
\end{description}

\begin{figure}
	\centering
	\includegraphics[width=\textwidth]{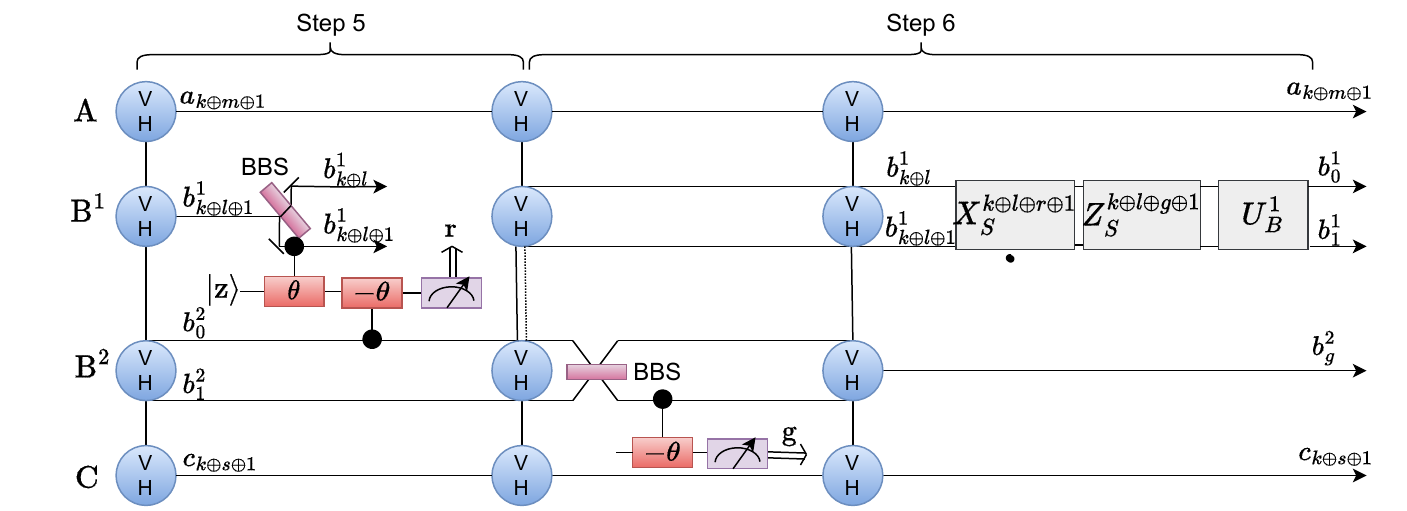}
	\caption{A schematic illustrating Step~5 and Step~6 of the CJRIO protocol is provided. Here, Bob$^1$ first initiates a new photon path using a BBS and enables the non-linear interaction $K_{b^1_{k\oplus l\oplus1}}(\theta)|z\rangle|b^1_{k\oplus l\oplus1}\rangle$ and forwards it to Bob$^2$ which enables the interaction $K_{b^2_0}(-\theta)|z\rangle|b^2_0\rangle$ and measures the CS, yielding the outcome $r$. Bob$^2$ then blends the two spatial paths of his photon and enables the cross-kerr interaction between one of his photon path and a CS followed by measuring the CS with outcome $g$ as depicted in the figure, that allows Bob$^1$ to implement a suitable unitary operation to achieve the state $\alpha_{B^1B^2}|b^1_0\rangle+\beta_{B^1B^2}|b^1_1\rangle$.}
	\label{Fig4.3:Step5-6_CJRIO}
\end{figure}

\begin{description}
	\item[Step~5] The spatial superposition of paths of photon B$^1$ that were diminished in the previous step can be retrieved using a BBS. Placing a BBS on path $|b^{1}_{k\oplus l\oplus1}\rangle$ generates a new path $|b^{1}_{k\oplus l}\rangle$. Bob$^1$ chooses one of the two paths (say, $|b^{1}_{k\oplus l\oplus1}\rangle$), and turn on the cross-Kerr interaction between the path $|b^{1}_{k\oplus l\oplus1}\rangle$ and an auxiliary CS $|z\rangle$ with shifting parameter $\theta$. The CS is then sent to Bob$^2$, who turns on the cross-Kerr interaction between the CS and his photon path $|b^{2}_{0}\rangle$ with phase shift of $-\theta$. Measurement on the CS gives an outcome $r=0$ $(1)$ corresponding to $|z\rangle$ $(|ze^{\pm i\theta})$. This process transforms the previous state in Equation \eqref{Eq4.19:Recover-Step4} to $|\Lambda_{kmslr}\rangle$, which can be written as
	\begin{equation}\label{Eq4.20:Measurement-Step5}
		|\Lambda_{kmslr}\rangle=|a_{k\oplus m\oplus1}\rangle(\alpha_{B^2}|b^1_{k\oplus l\oplus r\oplus 1}\rangle|b^2_0\rangle+(-1)^{k\oplus l\oplus 1}\beta_{B^2}|b^1_{k\oplus l\oplus r}\rangle|b^2_1)|c_{k\oplus s\oplus1}\rangle.
	\end{equation}
	Optical quantum circuit illustrating Step~5 of the \acrshort{CJRIO} protocol is shown in Figure \ref{Fig4.3:Step5-6_CJRIO}.
	\item[Step~6] For the implementation of the unitary operation, the cooperation between both Bob$^1$ and Bob$^2$ is required. In Step $4$, Bob$^1$ supports Bob$^2$ so that he can implement his operation $U^{2}_B$. This time, Bob$^2$ cooperates with Bob$^1$ by combining his photon paths $|b^{2}_{0}\rangle$ and $|b^{2}_{1}\rangle$ on a BBS. The cross-Kerr interaction is then allowed between $|b^{2}_{1}\rangle$ and an auxiliary CS $|z\rangle$ with phase shift of $\theta$. Measurement on the CS gives outcome $g=0$ $(1)$ corresponding to $|z\rangle$ $(|ze^{i\theta}\rangle)$. This process transforms the previous state in Equation \eqref{Eq4.20:Measurement-Step5} to $|\Lambda_{kmslrg}\rangle$, which can be written as
	\begin{equation}\label{Eq4.21:Measurement-Step6}
		|\Lambda_{kmslrg}\rangle=|a_{k\oplus m\oplus1}\rangle(\alpha_{B^2}|b^1_{k\oplus l\oplus r\oplus 1}\rangle+(-1)^{k\oplus l\oplus g\oplus 1}\beta_{B^2}|b^1_{k\oplus l\oplus r}\rangle)|b^2_g\rangle|c_{k\oplus s\oplus1}\rangle.
	\end{equation}
	It is clearly seen here that the coefficient $\alpha_{B^2}$ and $\beta_{B^2}$ reflects at $\text{B}^1$ end which was previously at the end of $\text{B}^2$. An appropriate unitary $Z^{k\oplus l\oplus g\oplus1}_{S}X^{k\oplus l\oplus r\oplus1}_{S}$, applied by Bob$^1$ on his qubit will produce the state $\alpha|b^1_0\rangle+\beta|b^1_1\rangle$. This process transforms the previous state in Equation \eqref{Eq4.21:Measurement-Step6} to
	\begin{equation}\label{Eq4.22:Recover-Step6}
		|\Lambda^{'}_{kmslrg}\rangle=|a_{k\oplus m\oplus1}\rangle(\alpha_{B^2}|b^1_{0}\rangle+\beta_{B^2}|b^1_{1}\rangle)|b^2_g\rangle|c_{k\oplus s\oplus1}\rangle.
	\end{equation}
	The intended operation $U^1_B$ can now be easily applied by Bob$^1$ on his photon $\text{B}^1$, which transforms the previous state to $|\Lambda^{''}_{kmslrg}\rangle$ as follows:
	\begin{equation}\label{Eq4.23:Recover-Step6}
		|\Lambda^{''}_{kmslrg}\rangle=|a_{k\oplus m\oplus1}\rangle(\alpha_{B^1B^2}|b^1_0\rangle+\beta_{B^1B^2}|b^1_1\rangle)|b^2_g\rangle|c_{k\oplus s\oplus1}\rangle.
	\end{equation}
	It is to be observed that $\alpha_{B^1B^2}|b^1_0\rangle+\beta_{B^1B^2}|b^1_1\rangle=U^{1}_{B}U^{2}_{B}(\alpha|b^{1}_{0}+\beta|b^{1}_{1})$. Still we haven't accomplished the \acrshort{CJRIO} task. To accomplish this task, all the parties involved in the \acrshort{CJRIO} will utilize \acrshort{P-DOF} of the quantum channel which has been explained in the next section.\\
	Optical quantum circuit illustrating Step~6 of the \acrshort{CJRIO} protocol is shown in Figure \ref{Fig4.3:Step5-6_CJRIO}.
\end{description}

\subsection{Utilizing P-DOF \label{Sec4.3.2:P-DOF_CJRIO}} 
\begin{figure}
	\centering
	\includegraphics[width=\textwidth]{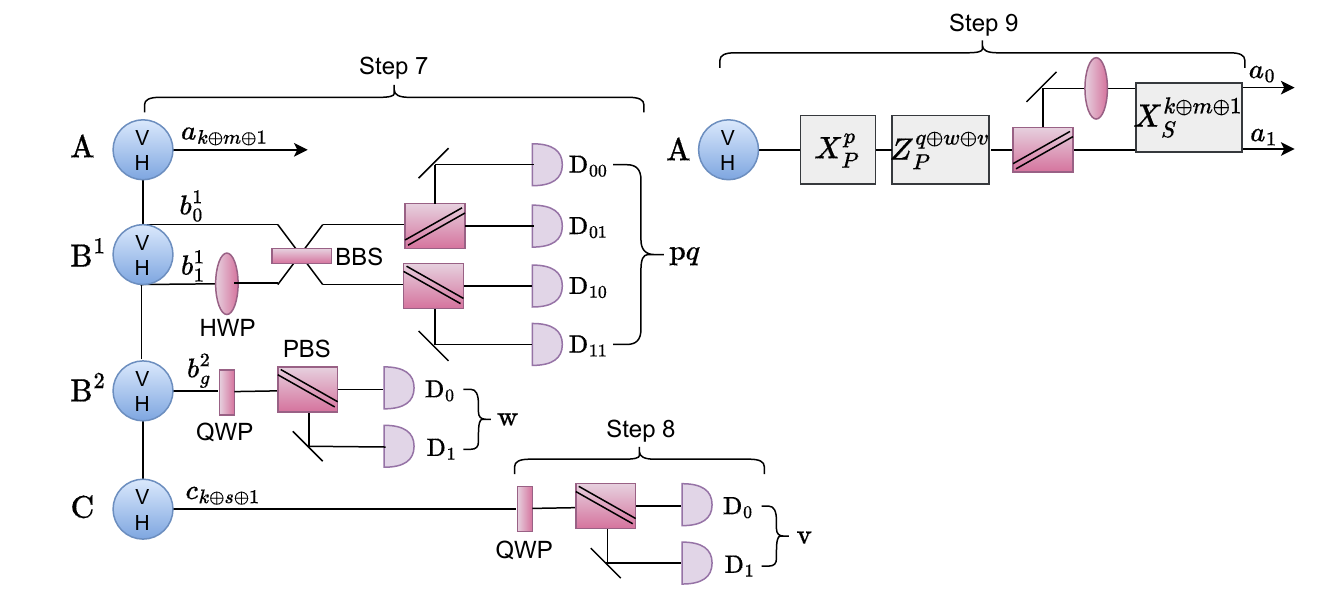}
	\caption{A schematic illustrating Step~7 to Step~9 of the CJRIO protocol is presented. Here, the joint parties Bob$^1$ and Bob$^2$ and the controller Charlie measure their respective photons in suitable basis. The measurement leads the collapse of photons $\text{B}^1$, $\text{B}^2$, and $\text{C}$ and we are only left with photon $\text{A}$. Alice then applies the suitable unitary operation followed by PBS and HWP along one of the paths to obtain the desired state $\alpha_{B^1B^2}|a_{0}\rangle+\beta_{B^1B^2}|a_{1}\rangle$.}
	\label{Fig4.4:Step7-9_CJRIO}
\end{figure}

The overall state when uses \acrshort{P-DOF} of the quantum channel used can be written as
\begin{equation}\label{eq:22}
	|\phi^{P}\rangle=|\Lambda^{''}_{kmslrg}\rangle|Q^{P}\rangle
\end{equation}
which is rewritten in its expanded form as
\begin{equation}
	\begin{split}
		|\phi^{P}\rangle &=|a_{k\oplus m\oplus1}\rangle[\alpha_{B^1B^2}|\text{H}\rangle_{A}|\text{H},b^1_0\rangle_{B^1}|\text{H},b^2_g\rangle_{B^2}|\text{H}\rangle_{C}+\alpha_{B^1B^2}|\text{V}\rangle_{A}|\text{V},b^1_0\rangle_{B^1}|\text{V},b^2_g\rangle_{B^2}|\text{V}\rangle_C\\
		&+\beta_{B^1B^2}|\text{H}\rangle_{A}|\text{H},b^1_1\rangle_{B^1}|\text{H},b^2_g\rangle_{B^2}|\text{H}\rangle_{C}+\beta_{B^1B^2}|\text{V}\rangle_{A}|\text{V},b^1_1\rangle_{B^1}|\text{V},b^2_g\rangle_{B^2}|\text{V}\rangle_C]|c_{k\oplus s\oplus1}\rangle.
	\end{split}
\end{equation}
Here, the state $|H,b^1_0\rangle$ represents the photon having horizontal polarization propagate along path $b^1_0$. The other states $|H,b^2_g\rangle$, $|V,b^1_0\rangle$, $|V,b^2_g\rangle$, $|H,b^1_1\rangle$, $|H,b^2_g\rangle$, $|V,b^1_1\rangle$ and $|V,b^2_g\rangle$ are also depicts in the similar fashion.\\
The primary objective is to jointly implement an operator on a qubit placed far apart. To accomplish the \acrshort{CJRIO} task, the coefficients $\alpha_{B^1B^2}$ and $\beta_{B^1B^2}$ must be shifted in the direction of Alice. To perform this task, the joint parties Bob$^1$ and Bob$^2$ and the controller Charlie must measure their photons in the proper bases.
\begin{description}
	\item[Step~7] A suitable measurement basis is chosen by Bob$^1$ and Bob$^2$ to measure their photons. Measurement basis is chosen by Bob$^1$ by placing a half-wave plate (HWP) on his photon path $b^1_1$ and then combine both the paths of his photon using the BBS. The HWP flips the polarization state of the photon as $|H,b^1_1\rangle\rightleftarrows|V,b^1_1\rangle$. This process transforms the previous state in Equation \eqref{Eq4.23:Recover-Step6} to
	\begin{equation}\label{Eq4.26:Basis-Step7}
		\begin{split}
			|\Omega\rangle& =|a_{k\oplus m\oplus1}\rangle[\alpha_{B^1B^2}|\text{H}\rangle_{A}|\text{H},b^1_0\rangle_{B^1}|\text{H},b^2_g\rangle_{B^2}|\text{H}\rangle_{C}+\alpha_{B^1B^2}|\text{H}\rangle_{A}|\text{H},b^1_1\rangle_{B^1}|\text{H},b^2_g\rangle_{B^2}|\text{H}\rangle_{C}\\
			&+\alpha_{B^1B^2}|\text{V}\rangle_{A}|\text{V},b^1_0\rangle_{B^1}|\text{V},b^2_g\rangle_{B^2}|\text{V}\rangle_C+\alpha_{B^1B^2}|\text{V}\rangle_{A}|\text{V},b^1_1\rangle_{B^1}|\text{V},b^2_g\rangle_{B^2}|\text{V}\rangle_C\\
			&+\beta_{B^1B^2}|\text{H}\rangle_{A}|\text{V},b^1_0\rangle_{B^1}|\text{H},b^2_g\rangle_{B^2}|\text{H}\rangle_{C}-\beta_{B^1B^2}|\text{H}\rangle_{A}|\text{V},b^1_1\rangle_{B^1}|\text{H},b^2_g\rangle_{B^2}|\text{H}\rangle_{C}\\
			&+\beta_{B^1B^2}|\text{V}\rangle_{A}|\text{H},b^1_0\rangle_{B^1}|\text{V},b^2_g\rangle_{B^2}|\text{V}\rangle_C-\beta_{B^1B^2}|\text{V}\rangle_{A}|\text{H},b^1_1\rangle_{B^1}|\text{V},b^2_g\rangle_{B^2}|\text{V}\rangle_C]|c_{k\oplus s\oplus1}\rangle.
		\end{split}
	\end{equation}
	This time Bob$^2$ chooses the measurement basis by placing a quarter wave plate (QWP) on his photon path $|b^2_g\rangle$, which acts like a Hadamard gate and transforms the polarization states as $|H,b^2_g\rangle\rightarrow(|H,b^2_g\rangle+|V,b^2_g\rangle)$ and $|V,b^2_g\rangle\rightarrow|H,b^2_g\rangle-|V,b^2_g\rangle$. This process transforms the previous state in Equation \eqref{Eq4.26:Basis-Step7} to
	\begin{equation}\label{Eq4.27:Operation-Step7}
		\begin{split}
			|\Omega^{'}\rangle&=|a_{k\oplus m\oplus1}\rangle[|\text{H},b^1_0\rangle_{B^1}|\text{H},b^2_g\rangle_{B^2}(\alpha_{B^1B^2}|\text{H}\rangle_{A}|\text{H}\rangle_{C}+\beta_{B^1B^2}|\text{V}\rangle_{A}|\text{V}\rangle_{C})\\
			&+|\text{H},b^1_0\rangle_{B^1}|\text{V},b^2_g\rangle_{B^2}(\alpha_{B^1B^2}|\text{H}\rangle_{A}|\text{H}\rangle_{C}-  \beta_{B^1B^2}\text{V}\rangle_{A}|\text{V}\rangle_{C})\\
			&+|\text{H},b^1_1\rangle_{B^1}|\text{H},b^2_g\rangle_{B^2}(\alpha_{B^1B^2}|\text{H}\rangle_{A}|\text{H}\rangle_{C}-\beta_{B^1B^2}|\text{V}\rangle_{A}|\text{V}\rangle_{C})\\
			&+|\text{H},b^1_1\rangle_{B^1}|\text{V},b^2_g\rangle_{B^2}(\alpha_{B^1B^2}|\text{H}\rangle_{A}|\text{H}\rangle_{C}+\beta_{B^1B^2}|\text{V}\rangle_{A}|\text{V}\rangle_{C})\\
			&+|\text{V},b^1_0\rangle_{B^1}|\text{H},b^2_g\rangle_{B^2}(\alpha_{B^1B^2}|\text{V}\rangle_{A}|\text{V}\rangle_{C}+\beta_{B^1B^2}|\text{H}\rangle_{A}|\text{H}\rangle_{C})\\
			&-|\text{V},b^1_0\rangle_{B^1}|\text{V},b^2_g\rangle_{B^2}(\alpha_{B^1B^2}|\text{V}\rangle_{A}|\text{V}\rangle_{C}-\beta_{B^1B^2}|\text{H}\rangle_{A}|\text{H}\rangle_{C})\\
			&+|\text{V},b^1_1\rangle_{B^1}|\text{H},b^2_g\rangle_{B^2}(\alpha_{B^1B^2}|\text{V}\rangle_{A}|\text{V}\rangle_{C}-\beta_{B^1B^2}|\text{H}\rangle_{A}|\text{H}\rangle_{C})\\
			&-|\text{V},b^1_1\rangle_{B^1}|\text{V},b^2_g\rangle_{B^2}(\alpha_{B^1B^2}|\text{V}\rangle_{A}|\text{V}\rangle_{C}+\beta_{B^1B^2}|\text{H}\rangle_{A}|\text{H}\rangle_{C})]|c_{k\oplus s\oplus1}\rangle.
		\end{split}
	\end{equation}
	Once the basis is chosen by both the joint parties, then they will do measurement of their qubits. The basis chosen by Bob$^1$ and Bob$^2$ are $\{|\text{H},b^1_0\rangle,|\text{H},b^1_1\rangle,|\text{V},b^1_0\rangle,|\text{V},b^1_1\rangle\}$ and $(\{|\text{H},b^2_g\rangle,|\text{V},b^2_g\rangle\})$, respectively. The measurement outcomes of Bob$^1$ and Bob$^2$ are $pq=00$, $01$, $10$, $11$ and $w=0$, $1$, respectively. Upon measurement, the state in Equation \eqref{Eq4.27:Operation-Step7} collapses to the following state:
	\begin{equation}
		|\Omega_{kmpqw}\rangle=|a_{k\oplus m\oplus1}\rangle\left\{\begin{matrix}
			(\alpha_{B^1B^2}|\text{H}\rangle_{A}|\text{H}\rangle_{C}+\beta_{B^1B^2}|\text{V}\rangle_{A}|\text{V}\rangle_{C})|c_{k\oplus s\oplus1}\rangle & \text{for} & pqw= 000,011\\ 
			(\alpha_{B^1B^2}|\text{H}\rangle_{A}|\text{H}\rangle_{C}-\beta_{B^1B^2}|\text{V}\rangle_{A}|\text{V}\rangle_{C})|c_{k\oplus s\oplus1}\rangle & \text{for} & pqw= 010,001\\ 
			(\alpha_{B^1B^2}|\text{V}\rangle_{A}|\text{V}\rangle_{C}+\beta_{B^1B^2}|\text{H}\rangle_{A}|\text{H}\rangle_{C})|c_{k\oplus s\oplus1}\rangle & \text{for} & pqw= 100,111\\ 
			(\alpha_{B^1B^2}|\text{V}\rangle_{A}|\text{V}\rangle_{C}-\beta_{B^1B^2}|\text{H}\rangle_{A}|\text{H}\rangle_{C})|c_{k\oplus s\oplus1}\rangle & \text{for} & pqw= 110,101
		\end{matrix}\right.
	\end{equation}
	Optical quantum circuit illustrating Step~7 of the \acrshort{CJRIO} protocol is shown in Figure \ref{Fig4.4:Step7-9_CJRIO}.
	\item[Step~8] Charlie, who acts a controller will now use her controlling power. She puts a QWP on path $c_{k\oplus s\oplus1}$ of her photon and then put a PBS whose outputs are measured with detector $D_0$ and $D_1$. The measurement outcomes are $v=0$, $1$ corresponding to $\{|\text{H},c_{k\oplus s\oplus1}\rangle, |\text{V},c_{k\oplus s\oplus1}\rangle\}$.
	\begin{equation}
		|\Omega_{kmpqwv}\rangle=|a_{k\oplus m\oplus1}\rangle\left\{\begin{matrix}
			(\alpha_{B^1B^2}|\text{H}\rangle+\beta_{B^1B^2}|\text{V}\rangle) & \text{for} & pqwv= 0000,0101, 0011, 0110\\ 
			(\alpha_{B^1B^2}|\text{H}\rangle-\beta_{B^1B^2}|\text{V}\rangle) & \text{for} & pqwv= 0001, 0100, 0010, 0111\\ 
			(\alpha_{B^1B^2}|\text{V}\rangle+\beta_{B^1B^2}|\text{H}\rangle) & \text{for} & pqwv= 1000, 1101, 1011, 1110\\ 
			(\alpha_{B^1B^2}|\text{V}\rangle-\beta_{B^1B^2}|\text{H}\rangle) & \text{for} & pqwv= 1001, 1100, 1010, 1111
		\end{matrix}\right.
	\end{equation}
	The coefficients $\alpha_{B^1B^2}$ and $\beta_{B^1B^2}$ ultimately reach at Alice's port, which is actually the desired task. To get the desired result of the \acrshort{CJRIO} task, Alice finally applies the unitary $Z^{q\oplus w\oplus v}_{P}X^{p}_{P}$ where $X_P=|H\rangle\langle V|+|V\rangle\langle H|$ and $Z_P=|H\rangle\langle H|-|V\rangle\langle V|$, to get the desired state in \acrshort{P-DOF} as
	\begin{equation}\label{Eq4.30:Recover-Step8}
		|\Omega_{km}\rangle=(\alpha_{B^1B^2}|H\rangle+\beta_{B^1B^2}|V\rangle)|a_{k\oplus m\oplus1}\rangle
	\end{equation}
	Optical quantum circuit illustrating Step~8 of the \acrshort{CJRIO} protocol is shown in Figure \ref{Fig4.4:Step7-9_CJRIO}.\\
	Till now, the desired state in \acrshort{P-DOF} has been achieved, which can be seen in Equation \eqref{Eq4.30:Recover-Step8}. The obtained state is needed to covert into \acrshort{S-DOF} to accomplish the complete \acrshort{CJRIO} task, which is described in the next step.\\
	\item[Step~9] The previously obtained state can be transformed into \acrshort{S-DOF} by placing a PBS and then HWP on the photon path $a_{k\oplus m\oplus1}$. Alice get the desired state $(\alpha_{B^1B^2}|a_{0}\rangle+\beta_{B^1B^2}|a_{1}\rangle)=U_{B}^1U_{B}^2|\psi\rangle_A$ by applying a unitary $X^{k\oplus m\oplus1}_{S}$. The \acrshort{CJRIO} task has now been accomplished successfully.\\
	Optical quantum circuit illustrating Step $7$ of the \acrshort{CJRIO} protocol is shown in Figure \ref{Fig4.4:Step7-9_CJRIO}.
\end{description}

\section{A Possible Generalization for CJRIO\label{Sec4.4:Generalization-CJRIO}}
The earlier proposed scheme for \acrshort{CJRIO} is generalized to an arbitrary number of joint parties and controllers. Assume there are $M$-joint parties (say Bob$^1$, Bob$^2$,..., Bob$^M$) and $N$-controllers (say Charlie$^1$, Charlie$^2$,...,Charlie$^N$). Each joint party has its corresponding operator, which is given as
\begin{equation}
	U^i_B=\begin{pmatrix}
		u^i_B & v^i_B\\ 
		-v^{*i}_B & u^{*i}_B
	\end{pmatrix}.
\end{equation}
The generalized task is that all $M$-parties should implement their operators jointly on an unknown qubit at remote node under the presence of all $N$-controllers. Mathematical description of the generalized task is given as follows
\begin{equation}
	\begin{split}
		|\psi_{B^1B^2...B^M}\rangle&=U^{1}_BU^{2}_B...U^{M}_B|\psi\rangle_X\\
		&=(\alpha_{B^1B^2...B^M}|x_0\rangle+\beta_{B^1B^2...B^M}|x_1\rangle).
	\end{split}
\end{equation}
To accomplish the generalized \acrshort{CJRIO} task, $1+M+N$ qubit hyperentangled state is used, which is given as follow
\begin{equation}
	|Q\rangle_{AB^1B^2...B^MC^1C^2...C^N}=|Q^{S}\rangle_{AB^1B^2...B^MC^1C^2...C^N}|Q^{P}\rangle_{AB^1B^2...B^MC^1C^2...C^N}
\end{equation}
where (up to the normalization factor),
\begin{equation}
	\begin{split}
	|Q^{S}\rangle_{AB^1B^2...B^MC^1C^2...C^N}=&|a_0\rangle_A|b^{1}_0\rangle_{B^1}|b^{2}_0\rangle_{B^2}...|b^{M}_0\rangle_{B^M}|c^{1}_0\rangle_{C^1}|c^{2}_0\rangle_{C^2}...|c^{N}_0\rangle_{C^N}\\
	&+|a_1\rangle_{A}|b^{1}_1\rangle_{B^1}|b^{2}_1\rangle_{B^2}...|b^{M}_1\rangle_{B^M}|c^{1}_1\rangle_{C^1}|c^{2}_1\rangle_{C^2}...|c^{N}_1\rangle_{C^N}
	\end{split}
\end{equation}
and
\begin{equation}
	\begin{split}
	|Q^{P}\rangle_{AB^1B^2...B^MC^1C^2...C^N}=&|H\rangle_A|H\rangle_{B^1}|H\rangle_{B^2}...|H\rangle_{B^M}|H\rangle_{C^1}|H\rangle_{C^2}...|H\rangle_{C^N}\\
	&+|V\rangle_A|V\rangle_{B^1}|V\rangle_{B^2}...|V\rangle_{B^M}|V\rangle_{C^1}|V\rangle_{C^2}...|V\rangle_{C^N}.
	\end{split}
\end{equation}
The various steps involved to accomplish the generalized \acrshort{CJRIO} task are described as follows:
\begin{description}
	\item[Step~1] Alice starts the protocol by entangling her unknown quantum state $|\psi\rangle_X$ with the \acrshort{S-DOF} of aforementioned channel $|Q^{S}\rangle_{AB^1B^2...B^MC^1C^2...C^N}$ in a similar fashion as done in Step~1 of Section \ref{Sec4.3:Protocol-CJRIO}. Their entanglement is shown in mathematical form as
	\begin{equation}
		|\Phi_k\rangle=\alpha|x_0\rangle|a_k\rangle\bigotimes_{i=1}^{M}|b^{i}_k\rangle\bigotimes_{j=1}^{N}|c^{j}_{k}\rangle+\beta|x_1\rangle|a_{k\oplus1}\rangle\bigotimes_{i=1}^{M}|b^{i}_{k\oplus1}\rangle\bigotimes_{j=1}^{N}|c^{j}_{k\oplus1}\rangle
	\end{equation}
	\item[Step~2] After entanglement of her unknown quantum state in the previous state, Alice disentangles her unknown photon from the quantum channel used. This can be done in a similar fashion as done in Step~2 of Section \ref{Sec4.3:Protocol-CJRIO}. This process leads to the collapse of the previous state to
	\begin{equation}
		|\Phi_{kmn}\rangle=|a_{k\oplus m\oplus1}\rangle(\alpha\bigotimes_{i=1}^{M}|b^{i}_k\rangle\bigotimes_{j=1}^{N}|c^{j}_{k}\rangle+\beta\bigotimes_{i=1}^{M}|b^{i}_{k\oplus1}\rangle\bigotimes_{j=1}^{N}|c^{j}_{k\oplus1}\rangle)
	\end{equation}
	\item[Step~3] Here, controllers use their powers through combining both paths of their photons on a BBS and allows cross-Kerr interaction between one of their photon path and an auxiliary CS with phase shift of $\theta$. Measurement on the CS gives outcome as $s_j=0$ $(1)$ corresponding to $|z\rangle$ $(|ze^{i\theta}\rangle)$. Their measurements collapses to the following state.
	\begin{equation}
		|\Phi_{kmns}\rangle=|a_{k\oplus m\oplus1}\rangle(\alpha\bigotimes^{M}_{i=1}|b^{i}_{k}\rangle-(-1)^{m\oplus n\oplus s_{1}\oplus s_{2}\oplus...\oplus s_{N}}\beta\bigotimes^{M}_{i=1}|b^{i}_{k\oplus1}\rangle)\bigotimes^{N}_{j=1}|c_{k\oplus s_{j}\oplus1}\rangle
	\end{equation}
	\item[Step~4] The role of controllers has been explained in the previous step. The joint parties will now operate their operations $U^{i}_{B}$, where $(i=1,2,..., M)$. The joint parties determine who will carry out the operation initially. Suppose, they decide that Bob$^M$ will first apply his operation. The rest of the joint parties follow Step~4 of the Section \ref{Sec4.3:Protocol-CJRIO}, which produces measurement results $l_i=0$ $(1)$. Bob$^M$ applies a suitable unitary operation to obtain $\alpha|b^{M}_0\rangle+\beta|b^{M}_1\rangle$. The desired unitary $U^{M}_{B}$ is now applied to obtain $\alpha_{B^M}|b^{M}_0\rangle+\beta_{B^M}|b^{M}_1\rangle$.
	\item[Step~5] This stage involves the gradual movement of the coefficients $\alpha_{B^M}$ and $\beta_{B^M}$, which are with Bob$^{M}$, towards Bob$^{M-1}$, Bob$^{M-2}$, and so forth. Here, the alternative joint partners collaborate simultaneously. Assume, Bob$^{M}$ and Bob$^{M-1}$ collaborate simultaneously. In order to create a new path, Bob$^{M-1}$ puts a BBS on his photon path. One of the paths then interacts with an auxiliary CS $|z\rangle$ through a cross-Kerr nonlinear interaction with phase shift $\theta$ and sends it to Bob$^{M}$, who then permits it to interact with his one photon path and measures it, yielding the measurement result $r_M=0$ $(1)$. Bob$^{M}$ blends both paths of his photon and turns on the cross-Kerr nonlinear interaction between an auxiliary CS and one of the photon paths with phase shift parameter $\theta$. Measurement on the CS gives the outcome as $g_{M}=0$ $(1)$. Based on the measurement result, Bob$^{M-1}$ will use the proper unitary operation to obtain $\alpha_{B^M}|b^{M-1}_0\rangle+\beta_{B^M}|b^{M-1}_1\rangle$, on which $U^{M-1}_{B}$ is  operated to obtain $\alpha_{B^{M-1}B^M}|b^{M-1}_0\rangle+\beta_{B^{M-1}B^M}|b^{M-1}_1\rangle$. Now, Bob$^{M-1}$ and Bob$^{M-2}$ go through this procedure again, and so on until Bob$^{1}$ and Bob$^{2}$. The ultimate state will now become
	\begin{equation}
		|\Phi_{kmns_{j}l_{i}r_{i}g_{i}}\rangle=|a_{k\oplus m\oplus1}\rangle(\alpha_{B^1B^2...B^M}|b^{1}_{0}\rangle+\beta_{B^1B^2...B^M}|b^{1}_{1}\rangle)\bigotimes^{M}_{i=2}|b^{i}_{g_{i}}\rangle\bigotimes^{N}_{j=1}|c_{k\oplus s_{j}\oplus1}\rangle
	\end{equation}
	\item[Step~6] The joint parties contribute at this stage. A suitable measurement basis is used to measure the photons of the joint parties. Bob$^{1}$ chooses his measurement basis by placing a HWP on his photon path $b^{1}_0$ and blends both the paths on a BBS. The remaining joint parties place a QWP on their respective photon path. After they choose a suitable measurement basis, they measure their photons.
	\item[Step~7] Once again controller uses the power by placing a QWP on their respective photon paths followed by PBS to measure their photons in a suitable measurement basis. Based on their measurement results, Alice applies a respective unitary operation to obtain $(\alpha_{B^1B^2...B^M}|\text{H}\rangle+\beta_{B^1B^2...B^M}|\text{V}\rangle)|a_{k\oplus m\oplus1}\rangle$, which is same as Equation \eqref{Eq4.30:Recover-Step8}. 
	\item[Step~8] The process involved in this step is identical to Step~9 of Section \ref{Sec4.3:Protocol-CJRIO}. 
\end{description}

\section{Existing variants of RIO as a sunset of CJRIO \label{Sec4.5:Variants-CJRIO}}
The proposed \acrshort{CJRIO} scheme can be viewed as a generalized scheme, and all current \acrshort{RIO} scheme variations can be derived as special cases of the proposed scheme. For instance, if Charlie and the associated actions are taken out of our proposed scheme, then it drops to the \acrshort{JRIO} scheme mentioned in Reference \cite{BaAn2022JRIO}. Eliminating Charlie, first simplify the resources used in Equation \eqref{Eq4.6:QCused} to a hyperentangled state of three qubit. This also removes the respective steps where the role of Charlie is involved such as, Step~3 and Step~8 of Section \ref{Sec4.3:Protocol-CJRIO}. In the similar fashion, if the action of either joint party is removed from the designed \acrshort{CJRIO} scheme, then it drops to the existing scheme of \acrshort{CRIO} \cite{BaAn2022CRIO}. Let's eliminate Bob$^2$, which simplify the resources used in Equation \eqref{Eq4.6:QCused} and also remove the respective steps where the role of Bob$^2$ is involved such as, Step~4, Step~5, Step~6 and Step~7 of Section \ref{Sec4.3:Protocol-CJRIO}.\\
Regarding the task efficiency, as far as known, no completely suitable formula for calculating efficiency of a quantum scheme has yet been developed. Though, efficiency is defined as the function of utilized resources and the quantity of task completed. For the \acrshort{RIO} task, efficiency can be quantified as 
\begin{equation}
	\eta=\frac{c}{b+e}
\end{equation}
where, $c$ represents the amount of $2\times2$ unitary operations used; this can be readily extended to the situation of multi-qubit operations; however, this manuscript is not concerned with multi-qubit operations. $b$ denotes the amount of classical communication (in bits) needed to complete the task; since the announcement of classical bit occurs only after each cross-Kerr interaction, this automatically accounts for the impact of the number of cross-Kerr interactions, while $e$ stands the quantity of e-bits needed to complete the assignment. This estimation of efficiency is similar to that introduced by Cabello \cite{Cabello2000QCrypt_Efficiency} and recently used in scenario of secure quantum communication (e.g., see \cite{Anindita2012DSQC} and \cite{Chitra2017SemiQC}). For the most general scenario of \acrshort{CJRIO} having $M$-joint parties and $N$-controllers, there are total $1+M+N$ parties including Alice with $c=M$, $b=4M+2N+1$, and $e=M+N+1$. This gives efficiency as
\begin{equation}
	\eta_{CJRIO}=\frac{M}{5M+3N+2}.
\end{equation}
It is now possible to gain a better understanding of the effects of increasing the number of joint parties and controllers by taking into account the special circumstances of the basic formula described above. For instance, if the case of no controller is considered then the obtained efficiency is $\eta_{JRIO}=\frac{M}{5M+2}$. Evidently, for higher $M$ values (i.e., for $M>>\frac{2}{5}=0.4$), $\eta_{JRIO}$ tends towards the value $\frac{1}{5}$ (implying $20\%$ efficiency) which is not influenced by $M$. Likewise, if we have $N$ controllers and only one Bob (the joint operation is irrelevant), the efficiency of the \acrshort{CRIO} scheme will be given as $\eta_{CRIO}=\frac{1}{7N+3}$. This shows that for intermediate value of $N$, $\eta_{CRIO}$ is reciprocal to $N$ with its highest value $\frac{1}{10}$ for the case $N=1$. Additionally, it is evident that $\eta_{CRIO}$ will disappear asymptotically (for higher $N$ value). The similar situation (i.e., the disappearing of efficiency) arises for the deterministic value of $M<<N$. However, such situations involving a huge number of people are impractical. In real-world scenarios, the efficiency of the designed scheme will be calculated using $\eta_{CJRIO}=\frac{M}{5M+3N+2}$. Lastly, it is important to note that this strategies continue to be successful with unit probability regardless of efficiency.

\section{The RIHO and RIPUO task\label{Sec4.6:Task-RIHO_RIPUO}}
Alice and Bob are two communicating parties in the both \acrshort{RIHO} and \acrshort{RIPUO} schemes, who are geographically separated from each other. Bob possesses an unknown unitary operator $U_B$ that can be represented differently depending on the situation (task). The unitary $U_B$ can be seen as a "lump operator" in one scenario, as mentioned in the Reference \cite{BaAn2007RIHO}. The mathematical form of unitary $U_B$ is given as
\begin{equation}\label{eq:U}
	U_B=\frac{1}{\sqrt{2}}\begin{pmatrix}
		u & v\\ 
		-v^{*} & u^{*}
	\end{pmatrix}
	\, = \frac{1}{\sqrt{2}}(\, U_0 \, + \, U_1),
\end{equation}
where,
\begin{equation}\label{eq:U_i}
	U_0=\begin{pmatrix}
		u & 0\\ 
		0 & u^{*}
	\end{pmatrix}
	\quad \& \quad 
	U_1=\begin{pmatrix}
		0 & v\\ 
		-v^{*} & 0
	\end{pmatrix}.
\end{equation}
Here, the operator $U_0$ commutes with the Pauli-$\sigma_z$ operator, while the operator $U_1$ anti-commutes with the Pauli-$\sigma_z$ operator. The operators $U_0$ and $U_1$ can be characterized as diagonal and anti-diagonal operators, respectively \cite{Wang06RIPUO}. When $U_B$ is assumed to be unimodular (i.e., satisfying $|u|^2+|v|^2=2$), the set of all such operators $U_B$ forms the $SU(2)$ group. Furthermore, any qubit rotation can be described by a matrix of this form. Consequently, the lump operator $U_B$ can be interpreted as a rotation and expressed as $U_B=\frac{1}{\sqrt{2}}(U_0+U_1)$. Here, the sub-operators $U_0$ and $U_1$ represent the restricted types of rotations, and imposing the conditions $|u|=|v|=1$ ensures the unitary of $U_0$ and $U_1$. This makes them convenient to visualize as physically realizable operators. To illustrate this further, consider the following examples. Recall that $R_{y}(\theta)$ and $R_{z}(\theta)$ represent single-qubit quantum gates that perform rotations of a qubit by an angle  $\theta$ around the y-axis and z-axis, respectively. These gates are expressed in matrix form as:
\begin{equation}\label{eq:ry}
	R_{y}(\theta)=\begin{pmatrix}
		\cos(\frac{\theta}{2}) & -\sin(\frac{\theta}{2})\\ 
		\sin(\frac{\theta}{2}) & \cos(\frac{\theta}{2})
	\end{pmatrix}
	\quad \& \quad 
	R_{z}(\theta)=\begin{pmatrix}
		\exp{(-\iota\frac{\theta}{2})} & 0\\ 
		0 & \exp{(\iota\frac{\theta}{2})}
	\end{pmatrix}.
\end{equation}
The operators $R_{z}(\theta)$ and $R_{y}(\theta=0)$ clearly belong to the category of $U_0$, while $R_{y}(\theta=\pi)$ falls under the category of $U_1$. These basic examples provide a clear physical visualization of the tasks intended for remote execution. Furthermore, by elaborating on these sub-operators, $U_0$ can be interpreted as arbitrary rotations around the z-axis, and $U_1$ as rotations by $\pi$ about any axis in the equatorial plane \cite{Huelga2002QRC}.\\
The first task involves Bob remotely applying $U_B$ to a single-qubit state $|\psi\rangle_X$ belonging to Alice. However, Bob cannot apply $U_m$ $(m=0,1)$ individually. By the end of the process, Alice's state $|\psi\rangle_X$ is transformed into either $U_0|\psi\rangle_X$ or $U_1|\psi\rangle_X$. This procedure is referred to as the remote implementation of the hidden operator because the specific sub-operator ($U_0$ or $U_1$) actually applied to Alice's state remains hidden within the lump operator $U_B$, which is available only to Bob.\\
In the second task, the sub-operator $U_0$ or $U_1$ can be viewed as a partially unknown operator as Bob is aware of the structure of the matrices or the location of non-zero matrix elements, but not the values of the matrix elements. In this case, Bob applies $U_m$ $(m=0,1)$ to Alice's state $|\psi\rangle_X$, which is transformed into $U_m|\psi\rangle_X$ by the end of the process.\\
The arbitrary quantum state held by Alice can be expressed as follows:
\begin{equation}\label{eq:Qs}
	|\psi\rangle_X=(\alpha|x_0\rangle+\beta|x_1\rangle)_X,
\end{equation}
where, $\alpha$ and $\beta$ are unknown coefficients that satisfy the normalization condition $|\alpha|^2+|\beta|^2=1$.\\
Based on the unitary visualization for the two scenarios, two distinct tasks of \acrshort{RIO} have been described. In the first task, Bob applies the unitary operator $U_m$ $(m=0,1)$ to Alice's unknown quantum state $|\psi\rangle_X$ using operator $U_B$. In the second task, Bob directly applies $U_m$ $(m=0,1)$ to Alice's quantum state $|\psi\rangle_X$ when he possesses the operator $U_m$. To accomplish both tasks, Alice and Bob share a two-qubit maximally entangled state, entangled in \acrshort{S-DOF} expressed as follows:
\begin{equation}\label{Eq4.46:Resource_RIHO}
	|\Omega^{+}\rangle_{AB}=|a_0\rangle_A |b_0\rangle_{B} + |a_1\rangle_A |b_1\rangle_{B}.
\end{equation}
In Equation \eqref{Eq4.46:Resource_RIHO}, the first qubit is with Alice, and the second qubit is with Bob. For simplicity, the normalization factor $1/\sqrt{2}$ has been dropped in Equation \eqref{Eq4.46:Resource_RIHO} and in the subsequent equations.\\
The procedures for accomplishing both tasks are outlined in Section \ref{Sec4.7:Protocols-RIHO_RIPUO}.

\section{Protocol for RIHO and RIPUO\label{Sec4.7:Protocols-RIHO_RIPUO}}
The unknown quantum state and the quantum channel utilized to accomplish both tasks can be jointly written as
\begin{equation}\label{Eq4.47:Joint-RIHO}
	|\psi\rangle_X|\Omega^{+}\rangle_{AB}=(\alpha|x_0\rangle+\beta|x_1\rangle)_X 
	\otimes(|a_0\rangle_A |b_0\rangle_{B} + |a_1\rangle_A |b_1\rangle_{B}).
\end{equation}
The two scenarios involving Bob’s unitary operators are: (i) Bob has control over $U_B$ but not $U_m$ $(m=0,1)$, meaning $U_m$ is hidden from him. (ii) Bob knows the structure of $U_m$ but does not know its exact values. In both cases, Bob is required to apply $U_m$ to Alice’s unknown quantum state $|\psi\rangle_X$. The protocols for scenario (i) and (ii) are outlined step by step in Section \ref{Sec4.7.1:Protocol-RIHO} and Section \ref{Sec4.7.2:Protocol-RIPUO}, respectively.
\subsection{Protocol for RIHO\label{Sec4.7.1:Protocol-RIHO}} 
\begin{figure}
	\centering
	\includegraphics[width=\textwidth]{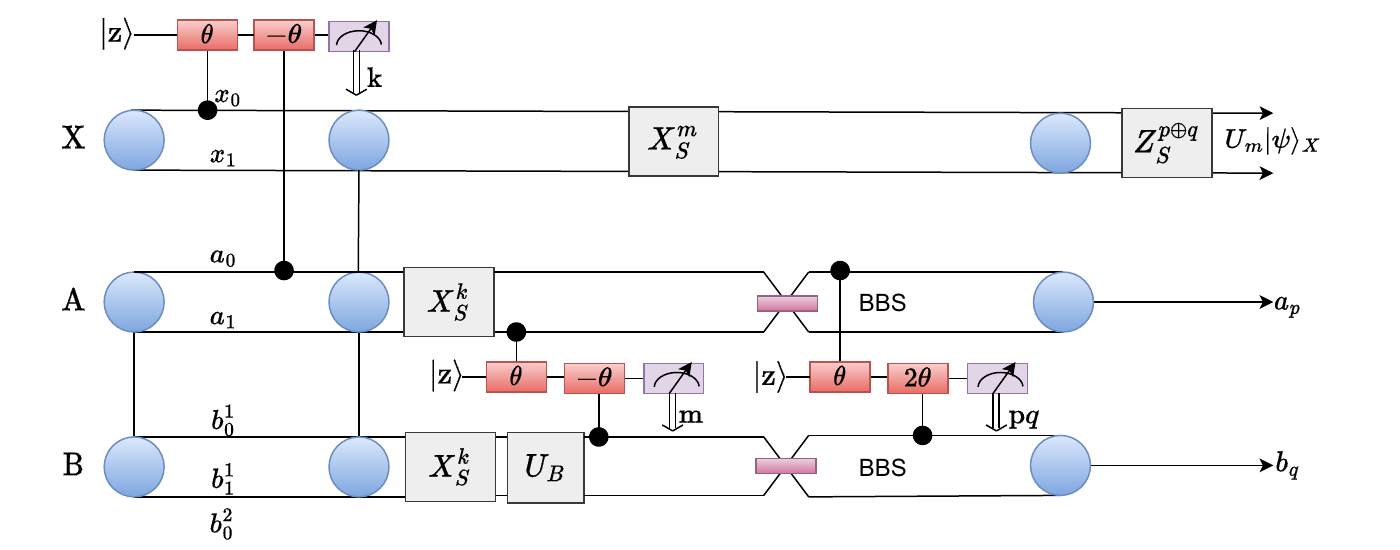}
	\caption{This figure illustrates the steps involved in the protocol for remote implementation of hidden operators. An unpolarized photon is represented by a circle.} 
	\label{Fig4.5:Fig01_RIHO}
\end{figure}

\begin{description}
	\item[Step~1] Firstly, Alice entangles her unknown state $|\psi\rangle_X$ with the quantum channel $|\Omega^{+}\rangle_{AB}$. To achieve this entanglement, she performs a cross-Kerr interaction between an auxiliary CS $|z\rangle$ and one path (say, $x_0$) of her photon $X$, followed by another cross-Kerr interaction between the $|z\rangle$ and path $a_0$ of her photon $A$. After the first cross-Kerr interaction between path $x_0$ and $|z\rangle$ with a phase shift parameter $\theta$, the joint state described in Equation \eqref{Eq4.47:Joint-RIHO} changes into $\alpha|x_0\rangle |a_0\rangle_A |b_0\rangle_{B}|z e^{i\theta}\rangle  +\beta|x_1\rangle |a_0\rangle_A |b_0\rangle_{B}|z\rangle+
	\alpha|x_0\rangle |a_1\rangle_A |b_1\rangle_{B} |z e^{i\theta}\rangle+\beta|x_1\rangle |a_1\rangle_A |b_1\rangle_{B} |z\rangle$. Due to the second cross-Kerr interaction between the path $a_0$ and $|z\rangle$ with phase shift parameter $-\theta$, the joint state described in Equation \eqref{Eq4.47:Joint-RIHO} changes into $\alpha|x_0\rangle_X |a_0\rangle_A |b_0\rangle_{B} + \beta|x_1\rangle_X |a_1\rangle_A |b_1\rangle_{B})|z\rangle+ \alpha|x_0\rangle_X |a_1\rangle_A |b_1\rangle_{B}|ze^{i\theta}\rangle + \beta|x_1\rangle_X |a_0\rangle_A |b_0\rangle_{B}|ze^{-i\theta}\rangle$. To obtain the desired entangled state, Alice performs a $\hat{\mathcal{X}}=a^\dagger+a$ quadrature homodyne measurement on the CS $|z\rangle$. Without loss of generality, $|z\rangle$ can be assumed to be real. This measurement is realized by projecting the coherent state $|z\rangle$ onto $|\mathcal{X}\rangle\langle\mathcal{X}|$, leading to
	\begin{equation}
		\begin{split}
			|\xi_h\rangle= & f(x,z)(\alpha|x_0\rangle_X |a_0\rangle_A |b_0\rangle_{B} + \beta|x_1\rangle_X |a_1\rangle_A |b_1\rangle_{B})\\
			& + f(x,z\cos{\theta})(\alpha|x_0\rangle_X |a_1\rangle_A |b_1\rangle_{B}e^{i \phi(x)} + \beta|x_1\rangle_X |a_0\rangle_A |b_0\rangle_{B}e^{-i \phi(x)}),
		\end{split}
	\end{equation}
	where, $f(x,z)=(1/2\pi)^{1/4}\exp[{-\frac{1}{4}(x-2z)^2}]$, $f(x,z\cos{\theta})=(1/2\pi)^{1/4}\exp[{-\frac{1}{4}(x-2z\cos{\theta})^2}]$ and $\phi(x)=(x-2z\cos{\theta})z\sin{\theta}$. It is important to note that the states $|ze^{ i\theta}\rangle$ and $|ze^{- i\theta}\rangle$ share the same Gaussian curve, $f(x,z\cos{\theta})$, and thus cannot be distinguished using homodyne measurement. In contrast, the states $|z\rangle$ and $|z e^{\pm i \theta}\rangle$ correspond to two distinct Gaussian curves. These states can be distinguished provided that the separation between their peaks is sufficiently large, satisfying the condition $(\mathcal{X}_d\sim z\theta^2\gg 1)$. The separation between the peaks and their midpoints is  given by $\mathcal{X}_d=2z(1-\cos{\theta})$ and $\mathcal{X}_0=z(1+\cos{\theta})$, respectively. If the measurement outcome, $\mathcal{X}$, satisfies $\mathcal{X}>\mathcal{X}_0$, then she obtains
	\begin{align}
		|\xi^{'} \rangle= \alpha|x_0\rangle_X |a_0\rangle_A |b_0\rangle_{B} + \beta|x_1\rangle_X |a_1\rangle_A |b_1\rangle_{B}.
		\label{eq:8}
	\end{align}
	Conversely, if the measurement outcome satisfies $\mathcal{X}<\mathcal{X}_0$, she obtains
	\begin{align}
		|\xi^{''} \rangle=\alpha|x_0\rangle_X |a_1\rangle_A |b_1\rangle_{B}e^{i \phi(x)} + \beta|x_1\rangle_X |a_0\rangle_A |b_0\rangle_{B}e^{-i \phi(x)}.
		\label{eq:9}
	\end{align}
	Since portions of the Gaussian curves $f(x,z)$ and $f(x,z\cos{\theta})$ overlap, there is a possibility of misidentifying the states $|\xi^{'} \rangle$ and $|\xi^{''} \rangle$. The likelihood of such misidentification is referred to as the error probability, which is given by
	\begin{align}
		P_{\rm{error}}= \frac{1}{2} \text{erfc}[\mathcal{X}_d/2\sqrt{2}],
		\label{eq:10}
	\end{align}
	where, $\text{erfc}$ denotes the complementary error function. In this case, the error probability is given by $P^1_{\rm{error}}= \frac{1}{2} \text{erfc}[z(1-\cos{\theta})/\sqrt{2}]$. If the state $|\xi^{''} \rangle$ is obtained, both Alice and Bob perform a classical feed-forward operation \cite{Munro_2005, Nemoto2004ckInteraction} to correct the phase shift $\phi(x)$ and apply the path-flip operator $X_S$, where $(X_S=|a_0\rangle\langle a_1|+|a_1\rangle\langle a_0|)$, to their respective photons $\text{A}$ and $\text{B}$. On the other hand, if $|\xi^{'} \rangle$ is obtained, no corrective action is required. After the measurement, the resulting joint state becomes entangled, as described by the following expression
	\begin{equation}
		|\xi^{'''}\rangle=\alpha|x_0\rangle_X |a_0\rangle_A |b_0\rangle_{B} + \beta|x_1\rangle_X |a_1\rangle_A |b_1\rangle_{B}.
		\label{Eq4.52:Measurement_Step1_RIHO}
	\end{equation}
	\item[Step~2] Bob applies an operator $U_B$ to his qubit $\text{B}$. Upon applying this operator, the state in Equation \eqref{Eq4.52:Measurement_Step1_RIHO} transforms into the state that can be written as 
	\begin{equation}
		|\xi^{''''}\rangle =(I\otimes I\otimes U_B)|\xi^{'''}\rangle=\frac{1}{\sqrt{2}}[\alpha |x_0\rangle_X |a_0\rangle_A (u|b_0\rangle-v^{*}|b_1\rangle)_{B} + \beta|x_1\rangle_X |a_1\rangle_A (v|b_0\rangle+u^{*}|b_1\rangle)_{B}].
		\label{Eq4.53:Operate_Stpe2_RIHO}
	\end{equation}
	where, $I$ represents $2\times 2$ identity matrix.
	\item[Step~3] Alice introduces an auxiliary CS denoted as $|z\rangle$, which she allows to interact with path $a_0$ of her photon $\text{A}$ through a cross-Kerr interaction with a phase shift $ \theta$. As a result, the state initially described by Equation \eqref{Eq4.53:Operate_Stpe2_RIHO} evolves into $\frac{1}{\sqrt{2}}[\alpha |x_0\rangle_X |a_0\rangle_A (u|b_0\rangle-v^{*}|b_1\rangle)_{B}|z e^{i\theta}\rangle + \beta|x_1\rangle_X |a_1\rangle_A (v|b_0\rangle+u^{*}|b_1\rangle)_{B}|z\rangle]$. Subsequently, Bob employs the same auxiliary CS $|z\rangle$, allowing it to interact with the path $b_0$ of his photon $\text{B}$ via the cross-Kerr interaction with a phase shift of $-\theta$. Following this interaction, the state evolves further as follows:
	\begin{align}
		|\xi^{'''''}\rangle=&\frac{1}{\sqrt{2}}[(\alpha u|x_0\rangle_X |a_0\rangle_A |b_0\rangle_{B} + \beta u^{*}|x_1\rangle_X |a_1\rangle_A |b_1\rangle_{B})|z\rangle+(-\alpha v^{*}|x_0\rangle_X |a_0\rangle_A |b_1\rangle_{B}|ze^{ i\theta}\rangle \nonumber \\
		&+ \beta v |x_1\rangle_X |a_1\rangle_A |b_0\rangle_{B}|ze^{- i\theta}\rangle)].
		\label{eq:13}
	\end{align}
	Next, Bob performs a $\hat{\mathcal{X}}$-quadrature measurement on the CS. As previously discussed, the states $|ze^{\pm i\theta}\rangle$ cannot be perfectly differentiated, however $|z\rangle$ and $|ze^{\pm i\theta}\rangle$ can be differentiated in the measurement with an error probability given by $ P^2_{\rm{error}}= \frac{1}{2} \text{erfc}[z(1-\cos{\theta})/\sqrt{2}]$. If the measurement outcome is $m = 0$, corresponding to the state $|z\rangle$, Bob takes no further action. Conversely, if the outcome is $m = 1$, corresponding to the state $|ze^{\pm i\theta}\rangle$, he applies a classical feed-forward operation to compensate for the phase shift $\pm\phi(x)$. He also applies the path flip operator $X_S$ to photon $\text{X}$. After these operations, the two states corresponding to the distinct outcomes are as follows:
	\begin{equation}
		|\xi_m\rangle = \frac{1}{\sqrt{2}}\left\{\begin{matrix}
			\alpha u|x_0\rangle_X |a_0\rangle_A |b_0\rangle_{B} + \beta u^{*}|x_1\rangle_X |a_1\rangle_A |b_1\rangle_{B} & \text{if}\, m=0\\ 
			-\alpha v^{*}|x_1\rangle_X |a_0\rangle_A |b_1\rangle_{B} + \beta v |x_0\rangle_X |a_1\rangle_A |b_0\rangle_{B} & \text{if}\, m=1
		\end{matrix}\right.
		\label{eq:14}
	\end{equation}
	\item[Step~4] A BBS is used by Alice and Bob to combine the two paths of their photons $\text{A}$ and $\text{B}$, respectively. The transformation rule for the BBS is given by $|a_i\rangle=\frac{1}{\sqrt{2}}[|a_i\rangle+(-1)^i|a_{i\oplus1}\rangle]$ $(i=0,1)$, where $\oplus$ denotes addition modulo 2 (for more details, refer to \cite{Kishore2019BS, BaAn2022CRIO}). As a result of this mixing, the paths $a_0$ and $a_1$ of photon $\text{A}$, and $b_0$ and $ b_1$ of photon $\text{B}$, transform as follows:
	\begin{equation}
		|\xi^{\prime}_m\rangle=\frac{1}{2}\left\{\begin{matrix}(|a_0\rangle|b_0\rangle+|a_1\rangle|b_1\rangle)(\alpha u|x_0\rangle+\beta u^{*}|x_1\rangle)+(|a_0\rangle|b_1\rangle+|a_1\rangle|b_0\rangle)(\alpha u|x_0\rangle-\beta u^{*}|x_1\rangle);&\\ \text{if}\, m=0&\\(|a_0\rangle|b_0\rangle+|a_1\rangle|b_1\rangle)(-\alpha v^{*}|x_1\rangle+\beta v |x_0\rangle)+((|a_0\rangle|b_1\rangle+|a_1\rangle|b_0\rangle)(-\alpha v^{*}|x_1\rangle-\beta v |x_0\rangle);&\\ \text{if}\, m=1&
		\end{matrix}\right.
		\label{eq:15}
	\end{equation}
	Alice initiates a cross-Kerr interaction between an auxiliary CS $|z\rangle$ and path $|a_1\rangle$, introducing a phase shift of $\theta$, and forwards the CS to Bob. Bob then initiates the cross-Kerr interaction between the CS and path $|b_1\rangle$, introducing a phase shift of $2\theta$, and performs a $\hat{\mathcal{X}}$-quadrature measurement on the CS. Various states obtained after the measurement are $|z\rangle, |ze^{i\theta}\rangle, |ze^{i2\theta}\rangle, |ze^{i3\theta}\rangle$ which gives corresponding outcomes as $pq = 00, 01, 10, 11$, respectively. Due to the overlap of adjacent Gaussian curves corresponding to $|z\rangle, |ze^{i\theta}\rangle, |ze^{i2\theta}\rangle, |ze^{i3\theta}\rangle$, misidentification may occur. The error probabilities for such misidentifications are $P^{31}_{\rm{error}}= \frac{1}{2} \text{erfc}[z(1-\cos{\theta})/\sqrt{2}]$, $ P^{32}_{\rm{error}}= \frac{1}{2} \text{erfc}[z(\cos{\theta}-\cos{2\theta})/\sqrt{2}]$ and $ P^{33}_{\rm{error}}= \frac{1}{2} \text{erfc}[z(\cos{2\theta}-\cos{3\theta})/\sqrt{2}]$ corresponding to distinguishing $|z\rangle$ from $|ze^{i\theta}\rangle$, $|ze^{i\theta}\rangle$ from $|ze^{i2\theta}\rangle$ and $|ze^{i2\theta}\rangle$ from $|ze^{i3\theta}\rangle$ respectively. If the measurement outcomes are $pq = 01$ or $10$, Alice applies the phase flip operator $(Z_S=|x_0\rangle\langle x_0|-|x_1\rangle\langle x_1|)$ on her photon $\text{X}$ and if outcomes are $pq = 00$ or $11$, no additional action is required. Finally, Bob successfully implements the unitary operation $U_m$ on Alice's arbitrary quantum state $|\psi\rangle_X$. The resulting normalized combined state is as follows:
	\begin{equation}\label{eq:16}
		|\varphi\rangle=|a_p\rangle_A |b_q\rangle_{B}(U_m |\psi\rangle_X)
	\end{equation}
	with probability of the \acrshort{RIHO} protocol being successful in terms of error occurs during the cross-Kerr interaction is
	\begin{align}
		P_{\rm{1Suc}}=1- P_{\rm{error}}^1P_{\rm{error}}^2(P_{\rm{error}}^{31}+P_{\rm{error}}^{32}+P_{\rm{error}}^{33}).
		\label{Eq4.58:Success_RIHO}
	\end{align}
\end{description}
Notably, Alice obtained $U_m|\psi\rangle_X$ $(m=0,1)$ despite Bob applying the operation $U_B$. Alice is now certain that she has received either $U_0|\psi\rangle_X$ or $U_1|\psi\rangle_X$ because the outcome $m$ is made public. Thus, implementing hidden operators into practice is a success. A visual representation of the several steps required to accomplish the \acrshort{RIHO} protocol can be found in Figure \ref{Fig4.5:Fig01_RIHO}. The \acrshort{RIHO} task can also be carried out with other quantum channels that are orthogonal to the channel employed in Equation \eqref{Eq4.46:Resource_RIHO}. Table \ref{Tab4.1:StepsRIHO} displays the variations in steps required when using various quantum channels. Essentially, a Bell state serves as a quantum channel in each of these cases. Bell states are the best candidates since entanglement is a necessary quantum resource for all \acrshort{RIO} tasks, and deterministic protocols would not function with unit fidelity without maximal entanglement state. The key point is that the ability to carry out the desired \acrshort{RIO} tasks with various Bell states with a slight modification to the particular step guarantees that Charlie, a third party, can control the protocols of our interest simply by preparing Bell states at random and allocating them to Alice and Bob without revealing which state is shared until a later stage. Some more detail about this topic shall be delved into in Section \ref{Sec4.8.2:Optimize-controlled-RIHO_RIPUO}. Nevertheless, the \acrshort{RIPUO} protocol shall be outlined first before proceeding.
\begin{table}
	\caption{Variations in operations at multiple stages when utilizing different quantum channels for the proposed RIHO protocol. Here, $|\Omega^{\pm}\rangle=|a_0\rangle|b_0\rangle\pm|a_1\rangle|b_1\rangle$, $|\Pi^{\pm}\rangle=|a_0\rangle|b_1\rangle\pm|a_1\rangle|b_0\rangle$.}
    \label{Tab4.1:StepsRIHO}
	\centering
	\begin{tabular}{|c|c|c|c|}
		\hline 
		Steps & $|\Omega^{-}\rangle$ & $|\Pi^{+}\rangle$ & $|\Pi^{-}\rangle$\\
		\hline 
		Step~1 & Same & $X_S^{k\oplus1}$ & $X_S^{k\oplus1}$\\
		\hline
		Step~2 & Same & Same & Same\\
		\hline
		Step~3 & Same & Same & Same\\
		\hline
		Step~4 & $Z^{p\oplus q\oplus1}_S$ & Same & $Z^{p\oplus q\oplus1}_S$\\
		\hline 
	\end{tabular}
\end{table}
 
\subsection{Protocol for RIPUO\label{Sec4.7.2:Protocol-RIPUO}}
\begin{figure}
	\centering
	\includegraphics[width=\textwidth]{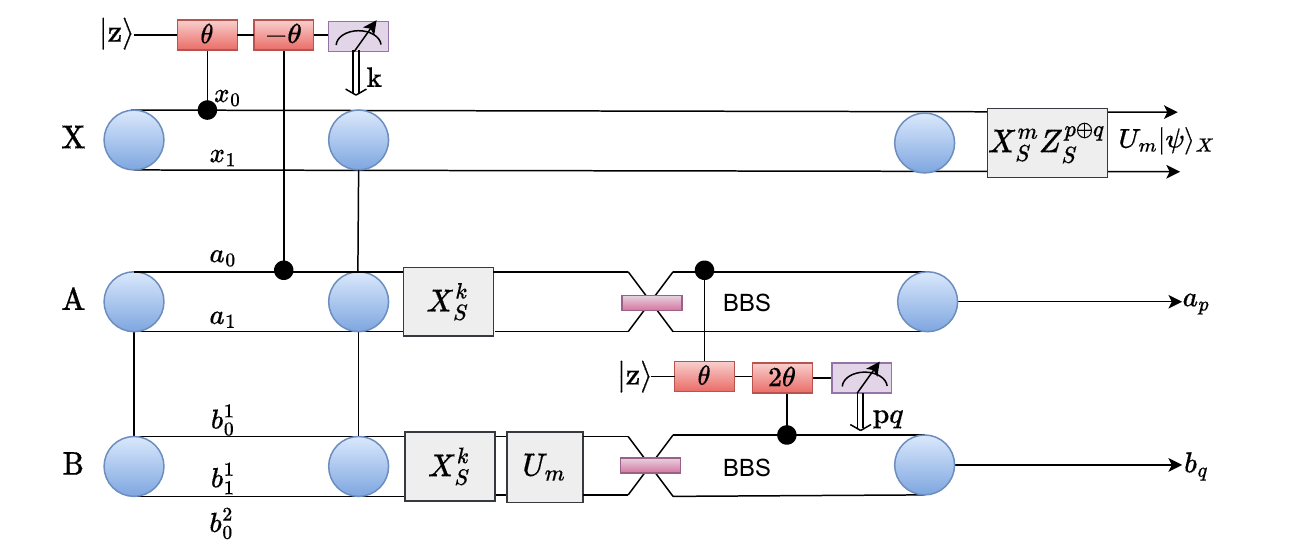}
	\caption{This figure illustrates the steps involved in the protocol for remote implementation of a partially unknown operator.}
	\label{Fig4.6:Fig02_RIPUO}
\end{figure}

\begin{description}
	\item[Step~1] The initial step is identical to Step $1$ outlined in Section \ref{Sec4.7.1:Protocol-RIHO}.
	\item[Step~2] Bob can choose to apply either $U_0$ or $U_1$ to his photon state. When Bob applies $U_0$, Equation \eqref{Eq4.52:Measurement_Step1_RIHO} changes to the following: 
	\begin{equation}\label{eq:5}
		|\zeta_{0}\rangle=(I\otimes I\otimes U_0)|\xi^{'''}\rangle=\alpha u|x_0\rangle_X |a_0\rangle_A |b_0\rangle_{B} + \beta u^{*}|x_1\rangle_X |a_1\rangle_A |b_1\rangle_{B}
	\end{equation}
	and when he applies $U_1$ then Equation \eqref{Eq4.52:Measurement_Step1_RIHO} changes to the following:
	\begin{equation}\label{eq:6}
		|\zeta_{1}\rangle=(I\otimes I\otimes U_1)|\xi^{'''}\rangle=-\alpha v^{*}|x_0\rangle_X |a_0\rangle_A |b_1\rangle_{B} + \beta v|x_1\rangle_X |a_1\rangle_A |b_0\rangle_{B}
	\end{equation}
	\item[Step~3] This step is identical to Step $4$ of Section \ref{Sec4.7.1:Protocol-RIHO}, with a minor modification after Bob measures the CS. Alice applies $X_{S}^{m}Z^{p\oplus q}_{S}$ if the measurement results are $pq$, where $m=0,1$ is the unitary $U_m$ choice that Bob applied in the preceding step. Ultimately, Bob can apply $U_m$ to Alice's state $|\psi\rangle_X$, resulting in   
	\begin{equation}\label{eq:7}
		|\varsigma\rangle=|a_p\rangle_A |b_q\rangle_{B}(U_m |\psi\rangle_X) 
	\end{equation}
	with probability of the \acrshort{RIPUO} protocol being successful in terms of error occurs during the cross-Kerr interaction is
	\begin{align}
		P_{\rm{2Suc}}=1- P_{\rm{error}}^1(P_{\rm{error}}^{31}+P_{\rm{error}}^{32}+P_{\rm{error}}^{33}),
		\label{Eq4.62:Success_RIPUO}
	\end{align}
	where, $P_{\rm{error}}^1$ and $P_{\rm{error}}^{3i}$ ($i=1,2,3$) represent the same error probabilities as those encountered in Step $1$ and Step $4$ of the \acrshort{RIHO} protocol, respectively.
\end{description}
It is evident that Bob applied the operator $U_m$ $(m=0,1)$, and Alice obtained the $U_m|\psi\rangle_X$ after completing all steps of the suggested protocol in Section \ref{Sec4.7.2:Protocol-RIPUO}. Thus, a partially unknown operator is successfully implemented. The several steps that have been taken to accomplish the \acrshort{RIPUO} protocol is illustrated in Figure \ref{Fig4.6:Fig02_RIPUO}. Similar to \acrshort{RIHO}, \acrshort{RIPUO} can also be carried out with various quantum channels that are orthogonal to the channel in Equation \eqref{Eq4.46:Resource_RIHO}. Table \ref{Tab4.2:stepsRIPUO} display the variations in steps required when using various quantum channels.
\begin{table}
	\caption{Variations in operations at multiple stages when utilizing different quantum channels for the proposed RIPUO protocol. Here, $|\Omega^{\pm}\rangle=|a_0\rangle|b_0\rangle\pm|a_1\rangle|b_1\rangle$, $|\Pi^{\pm}\rangle=|a_0\rangle|b_1\rangle\pm|a_1\rangle|b_0\rangle$.}
	\label{Tab4.2:stepsRIPUO}
	\centering
	\begin{tabular}{|c|c|c|c|}
		\hline 
		Steps & $|\Omega^{-}\rangle$ & $|\Pi^{+}\rangle$ & $|\Pi^{-}\rangle$\\
		\hline 
		Step~1 & Same & $X_S^{k\oplus1}$ & $X_S^{k\oplus1}$\\
		\hline
		Step~2 & Same & Same & Same\\
		\hline
		Step~3 & $X_{S}^{m}Z^{p\oplus q\oplus1}_S$ & Same & $X_{S}^{m}Z^{p\oplus q\oplus1}_S$\\
		\hline 
	\end{tabular}
\end{table}

\subsection{Dissipation of auxiliary coherent state\label{Sec4.7.3:Dissipation-RIHO_RIPUO}}
\begin{figure}
	\centering
	\begin{subfigure}{0.5\textwidth}
		\centering
		\includegraphics[width=0.8\linewidth]{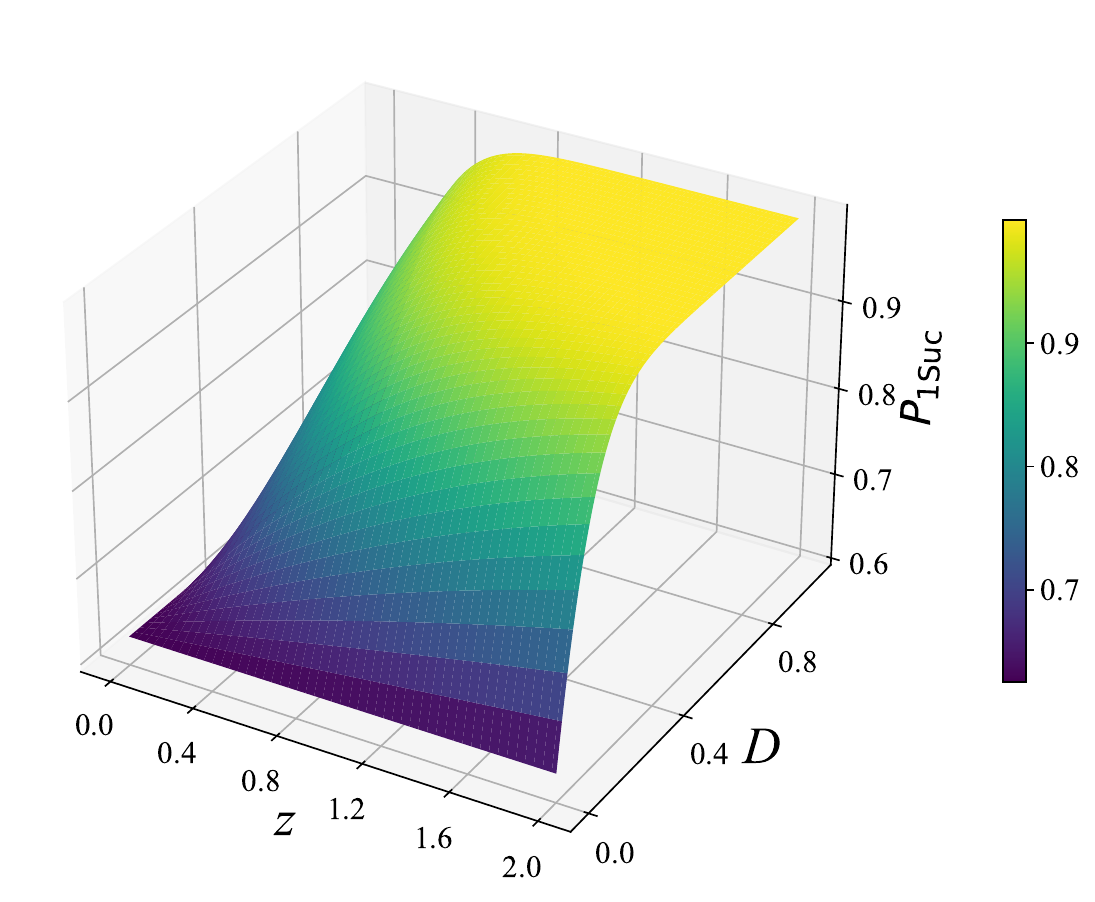}
		\caption{}
		\label{fig:sub-fig-a}
	\end{subfigure}
	\begin{subfigure}{0.5\textwidth}
		\centering
		\includegraphics[width=0.8\linewidth]{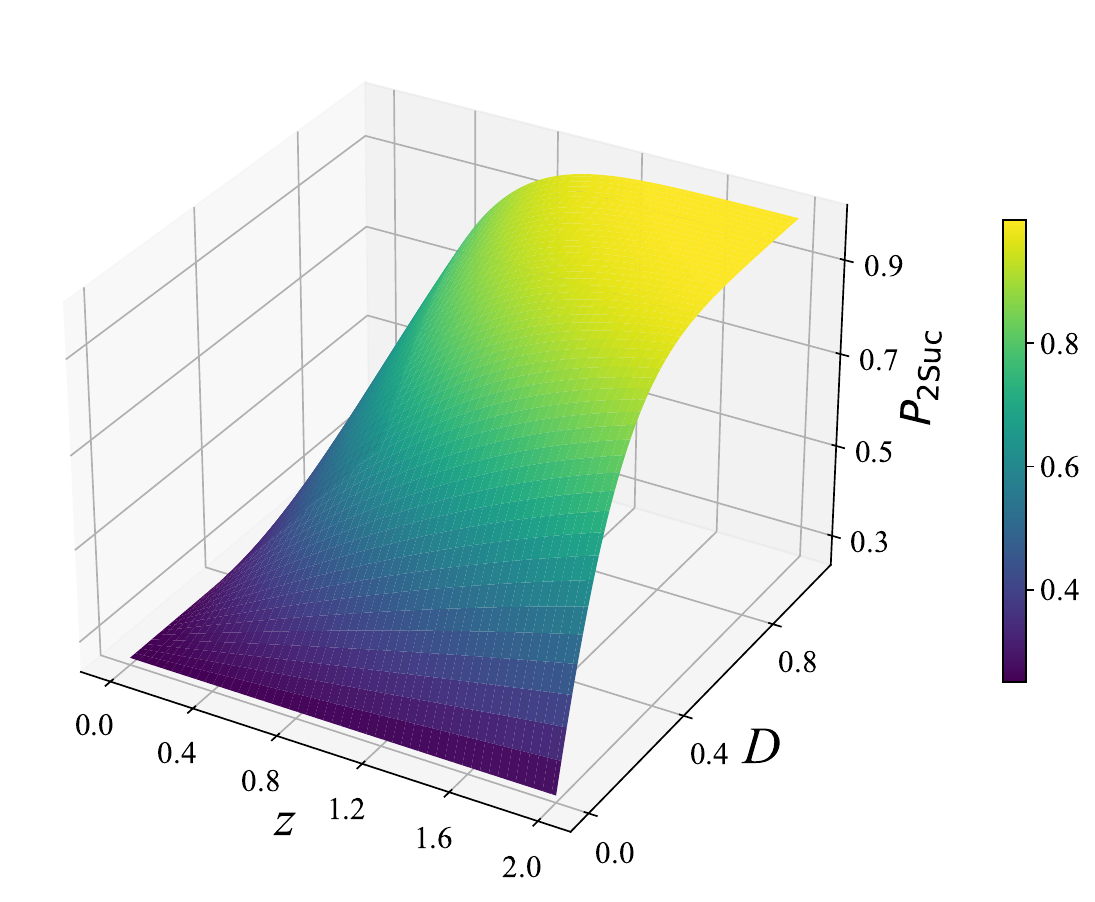}
		\caption{}
		\label{fig:sub-fig-b}
	\end{subfigure}
	\\[1ex] 
	\begin{subfigure}{0.5\textwidth}
		\centering
		\includegraphics[width=0.8\linewidth]{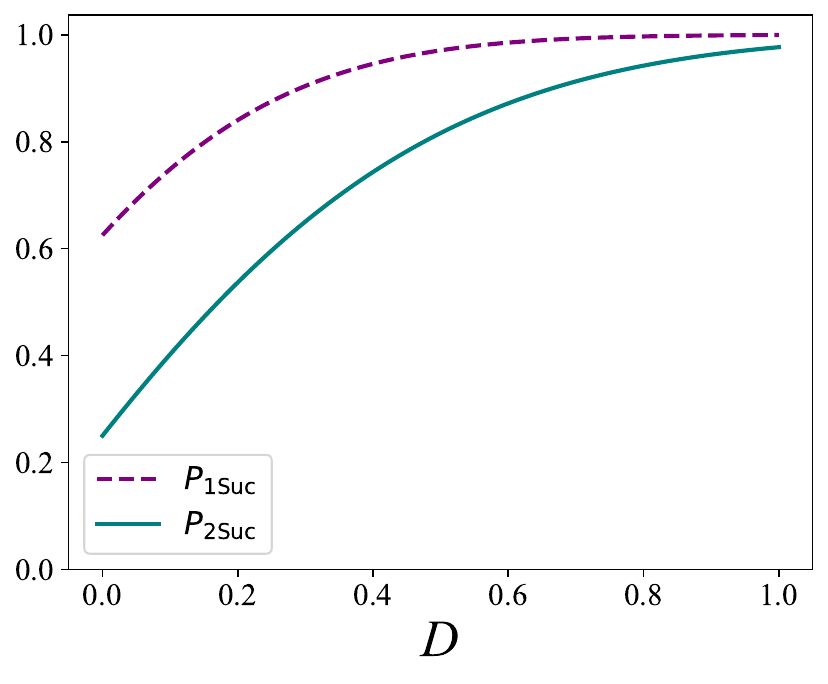}
		\caption{}
		\label{fig:sub-fig-c}
	\end{subfigure}
	\caption{The modified success probabilities is shown in (a) and (b) as functions of dissipative parameter $D$  and initial amplitude $z$ of coherent state, with $\theta=\pi$ radians phase shift for both the RIHO ($P_{\rm{1Suc}}$) and the RIPUO ($P_{\rm{2Suc}}$) protocols. In (c), the behavior of the same probabilities $P_{\rm{1Suc}}$ and $P_{\rm{2Suc}}$ as a function of $D$ is shown, with phase shift $\theta=\pi$ radians and an initial amplitude of the coherent state $z=1$.}
	\label{fig:Fig03}
\end{figure}

In real physical scenario, the environment is associated with the CS $|z\rangle$. Some photons may be lost to the environment as a result of this undesirable coupling with the surrounding environment. Consequently, it is crucial to investigate how the dissipative coherent state $|z\rangle$ affects the $\hat{\mathcal{X}}$ quadrature measurement. A master equation can explain the dynamics of interactions between a quantum system and its surroundings. The master equation offers in accounting the evolution of the system as a result of its coupling with the environment, with an emphasis on the impact of this interaction on the density matrix of the system. $\rho(t)$. The master equation in the interaction picture, as described in \cite{SB, BP07, GZ04, l20}, is given by
\begin{align}
	\partial\rho(t)=\frac{\gamma}{2}[2b\rho(t)b^\dagger-\rho(t)b^\dagger b-b^\dagger b \rho(t)].
	\label{Eq4.63:MasterEq}
\end{align}
Here, $\rho(t)$ represents the density matrix of the coherent state $|z\rangle$ at any given time $t$. The operators $b$ and $b^\dagger$ denote the annihilation and creation operators of the coherent state, respectively, and $\gamma$ is the dissipation constant associated with the coherent state.\\
Equation \eqref{Eq4.63:MasterEq} can be solved using the method outlined in References \cite{P90, BK, J06}. The Gaussian curves are adjusted to take the form $f(x, Dz\cos{(n\theta)})=(1/2\pi)^{1/4}\exp[{-\frac{1}{4}(x-2Dz\cos{(n\theta)})^2}]$, where $D=e^{-\gamma t}$ and $n=0,\pm1,\pm2,\pm3$. This modification to the Gaussian curves also impacts the error probabilities, which are consequently altered to $P^{1d}_{\rm{error}}= \frac{1}{2} \text{erfc}[Dz(1-\cos{\theta})/\sqrt{2}]$, $P^{2d}_{\rm{error}}= \frac{1}{2} \text{erfc}[Dz(1-\cos{\theta})/\sqrt{2}]$, $P^{31d}_{\rm{error}}= \frac{1}{2} \text{erfc}[Dz(1-\cos{\theta})/\sqrt{2}]$, $P^{32d}_{\rm{error}}= \frac{1}{2} \text{erfc}[Dz(\cos{\theta}-\cos{2\theta})/\sqrt{2}]$ and $P^{33d}_{\rm{error}}= \frac{1}{2} \text{erfc}[Dz(\cos{2\theta}-\cos{3\theta})/\sqrt{2}]$. As a result, the success probabilities given by Equation \eqref{Eq4.58:Success_RIHO} and Equation \eqref{Eq4.62:Success_RIPUO} are modified. To illustrate the effect of the dissipation process on the success probabilities, their variation with respect to different physical parameters has been plotted, as shown in Figure \ref{fig:Fig03}. 

Figure \ref{fig:Fig03} (a,b) shows the modified success probabilities of \acrshort{RIHO} ($P_{\rm{1Suc}}$) and \acrshort{RIPUO} ($P_{\rm{2Suc}}$) as functions of the dissipative parameter $D=e^{-\gamma t}$ and the initial amplitude $z$ of the coherent state $|z\rangle$, considering a phase shift $\theta$. As we can see, both protocols have the high success probability when initial amplitude of the CS is $1$ and coupling with the environment is weak, i.e., a dissipation constant $\gamma$ is small. Motivated by the earlier observation, a plot is provided shown in Figure \hyperref[fig:Fig03]{3(c)}. It can seen that both protocols have high success probability when the dissipation constant is small.

\section{A general method for selecting quantum channels for different variants of the proposed protocols\label{Sec4.8:General-RIHO_RIPUO}}
There exist several variations of the well known protocols for quantum communication. In particular, a number of \acrshort{RSP} variations have been thoroughly investigated. For instance, controlled-\acrshort{RSP}, joint-\acrshort{RSP}, controlled-joint-\acrshort{RSP}, bidirectional-\acrshort{RSP}, and similar approaches (see Ref. \cite{vishal2015cb_RSP_NC} and references therein for details) are all variations of the \acrshort{RSP} scheme. Similarly, the proposed protocols for the \acrshort{RIHO} or \acrshort{RIPUO} may have multiple variations. Various potential variations of the proposed protocols are covered in Section \ref{Sec4.8.1:variants-riho-ripuo}. Additionally, the method for optimizing the controller's qubit is covered in Section \ref{Sec4.8.2:Optimize-controlled-RIHO_RIPUO}.

\subsection{Different possible variants of the RIHO and RIPUO protocols\label{Sec4.8.1:variants-riho-ripuo}}
It should be noticed that the quantum channel, which is provided by Equation \eqref{Eq4.46:Resource_RIHO}, is the same for both the proposed protocols. Remarkably, several variations of the proposed protocols can be executed with the similar quantum channel. Subsequently, it will be demonstrated that the proposed protocols can be made more general by changing just a few steps of the original protocols.\\
It can be assumed that an arbitrary quantum state situated at a distance is intended to be controlled by an operator, which is shared among $M$-parties. This might be seen as the multiparty joint \acrshort{RIPUO} or \acrshort{RIHO}. A quantum channel that can be used in general for the same purpose is written as follows:
\begin{equation}\label{Eq4.64:J-QC}
	|\Upsilon\rangle_{AB^1B^2...B^M}=(|a_0\rangle_A \otimes_{i=1}^{M}|b_0^{i}\rangle_{B^i} + |a_1\rangle_A \otimes_{i=1}^{M}|b_1^{i}\rangle_{B^i}).
\end{equation}
Similarly, if there are $N$ controllers along with $M$ joint parties, the quantum channel required to accomplish the controlled-joint task is described as follows:
\begin{equation}\label{Eq4.65:JC-QC}
	|\Theta\rangle_{AB^1B^2...B^MC^1C^2...C^N}=(|a_0\rangle_A \otimes_{i=1}^{M}|b_0^{i}\rangle_{B^i}\otimes_{j=1}^{N}|c_0^{j}\rangle_{C^j} + |a_1\rangle_A \otimes_{i=1}^{M}|b_1^{i}\rangle_{B^i}\otimes_{j=1}^{N}|c_1^{j}\rangle_{C^j}).
\end{equation}
Note that the quantum channel described in Equation \eqref{Eq4.65:JC-QC} represents the most general form for the \acrshort{RIHO} and \acrshort{RIPUO} protocols occurs unidirectional, which can derive the channels for other variants. It is also possible to extend the proposed generalized protocol from unidirectional to multiple directions. For example, several parties can simultaneously remotely implement operators on each other's qubit. Such variant of \acrshort{RIO} is known as multi-directional \acrshort{RIHO} and \acrshort{RIPUO}. If there are $m$ possible direction of implementing hidden or partially unknown operators remotely then the resources used will be $|\Theta\rangle^{\otimes m}$. The protocols and strategies will follow a similar approach.\\
One could also consider reducing the number of qubits used so far in the controlled variants of the protocols offered. The technique for optimizing the controller qubit is covered in the next section.

\subsection{Optimizing controller qubit in the controlled variants of the RIHO and RIPUO protocol\label{Sec4.8.2:Optimize-controlled-RIHO_RIPUO}}
So far, different variants such as, controlled, cyclic controlled and bidirectional cyclic controlled, of the proposed protocol (i.e., \acrshort{RIHO} and \acrshort{RIPUO}) have been reported \cite{BaAn2007RIHO,FanQ2008CRIPUO,PengJ2019CCRIPUO,Peng22MPCRIPUO,PengJ2022BCCRIPUO}. In order to implement such schemes, the controller must utilizes a multi-partite entangled states as a quantum resource. Combining this with the state intended for the remote operation results in a complex state of the form
\begin{equation}
	|\Psi\rangle =\frac{1}{\sqrt{2}}(|\chi_\mu\rangle|\mu\rangle+|\chi_\nu\rangle|\nu\rangle,
\end{equation}
where, $\langle \chi_\mu|\chi_\nu\rangle = 0$ and $\langle \mu|\nu\rangle = 0$ (refer to Equation ($5$) of Reference \cite{PengJ2019CCRIPUO} and Equation ($9$) of Reference \cite{FanQ2008CRIPUO, PengJ2022BCCRIPUO}), along with the subsequent strategy as an example \footnote{Extending to the multi-controller case is straightforward. In that case, the combined state is represented as $|\psi\rangle =\sum_{\mu=0}^N\frac{1}{\sqrt{N}}(|\chi_\mu\rangle|\mu\rangle$, where $|\mu\rangle$ serves as a product state that forms the basis set $\{|\mu\rangle\}$ in an $N$-dimensional space.}. At this point, Charlie typically measures in $\{|\mu\rangle,|\nu\rangle\}$ basis (computational basis is the frequent choice, but generally speaking, it can be any basis) and reveals the measurement result. Knowing Charlie's measurement result enables Alice and Bob(s) to determine their shared state and take appropriate action. In the scenario described above, based on disclosure of the state $|\mu\rangle  (|\nu\rangle)$, Alice and Bob are aware of their shared state $|\chi_\mu\rangle$  ($|\chi_\nu\rangle$) and will carry out the remaining steps of the protocol as necessary. At this point, there are two key conceptual aspects to consider. Firstly, Alice and Bob(s) are presumed to be semi-honest in any such schemes for controlled computing or communication. Otherwise, they could independently prepare states such as $|\chi_\mu\rangle$  or $|\chi_\nu\rangle$ and carry out the protocol on their own. Secondly, Charlie shares only a small piece of information to reveal the outcome of his measurement, thereby maintaining control over the other semi-honest participants. To achieve this in the standard approach, she must possess at least one qubit, which needs to be entangled with the states of Alice and Bob. Only in this case her measurement results in different shared states, thereby granting him the power to control. This approach necessitates the creation and maintenance of large entangled states, which are challenging to generate and sustain. Here, we observe that Charlie can maintain his control even without keeping a qubit. In particular, he might choose the initial state at random from a group of orthogonal states, implies that, depending on whether Bob began with a state $|\Psi_\mu\rangle=|\Omega^{+}\rangle^{\otimes n}$ or $|\Psi_\nu\rangle=|\Omega^{-}\rangle^{\otimes n}$, Alice and Bob's states becomes $|\chi_\mu\rangle$ or $|\chi_\nu\rangle$. Here, $|\Omega^{\pm}\rangle=|a_0\rangle|b_0\rangle\pm|a_1\rangle|b_1\rangle$, and $n$ is an integer. In the case of most basic unidirectional protocols (\acrshort{RIHO} and \acrshort{RIPUO}) and those with a single controller (CRIHO and CRIPUO), the state with $n=1$ described above would be adequate. Instead of revealing the measurement result in the case of CRIHO and CRIPUO, Charlie would provide information of one bit, whether he began with $|\Psi_\mu\rangle$ or $|\Psi_\nu\rangle$. She would have the same level of control over Alice and Bob(s), who are semi-honest in the sense they try to steal the information without disturbing the protocol. The only difference is that Charlie wouldn't have to prepare or maintain massive entangled states. A similar result is achieved if Charlie begins from $|\Psi_\mu\rangle=|\Pi^{+}\rangle^{\otimes n}$ or $|\Psi_\nu\rangle=|\Pi^{-}\rangle^{\otimes n}$ where, $|\Pi^{\pm}\rangle=|a_0\rangle|b_1\rangle\pm|a_1\rangle|b_0\rangle$. A closer examination of the scheme presented here reveals that the majority of such schemes may be implemented with one or more Bell states, which are comparatively simple to generate and maintain. For instance, a scheme for bidirectional \acrshort{RIPUO} (BRIPUO), bidirectional \acrshort{RIHO} (BRIHO), and their controlled versions (i.e., CBRIHO and CBRIPUO) can be realised with two Bell states (i.e., by taking into account $n=2$ in the states mentioned above) and the aforementioned strategy. In contrast, a unidirectional (bidirectional) cyclic implementation of RIHO and RIPUO involving $m$ parties different from the controller would need $m\,(2m)$ Bell states. This fact is summed up in the Table \ref{Tab4.3:Optimize_RIHO_RIPUO}, where the minimal resources needed to do various \acrshort{RIO} tasks have been stated using the proposed concepts and have contrasted them with the quantum resources used in previous research to complete the comparable jobs. Benefit of the current strategy is evident from this comparison. Here, the most basic form of the cyclic version, $m=3$ case is considered with three users, Alice, Bob, and David. In this case, Alice wants to remotely apply a partially unknown or hidden quantum operator to Bob's arbitrary quantum state, while Bob (David) wish to carry out comparable operations on David's (Alice's) states. To accomplish the task, three Bell states ($(|\Omega^{+}\rangle_{AB}|\Omega^{+}\rangle_{B'D}|\Omega^{+}\rangle_{D'A'})$) are required as a quantum resource. The qubits labeled as $A$ and $A'$ belong to Alice, $B$ and $B'$ belong to Bob, and $D$ and $D'$ belong to David. The controlled version of this protocol would use the same number of quantum channel, while the bidirectional and controlled bidirectional versions of cyclic RIPUO and cyclic RIHO would use twice as many Bell states.
\begin{table}
	\caption{Optimal quantum resources for different variants of the proposed protocols for RIHO and RIPUO. Here, $|\Omega^{\pm}\rangle=|a_0\rangle|b_0\rangle\pm|a_1\rangle|b_1\rangle$, $|\Pi^{\pm}\rangle=|a_0\rangle|b_1\rangle\pm|a_1\rangle|b_0\rangle$.}
	\label{Tab4.3:Optimize_RIHO_RIPUO}
	\centering
	\begin{tabular}{|c|c|c|c|c|}
		\hline 
		S.No & Quantum Channel Used & Purpose & Party & Optimal Channel\\
		\hline 
		1 & $\frac{1}{\sqrt{2}}(|000\rangle+|111\rangle)_{AA^{'}B}$ \cite{BaAn2007RIHO} & RIHO & 2 & $|\Omega^{\pm}\rangle$ or $|\Pi^{\pm}\rangle$\\
		\hline
		2 & $\frac{1}{\sqrt{2}}(|0000\rangle+|1111\rangle)_{AA^{'}BC}$ \cite{BaAn2007RIHO} & CRIHO & 2+1 & $|\Omega^{\pm}\rangle$ or $|\Pi^{\pm}\rangle$\\
		\hline
		3 & $\frac{1}{\sqrt{2}}(|000\rangle+|111\rangle)_{ABC}$ \cite{FanQ2008CRIPUO} & CRIPUO & 2+1 & $|\Omega^{\pm}\rangle$ or $|\Pi^{\pm}\rangle$\\
		\hline
		4 & $|+\rangle|\phi^{+}\rangle^{\otimes3}+|-\rangle|\phi^{-}\rangle^{\otimes3}$ \cite{PengJ2019CCRIPUO} & CCRIPUO & 3+1 & $|\Omega^{\pm}\rangle^{\otimes3}$ or $|\Pi^{\pm}\rangle^{\otimes3}$\\
		\hline
		5 & $|0\rangle|\phi^{+}\rangle^{\otimes6}+|1\rangle|\phi^{-}\rangle^{\otimes6}$ \cite{PengJ2022BCCRIPUO} & BCCRIPUO & 3+1 & $|\Omega^{\pm}\rangle^{\otimes6}$ or $|\Pi^{\pm}\rangle^{\otimes6}$\\
		\hline
	\end{tabular}
\end{table}
It is important to note that, in relation to the previously discussed strategy of quantum resource optimization, let's attempt to optimize the controllers' qubit provided in Equation \eqref{Eq4.65:JC-QC}, since it is the universal quantum resource for the \acrshort{RIHO} and \acrshort{RIPUO}. Among $N$ controllers, one controller (say $n^{th}$ controller) may maintain his or her capacity to regulate Alice and Bob execution of an operator remotely, in a traditional way. Specifically, $n^{th}$ controller selects a random bit $r_n\in\{0,1\}$ and prepares $M+N$-partite entangled state (as opposed to the $M+N+1$ state described in Equation \eqref{Eq4.65:JC-QC}).    
\begin{equation}\label{eq:JC-QC1}
	\begin{split}
	|\Theta^{'}\rangle_{AB^1B^2...B^MC^1...C^{n-1}...C^{n+1}...C^N}=&(|a_0\rangle_A \otimes_{i=1}^{M}|b_0^{i}\rangle_{B^i}\otimes_{j=1, j\neq n}^{N}|c_0^{j}\rangle_{C^j}\\
	&+ (-1)^{r_n} |a_1\rangle_A \otimes_{i=1}^{M}|b_1^{i}\rangle_{B^i}\otimes_{j=1, j\neq n}^{N}|c_1^{j}\rangle_{C^j}).
   \end{split}
\end{equation}
The controller then allows all other parties to share the state that has been prepared. The controller $n$ broadcasts the exact value of $r_n$ at the finishing stage of the protocol. In qualitative terms, the controller's capability remains unaffected, but the quantum resource requirement is saved with one qubit and also saved the expense of measurement on controller's qubit. However, the most effective strategy can be achieved if the $N$ controllers collaborate in the following manner. Consider a scenario where one controller among all, say controller $1$, generates an entangled state involving $M + 2$ parties, structured as follows:
\begin{equation}\label{Eq4.68:JC-QC2}
	|\Theta^{''}\rangle_{AB^1B^2...B^MC}=(|a_0\rangle_A \otimes_{i=1}^{M}|b_0^{i}\rangle_{B^i}|c_0\rangle_{C} + (-1)^{r_1} |a_1\rangle_A \otimes_{i=1}^{M}|b_1^{i}\rangle_{B^i}|c_1\rangle_{C}).
\end{equation}
using a randomly chosen bit $r_1\in\{0,1\}$. Controller $1$ then allows the sharing of the state (Equation \eqref{Eq4.68:JC-QC2}) among Alice, $M$ Bobs, and controller $2$, with qubit $C$ being held by the controller $2$. Controller $2$ decides whether or not to apply the phase-flip operation to qubit $C$ (mathematically, this means that controller $2$ applies the operator $Z^{r_2}$ to qubit $C$, where $r_2$ is a randomly chosen bit). Afterward, controller $2$ forwards qubit $C$ to controller $3$, who performs a similar operation using a new random bit $r_3$. Controller $3$ then sends qubit $C$ to controller $4$, and so on, until controller $N$. As a result, the total state shared among all the parties becomes:
\begin{equation}\label{eq:JC-QC3}
	|\Theta^{'''}\rangle_{AB^1B^2...B^MC}=(|a_0\rangle_A \otimes_{i=1}^{M}|b_0^{i}\rangle_{B^i}|c_0\rangle_{C} + (-1)^{r} |a_1\rangle_A \otimes_{i=1}^{M}|b_1^{i}\rangle_{B^i}|c_1\rangle_{C}).
\end{equation}
where, $r=\sum_{j=1}^{N}r_j$. During execution of the protocol, controller $1$ must measure qubit $C$, yielding an outcome $r_0\in\{0,1\}$. At a suitable point near the final step of the protocol, controller $1$ reveals both $r_0$ and $r_1$, while each of the other controllers reveal respective random bit $r_j$. This allows all the parties (Alice and all $M$ Bobs) to determine the absolute values of $r_0$ and $r$, enabling them to correctly finish the protocol. Each controller retains their individual autonomy, and the quantum resource cost is significantly reduced. 

\section{Conclusion\label{Sec4.9:Conclusion-RIHO_RIPUO}}
In this chapter, schemes for three variants of \acrshort{RIO} are proposed. The proposed variants of \acrshort{RIO} are named as \acrshort{CJRIO}, \acrshort{RIHO} and \acrshort{RIPUO}. Different quantum resources are utilized in the proposed schemes. For example, \acrshort{CJRIO} uses a four-qubit hyperentangled state (entangled in two \acrshort{DOF}, \acrshort{S-DOF} and \acrshort{P-DOF}) while \acrshort{RIHO} and \acrshort{RIPUO} uses a two-qubit maximally entangled state (entangled in \acrshort{S-DOF}). The proposed protocols uses photon as a qubit and due to the least interacting properties of photons, cross-Kerr interaction method is used to allow interaction between photons.\\
The \acrshort{CJRIO} scheme involves total four parties, a receiver (Alice), two senders ($\text{Bob}^1$ \& $\text{Bob}^2$), and one controller (Charlie). The two senders jointly teleport an operator to the receiver under the presence of a controller Charlie. This scheme is also generalized to arbitrary senders and controllers. Efficiency of the scheme is estimated in terms of number of operators to be teleported remotely, amount of quantum resources utilized and classical bits disclosed.\\
The \acrshort{RIHO} and \acrshort{RIPUO} schemes involve two parties, one sender (Bob), and one receiver (Alice). In the \acrshort{RIHO} scheme, a hidden operator is teleported to Alice. The operator is hidden in the sense that intended operator is dissolved inside a super-operator. In the \acrshort{RIPUO} scheme, a partially unknown operator is teleported. The operator is partially known or unknown in the sense that its matrix structure is known but not the values of its matrix elements. Both the schemes are then extended to their other possible variants. One important method is introduced, which gives controllers a power to control the protocols even without keeping qubits. Most interestingly, the dissipation of CS used for the photon-photon interaction is well studied. The success probability of both the schemes is estimated in terms of errors occurs during the interaction.\\
It is to be noted that the all proposed schemes of \acrshort{RIO} can be experimentally realized with the currently available technologies. In particular, control power may be calculated by expanding the work done in References \cite{ShohiniG2015ControlPower,BaAn2019ControlPower}. Furthermore, the teleportation of an operator between two network nodes that share a maximally entangled state has been discussed so far. The following question arises here: What happens if Alice and Bob exist in the same network, but the entangled state is not shared? This question is explored in Reference \cite{WuFan22RIO_QC} within the framework of quantum multihop networks. However, without limiting ourselves to quantum multihop networks and basic \acrshort{RIO} protocols, various \acrshort{RIO} variants in a universal network can be explored, as the proposed schemes rely solely on Bell states. An efficient method for entanglement routing in such broader network scenarios is recently reported in the Reference \cite{Vaisakh22QN}. This chapter ends with the expectation that the suggested protocols will soon be realized experimentally and will lead to numerous applications.
\newpage

\chapter{QUANTUM ANONYMOUS VETO}\label{Ch5:QV}
\graphicspath{{Chapter5/Chapter5Figs/}{Chapter5/Chapter5Figs/}}

\section{Introduction}\label{Sec5.1:Intro-QV}
We often come across certain concepts or laws of quantum physics that are counterintuitive and consequently difficult to accept as we live in a classical world. Interestingly, these quantum physics laws, with no classical counterpart, resulted in several examples of quantum advantages that led to the second quantum revolution. This revolution has made us capable of developing quantum technologies. Some of such technologies have been built and still developing some are futuristic technologies. To be specific, we can think of products and devices recently developed in the domains of quantum computing \cite{Arute2019qSupremacy,Montanaro2016qAlgo}, quantum communication \cite{Pathak2017QCrypt}, and quantum metrology \cite{Arvidsson2020qMetrology,Giovannetti2011qMetrology}. It is to be noted that the quantum advantages are not restricted to these domains only. Consequently, it is generally accepted that the second quantum revolution is almost certain to impact our daily life in the near future \cite{dowling2003qRevolution2}. However, different possible facets of technologies where we may use quantum resources for advancement, are not yet fully explored. Among different facets of quantum technologies, quantum cryptography \cite{Pathak2017QCrypt}, which provides unconditional security, is the most mature field. The characteristic task 
 of quantum cryptography is \acrfull{QKD}, where a secure key is distributed securely between two legitimate parties using the quantum resources. The first-ever \acrshort{QKD} protocol was introduced by Charles Bennett and Gilles Brassard in 1984, popularly known as BB84 protocol \cite{Bennett1984QCrypt}. Since then, significant advancements have been made. Various schemes for \acrshort{QKD} have been designed and implemented \cite{Ekert1991QCrypt,Bennett1992QKD}. Building on the foundational principles of \acrshort{QKD}, the concept of achieving unconditional security using quantum resources has been extended to numerous innovative protocols. These include one-way and two-way protocol for secure direct quantum communication (see References \cite{Pathak2015DSQC,Anindita2012DSQC,Srikara2020cvDSQC} and Chapter 8 of Reference \cite{Pathak2013book}) as well as protocols for \acrfull{SMQC} \cite{Ashwin2020CVqDialogue,Crepeau2002bookQC}. Various tasks where \acrshort{SMQC} is needed are auction, voting, private comparison, etc. Since these tasks are extremely useful in our daily lives, \acrshort{SMQC} schemes have received considerable amount of attention. One of the most crucial \acrshort{SMQC} tasks is voting as it is a necessary component of any democratic system in which choices are made collectively. The potential uses of quantum resources for the anonymous voting scenario was first demonstrated in 2006 by Hillery et al. \cite{Hillery2006qVoting}, and almost concurrently by Vaccaro et al. \cite{VaccaroJ2007qVoting} (for a historical perspective, see \cite{Kishore2017qVoting}). Since then, significant progress has been made in this field, with the introduction of numerous new protocols in recent years (\cite{Kishore2017qVoting,Parakh2022Qvoting,SandeepM2021QV} and references therein).\\
A notable subclass within the broad category of voting schemes focuses on unanimous decision-making, where a proposal is rejected if even a single member disagrees. The United Nation (UN) Security Council is a notable example of an organization that often performs this task, as when a proposal is submitted to UN for voting, it is immediately rejected if one or more council members disagree. This type of scheme is referred to as a veto, where maintaining the confidentiality of individual votes has been often a critical requirement. The first quantum-based solution to this was reported by Rahman and Kar using the GHZ type entangled state \cite{RahamanR2015QV}. Subsequently, Wang et al. proposed a comprehensive solution utilizing the same entangled state (GHZ state) \cite{WangQ2021QV} and also conducted experimental testing of the protocol on cloud-based IBM quantum computer for a specific situation where 4 voters participated in a veto process. A subsequent research work by Mishra et al. \cite{SandeepM2021QV} suggested a number of novel \acrfull{QAV} schemes, which are categorized on the basis of the quantum resources they employed and their viability in practical physical implementations. Additionally, the schemes of Mishra et al. have been divided into three categories: deterministic, iterative, and probabilistic. Among the proposed schemes, two (identified as QAV-6 and QAV-7 in \cite{SandeepM2021QV}, referred to as Protocol A and Protocol B in this thesis) are focused here due to their superior efficiency. In this chapter, the two most efficient \acrshort{QAV} protocols designed by Mishra et al. \cite{SandeepM2021QV}, are experimentally verified on IBM quantum computer accessible on the cloud. The impact of different types of noisy environments such as bit-flip, phase damping, amplitude damping, and depolarizing noise have also been examined on these protocols.\\
Before moving forward, it is important to note that the presence of decoherence has so far prevented the development of a fully scalable quantum computer. Nevertheless, a range of \acrfull{NISQ} computers have been developed in recent years. Moreover, several organizations have enabled cloud-based access to these quantum computers, making them widely available for use. IBM is the leading organization as it has provided cloud access to the NISQ computers to common users since 2016. Quantum computers available at IBM Quantum platform are comprised of transmon qubits and are based on superconductivity. However, each of the quantum computer available on the cloud is unique, as different quantum computers have different topologies and different number of qubits. Without diving into more details, it is worth noting that numerous quantum communication and computing tasks have previously been implemented using IBM quantum computers \cite{Ashwin2021factorIBM,Sisodia2017bmIBM,Sisodia2017OpticalDesign_QT,Harper2019FT_IBM,AcasieteF2020qWalkIBM,Britant2024IBM}. For instance, IBM quantum computers have been utilized in various applications, including solving factorization problem using hybrid classical-quantum algorithm \cite{Ashwin2021factorIBM}, nondestructive Bell states discrimination \cite{Sisodia2017bmIBM}, experimental implementation of an optimal teleportation scheme \cite{Sisodia2017OpticalDesign_QT}, fault-tolerant quantum gates implementation \cite{Harper2019FT_IBM}, designing quantum circuit to model quantum walks \cite{AcasieteF2020qWalkIBM}, and few more tasks (see \cite{Bikash2019qRouterIBM} and its references). Building on this extensive list, the aim of this chapter is to report the realization of two effective \acrshort{QAV} schemes using a quantum computer available at IBMQ. The motivation for conducting this experiment is twofold: first, anonymous veto has numerous applications in social contexts, and second, since veto systems typically involve a small amount of voters therefore it can be implemented commercially (using the technologies available today) if realized successfully.

\section{Quantum anonymous veto protocols\label{Sec5.2:Protocols-QAV}}
As mentioned above, Mishra et al. \cite{SandeepM2021QV} recently introduced a collection of schemes for \acrshort{QAV}. Two of the proposed protocols are notably more effective in terms of qubit usage than other schemes that they have proposed. In the subsections of this section, a step-by-step description of these two protocols will be provided, referring to them as Protocol A and Protocol B. Both the protocols are facilitated by a \acrfull{VA} Alice, who is considered as semi-honest. Alice is semi-honest in the sense that while she adheres strictly and honestly to the protocol, she may attempt to gain additional information about the strategy (e.g., votes) of other participants (voters). The voting process involves $n$ voters, denoted as $\left\{ V_{i},V_{2},\cdots,V_{n}\right\}$.

\subsection{Protocol A}
Protocol A is iterative in nature and employs a Bell state for its execution, with one qubit serving as the home qubit and the other as the travel qubit. Protocol A can be outlined through the following sequence of steps:
\begin{description}
	\item [Step~A1] The VA prepares a two-qubit maximally entangled state ($\vert\phi^{+}\rangle=\frac{1}{\sqrt{2}}(|00\rangle+|11\rangle)$) i.e. the Bell state. First qubit of the Bell state is kept by the VA itself, and the second qubit is sent to the voter $V_{1}$ (the first voter).  
	\item [Step~A2] $V_{1}$ applies the unitary operator $\sigma_{z}(t)=\left[\begin{array}{cc}
		1 & 0\\
		0 & e^{i\frac{\pi}{2^{t}}}
	\end{array}\right]$ with $t=0$ if $V_{1}$ choose to veto; otherwise, the identity operation $I=\left[\begin{array}{cc}
		1 & 0\\
		0 & 1
	\end{array}\right]$ is applied. After performing the unitary operation, $V_1$ sends the travelling qubit to $V_2$, who can cast his vote in the identical manner. This process continues sequentially, with each voter encoding their decision, until $V_n$, who then sends the travelling qubit back to the VA after casting his vote.\\
	It is important to note that $(t+1)$ represents the number of iterations of the protocol. Thus, $t=0$ corresponds to the first iteration \footnote{A voter applies $\sigma_{z}(n-1)=\left[\begin{array}{cc}
			1 & 0\\
			0 & e^{i\frac{\pi}{2^{n-1}}}
		\end{array}\right]$ in the $n^{th}$ iteration to cast his vote.} and a voter applies $\left[\begin{array}{cc}
		1 & 0\\
		0 & e^{i\pi}
	\end{array}\right]=\left[\begin{array}{cc}
		1 & 0\\
		0 & -1
	\end{array}\right]=\sigma_{z}$ in the first iteration.
	\item [Step~A3] The VA measures both the home qubit in his possession and the travelling qubit acquired from $V_n$ together in the Bell basis.\\ 
	If odd (or even, comprising 0) amount of voters applied veto, the VA's measurement will yield the state $\vert\phi^{-}\rangle=\frac{1}{\sqrt{2}}(|00\rangle-|11\rangle)$ $\left(\vert\phi^{+}\rangle=\frac{1}{\sqrt{2}}(|00\rangle+|11\rangle)\right)$, and the protocol aborts (proceed to the next step, as the result obtained is inconclusive).
	\item [Step~A4] Steps~A1 to A3 are iterated for $t = 1$, and so on, with each iteration incrementing the $t$ value by one, until a conclusive result is obtained.
\end{description}
If $n$ voters participate in the voting process, the maximum iterations that are needed to reach a conclusive outcome would be $1+\log_{2}n$, with each iteration reducing the number of voting possibilities by half. It is important to note that this protocol is iterative, and the amount of iterations could potentially reveal the amount of voters who utilized their veto right (for instance, when parties vetoed number is even). However, this leakage does not reveal the individuality of voters who have performed the veto. Therefore, it does not compromise the anonymity of voters. Additionally, to prevent such partial leakage, a protocol can be implemented where a conclusive result is achieved in just one iteration. In the following, another protocol, which is referred as Protocol B will be described.

\subsection{Protocol B}
Unlike Protocol A, this protocol (Protocol B) is deterministic and yields a conclusive result in just a single iteration. Additionally, each voter is assigned a set of unitary operations that can be utilized to cast a veto. Specifically, the $i$th user holds a unitary $U_{i}$ that they can apply to perform a veto. For instance, Table \ref{Tab5.1:unitary-veto} presents the possible unitary and various maximally entangled states for four voters. The procedure for executing the Protocol B for a general $n$ voters scenario involves the following steps:
\begin{description}
	\item [Step~B1] The verifier VA, locally prepares a $m$-qubit maximally entangled state $\mid\psi_{in}\rangle$, where $m\geq(n-1)c$. Here, $c=1$ since each voter can encode the information of one bit only, either he agree or disagree with the proposal. VA assign $l$ qubits, such that $l<m$, as the travelling qubits, while retaining $\left(m-l\right)$ qubits with herself as the home qubits. The $l$ travelling qubits are sent to each voter and eventually returned to the VA.
	\item [Step~B2] The voter $V_{i}$ $(1\leq i\leq n-1)$ performs the identity operation if he agree with the proposal, otherwise he applies the operator $U_{i}$ if he disagree. Following the operation, all $l$ travelling qubits are forwarded to voter $V_{i+1}$. This sequence continues until it reaches voter $V_{n}$, who, after completing his operation, returns the travelling qubits to the VA.
	\item [Step~B3] The VA performs a measurement on the final state ($|\psi_{fin}\rangle$) in the identical basis used to prepare the initial state. If $\langle\psi_{in}|\psi_{fin}\rangle=1$, it indicates that either no voter has applied veto or all voters have applied veto. If $\langle\psi_{in}|\psi_{fin}\rangle\neq1$, it indicates that at least a single voter has done veto. This approach enables the voters to determine whether a consensus has been achieved.
	\begin{table}
		\caption{An examples of quantum states and their corresponding quantum operations that can be utilized to implement Protocol B.\label{Tab5.1:unitary-veto}}
		\begin{centering}
			\begin{tabular}{|>{\centering}p{2cm}|>{\centering}p{2cm}|>{\centering}p{2cm}|>{\centering}p{9cm}|}
				\hline 
				Number of voters  & Number of travel qubits  & Quantum state  & Encoding operation ($O_{i}$)\tabularnewline
				\hline 
				Four  & Two  & Four-qubit cluster state  & $U_{1}:\{X\otimes iY\},U_{2}:\{X\otimes Z\},U_{3}:\{iY\otimes Z\},U_{4}:\{iY\otimes iY\}$\tabularnewline
				\hline 
				Four  & Two  & GHZ state  & $U_{1}:\{X\otimes I\},U_{2}:\{X\otimes X\},U_{3}:\{iY\otimes X\},U_{4}:\{iY\otimes I\}$\tabularnewline
				\hline 
			\end{tabular}
			\par\end{centering}
	\end{table}
\end{description}

\section{Quantum veto protocols using IBM quantum computer\label{Sec5.3:Experimet-QAV}}
\begin{center}
	\begin{figure}[h!]
		\centering
		\includegraphics[width=0.8\paperwidth]{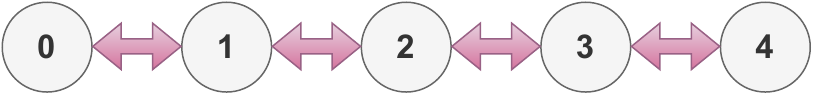} 
	\caption{Topology of the IBMQ Manila\label{Fig5.1:Topology-Manila}.}
\end{figure}
\end{center}

As already noted, IBM offers cloud access to different quantum computers, each differing in size (amount of qubits) and topology \cite{Ibm2021}. For the experimental implementation of the protocols described above, it have been assumed that there are four voters. Furthermore, Protocol A relies solely on the Bell states, making it feasible to implement using even a small quantum computer for a prototype demonstration. In comparison, a diverse range of entangled states can be used to implement the Protocol B. Naturally, the selection of unitary operations depends on the entangled state chosen. Table \ref{Tab5.1:unitary-veto} is shown the unitaries for voters for implementing Protocol B using either a 4-qubit (cluster) or 3-qubit (GHZ) entangled state. Here, the focus is limited to implementing Protocol B exclusively with these maximally entangled states. Consequently, a quantum computer with at least four qubits, capable of handling quantum computing tasks, will suffice for the requirements. A IBMQ Manila has been chosen due to its availability during the time experiments were conducted. IBMQ Manila is a quantum computer having five superconducting based qubit, with its topology illustrated in Figure \ref{Fig5.1:Topology-Manila}. The following sections provide a brief overview of the quantum circuits designed for implementing both protocols for quantum veto and its results obtained after executing these circuits on IBMQ Manila. A two-qubit quantum gate, such as a CNOT gate, can directly be implemented only between connecting qubits. Two qubits are connected by a bidirectional arrow that indicates either qubit appeared at the end of the bidirectional arrow can serve as the control qubit. Nevertheless, the accuracy of each CNOT gate may vary. The errors caused by implementing a CNOT gate are influenced by specific choice of qubit pair on which it is executed. To clarify this point, Table \ref{Tab5.2:Calibration-Manila} provides a list of parameters which indicates errors associated with different gate, while their execution on IBMQ Manila during the time experiment is performed. The table is limited to qubits 0, 1, 2, and 3, as shown in Figure \ref{Fig5.1:Topology-Manila} and labeled as $Q_{0}$, $Q_{1}$, $Q_{2}$, and $Q_{3}$ in Table \ref{Tab5.2:Calibration-Manila}, since these are the only qubits used in the experiment. The Bell state is prepared on qubit $Q_{0}$ and $Q_{1}$ for Protocol A while qubit $Q_{0},Q_{1}$ and
$Q_{2}$ are used to prepare GHZ state and the entire qubits $\{Q_{0},Q_{1},Q_{2},Q_{3}\}$ are used to prepare the four qubit cluster state.

\begin{table}
	\caption{The calibration data of IBMQ Manila on March 26, 2022. Here, cxi\_j, representing a $\rm{CNOT}$ gate with qubit i as the control and qubit j as the target.\label{Tab5.2:Calibration-Manila}}
	\begin{centering}
		\begin{tabular}{|c|c|c|>{\centering}p{1.6cm}|>{\centering}p{2cm}|>{\centering}p{2cm}|>{\centering}p{4.4cm}|}
			\hline 
			Qubit  & T1 ($\mu s$)  & T2 ($\mu s$)  & \centering{}Frequency (GHz)  & \centering{}Readout assignment error  & \centering{}Single-qubit Pauli-X-error  & \centering{}CNOT error\tabularnewline
			\hline 
			$Q_{0}$  & 138.5 & 96.57  & \centering{}4.962  & \centering{}$4.91\times10^{-2}$  & \centering{}$3.47\times10^{-4}$  & \centering{}cx0\_1: $6.312\times10^{-3}$\tabularnewline
			\hline 
			$Q_{1}$  & 138.98  & 70.49  & \centering{}4.838  & \centering{}$3.66\times10^{-2}$  & \centering{}$2.92\times10^{-4}$  & \centering{} cx1\_2: $6.312\times10^{-3}$, cx1\_0: $1.081\times10^{-2}$\tabularnewline
			\hline 
			$Q_{2}$  & 147.72 & 26.01 & \centering{}5.037  & \centering{}$2.32\times10^{-2}$  & \centering{}$2.51\times10^{-4}$  & \centering{}cx2\_3: $7.329\times10^{-3}$, cx2\_1:$1.112\times10^{-2}$\tabularnewline
			\hline 
			$Q_{3}$  & 186.97 & 53.8  & \centering{}4.951  & \centering{}$2.37\times10^{-2}$  & \centering{}$3.23\times10^{-4}$  & \centering{}cx3\_2: $7.329\times10^{-3}$\tabularnewline
			\hline 
		\end{tabular}
		\par\end{centering}
\end{table}

\subsection{Experimental realization of Protocol A}
\begin{figure}
	\begin{centering}
		\includegraphics[width=0.8\paperwidth]{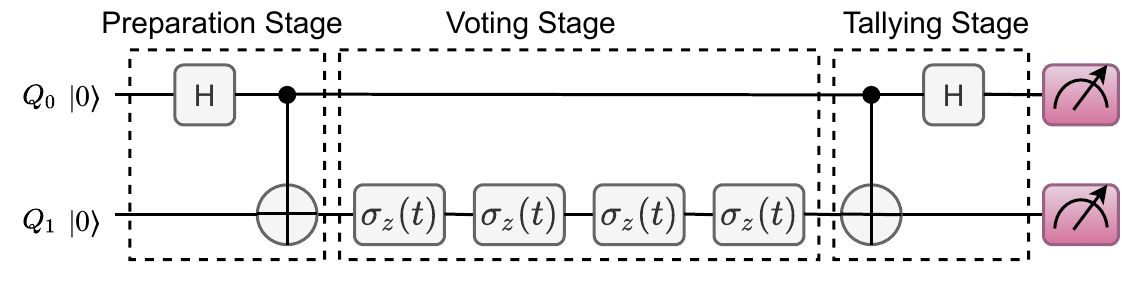} 
		\par\end{centering}
	\caption{A quantum circuit designed for the experimental realization
		of Protocol A. \label{Fig5.2:Ckt-ProtocolA}}
\end{figure}

Protocol A starts with the preparation of Bell state $|\phi_{int}\rangle=\frac{1}{\sqrt{2}}(|00\rangle+|11\rangle)$ by the VA, which is created using a Hadamard gate followed by a CNOT gate, as shown in the leftmost block of Figure \ref{Fig5.2:Ckt-ProtocolA}. Subsequently, the voters and VA adhere to the protocol and to accomplish this, the four voters sequentially apply unitary operations. The voter $V_{i}$ ($i\in\{0,1\}$) applies $\sigma_{z}(t)$. The choice of unitary $\sigma_{z}(t)$ is based on the iteration number and decision of voters to veto. A quantum circuit illustrating voting stage of the protocol is depicted in the central part of Figure \ref{Fig5.2:Ckt-ProtocolA}. This step results in the final state $|\phi_{fin}\rangle$, generated after all the four voters have cast their votes during a specific round of iteration. If $\langle\phi_{int}|\phi_{fin}\rangle=1$, then the result is inconclusive. Otherwise, the VA can determine that there is no unanimous agreement on the proposal, indicating that a veto has been made. To determine whether the final state fulfils the condition $\langle\phi_{int}|\phi_{fin}\rangle=1$, $|\phi_{fin}\rangle$ must be measured in the Bell basis. However, IBM quantum computer only allows measurement in computational basis. As a result, an attempt is made to perform measurement in Bell basis. For that, a reverse EPR circuit is used, which is constructed by applying CNOT gate followed by Hadamard gate (refer to block at right-side of Figure \ref{Fig5.2:Ckt-ProtocolA}). Figure \ref{Fig5.2:Ckt-ProtocolA} illustrates the complete quantum circuit design for Protocol A to execute on IBMQ Manila. Given that we have four voters, the protocol requires a maximum of $(1+\log_{2}4=3)$ iterations to reach a conclusive result.

There are total five different possibilities of voting patterns in this scenario: no voter has vetoed, one voter among four has vetoed, two voters among four have vetoed, three voters among four have vetoed, and all voters have done veto. The expected result outcomes for all different possible voting patterns for each iteration are presented in the sixth column of Table \ref{Tab5.3:Outcome_Bell_QV}. The remaining task is to verify whether the experimental result obtained after executing the circuit depicted in Figure \ref{Fig5.2:Ckt-ProtocolA}, on IBMQ Manila produces a consistent result (reported in the sixth column as the final state of Table \ref{Tab5.3:Outcome_Bell_QV}) with the expected result outcomes (reported in the eighth column as the simulator result of Table \ref{Tab5.3:Outcome_Bell_QV}).
\begin{table}
	\caption{Comparison between the theoretically expected results for implementing the quantum veto protocol using Bell state (Protocol A) and the experimentally obtained results. Here, $|\phi^{\pm}\rangle=\frac{1}{\sqrt{2}}\left(|00\rangle+|11\rangle\right)$
		and simulator results corresponds to the measurement outcomes obtained by executing the circuit shown in Figure \ref{Fig5.2:Ckt-ProtocolA} on IBM
		Qasm simulator. \label{Tab5.3:Outcome_Bell_QV}}
	\centering{}%
	\begin{tabular}{|>{\centering}p{0.5cm}|>{\centering}p{1cm}|>{\centering}p{0.7cm}|>{\centering}p{2.2cm}|>{\centering}p{1.5cm}|>{\centering}p{0.9cm}|>{\centering}p{1.85cm}|>{\centering}p{1.47cm}|>{\centering}p{1.59cm}|>{\centering}p{1cm}|}
		\hline 
		S. No.  & Initial state  & Veto done  & Which voter(s) has (have) vetoed  & Iteration No.  & Final state  & Result  & Expected measurement outcome on simulator  & Probability of obtaining the expected result on real device  & Fidelity (\%)\tabularnewline
		\hline 
		1  & $|\phi^{+}\rangle$  & zero  & No one  & Iteration 1  & $|\phi^{+}\rangle$  & Inconclusive  & 00  & 0.971 & 97.85\tabularnewline
		\hline 
		2  & $|\phi^{+}\rangle$  & one  & Any one voter among the four voters  & Iteration 1  & $|\phi^{-}\rangle$  & Conclusive  & 10  & 0.940 & 94.54\tabularnewline
		\hline 
		\multirow{2}{0.5cm}{3} & \multirow{2}{1cm}{$|\phi^{+}\rangle$} & \multirow{2}{1.1cm}{two} & \multirow{2}{2.5cm}{Any two of the four voters (e.g., $1^{st}$ \& $3^{rd}$ or $3^{rd}$
			\& $4^{th}$)} & Iteration 1  & $|\phi^{+}\rangle$  & Inconclusive  & 00  & 0.963 & 96.70\tabularnewline
		\cline{5-10} \cline{6-10} \cline{7-10} \cline{8-10} \cline{9-10} \cline{10-10} 
		&  &  &  & Iteration 2  & $|\phi^{-}\rangle$  & Conclusive  & 10  & 0.939 & 94.40\tabularnewline
		\hline 
		4  & $|\phi^{+}\rangle$  & three  & Any three of the four voters (e.g., $1^{st}$, $2^{nd}$ \& $4^{th}$
		or$1^{st}$, $3^{rd}$ \& $4^{th}$)  & Iteration 1  & $|\phi^{-}\rangle$  & Conclusive  & 10  & 0.934 & 95.55\tabularnewline
		\hline 
		\multirow{3}{0.5cm}{5} & \multirow{3}{1cm}{$|\phi^{+}\rangle$ } & \multirow{3}{1.1cm}{four} & \multirow{3}{2.5cm}{All the four voters } & Iteration 1  & $|\phi^{+}\rangle$  & Inconclusive  & 00  & 0.964 & 97.21\tabularnewline
		\cline{5-10} \cline{6-10} \cline{7-10} \cline{8-10} \cline{9-10} \cline{10-10} 
		&  &  &  & Iteration 2  & $|\phi^{+}\rangle$  & Inconclusive  & 00  & 0.969 & 96.54\tabularnewline
		\cline{5-10} \cline{6-10} \cline{7-10} \cline{8-10} \cline{9-10} \cline{10-10} 
		&  &  &  & Iteration 3  & $|\phi^{-}\rangle$  & Conclusive  & 10  & 0.927 & 92.36 \tabularnewline
		\hline 
	\end{tabular}
\end{table}

\begin{figure}
	\begin{centering}
		\begin{tabular}{c}
			\includegraphics[scale=1]{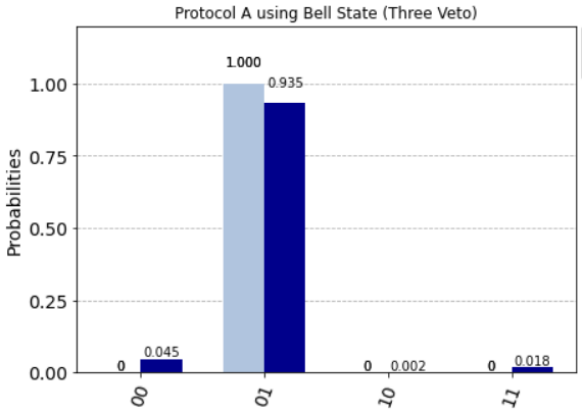}\\
			(a)\\
			\includegraphics[scale=1]{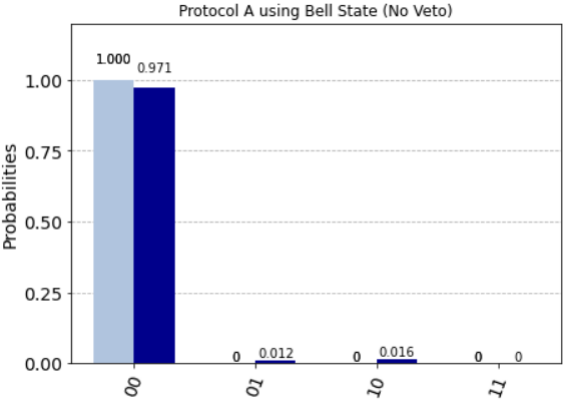}\\
			(b)
		\end{tabular}
		\par\end{centering}
	\caption{Results obtained from the real device and simulator for Protocol A using Bell state for the (a) conclusive and (b)  inconclusive outcomes. \label{Fig5.3:Result_ProtocolA}}
\end{figure}

The quantum circuit depicted in Figure \ref{Fig5.2:Ckt-ProtocolA} has been experimentally implemented on IBMQ Manila. Additionally, the circuit was executed using the Qasm simulator available at IBM Quantum platform. Each experiment on the real device was conducted with total $8192$ shots. As anticipated, the result obtained through simulation is perfectly aligned with the theoretical expectations reported in the sixth column of Table \ref{Tab5.3:Outcome_Bell_QV}. However, because of the intrinsic error while implementing quantum gates which is shown in Table \ref{Tab5.2:Calibration-Manila}), the experimental results differ slightly from the expected result outcomes, as shown in the last two columns of Table \ref{Tab5.3:Outcome_Bell_QV}. In particular, the probability of obtaining the expected result is provided in the last two columns of Table \ref{Tab5.3:Outcome_Bell_QV}. In the actual experimental scenario, a slightly different outcome (in comparison to the expected outcome) is obtained and the difference is quantified via fidelity. The term fidelity is given as $F(\sigma,\rho)=Tr\left[\sqrt{{\sqrt{{\sigma}}\rho\sqrt{{\sigma}}}}\right]^{2}$, where $\sigma$ is theoretical density state matrix and $\rho$ is the experimental density state matrix. To calculate fidelity, the density matrix state is needed, which is estimated using a technique called \acrfull{QST}. A detailed discussion on the \acrshort{QST} technique is already mentioned in Chapter \ref{Ch1: Introduction}. This technique is available as a built-in feature of IBM Quantum experience. The fidelity variation from 92.36\% to 97.85\% is shown in the last column of Table \ref{Tab5.3:Outcome_Bell_QV}. This indicates that the experimental results obtained are very close to the theoretical result, demonstrating that the schemes of quantum veto can be implemented with quite high probability of success. To demonstrate this idea, Figure \ref{Fig5.3:Result_ProtocolA} presents the experimental outcomes of Protocol A for the case of zero and three veto, alongside the obtained result from both simulator and real quantum device. The closeness of the results shows that the \acrshort{QAV} was successfully implemented.\\
Furthermore, the consistency of the results obtained was checked by conducting $10$ run of experiment independently, each with $8192$ shots, for the scenario where three of the four voters have applied veto. The obtained fidelities (in \%) are 95.13, 95.65, 95.53, 95.91, 96.88, 95.84,
93.87, 95.73, 94.96 and 95.39 for $10$ runs of the experiment. This dataset exhibits a standard deviation of approximately $0.77$, demonstrating a high level of consistency in the results.

It is worth noting that as the number of voters increases, the quantum circuit size (directly depends on the number of gates used in the circuit) will also grow since each voter applies their respective unitary operation. The increase in gate count results in introducing an error to some degree, leading to a natural decline in fidelity. However, the \acrshort{QAV} scheme can be realized in practical scenarios with currently available technology, such as the UN Security Council where the right to veto is given to five members only.

\subsection{Experimental realization of Protocol B}
\begin{figure}
	\begin{centering}
		\includegraphics[scale=0.6]{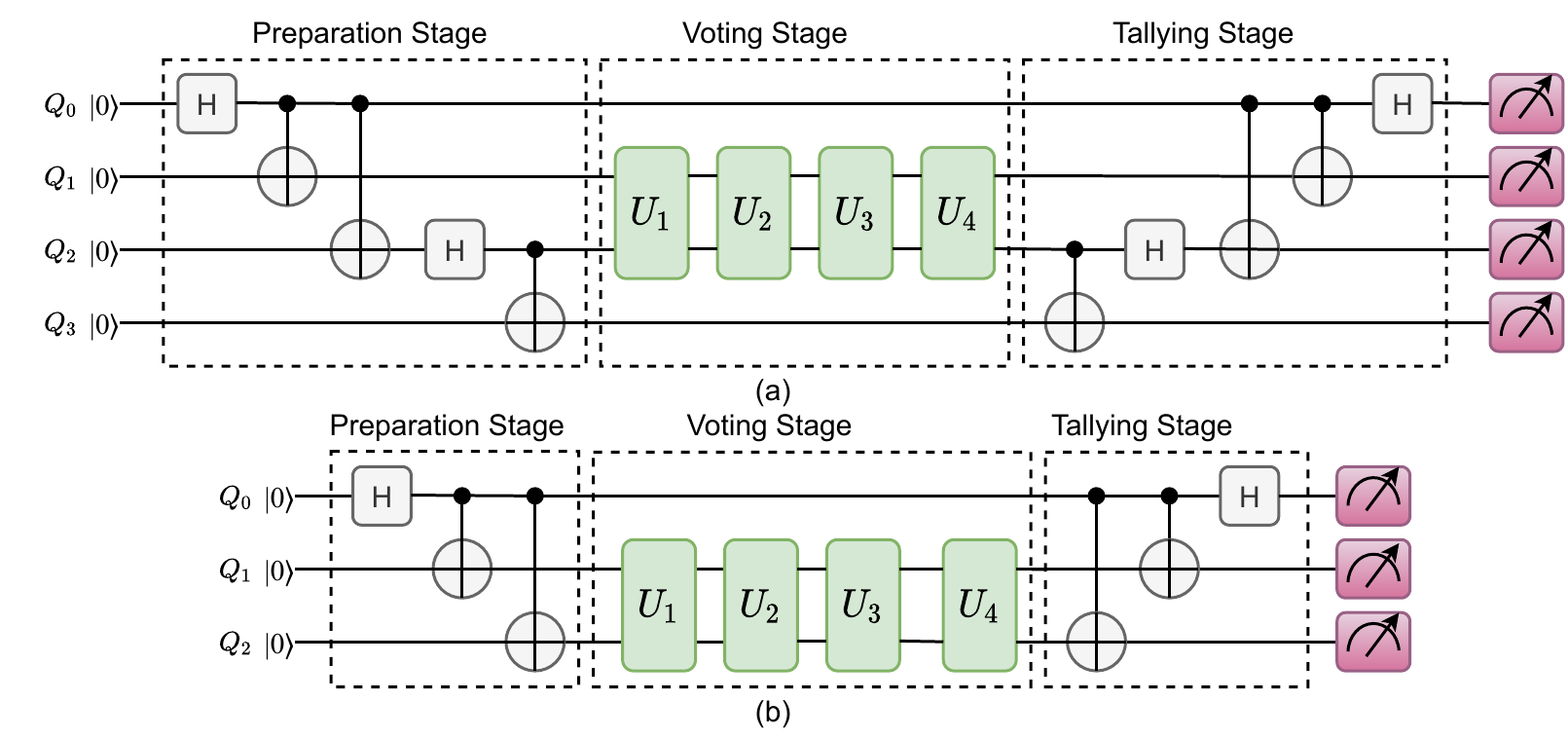} 
	\end{centering}
	\caption{A quantum circuit for implementing Protocol B using (a) the cluster state and (b) the GHZ state. \label{Fig5.4:Ckt-ProtocolB}}
\end{figure}

\begin{figure}
	\begin{centering}
		\begin{tabular}{cc}
			\includegraphics[scale=0.65]{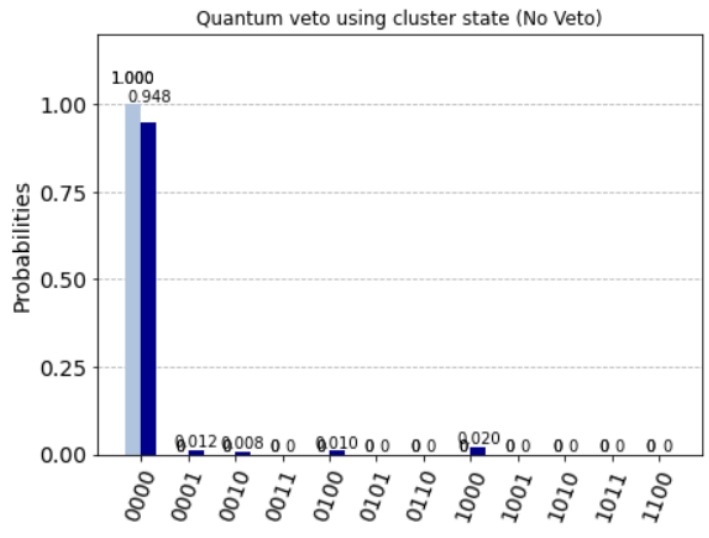} &
			\includegraphics[scale=0.65]{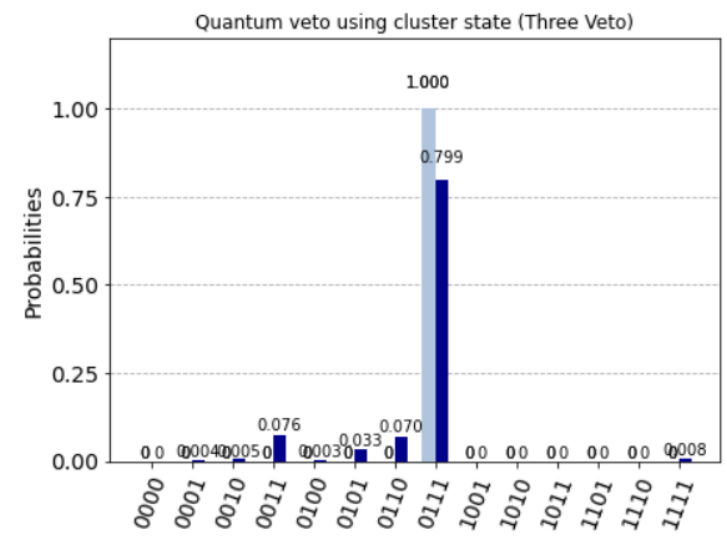}\\
			(a) & (b)\\
			\includegraphics[scale=0.65]{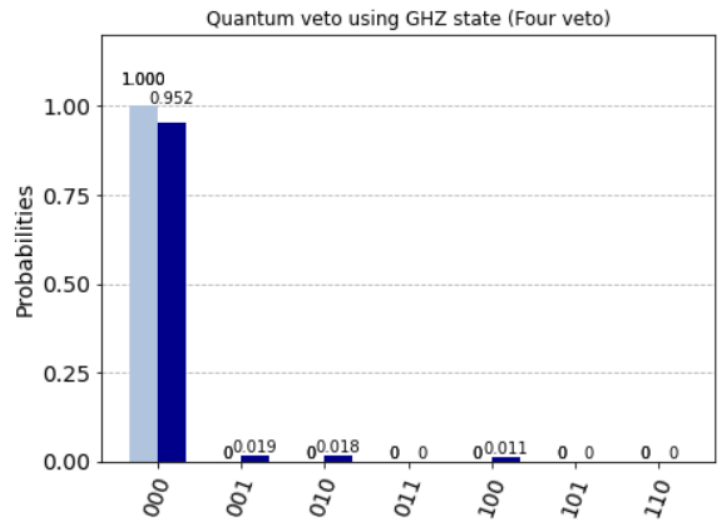} &
			\includegraphics[scale=0.65]{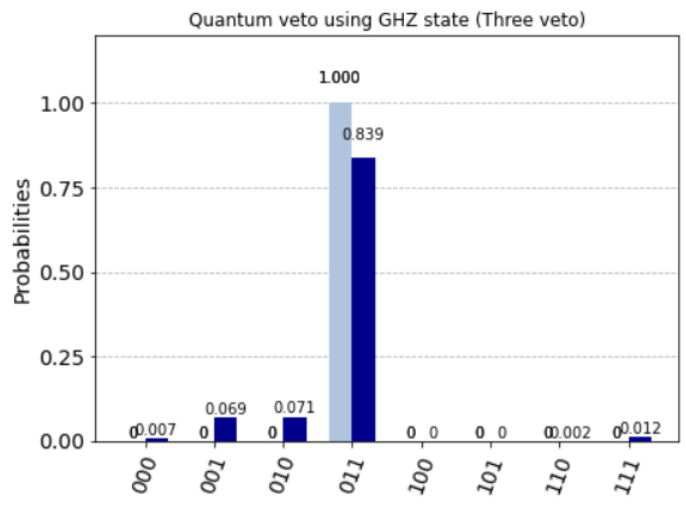}\\
			(c) & (d)
		\end{tabular}
		\par\end{centering}
	\caption{Result obtained from the real device and simulator for Protocol B using cluster state and GHZ state, showing the (a,c) inconclusive and (b,d) conclusive outcomes. \label{Fig5.5:Result-ProtocolB}}
\end{figure}

Protocol B is also realized for the case of four voters. The quantum circuit designed to realize Protocol B is classified into three parts. First is the preparation stage which prepares an entangled quantum state (see leftmost part of Figure \ref{Fig5.4:Ckt-ProtocolB} (a) and (b), which can generate a four-qubit cluster state $|\psi_{c}\rangle=\frac{1}{\sqrt{2}}(|0000\rangle+|0011\rangle+|1100\rangle-|1111\rangle)$
and three-qubit GHZ state $|\psi_{GHZ}\rangle=\frac{1}{\sqrt{2}}(|000\rangle+|111\rangle)$,
respectively). Next is the voting stage where all the voters cast their votes. The voting is done in accordance with the rule described in Table \ref{Tab5.1:unitary-veto}. All voters sequentially cast their votes and finally send their traveling qubits to the VA. In the final stage (also called as tallying stage), where the VA tallies all the votes by measuring all the qubits in the basis used for the preparation of the entangled state in the preparation stage. However, IBM only allows to perform measurement in the computational basis. To perform a measurement in some other basis, an inverse circuit is used to map into the computational basis measurement. In this case, an inverse of the circuit used in the preparation stage is used to map into the computational basis measurement outcome (see the rightmost part of Figure \ref{Fig5.4:Ckt-ProtocolB}). The seventh columns of Table \ref{Tab5.4:Result-Cluster-ProtocolB} and \ref{Tab5.5:Result-GHZ-ProtocolB} show the expected results for the five different possible ways of the voting scenario in the case of four voters. The result obtained from executing the circuit designed for Protocol B using the cluster and GHZ state on IBM Qasm simulator is shown in the eighth column of the Table \ref{Tab5.4:Result-Cluster-ProtocolB} and \ref{Tab5.5:Result-GHZ-ProtocolB} respectively. The obtained result is found to be consistent with the expected outcomes. The last two columns of both the tables show the experimental result obtained after executing the circuits using a real quantum computer (IBMQ Manila). The experimental results are found to be slightly different from the expected result due to intrinsic errors of the system use. Figure \ref{Fig5.5:Result-ProtocolB} also illustrates the result obtained from both the simulator and real quantum device for no veto and three veto cases. In particular, Figure \ref{Fig5.5:Result-ProtocolB} (a, b) and (c, d) illustrates the result obtained from implementing Protocol B using four-qubit cluster state and three-qubit GHZ state, respectively. This clearly shows the successful demonstration of Protocol B on for \acrshort{QAV}.
\newpage
\begin{longtable}{|>{\centering}p{0.5cm}|>{\centering}p{1cm}|>{\centering}p{0.7cm}|>{\centering}p{1.5cm}|>{\centering}p{3cm}|>{\centering}p{1cm}|>{\centering}p{1.5cm}|>{\centering}p{1.7cm}|>{\centering}p{1.2cm}|}\caption{Results obtained from the implementation of quantum veto protocol using the cluster state (Protocol B). Here, $|\psi_{c}\rangle=\frac{1}{2}(|0000\rangle+|0011\rangle+|1100\rangle-|1111\rangle)$ \label{Tab5.4:Result-Cluster-ProtocolB}}\\
		\hline 
		{S. No.} & {Initial State } & {Veto done} & {Which voter(s) has (have) vetoed} & {Final State } & {Is the result conclusive? } & {Expected measurement outcome on simulator} & {Probability of getting expected result on real device } & {Fidelity (\%)}\tabularnewline
		\hline 
		{1 } & {$|\psi_{c}\rangle$ } & {zero} & {No one } & {$\frac{1}{2}(|0000\rangle+|0011\rangle+|1100\rangle-|1111\rangle)$ } & {No } & {0000 } & 0.948 & 94.65\tabularnewline
		\hline 
		\multirow{4}{0.5cm}{{\centering{2}}} & \multirow{4}{1cm}{{\centering{$|\psi_{c}\rangle$}}} & \multirow{4}{1cm}{{\centering{one}}} & {$1^{st}$ } & {$\frac{1}{2}(|0101\rangle-|0110\rangle-|1001\rangle-|1010\rangle)$ } & {Yes } & {1111 } & 0.837 & 86.92\tabularnewline
		\cline{4-9} \cline{5-9} \cline{6-9} \cline{7-9} \cline{8-9} \cline{9-9} 
		&  &  & {$2^{nd}$ } & {$\frac{1}{2}(|0100\rangle-|0111\rangle+|1000\rangle+|1011\rangle)$ } & {Yes } & {0110 } & 0.907 & 91.32\tabularnewline
		\cline{4-9} \cline{5-9} \cline{6-9} \cline{7-9} \cline{8-9} \cline{9-9} 
		&  &  & {$3^{rd}$ } & {$\frac{1}{2}(-|0100\rangle+|0111\rangle+|1000\rangle+|1011\rangle)$ } & {Yes } & {1110 } & 0.860 & 88.38\tabularnewline
		\cline{4-9} \cline{5-9} \cline{6-9} \cline{7-9} \cline{8-9} \cline{9-9} 
		&  &  & {$4^{th}$ } & {$\frac{1}{2}(-|0101\rangle+|0110\rangle-|1001\rangle-|1010\rangle)$ } & {Yes } & {0111 } & 0.857 & 87.74\tabularnewline
		\hline 
		\multirow{6}{0.5cm}{{\centering{3}}} & \multirow{6}{1cm}{{\centering{$|\psi_{c}\rangle$}}} & \multirow{6}{1cm}{{\centering{two}}} & {$1^{st}$ \& $2^{nd}$ } & {$\frac{1}{2}(-|0001\rangle-|0010\rangle+|1101\rangle-|1110\rangle)$ } & {Yes } & {1001 } & 0.878 & 91.76\tabularnewline
		\cline{4-9} \cline{5-9} \cline{6-9} \cline{7-9} \cline{8-9} \cline{9-9} 
		&  &  & {$1^{st}$ \& $3^{rd}$ } & {$\frac{1}{2}(|0001\rangle+|0010\rangle+|1101\rangle-|1110\rangle)$ } & {Yes } & {0001 } & 0.917 & 93.65\tabularnewline
		\cline{4-9} \cline{5-9} \cline{6-9} \cline{7-9} \cline{8-9} \cline{9-9} 
		&  &  & {$1^{st}$ \& $4^{th}$ } & {$\frac{1}{2}(|0000\rangle+|0011\rangle-|1100\rangle+|1111\rangle)$ } & {Yes } & {1000 } & 0.920 & 93.12\tabularnewline
		\cline{4-9} \cline{5-9} \cline{6-9} \cline{7-9} \cline{8-9} \cline{9-9} 
		&  &  & {$2^{nd}$ \& $3^{rd}$ } & {$\frac{1}{2}(-|0000\rangle-|0011\rangle+|1100\rangle-|1111\rangle)$ } & {Yes } & {1000 } & 0.924 & 94.12\tabularnewline
		\cline{4-9} \cline{5-9} \cline{6-9} \cline{7-9} \cline{8-9} \cline{9-9} 
		&  &  & {$2^{nd}$ \& $4^{th}$ } & {$\frac{1}{2}(-|0001\rangle-|0010\rangle-|1101\rangle+|1110\rangle)$ } & {Yes } & {0001 } & 0.925 & 94.13\tabularnewline
		\cline{4-9} \cline{5-9} \cline{6-9} \cline{7-9} \cline{8-9} \cline{9-9} 
		&  &  & {$3^{rd}$ \& $4^{th}$ } & {$\frac{1}{2}(-|0001\rangle-|0010\rangle+|1101\rangle-|1110\rangle)$ } & {Yes } & {1001 } & 0.898 & 90.63\tabularnewline
		\hline 
		\multirow{4}{0.5cm}{{\centering{4}}} & \multirow{4}{1cm}{{\centering{$|\psi_{c}\rangle$}}} & \multirow{4}{1cm}{{\centering{three}}} & {$1^{st}$, $2^{nd}$ \& $3^{rd}$ } & {$\frac{1}{2}(-|0101\rangle+|0110\rangle-|1001\rangle-|1010\rangle)$ } & {Yes } & {0111 } & 0.825 & 89.56\tabularnewline
		\cline{4-9} \cline{5-9} \cline{6-9} \cline{7-9} \cline{8-9} \cline{9-9} 
		&  &  & {$1^{st}$, $2^{nd}$ \& $4^{th}$ } & {$\frac{1}{2}(-|0100\rangle+|0111\rangle+|1000\rangle+|1011\rangle)$ } & {Yes } & {1110 } & 0.799 & 89.15\tabularnewline
		\cline{4-9} \cline{5-9} \cline{6-9} \cline{7-9} \cline{8-9} \cline{9-9} 
		&  &  & {$1^{st}$, $3^{rd}$ \& $4^{th}$ } & {$\frac{1}{2}(-|0100\rangle+|0111\rangle-|1000\rangle-|1011\rangle)$ } & {Yes } & {0110 } & 0.791 & 91.69\tabularnewline
		\cline{4-9} \cline{5-9} \cline{6-9} \cline{7-9} \cline{8-9} \cline{9-9} 
		&  &  & {$2^{nd}$, $3^{rd}$ \& $4^{th}$ } & {$\frac{1}{2}(-|0101\rangle+|0110\rangle+|1001\rangle+|1010\rangle)$ } & {Yes } & {1111 } & {0.792} & 86.78\tabularnewline
		\hline 
		{5} & {$|\psi_{c}\rangle$ } & {four} & {$1^{st}$, $2^{nd}$, $3^{rd}$ \& $4^{th}$ } & {$\frac{1}{2}(|0000\rangle+|0011\rangle+|1100\rangle-|1111\rangle)$ } & {No } & {0000 } & 0.949 & 95.65\tabularnewline
	  \hline
\end{longtable}

\begin{table}
	\caption{Results obtained from the implementation of quantum veto protocol using the GHZ state (Protocol B). Here, $|\psi_{GHZ}\rangle=\frac{1}{\sqrt{2}}(|000\rangle+|111\rangle)$. }
	\label{Tab5.5:Result-GHZ-ProtocolB}
	\begin{centering}
		\begin{tabular}{|>{\centering}p{0.5cm}|>{\centering}p{1.2cm}|>{\centering}p{0.7cm}|>{\centering}p{1.5cm}|>{\centering}p{2.8cm}|>{\centering}p{1cm}|>{\centering}p{1.5cm}|>{\centering}p{1.7cm}|>{\centering}p{1.2cm}|}
			\hline 
			{S. No.} & {Initial State } & {Veto done} & {Which voter(s) has (have) vetoed} & {Final State } & {Is the result conclusive? } & {Expected measurement outcome on simulator} & {Probability of getting expected result on real device } & {Fidelity (\%)}\tabularnewline
			\hline 
			{\centering{1}} & {$\centering{|\psi_{GHZ}\rangle}$ } & {\centering{zero} } & {No one } & {$\frac{1}{\sqrt{2}}(|000\rangle+|111\rangle)$} & {No} & {000} & 0.924 & 95.98\tabularnewline
			\hline
			\multirow{4}{0.5cm}{\centering{2}} & \multirow{4}{1cm}{\centering{$|\psi_{GHZ}\rangle$}} & \multirow{4}{1cm}{one}
			& {$1^{st}$} & {$\frac{1}{\sqrt{2}}(|010\rangle+|101\rangle)$} & {Yes} & {010} & 0.891 & 93.54\tabularnewline
			\cline{4-9} \cline{5-9} \cline{6-9} \cline{7-9} \cline{8-9} \cline{9-9}
			&  &  & {$2^{nd}$} & {$\frac{1}{\sqrt{2}}(|011\rangle+|100\rangle)$} & {Yes} & {011} & 0.905 & 91.27\tabularnewline
			\cline{4-9} \cline{5-9} \cline{6-9} \cline{7-9} \cline{8-9} \cline{9-9} 
			&  &  & {$3^{rd}$} & {$\frac{1}{\sqrt{2}}(-|011\rangle+|100\rangle)$} & {Yes} & {111} & 0.870 & 89.19\tabularnewline
			\cline{4-9} \cline{5-9} \cline{6-9} \cline{7-9} \cline{8-9} \cline{9-9} 
			&  &  & {$4^{th}$} & {$\frac{1}{\sqrt{2}}(-|010\rangle+|101\rangle)$} & {Yes} & {110} & 0.864 & 91.23\tabularnewline
			\hline 
			\multirow{6}{0.5cm}{\centering{3}} & \multirow{6}{1cm}{$|\psi_{GHZ}\rangle$} & \multirow{6}{1cm}{two} &
			$1^{st}$ \& $2^{nd}$ & {$\frac{1}{\sqrt{2}}(|001\rangle+|110\rangle)$ } & {Yes} & {001} & 0.923 & 95.02\tabularnewline
			\cline{4-9} \cline{5-9} \cline{6-9} \cline{7-9} \cline{8-9} \cline{9-9}
			&  &  & {$1^{st}$ \& $3^{rd}$} & {$\frac{1}{\sqrt{2}}(-|001\rangle+|110\rangle)$} & {Yes} & {101} & 0.899 & 90.33\tabularnewline
			\cline{4-9} \cline{5-9} \cline{6-9} \cline{7-9} \cline{8-9} \cline{9-9}
			&  &  & {$1^{st}$ \& $4^{th}$} & {$\frac{1}{\sqrt{2}}(-|000\rangle+|111\rangle)$} & {Yes} & {100} & 0.948 & 94.81\tabularnewline
			\cline{4-9} \cline{5-9} \cline{6-9} \cline{7-9} \cline{8-9} \cline{9-9} 
			&  &  & {$2^{nd}$ \& $3^{rd}$} & {$\frac{1}{\sqrt{2}}(-|000\rangle+|111\rangle)$} & {Yes} & {100} & 0.918 & 93.75\tabularnewline
			\cline{4-9} \cline{5-9} \cline{6-9} \cline{7-9} \cline{8-9} \cline{9-9}
			&  &  & {$2^{nd}$ \& $4^{th}$} & {$\frac{1}{\sqrt{2}}(-|001\rangle+|110\rangle)$} & {Yes} & {101} & 0.904 & 90.92\tabularnewline
			\cline{4-9} \cline{5-9} \cline{6-9} \cline{7-9} \cline{8-9} \cline{9-9}
			&  &  & {$3^{rd}$ \& $4^{th}$} & {$\frac{1}{\sqrt{2}}(-|001\rangle-|110\rangle)$} & {Yes} & {001} & 0.904 & 94.04\tabularnewline
			\hline 
			\multirow{4}{0.5cm}{\centering{4}} & \multirow{4}{1cm}{\centering{$|\psi_{GHZ}\rangle$}} & \multirow{4}{1cm}{three} & {$1^{st}$, $2^{nd}$ \& $3^{rd}$} & {$\frac{1}{\sqrt{2}}(-|010\rangle+|101\rangle)$} & {Yes} & {110} & 0.839 & 91.00\tabularnewline
			\cline{4-9} \cline{5-9} \cline{6-9} \cline{7-9} \cline{8-9} \cline{9-9}
			&  &  & {$1^{st}$, $2^{nd}$ \& $4^{th}$} & {$\frac{1}{\sqrt{2}}(-|011\rangle+|100\rangle)$} & {Yes} & {111} & 0.832 & 87.84\tabularnewline
			\cline{4-9} \cline{5-9} \cline{6-9} \cline{7-9} \cline{8-9} \cline{9-9}
			&  &  & {$1^{st}$, $3^{rd}$ \& $4^{th}$} & {$\frac{1}{\sqrt{2}}(-|011\rangle-|100\rangle)$} & {Yes} & {011} & 0.875 & 90.01\tabularnewline
			\cline{4-9} \cline{5-9} \cline{6-9} \cline{7-9} \cline{8-9} \cline{9-9}
			&  &  & {$2^{nd}$, $3^{rd}$ \& $4^{th}$} & {$\frac{1}{\sqrt{2}}(-|010\rangle-|101\rangle)$} & {Yes} & {010} & 0.905 & 92.86\tabularnewline
			\hline 
			{5} & {$|\psi_{GHZ}\rangle$} & {four} & {$1^{st}$, $2^{nd}$, $3^{rd}$ \& $4^{th}$} & {$\frac{1}{\sqrt{2}}(-|000\rangle-|111\rangle)$} & {No} & {000} & 0.952 & 95.77\tabularnewline
			\hline 
		\end{tabular}
		\par\end{centering}
\end{table}

\section{A comparative analysis of Protocol A and Protocol B \label{Sec5.4:Comparison-Protocols-QAV}}
In the previous section, it has been clearly demonstrated that both the protocols (Protocol A and Protocol B) can be realized experimentally using IBMQ Manila with a few errors occurring while experimenting. However, a comparative analysis of the relative performance of these protocols has not yet been conducted. A comparative assessment can be made after examining the last two columns of the tables presented in this chapter (Table \ref{Tab5.3:Outcome_Bell_QV}, \ref{Tab5.4:Result-Cluster-ProtocolB} and \ref{Tab5.5:Result-GHZ-ProtocolB}). One can clearly observe that the fidelity for implementing Protocol A using the Bell state ranges from  92.36\% and 97.85\%, whereas the fidelity for implementing Protocol B using the cluster (GHZ) state ranges from 86.78\% and 95.65\% (87.84\% and 95.98\%). The fidelity ranges obtained clearly indicate that protocols using Bell state outperform those using the GHZ state, which in turn, perform better than the cluster state based schemes. This result is expected as preparing a complex entangled state leads to a greater challenge in creating and maintaining it. Till now, the performance of both protocols (Protocol A and B) has been analyzed under ideal conditions where there is no noise in the quantum channels. Now, the protocols are analyzed under a more realistic scenario that includes various noises in the channels which is discussed in the next section.
\begin{figure}[h!]
	\begin{centering}
		\begin{tabular}{c}
			\includegraphics[scale=0.8]{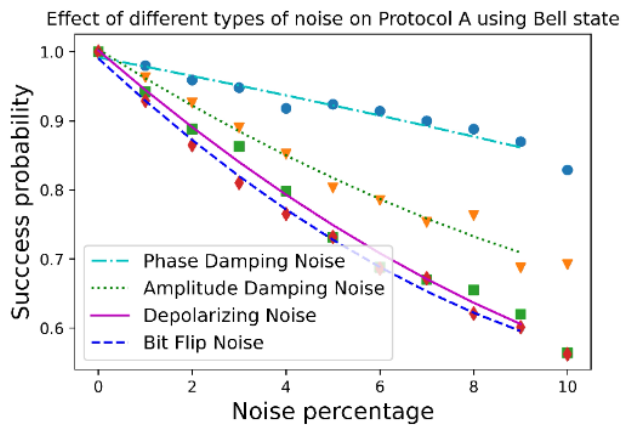}\\
			(a)\\
			\includegraphics[scale=0.8]{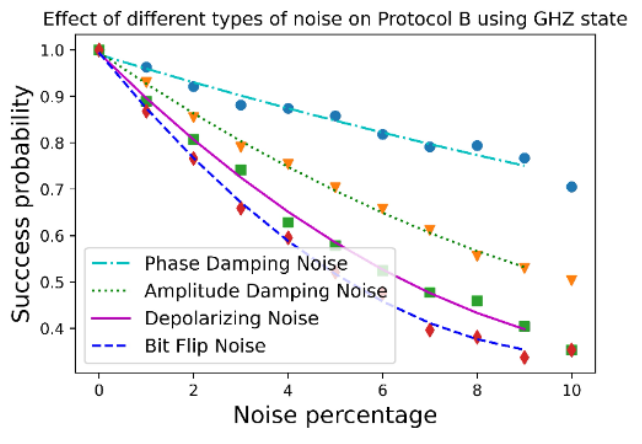}\\
			(b)\\
			\includegraphics[scale=0.8]{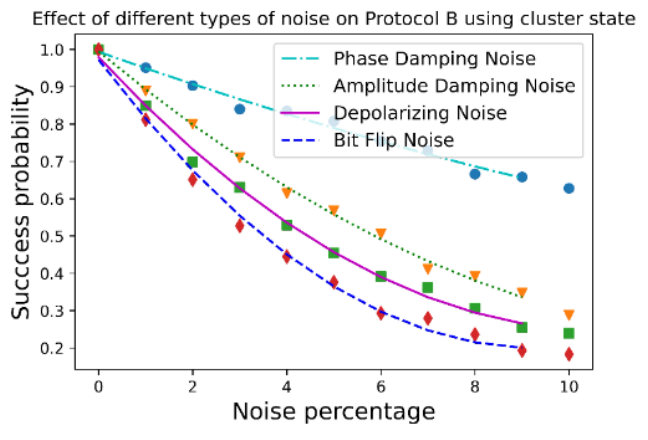}\\
			(c)
		\end{tabular}
		\par\end{centering}
	\caption{Impact of phase damping, amplitude damping, depolarization, and bit
		flip noise on: (a) Protocol A with the Bell state (b) Protocol B with
		the GHZ state (c) Protocol B with the cluster state. \label{Fig5.6:Noise Effect-QAV}}
\end{figure}

\subsection{Effect of noise\label{sec:Effect-of-noise}}
\begin{figure}[h!]
	\begin{centering}
		\begin{tabular}{cc}
			\includegraphics[scale=0.7]{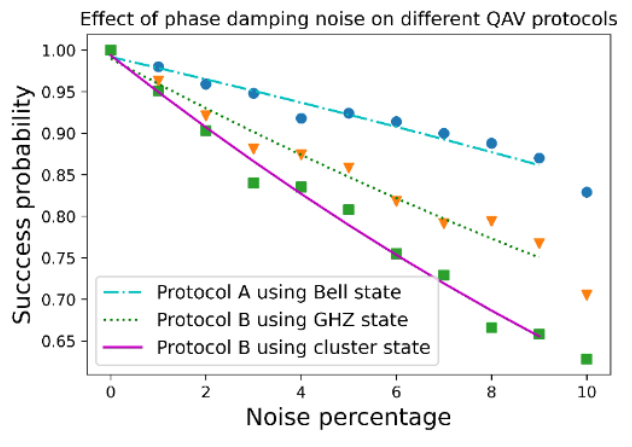} &
			\includegraphics[scale=0.7]{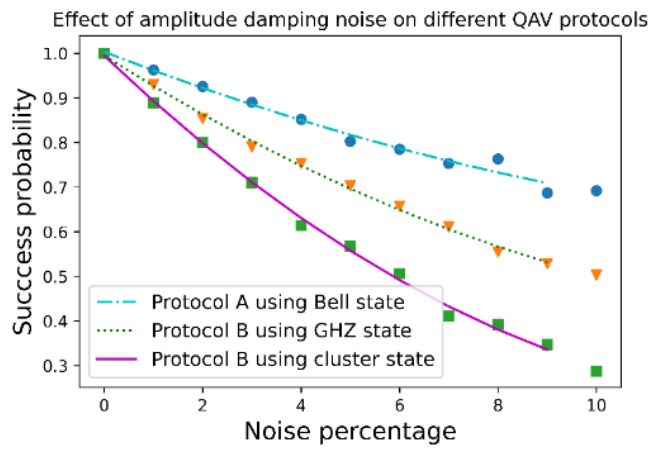}\\
			(a) & (b)\\
			\includegraphics[scale=0.7]{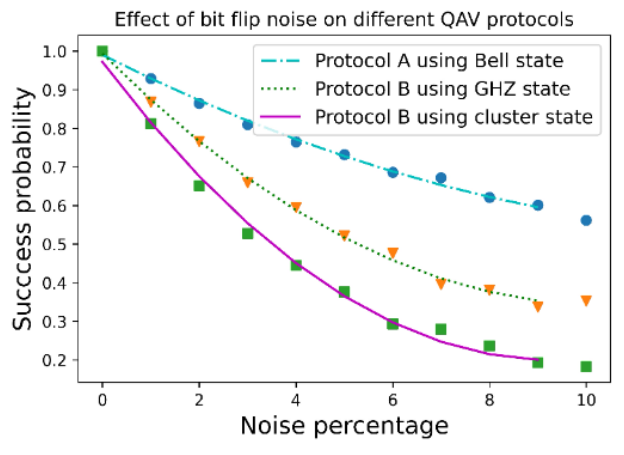} &
			\includegraphics[scale=0.7]{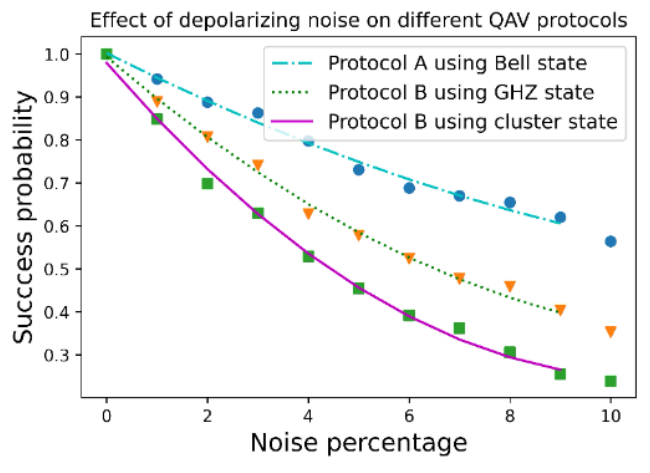}\\
			(c) & (d)
		\end{tabular}
		\par\end{centering}
	\caption{Impact of (a) phase damping (b) amplitude damping (c)
		bit flip and (d) depolarizing noise on Protocol A and Protocol B. \label{Fig5.7:Comparision-Noise-QAV}}
\end{figure}

An isolated quantum system without any environmental affect can not be physically possible. The environmental effect on a quantum system results in loss of its quantum mechanical features. Thus, it is important to study the effect of unwanted noise on the quantum system. Consequently, any quantum protocol is considered to be physically useful when it is robust against noise (for more details, see \cite{Anindita2017qDialogueNC, Anindita2018qConferenceNC, RishiD2016verifyQubitNC, Kishore2018qPrivateComparisionNC}).\\
Modeling of noise is already discussed in detail in Section \ref{sec1.5:Noise Model} of Chapter \ref{Ch1: Introduction} and further described in Section \ref{sec2.3.2:Effect-of-noise-MQT} of Chapter \ref{Ch2:MQT} and \ref{sec3.4:Effect-of-noise-QB} of Chapter \ref{Ch3:QB}. In Chapter \ref{Ch2:MQT} \& \ref{Ch3:QB}, the techniques to model noise in quantum circuits have already been discussed and also mentioned that the important noise models are available on qiskit \cite{NoiseQiskit}, an open source python library. Here, some relevant noises such as, bit-flip, amplitude damping, phase damping and depolarizing noise will mainly be discussed. The impact of these noises on the protocols (Protocol A and B) can be analyzed by examining the corresponding variations in fidelity compared to their ideal scenario. An approach different from the noise modeling technique available on qiskit was used to study the effect of two noisy conditions (amplitude and phase damping) on the \acrshort{QAV} protocols \cite{SandeepM2021QV}. Using the technique available on qiskit for modeling various noisy environments, Protocol A that uses Bell state and Protocol B that uses both the cluster state and the GHZ state are studied in various noisy conditions. To check the effect of these noises, success probability of the protocols stated in Section \ref{Sec5.2:Protocols-QAV} is plotted as function of noise percentage which is illustrated in Figure \ref{Fig5.6:Noise Effect-QAV}. The results obtained are found to be consistent with the findings of Reference \cite{SandeepM2021QV} and demonstrates that the effects of various noises are ranked in descending order (on basis of obtained fidelity) as bit-flip, depolarizing, amplitude damping and phase damping noise.\\
For a comparative study of protocols discussed in Section \ref{Sec5.2:Protocols-QAV}, success probability of the protocols are plotted as function of noise percentage for each noisy condition illustrated in Figure \ref{Fig5.7:Comparision-Noise-QAV}. Further, it is easily observed that schemes that uses Bell state as a quantum resource performs better than the schemes that uses a higher qubit entangled state as a quantum resource.

\section{Conclusion\label{Sec5.5:Conclusion-QAV}}
In real-life scenarios, voting plays a significant role in the decision-making process. Several schemes where voting is mandatory needs the anonymity of voters and also its security. As quantum mechanical features guarantee security, researchers have proposed several protocols for quantum voting since 2006. A particular type of quantum voting scenario known as \acrshort{QAV} where voters have given right to perform veto in anonymous sense, has drawn a special attention to the UN security council. A set of protocols for the \acrshort{QAV} have been recently proposed \cite{SandeepM2021QV}. Here, two efficient \acrshort{QAV} protocols (referred as Protocol A and B) have been implemented using a quantum computer (IBMQ Manila) available on cloud at IBM Quantum platform. Protocol A is realized using the Bell state whereas, Protocol B is realized using the cluster state and three-qubit GHZ state. Both protocols were successfully implemented with high success probabilities, and these are physically realizable for a small number of voters (as in the case of UN security council) with currently available technology. Additionally, it is noticed that Protocol A that uses Bell state outperforms Protocol B that uses higher qubit entangled state under both ideal and noisy conditions. Furthermore, on the basis of decreasing value of the fidelity, different noise models examined can be ranked in descending order as bit-flip, depolarizing, amplitude damping and phase damping noise.
\newpage

\chapter{EXPERIMENTAL REALIZATION OF DISTRIBUTED PHASE REFERENCE QKD PROTOCOLS}\label{Ch6:COW-DPS}
\graphicspath{{Chapter6/Chapter6Figs/}{Chapter6/Chapter6Figs/}}
\section{Introduction}\label{sec:intro-COW-DPS}
\begin{figure}[h!]
	\centering{}
	\includegraphics[scale=1]{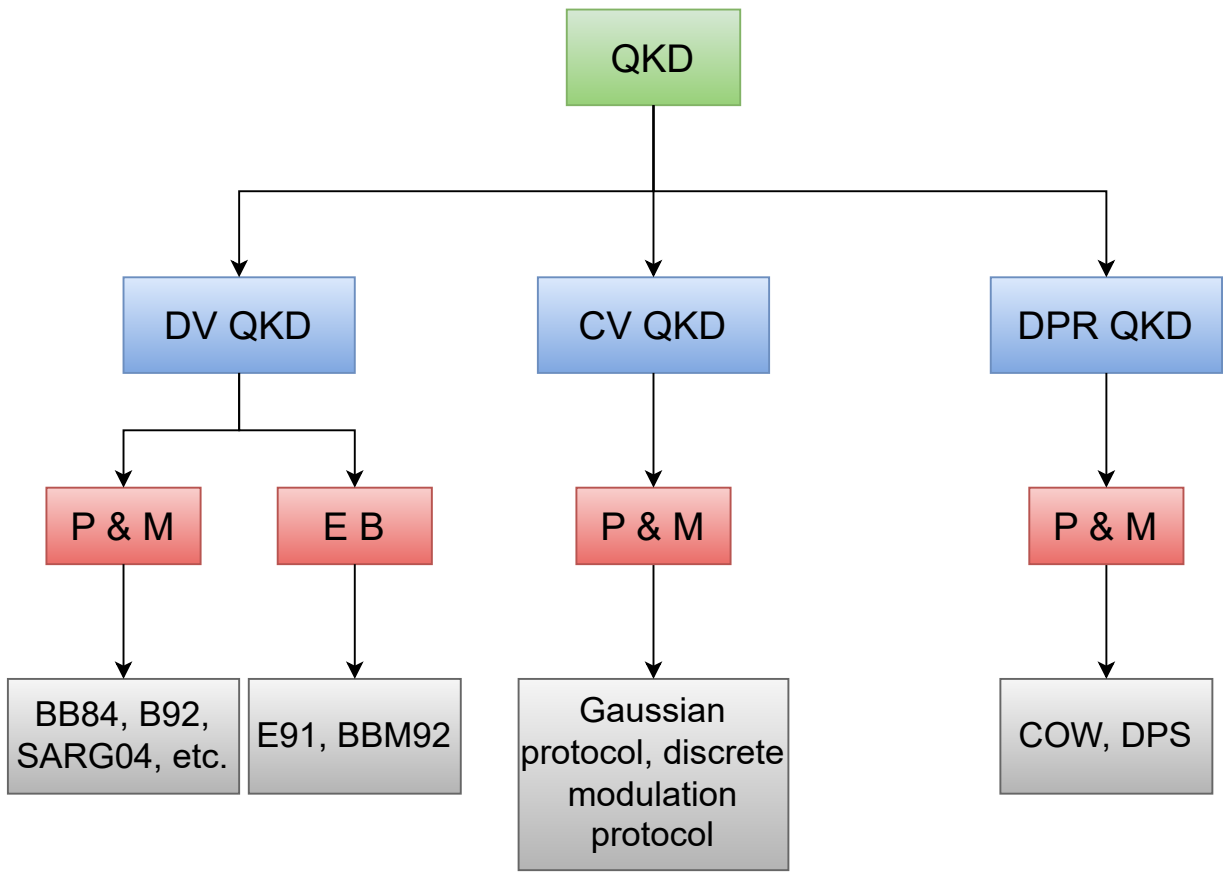}
	\caption{\label{fig6.1:QKD_structure} Classification of QKD where
		P \& M refers to prepare-and-measure-based and EB refers to entanglement based QKD protocol \cite{sharma2021quantum}.}
\end{figure}

Since ancient times, humans have made many efforts to securely exchange their secret information. A piece of information is said to be secret when the information is accessible to legitimate parties only. Ideally, an illegitimate party would not be able to extract the secret information. A field of research where protocols for exchanging such secret information between two or more authenticated parties are studied is known as cryptography. A process to generate a secret key using classical features (resources) is known as classical cryptography. The secrecy of the schemes used in modern classical cryptography relies on encoding or decoding through mathematically complex problems \cite{rothe2005complexity}. However, the advent of new algorithms or powerful computing devices leading to the capability of solving these complex problems within a certain time limit can break the secrecy of classical cryptography. Consequently, all schemes for classical cryptography are conditionally secure (as the security is conditioned on the computational capabilities of the adversaries). Thus, there is a need for an unconditionally secure cryptographic scheme. Quantum cryptography aims to address that need. Quantum cryptography can provide the capability to perform unconditionally secure communication as its security is guaranteed by quantum mechanical laws \cite{mayers2001unconditional} and is independent of the computational power of the adversaries. A particular process of generating (more specifically distributing) a secure key between two legitimate users with the help of quantum mechanical features (resources) is known as \acrfull{QKD}. The first-ever \acrshort{QKD} protocol was proposed by Charles Bennett and Gilles Brassard in 1984 \cite{Bennett1984QCrypt}, and since then, the field has advanced significantly \cite{Gisin2002QCrypt,pirandola2020advances,Pathak2017QCrypt,Mandeep2025QCryptRev,Curty2024QCryptSecurity,Soubusta2001QCrypt,Soubusta2013ExpEve,Genovese2010QKD}. With advancements in technology, \acrshort{QKD} has transitioned from a theoretical concept to experimental demonstration leading to commercial solutions with higher key generation rates and transmission distances. The most frequently used transmission channels in \acrshort{QKD} systems are free-space and optical fibers. Several countries have done field trials of the \acrshort{QKD} systems over the fiber-based networks in their metropolitan cities \cite{peev2009secoqc, sasaki2011field, wang2014field, tang2016measurement}. Recently, significant advancements have also been reported in the field of satellite-assisted quantum communications \cite{bedington2017progress,liao2017satellite,dai2020towards,Zeilinger2021satellite,Pirandola2021satellite}. The quantum cryptography as a field seems to be a very promising research area for many leading government and private organizations \cite{stanley2022recent}.\\
The existing \acrshort{QKD} protocols can be broadly classified into three main categories: (i) Discrete variable \acrshort{QKD} (DV QKD)
protocols \cite{Bennett1984QCrypt,Bennett1992QKD,Scarani2004QCrypt,Ekert1991QCrypt}, 
(ii) Continuous variable \acrshort{QKD} (CV QKD) protocols \cite{ralph1999continuous},
and (iii) \acrfull{DPR} \acrshort{QKD} protocols \cite{Inoue2002DPS-QKD,Stucki2005COW-QKD,Lucamarini2018TF-QKD,tamaki2012phase} (refer to Figure \ref{fig6.1:QKD_structure}). However, this classification is not unique, as both DV and CV version of many \acrshort{QKD} protocols (such as, BB84 and B92) does exist. Encoding processes are different for all these different categories. In DV QKD, information is encoded in discrete properties of a quantum state, such as the polarization of single photons. Examples include the BB84 protocol \cite{Bennett1984QCrypt}, the B92 protocol \cite{Bennett1992QKD}, and the SARG04 protocol \cite{Scarani2004QCrypt}. In CV QKD, non-classical features of squeezed or coherent states, such as their quadratures is used for encoding information. Examples include the discrete modulation protocol \cite{hillery2000quantum}, the Gaussian protocol \cite{cerf2001quantum}, and the CV-B92 protocol \cite{srikara2020continuous}. In \acrshort{DPR} \acrshort{QKD}, the encoding is done using the time arrival of weak pulse train or phase difference between the successive weak pulses. Examples include the \acrfull{COW} \acrshort{QKD} \cite{stucki2009continuous}, the \acrfull{DPS} \acrshort{QKD} \cite{inoue2005dpsPNS}, the \acrfull{DPTS} \acrshort{QKD} \cite{Bacco2016DPR-QKD}. It is to be noted that \acrshort{DPR} \acrshort{QKD} does not require an ideal single photon source for encoding instead a \acrfull{WCP} is used. Looking at the current state of technology, \acrshort{DPR} schemes are regarded as the best practical solution to realize \acrshort{QKD}. In other words, protocols for \acrshort{DPR} \acrshort{QKD} are easier to implement experimentally than others with higher key generation rates. Among the \acrshort{DPR} category \acrshort{QKD} protocols, \acrshort{COW} and \acrshort{DPS} have been studied most. The \acrshort{DPS} protocol for \acrshort{QKD} was initially introduced in the year $2002$ \cite{Inoue2002DPS-QKD}, followed by the introduction of the \acrshort{COW} protocol in the year $2005$ \cite{Stucki2005COW-QKD}. Since then, a significant advancement has been made in their experimental realization across various distances. An important thing to mention here is that the \acrshort{DPR} \acrshort{QKD} still faces a problem in achieving unconditional security \cite{gonzalez2020upper, mizutani2023finite,Bacco2016DPR-QKD,DaLio2019DPR-QKD}. However, in recent years, \acrfull{MDI} \acrshort{QKD} and \acrfull{TF} \acrshort{QKD} have gained prominence due to their improved security and enhanced key generation rates over long distances \cite{fan2022robust,Lucamarini2018TF-QKD,zhou2023experimental,xie2022breaking,Wang2022TF-QKD,Genovese2022TF-QKD,Genovese2024TF-QKD}. The ease of implementation of the \acrshort{DPR} family \acrshort{QKD} makes it popular for the field deployment. A comprehensive analysis in the major developments in the experimental implementation of \acrshort{DPR} \acrshort{QKD} will be presented later in this chapter. However, a focus will be solely on the experimental advancements in key rates for \acrshort{DPR} \acrshort{QKD} protocols.

Although the \acrshort{DPR} \acrshort{QKD} systems have been successfully implemented in various locations, several issues remain unaddressed, as the focus has primarily been on key generation rates and communication distances. During the implementation process, some of the key parameters such as \acrfull{DR}, \acrfull{CR} and \acrfull{DT} of the detector are often overlooked. The details of these key parameters will be discussed in Section \ref{sec6.2.3:postprocessing}. An important thing to mention here is that during any \acrshort{QKD} implementation process, the key generation rate across various distances is heavily influenced by these parameters. This chapter aims to analyze the influence of several parameters such as \acrshort{DR}, \acrshort{CR} and \acrshort{DT} of the detector on key generation rates over various distances. A similar work was performed earlier, where key rate analysis of \acrshort{COW} \acrshort{QKD} has been done in accordance with several parameters across various distances \cite{Malpani2024ExpCOW-QKD}. An identical analysis is done for \acrshort{DPS} \acrshort{QKD} in this chapter. In addition, an analysis of \acrshort{COW} \acrshort{QKD} is also done considering monitoring line to check eavesdropping, which was missing in the earlier work reported in Reference \cite{Malpani2024ExpCOW-QKD}. Furthermore, a comparison has also been made between both the \acrshort{COW} and the \acrshort{DPS} \acrshort{QKD} protocols.

\section{DPS and COW QKD protocols \label{sec:COW-protocol-definition}}
Two well-known protocols from \acrshort{DPR} family are \acrshort{COW} and \acrshort{DPS}. The similarities and differences between both protocols are summarized in Table \ref{tab6.1:COW-DPS}. Encoding in \acrshort{COW} \acrshort{QKD} is done using time bin information of pulses with average photon number $(\mu=\left|\alpha\right|^{2}=0.5)$, while encoding in \acrshort{DPS} \acrshort{QKD} is done using phase information of consecutive pulses with average photon number $(\mu=\left|\alpha\right|^{2}=0.2)$. The major advantage of both protocols is their robustness against the \acrfull{PNS} attack and their insensitivity towards polarization. Another difference between both the protocols is their usage of detector numbers. In \acrshort{COW} \acrshort{QKD}, three detectors are needed with one in the data line to collect key information and the other two in the monitoring line to check eavesdropping. It is to be noted that the \acrshort{COW} \acrshort{QKD} can also be implemented using two detectors as one detector can check eavesdropping on the monitoring line (counts in ${\rm D}_{M1}$ is reduced in case of eavesdropping). However, in \acrshort{DPS} \acrshort{QKD}, two detectors are needed.
\begin{table}
	\caption{\label{tab6.1:COW-DPS} A table to compare the COW and DPS QKD protocols.}
	\begin{centering}
		\begin{tabular}{|p{1.5cm}|p{3cm}|p{4cm}|p{4cm}|}
			\hline 
			S No. & Properties & DPS & COW\tabularnewline
			\hline 
			1 & Encoding & Phase difference between consecutive coherent pulse & Combining vacuum and coherent pulse \tabularnewline
			\hline 
			2 & Source & WCP & WCP\tabularnewline
			\hline 
			3 & $\mu$ & $0.5$ & $0.2$\tabularnewline
			\hline 
			4 & No of detectors & $2$ & $3$\tabularnewline
			\hline 
			5 & PNS attack effect & No & No\tabularnewline
			\hline 
			6 & Intensity & Constant & Modulated\tabularnewline
			\hline 
			7 & Phase & Modulated & Constant\tabularnewline
			\hline 
			8 & Polarization & Insensitive & Insensitive\tabularnewline
			\hline 
		\end{tabular}
		\par\end{centering}
\end{table}
A brief explanation of the steps involved in implementing both \acrshort{COW} and \acrshort{DPS} \acrshort{QKD} protocols is provided.

\subsection{DPS QKD}
\begin{figure}
	\begin{centering}
		\begin{tabular}{c}
			\includegraphics[scale=0.5]{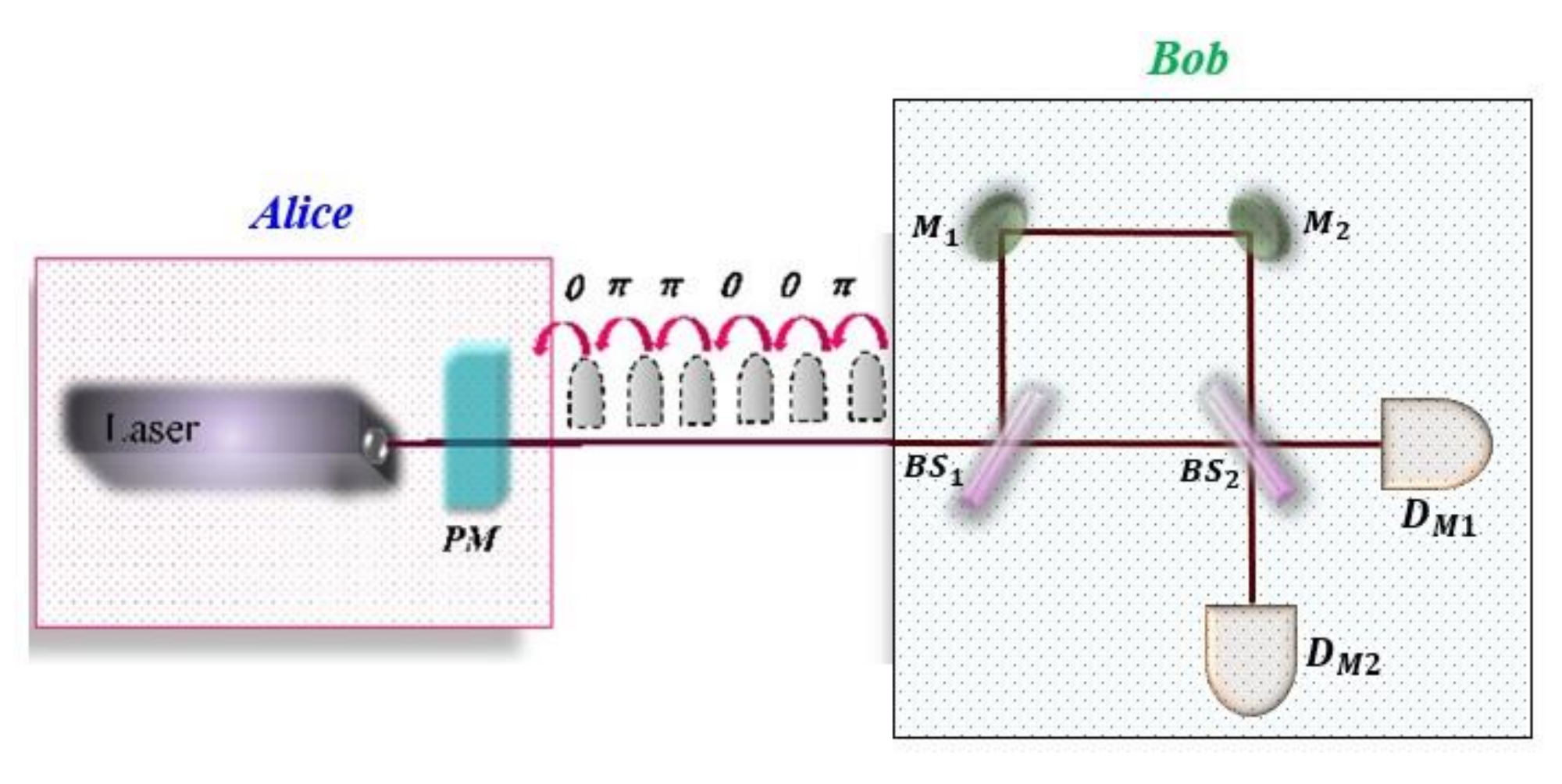}\tabularnewline
		\end{tabular}
		\par\end{centering}
		\caption{\label{fig6.2:DPS} A block diagram illustrating the DPS protocol. PM: phase
		modulator, ${\rm BS_{1}}$, ${\rm BS_{2}}$ are beam splitters and
		$M_{1}$, $M_{2}$ are mirrors.}
\end{figure}

The concept of \acrshort{DPS} \acrshort{QKD} protocol was first introduced in the year $2002$ by Inoue et al. \cite{Inoue2002DPS-QKD}. Alice starts the protocol by creating a \acrshort{WCP} with mean photon number $0.2$ and modulated their phase using a phase modulator to either $0$ or $\pi$. The modulated pulses are sent to Bob through an optical fiber. Bob passes the incoming pulses through a \acrfull{MZI} with $1$ bit delay. The pulses coming out from the \acrshort{MZI} are detected either using \acrfull{SPD}s or \acrfull{SNSPD}s. Among the pulse train, two consecutive pulses interfere within the \acrshort{MZI} which leads to the detection of one of the two detectors gives the phase difference information of the two pulses. In Figure \ref{fig6.2:DPS}, a click on $D_{M1}\,\left(D_{M2}\right)$ indicates a phase difference of $0\,(\pi)$ between two consecutive pulses. The step-wise explanation of the \acrshort{DPS} \acrshort{QKD} protocol are outlined as follows:
\begin{enumerate}
	\item Alice transmits a sequence of attenuated pulses with mean photon number below one (usually $0.2$). These attenuated pulses are modulated in phase $(0,\pi)$ using phase modulator.
	\item Bob utilizes photon detectors and \acrshort{MZI} with one bit delay to detect the phase modulated pulses and gain the phase information between two successive pulses and their arrival time.
	\item Bob communicates over the public classical channel about the click time of either of two detectors.
	\item Alice gains the click time information of Bob's detector.
	\item Alice and Bob assign bit $0 (1)$ in accordance to phase information between consecutive pulses $0 (\pi)$ and generate a shifted key at both ends.
	\item Final secret key is obtained after post-processing processes including error correction and privacy amplification on the shifted key.
\end{enumerate}

\subsection{COW QKD}
\begin{figure}
	\begin{centering}
		\begin{tabular}{c}
			\includegraphics[scale=0.5]{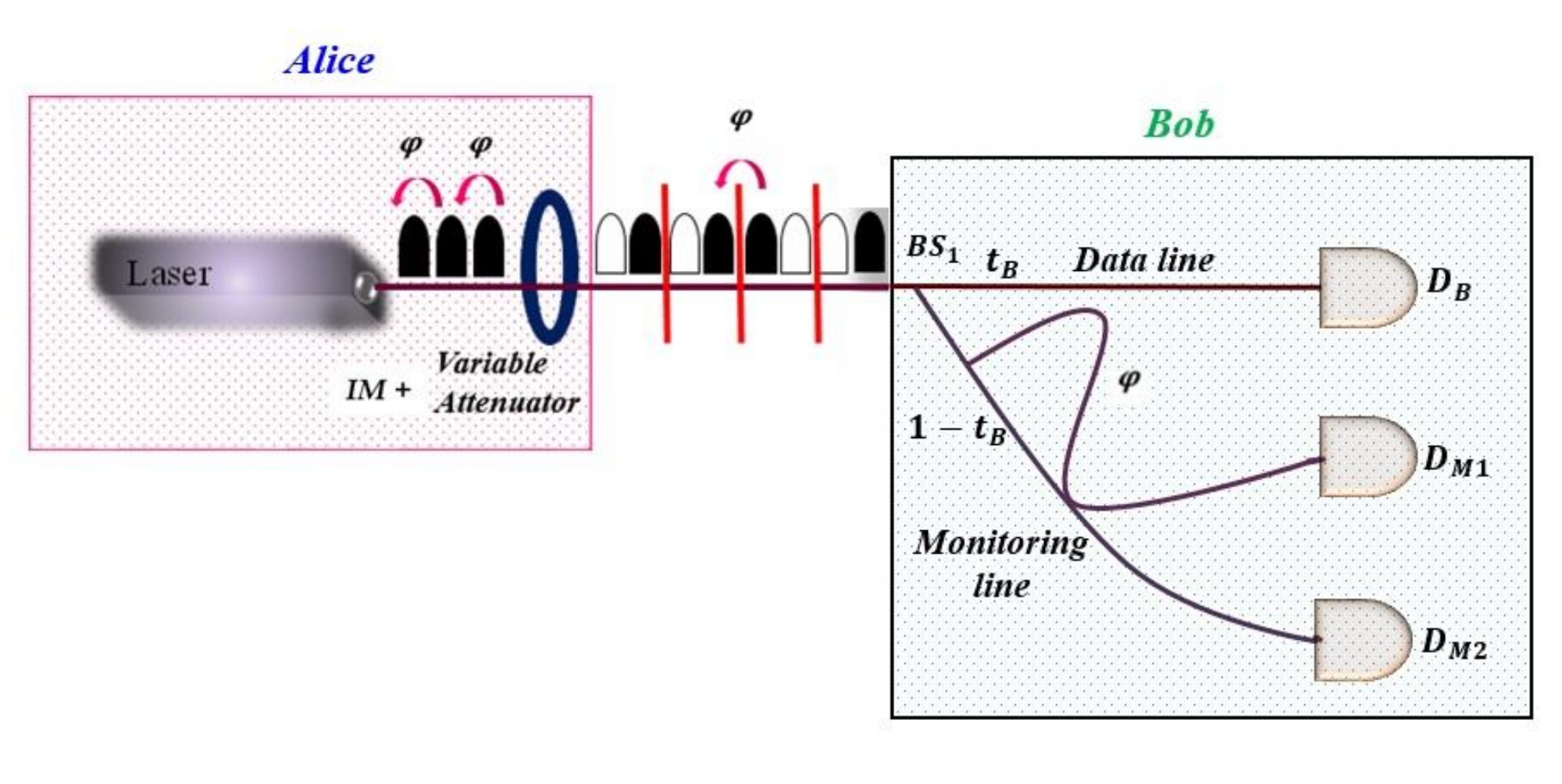}\tabularnewline
		\end{tabular}
		\par\end{centering}
	\caption{\label{fig6.3:COW} A block diagram illustrating the COW protocol. IM: intensity modulator, BS1: beamsplitter.}
\end{figure}

Encoding in the \acrshort{COW} \acrshort{QKD} protocol is done using a pair consisting of one empty pulse and one non-empty pulse. Steps involved in the protocol are outlined as follows:
\begin{enumerate}
	\item Alice generates a sequence of empty and non-empty pulses in different orders $\left|0\right\rangle \left|\alpha\right\rangle$ and $\left|\alpha\right\rangle \left|0\right\rangle$ corresponds to classical bit $1$, $0$ with each probability $\frac{\left(1-f\right)}{2}$. She adds two non-empty pulse sequences $\left|\alpha\right\rangle \left|\alpha\right\rangle$ with probability $f$ in between which are referred as decoy pulses. It is to be noted that the average photon number should be below $1$ $\left(\left|\alpha\right|^{2}<1\right)$. These sequences of pulses are sent to Bob over the quantum channel.
	\item Bob measures the arrival time of the $90\%$ pulse sequence using his detector $D_{{\rm B}}$ and record it, while the remaining $10\%$ are detected in the monitoring arm to check eavesdropping (refer to Figure \ref{fig6.3:COW}).\footnote{A $90:10$ distribution of pulses among the data and monitoring arm which is standard have been implemented. However, researchers have employed a $95:5$ distribution \cite{shaw2022optimal}. This approach results in the enhancement of the key generation rate, compromising security because of the significant reduction of pulses in the monitoring line.}
	\item Bob checks coherence between two consecutive non-empty pulses using detectors $D_{{\rm M1}}$ and $D_{{\rm M2}}$, presents in the monitoring arm. The \acrshort{MZI} is arranged in the monitoring arm such that only $D_{{\rm M1}}$ clicks in the absence of any eavesdropping (refer to Figure \ref{fig6.3:COW}). The protocol is aborted if detection at $D_{{\rm M2}}$ exceeds the threshold.
	\item A shifted key is obtained after arrival time matching of pulses at both ends Alice and Bob. Post-processing processes such as, error correction and privacy amplification on the shifted key produce a private secret key for both encryption and decryption.
\end{enumerate}

\subsection{Post-processing}\label{sec6.2.3:postprocessing}
The procedure described above for both \acrshort{COW} and \acrshort{DPS} \acrshort{QKD} helps in obtaining the raw key which cannot be used for encryption and decryption. The raw key should go through post-processing processes to obtain the secret key which can be used for encryption and decryption. Error correction and privacy amplification are the two major steps of post-processing processes. Errors in the raw key occur due to imperfection in the used devices and losses in data transmission. Eavesdropping in the channel may also cause errors. Error correction protocols are able to correct the errors introduced into the raw key. Privacy amplification protocols amplify the security of the final key with the cost of key size reduction. The main post-processing steps for \acrshort{QKD} are described as follows:
\begin{enumerate}
	\item {\bf Parameter estimation:} Error in the raw key arises due to various factors like, imperfection in optical device used, channel transmission loss, eavesdropping. It may be noted that Eve may hide behind the errors if we unable to distinguish the cause of error. A threshold is set to distinguish whether the error is due to eavesdropping or device imperfection. A portion of raw key is extracted to estimate the \acrfull{QBER} is known as \acrshort{DR}. The communication protocol aborted when the estimated \acrshort{QBER} is above the threshold value otherwise the protocol continues. Randomly picked samples from the raw key gives a good estimation of \acrshort{QBER}. In general \acrshort{DR} is taken as 50\% of the raw key, but it reduces the key rate. So, for larger raw key size with randomly picked samples, a 3–10\% \acrshort{DR} is statistically sufficient to detect a trace of eavesdropping \cite{israel1992determining}.	
	\item {\bf Error correction:} The communicating parties (Alice and Bob) identify the location of errors in the raw key and follow some error correction protocol to correct them. The most widely used error correction protocol in \acrshort{QKD} is low-density parity-check (LDPC) \cite{mackay1999good}. During the error correction process, some information is exchanged over the classical channel. This process ends with the generation of identical key at both ends of the communicating parties.
	\item {\bf Privacy amplification:} During the information exchange in the error correction process, some leakage may happen which is beneficial for Eve to extract some authentic information. The key size is compressed using some compression function to make the information leakage negligible to Eve. The factor with which the error corrected key is compressed is known as \acrshort{CR}. Several schemes for privacy amplification are available which can be executed.
\end{enumerate}

\section{Experimental realization of DPR QKD protocols and comparison of key rates with existing realizations\label{sec:Advances}}
\begin{figure}
	\begin{centering}
		\begin{tabular}{c}
			\includegraphics[scale=0.75]{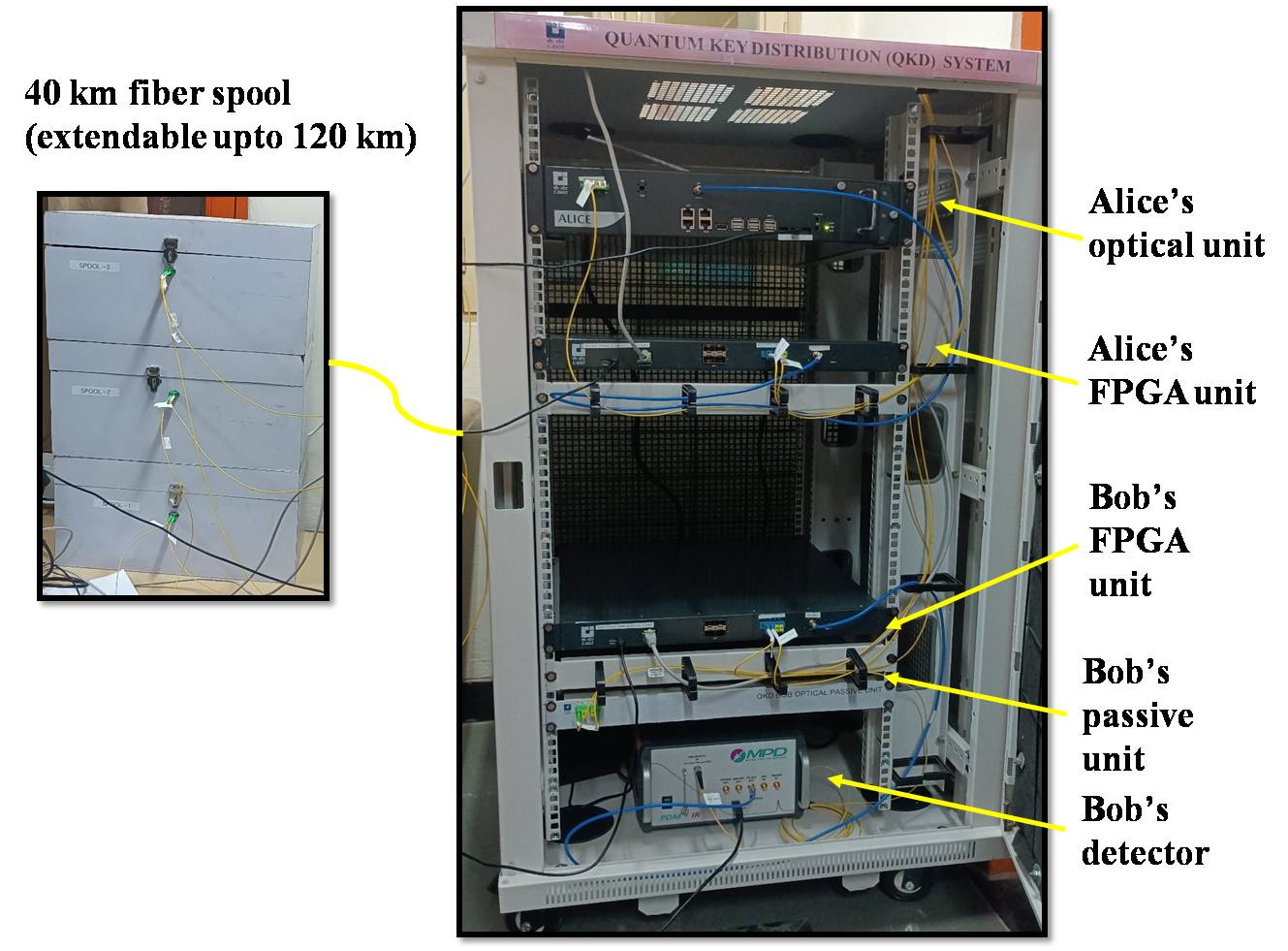}\tabularnewline
		\end{tabular}
		\par\end{centering}
	\caption{\label{fig6:QKD_set_up}A snapshot of the experimental set-up used for the DPR QKD implementation.}
\end{figure}

Over the past two decades, a significant advancement have been made in experimental realization of both \acrshort{COW} and \acrshort{DPS} \acrshort{QKD} protocols which is presented in Tables \ref{tab6.2:Survey-DPS} and \ref{tab6.3:Survey-COW}. The realization of both the protocols have been made possible in cooperation with Center for Development of Telematics (C-DOT) \cite{Cdot}. This set-up consists of three 19 inches rack mountable boxes and wooden boxes containing optical fibers spools. A snapshot of the set-up used for the experiment is illustrated in Figure \ref{fig6:QKD_set_up}. Two boxes are assigned to Alice, one for her optical units an other for her controlling unit (FPGA board). Bob boxes contain the detecting units. Alice and Bob does communication over the optical fiber channels. Further, the effect of several classical shifting parameters such as \acrshort{CR}, \acrshort{DR} and detector's \acrshort{DT} on the key generation rate have been studied. A recent study reported such effects for \acrshort{COW} \acrshort{QKD} over a wide range of communication distance \cite{Malpani2024ExpCOW-QKD}, but without monitoring arm. However, the \acrshort{COW} \acrshort{QKD} realization with monitoring arm is also reported in this chapter. The components along with its specification used for the experimental realization is listed in Table \ref{tab6.4:Componenets-Exp}.
\begin{table}
\caption{\label{tab6.2:Survey-DPS}The table outlines the
	advancements in experimental realization of the DPS QKD protocol. The symbol \# indicates that the information
	is not provided in the mentioned paper while {*} denotes dispersion shifted
	fiber.}
	\begin{centering}
		\begin{tabular}{|p{1.5cm}|p{1.5cm}|p{2.5cm}|p{1.5cm}|p{1.5cm}|p{2cm}|}
			\hline 
			QKD & Mode & Detector Type & Distance (km) & KR (bps) & References\tabularnewline
			\hline 
			DPS & Fiber &   APD & $80$ & $21352$ & this work\tabularnewline \hline
			DPS & Fiber{*} &  up-conversion-assisted hybrid photon detector (HPD) & $10$ & $1.3$Mbps & \cite{zhang2009megabits}\tabularnewline
			\hline 
			DPS & Fiber &  SNSPD & $90$ & $2100$bps & \cite{sasaki2011field}\tabularnewline
			\hline
			DPS & Fiber &  APD & $20$ & $3076$ & \cite{honjo2004differential}\tabularnewline
			\hline 
			DPS & Fiber & InGaAs APD & $25$ & $9000$ & \cite{honjo2009differential}\tabularnewline
			\hline 
			DPS & Fiber &  Low jitter up-conversion detector & $100$ & $106$ & \cite{diamanti2006100}\tabularnewline
			\hline 
			DPS & Fiber &  low-jitter up-conversion detector
			$10$ GHz clock & $105$ & $3700$ & \cite{takesue200610}\tabularnewline
			\hline 
			DPS & Fiber &  InGaAs/InP APD ($6\%$ eff) & $100-160$ & $24000-490$ & \cite{namekata2011high}\tabularnewline
			\hline 
			DPS & Fiber & SNSPD & $200$ & $12.1$ & \cite{Takesue2007DPS-QKD}\tabularnewline
			\hline 
			DPS & Fiber &  Ultra low noise SNSPD & $260$ & $1.85$ bps & \cite{wang20122}\tabularnewline
			\hline 
			DPS & Fiber{*} & SNSPD with ultra low DCR $0.01$counts per second & $336$  & \# & \cite{shibata2014quantum}\tabularnewline
			\hline 
			DPS & Fiber & SNSPD  with $0.01$counts per second & $265$  & $192$ & \cite{pathak2023phase}\tabularnewline
			\hline
		\end{tabular}
		\par\end{centering}
\end{table}

\begin{table}
	\caption{\label{tab6.3:Survey-COW} The table outlines the
	advancements in experimental realization of the COW QKD protocol. The symbol \# indicates that the information is not
	provided clearly in the mentioned paper.}
	\begin{centering}
		\begin{tabular}{|p{1.5cm}|p{1.5cm}|p{2.5cm}|p{1.5cm}|p{1.5cm}|p{2cm}|}
			\hline 
			QKD & Mode  & Detector Type & Distance (km) & KR (bps) & References\tabularnewline
			\hline 
			COW & Fiber & InGaAs SPD (free running mode) & (a)$120$
			
			(b) $145$ & (a) $241-2410$
			
			(b)$154-1184$ & \cite{Malpani2024ExpCOW-QKD}\tabularnewline
			\hline 
			COW & Fiber & InGaAs SPD (Lab) & $150$ & $>50$ & \cite{stucki2009continuous}\tabularnewline
			\hline 
			COW & Fiber & SNSPD (Field) & $150$ & $2.5$ & \cite{stucki2009continuous}\tabularnewline
			\hline 
			COW & Fiber & SNSPD & $100-250$ & $6000-15$ & \cite{Stucki2009ExpCOW-QKD}\tabularnewline
			\hline 
			COW & Fiber & InGaAs SPD & (a) $14.2$
			
			(b) $45.6$
			
			(c) $24.2$
			
			(d) $36.6$
			
			(total 121 km with three trusted nodes) & (a) $1956$
			
			(b) $1314$
			
			(c) $763$
			
			(d) $1906$ & \cite{wonfor2019field}\tabularnewline
			\hline 
			COW & Fiber & InGaAs/InP negative feedback avalanche diodes (NFADs) & $307$ & $3.18$ & \cite{korzh2015COW}\tabularnewline
			\hline 
			COW & Fiber & InGaAs SPD & $25$ & $22500$ & \cite{walenta2014COW}\tabularnewline
			\hline 
			COW & Fiber & InGaAs SPD (gated mode) & $150$ & $<1000$ & \cite{shaw2022optimal}\tabularnewline
			\hline 
		\end{tabular}
		\par\end{centering}
\end{table}

\begin{table}
	\caption{\label{tab6.4:Componenets-Exp} Details of the
	various components that are utilized in the experiment.}
	\begin{centering}
		\begin{tabular}{|c|c|}
			\hline 
			\textbf{Component/ Technique} & \textbf{Property/ Type}\tabularnewline
			\hline 
			Laser & PS-NLL laser from Teraxion\tabularnewline
			\hline 
			Fiber & SMF-28 (ITU-TG652D) (loss-0.2 dB/km)\tabularnewline
			\hline 
			Intensity modulator (COW) & Lithium Niobate based\tabularnewline
			\hline 
			Phase modulator (DPS) & MPZ-LN series\tabularnewline
			\hline 
			Random number generator & TRNG\tabularnewline
			\hline 
			Operating Temperature & $ 10^{\rm{o}}{\rm C}-2\ensuremath{8^{\rm{o}}}{\rm C}$\tabularnewline
			\hline 
			Detector & SPD\_OEM\_NIR from Aurea Technology\tabularnewline
			\hline 
			Error correction technique & LDPC\tabularnewline
			\hline 
			Privacy amplification & Toeplitz based hashing\tabularnewline
			\hline 
		\end{tabular}
		\par\end{centering}
\end{table}

Experiment performed to realize the \acrshort{DPS} \acrshort{QKD} has been tested for altered distances $80$ km, $100$ km and $120$ km with help of fiber spool. The testing is done by observing the key rate variations for different allowed values of \acrshort{DR}, \acrshort{CR} and \acrshort{DT}. \acrshort{DR} can vary from  $3.125\%$ to $50\%$, \acrshort{CR} can vary from $50\%$ to $95\%$, the variation in detector's \acrshort{DT} can be done from $20\,\mu$s to $50\,\mu$s. A higher \acrfull{KR} of $21352$ bps at DR $=3.125\%$, CR $=50\%$, DT $=20\,\mu$s for $80$ km communication through \acrshort{DPS} \acrshort{QKD} protocol is achieved. The key generation rate in the experiment aligns closely with those reported by other groups. For instance, Namekata et al. in the year $2011$ reported a \acrshort{KR} of $24$ kbps over a distance of $100$ km. A slightly better \acrshort{KR} was obtained because Avalanche Photodiodes (APDs) were used in gated mode and instead of standard telecom grade fiber, dispersion-shifted fiber (DSF) was used. They employed 2-GHz sinusoidally APDs in gated mode with a probability of dark count $2.8\times10^{-8}$ (55 counts per second), demonstrating that APDs operating in gated mode outperform the \acrshort{SNSPD}s. In $2011$, Sasaki et al. achieved a \acrshort{KR} of $2.1$ kbps using \acrshort{SNSPD}s and standard fiber during the field trail in Tokyo \cite{sasaki2011field}. It has also been shown that the use of \acrshort{SNSPD}s along with \acrfull{ULL} fiber, can significantly enhance key rates. For example, Wang et al. \cite{wang20122} demonstrated \acrshort{QKD} over $260$ km using \acrshort{ULL} telecom fiber with a loss coefficient of $0.164$ dB/km. Similarly, a higher key rate and longer communication distance using \acrshort{ULL} fiber can be achieved. Another experimental demonstration by Shibata et al. \cite{shibata2014quantum} achieved a \acrshort{KR} of $0.03$ bps over a communicating distance $336$ km through \acrshort{ULL} fiber. In the recent work of Pathak et al., a \acrshort{KR} of $193$ bps has been achieved, maintaining \acrshort{QBER} below $1\%$ while implementing \acrshort{DPS} \acrshort{QKD} \cite{pathak2023phase}. Their implementation utilized a standard telecom fiber and \acrshort{SNSPD}s. They further extrapolated that a secure \acrshort{KR} of $0.11$ bps can be achieved for a distance of $380$ km if \acrshort{ULL} fiber is used. A secure \acrshort{KR} of $10-25$ kbps over a distance of $80-120$ km using standard telecom fiber and APDs is achieved from the experiment reported in this chapter. These results are consistent with those reported by other groups. Also, light is shed on the variation of \acrshort{KR}s in accordance with \acrshort{DR}, \acrshort{CR} and \acrshort{DT} which was often overlooked in previous works. The effect of these classical shifting parameters (\acrshort{DR}, \acrshort{CR} and \acrshort{DT}) on the secure \acrshort{KR} is discussed in the next section.\\
The key factors that affect the \acrshort{COW} and \acrshort{DPS} protocol are listed in Table \ref{tab6.5:COW-DPS-key} along with their corresponding values chosen for experimental realization. Based on these factors, a theoretical formulation has been provided to estimate clicks and secret \acrshort{KR}. Detectors at Bob's side encounter clicks per second for both \acrshort{COW} and \acrshort{DPS} protocols which are formulated as:
\begin{equation}\label{eq:ClickCOW}
	C^{COW}_{total}=\frac{\tau}{1+\tau}
\end{equation}
\begin{equation}\label{eq:CLickDPS}
	C^{DPS}_{total}=\frac{\tau\times10^{-{l_m}/10}}{1+\tau\times10^{-{l_m}/10}}
\end{equation}
where, $\tau=\eta\times\mu\times f\times10^{{-l_f}d/10}$. Details of parameters affecting clicks on detector are summarized in Table \ref{tab6.5:COW-DPS-key}. It is clearly seen that for \acrshort{DPS}, an additional factor $10^{-{l_m}/10}$ is incorporated due to the \acrshort{MZI} loss.\\
The final secure key rate is obtained based on the clicks, \acrshort{DR} and \acrshort{CR}. The key rate formulation for both the protocols are mapped as follows:
\begin{equation}\label{eq:Key_COW}
	R_k^{COW}=C^{COW}_{total}\times(1-DR)\times(1-CR)
\end{equation}
\begin{equation}\label{eq:Key_DPS}
	R_k^{DPS}=2\times C^{DPS}_{total}\times(1-DR)\times(1-CR).
\end{equation}
The key generation rate for \acrshort{DPS} protocol is almost double to \acrshort{COW} because two detectors are used in the \acrshort{DPS} protocol for the key generation. It is observed that the final secure key rate depends on several experimental parameters. Further, the effect of several parameters such as, \acrshort{DR}, \acrshort{CR}, and detector's \acrshort{DT} on the final \acrshort{KR} will be studied.
\begin{table}
	\caption{\label{tab6.5:COW-DPS-key} Parameters that are influencing the key rate of the COW and DPS protocols.}
	\begin{center}
		\begin{tabular}{|p{1cm}|p{5cm}|p{3cm}|p{3cm}|}
			\hline 
			S No. & Parameters & COW & DPS\\
			\hline 
			1 & Average photon per pulse ($\mu$) & 0.5 & 0.2\\
			\hline 
			2 & Loss in fibre ($l_f$) & $0.5$ dB/km & $0.2$ dB/km\\
			\hline 
			3 & Distance ($d$) & $40-120$ km & $40-120$\\
			\hline 
			4 & Pulse frequency ($f$) & $500$ MHz & $1$ GHz\\
			\hline 
			5 & Optical coupler $t_c$ & $90:10$ & $-$\\
			\hline 
			6 & Detector efficiency $\eta$ & $10$ \% & $10$ \%\\
			\hline 
			7 & Dead time ($t_d$) & $20-50$ $\mu s$ & $20-50$ $\mu s$\\
			\hline 
			8 & MZI loss ($l_m$) & $-$ & $2$ dB\\
			\hline 
			9 & Disclose rate ($DR$) & $3.125-50$ \% & $3.125-50$ \%\\
			\hline
			10 & Compression ratio ($CR$) & $50-95$ \% & $50-95$ \%\\
			\hline
		\end{tabular}
	\end{center}
\end{table}

\section{Analysis of key rates with different parameters\label{sec:Observation}}
This section mainly focuses on key rate analysis as a function of different classical shifting parameters for the \acrshort{DPS} \acrshort{QKD} protocol. Further, a comparative analysis of both \acrshort{COW} and \acrshort{DPS} protocols has been made.
\begin{figure}
	\begin{centering}
		\begin{tabular}{cc}
			\includegraphics[scale=0.85]{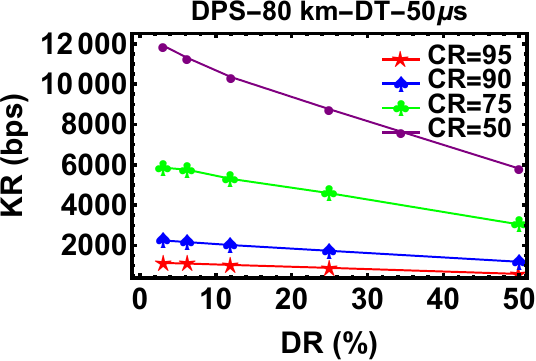} & \includegraphics[scale=0.85]{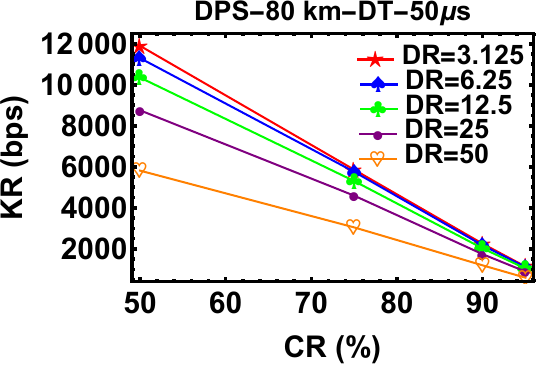}\\
			(a) & (d)\\
			\includegraphics[scale=0.85]{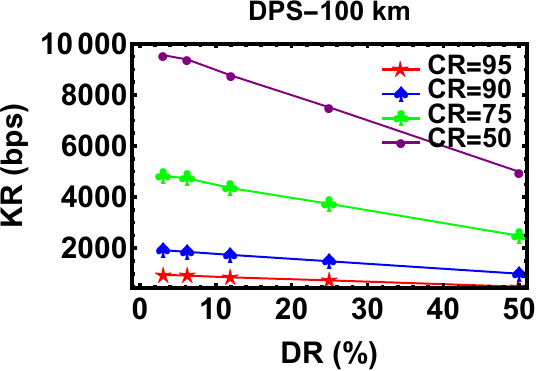} & \includegraphics[scale=0.85]{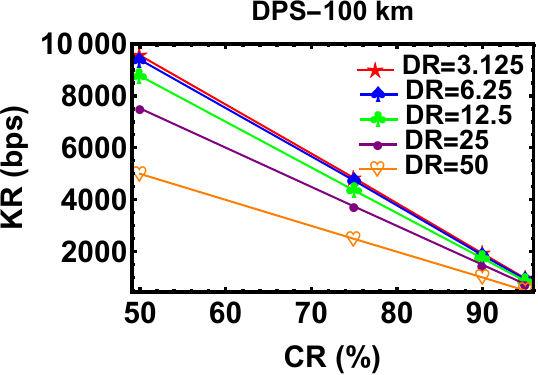}\\
			(b) & (e)\\
			\includegraphics[scale=0.85]{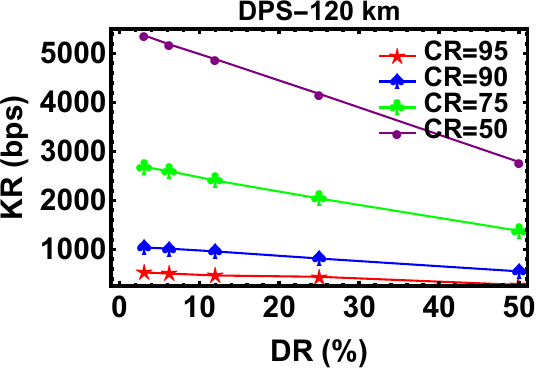} & \includegraphics[scale=0.85]{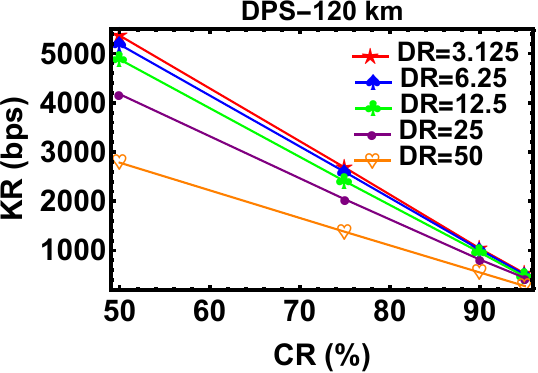}\\
			(c) & (f)
		\end{tabular}
		\par\end{centering}
	\caption{\label{fig6.4:DPS-KR-CR} Plots for the obtained KR of the DPS protocol as a function of CR for various DR (a) at $80$ km (b) at $100$ km,
		(c) at $120$, with detector's DT $=50\,\mu$s. The KR decreases with increasing distance and increases with higher DR.}
\end{figure} 

\subsection{Variation of key rate with respect to disclose rate and compression ratio for DPS QKD}
The obtained data of \acrshort{KR} at different allowed values of \acrshort{DR}, \acrshort{CR} and \acrshort{DT} are plotted for various distances $80$ km, $100$ km and $120$ km in Figure \ref{fig6.4:DPS-KR-CR} keeping the fixed value of \acrshort{DT} at $50\,\mu$s. Specially, Figure \ref{fig6.4:DPS-KR-CR} (a), (b) and (c) represents the \acrshort{KR} variation with respect to \acrshort{DR} for different values of \acrshort{CR} at distances $80$, $100$
and $120$ km, respectively. Similarly, Figure \ref{fig6.4:DPS-KR-CR} (d), (e), and (f) represents the \acrshort{KR} variation with respect to \acrshort{CR} for different values of \acrshort{DR} at distances $80$, $100$ and $120$ km, respectively. As discussed in Section \ref{sec6.2.3:postprocessing}, channel error is estimated using a portion of signals received at Bob's end. These signals are disclosed and do not contribute to the generated key, meaning a higher \acrshort{DR} reduces the \acrshort{KR}. Similarly, the \acrshort{CR} determines the portion with which the key is compressed to amplify its security. An increase in \acrshort{CR} naturally lowers the \acrshort{KR}. In this experimental implementation, as anticipated, the \acrshort{KR} increases as the \acrshort{DR} and \acrshort{CR} decrease across various distances.

 \begin{figure}
 	\begin{centering}
 		\begin{tabular}{cc}
 			\includegraphics[scale=0.85]{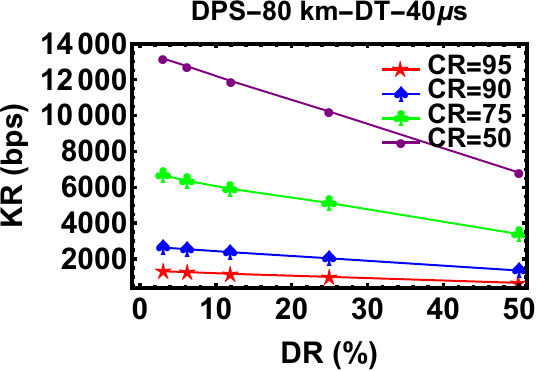} & \includegraphics[scale=0.85]{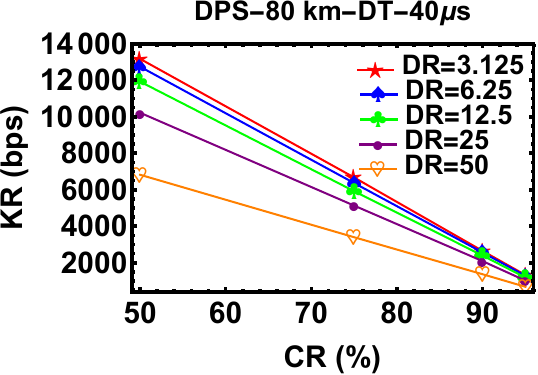}\\
 			(a) & (d)\\
 			\includegraphics[scale=0.85]{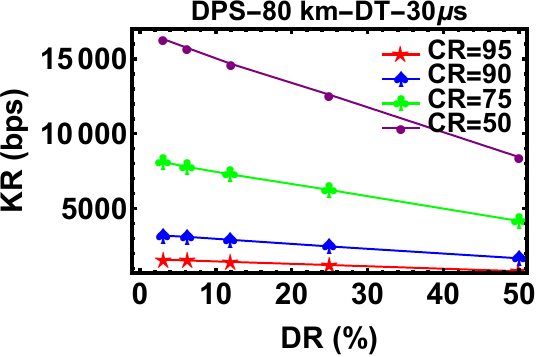} & \includegraphics[scale=0.85]{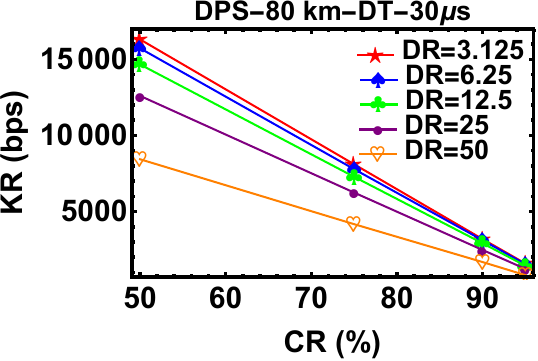}\\
 			(b) & (e)\\ 
 			\includegraphics[scale=0.85]{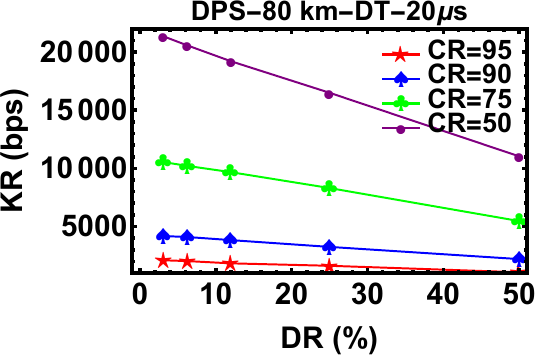} & \includegraphics[scale=0.85]{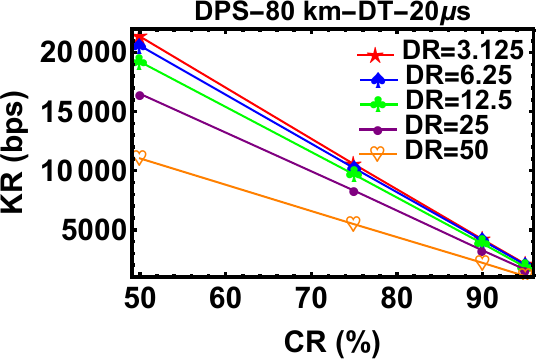}\\
 			(c) & (f)
 		\end{tabular}
 		\par\end{centering}
 	\caption{\label{fig6.5:DPS-KR-CR-PA} Plots for the obtained KR of the DPS protocol as a function of DR (CR) for various CR (DR) at detector's DT $=40\,\mu$s, $=30\,\mu$s and $=20\,\mu$s for $80$ km communication distance.}
 \end{figure} 

\subsection{Variation of key rate with respect to detector dead time for DPS QKD}
\acrshort{DT} of detector play a crucial role in implementing any \acrshort{QKD} experiment. \acrshort{DT} of a detector is interpreted as the time interval between two consecutive detections for which detector becomes inactive or unable to detect signals. A longer \acrshort{DT} means the detector cannot detect photons for an extended period. Ideally, \acrshort{DT} of detector should be kept to its minimum possible value. Keeping these facts in mind, an experiment has been performed for different possible values of \acrshort{DT} of detector. Based on the specification of the detector used, an experiment at \acrshort{DT} values $20\,\mu s-50\,\mu s$ has been performed. Figure \ref{fig6.4:DPS-KR-CR} already depicts the results at \acrshort{DT}=$50\,\mu s$. Figure \ref{fig6.5:DPS-KR-CR-PA} (a)-(c) ((d)-(f)) represents the \acrshort{KR} variation with respect to \acrshort{DR} (\acrshort{CR}) for different values of \acrshort{CR} (\acrshort{DR}) at \acrshort{DT} $40\,\mu s$, $30\,\mu s$ and $20\,\mu s$, respectively keeping fixed communication distance of $80$ km using the \acrshort{DPS} protocol. As expected, the results demonstrate that \acrshort{KR} increases as \acrshort{DT} decreases. It should be noted that this study focuses on the variation of \acrshort{KR} with \acrshort{DT}. However, \acrshort{DT} of a detector is also correlated to several other parameters such as, its efficiency, dark count rate and after pulse probability. But the detector used in this experiment has a freedom to tune its \acrshort{DT} only. Therefore, the analysis is limited to the impact of \acrshort{DT} on \acrshort{KR} among these interrelated parameters.

\begin{figure}
	\begin{centering}
		\begin{tabular}{c}
			\includegraphics[scale=1]{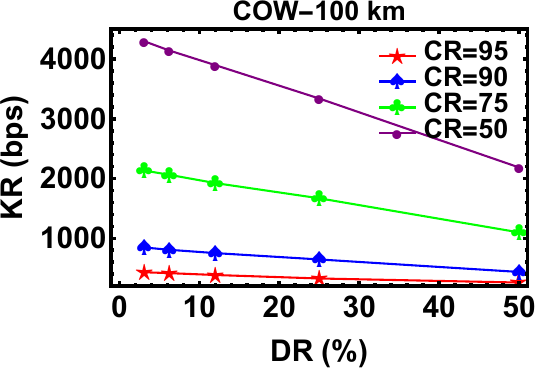}\\
			(a)\\
			\includegraphics[scale=1]{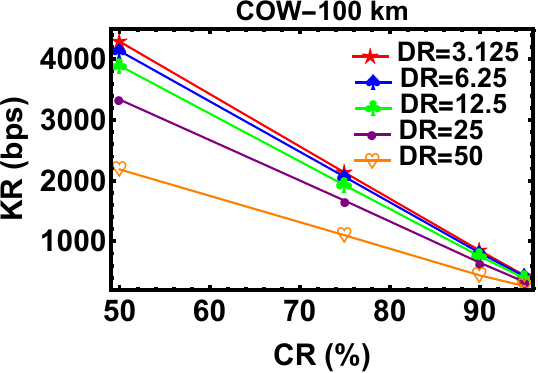}\\
			(b)
		\end{tabular}
		\par\end{centering}			
	\caption{\label{fig6.6:COW-KR-CR-PA-1} Plots for the obtained KR of the
		COW QKD protocol as a function of (a) DR for various CR and (b) CR for various DR, for $100$ km distance with detector's DT $=50\,\mu$s.}
\end{figure}

\subsection{Variation of key rate with respect to disclose rate and compression ratio for COW QKD}
An earlier work reported in Reference \cite{Malpani2024ExpCOW-QKD} shown the implementation of the \acrshort{COW} \acrshort{QKD} protocol without monitoring arm. Here, it will be extend and report \acrshort{COW} \acrshort{QKD} implementation with the monitoring arm.  Figure \ref{fig6.6:COW-KR-CR-PA-1} (a)-(b) represents the \acrshort{KR} variation of \acrshort{COW} \acrshort{QKD} protocol with respect to \acrshort{DR} (\acrshort{CR}) for different values of \acrshort{CR} (\acrshort{DR}) for $100$ km communication distance, keeping fixed value of \acrshort{DT} to $50\,\mu s$. As observed, the \acrshort{KR} value increases for lower value of \acrshort{CR} and \acrshort{DR}. A comparative analysis between both \acrshort{COW} and \acrshort{DPS} is studied in the next section.

\begin{figure}[h!]
	\begin{centering}
		\begin{tabular}{c}
			\includegraphics[scale=1]{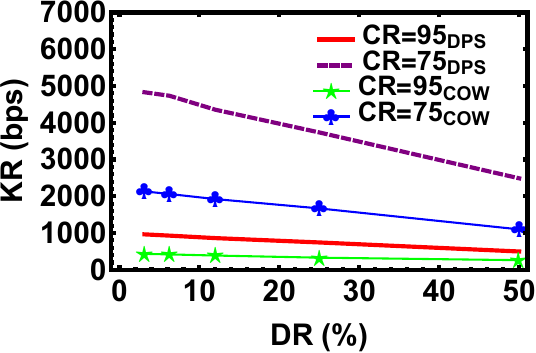}\tabularnewline
		\end{tabular}
		\par\end{centering}
	\caption{\label{fig6.7:COW-DPS} Plots for obtained KR as a function of DR for various CR for both DPS (Purple dashed and Red solid line) and COW (Blue and Green line with plot markers) QKD protocol. It is obvious that KR for DPS is higher than COW.}
\end{figure}

\begin{figure}[h!]
	\begin{centering}
		\begin{tabular}{c}
			\includegraphics[scale=1]{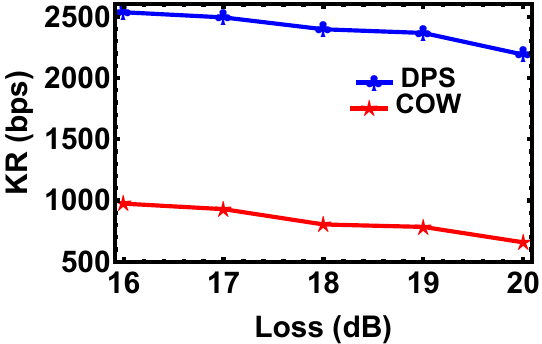}\tabularnewline
		\end{tabular}
		\par\end{centering}
	\caption{\label{fig6.8:COW-DPS-Loss} Plots for obtained KR as function of channel losses for both COW and DPS QKD protocol.}
\end{figure}
\subsection{A comparative analysis of key rate of DPS and COW QKD}
Both \acrshort{COW} and \acrshort{DPS} \acrshort{QKD} scheme belongs to the same \acrshort{DPR} category \acrshort{QKD}, making it relevant to compare their performance under the identical parameters, except laser repetition rate which is $500$ MHz for \acrshort{COW} while $1$ GHz for \acrshort{DPS}. The reason behind different repetition rate is the nature of pulse uses for encoding, \acrshort{COW} uses sequence of empty and non-empty pulses while \acrshort{DPS} uses all non-empty pulses. Here, comparison is done for key generation over $100$ km. Figure \ref{fig6.7:COW-DPS} illustrates the \acrshort{KR} variation for both \acrshort{COW} and \acrshort{DPS} protocols with respect to \acrshort{DR} for different \acrshort{CR} keeping fixed \acrshort{DT} value of the detector to $=50\,\mu$s. A significant higher \acrshort{KR} is observed in case of \acrshort{DPS} \acrshort{QKD} protocol for all cases. Additionally, a plot for \acrshort{KR} variation with channel loss is also provided in Figure \ref{fig6.8:COW-DPS-Loss}.

\begin{figure}[h!]
	\begin{centering}
		\begin{tabular}{c}
			\includegraphics[scale=1]{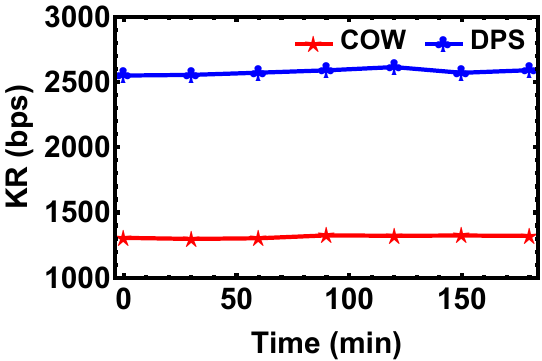}\\
			(a)\\ 
			\includegraphics[scale=1]{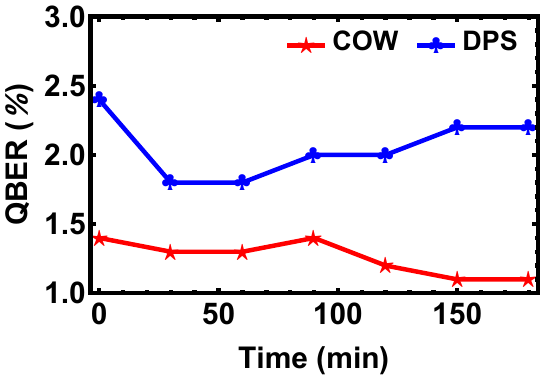}\\
			(b)
		\end{tabular}
		\par\end{centering}
	\caption{\label{fig6.9:stability plot} The stability plot for (a) KR and (b) QBER for both COW and DPS QKD protocol with fixed value of DS to $3.125$ \%, CR to $90$ \% and DT to $=50\,\mu$s for $80$ km communication distance.}
\end{figure}
\subsection{Key rate and QBER stability}
To check the stability of \acrshort{KR} and \acrshort{QBER} value which is obtained, the \acrshort{QKD} set-up run for a longer period keeping all parameters fixed. Figure \ref{fig6.9:stability plot} (a) and (b) illustrates the \acrshort{KR} and \acrshort{QBER} stability plot upto $3$ hour for both \acrshort{COW} and \acrshort{DPS} protocols for $80$ km communication distance, keeping fixed value of \acrshort{DR} to $3.125$ \%, \acrshort{CR} to $90$ \% and \acrshort{DT} to $50$ $\mu s$. It is found that the \acrshort{KR} and \acrshort{QBER} for both protocols remains stable throughout the testing period.

\section{Conclusion}
In the current era, quantum technology is considered as one of the major players for secure communication and faster computation. \acrshort{QKD} is a mature field where a secure key is generated among two legitimate parties. Several protocols for \acrshort{QKD} have been proposed and analyzed, most use either single photon source or entangled photons for communication. But with currently available technology, it is difficult to generate single photon on demand and maintaining entangled photons is also quite challenging. So, researchers have used attenuated laser sources for \acrshort{QKD} implementation. \acrshort{DPR} category \acrshort{QKD} such as \acrshort{COW} and \acrshort{DPS} uses such attenuated coherent pulse for encoding and decoding. In this Chapter, an experimental realization of both \acrshort{COW} and \acrshort{DPS} \acrshort{QKD} protocols have been reported over various distances. Specially, the effect on \acrshort{KR} due to changes in various classical shifting parameters such as, \acrshort{DR}, \acrshort{CR} and detector's \acrshort{DT} have been analyzed. The \acrshort{KR} obtained aligns with the results reported by other research groups for the same protocol. This study provides valuable insights into optimizing these shifting parameters to enhance key rates over longer distances, marking another step toward the realization of secure quantum communication networks.
\newpage
\chapter{CONCLUSION AND SCOPE FOR THE FUTURE WORKS}\label{Ch7:Conclusion}
\graphicspath{{Chapter7/Chapter7Figs/}{Chapter7/Chapter7Figs/}}
\section{Structure of the thesis}
Quantum mechanics has some intrinsic features that have no classical analogue. These non-classical features led to several no-go theorems (see Chapter 3 of Reference \cite{Pathak2013book}.). Initially, these no-go theorems were viewed as limitations of quantum mechanics, but after recent developments in quantum computation, quantum communication and quantum foundation, these are seen as advantages of quantum mechanics. Research in quantum information science is broadly divided into four categories: quantum computation, quantum communication, quantum sensing and quantum materials. The major motivations of doing research in the domain of quantum information science are: supremacy (quantum computers can solve certain computation problems much faster than classical computer), security (quantum communications are unconditionally secure) and accessibility (quantum computers are easily accessible on cloud at various platforms). This thesis covers recent progress in quantum computation and quantum communication along with several new results. Chapter $1$ covers some preliminary concepts required to understand the thesis work with historical developments in the domain of quantum computing and quantum communication. Algorithms designed for quantum computing tasks are further classified into two categories: algorithms that need to speed up and algorithm that needs security. Both faces problems due to difficulties in scaling a quantum computation power in NISQ era which can be solved using an approach of distributed quantum computing or some different approach. Three protocols are designed in Chapter $4$ which have direct applications in distributed photonic quantum computing.\\
Quantum communication schemes are classified into two categories: schemes that require security and schemes that don't essentially require security. Example for the schemes that require security is quantum cryptography (e.g., \acrfull{QKD}) which has both theoretical and experimental aspects. The experimental aspect of \acrshort{QKD} is covered in Chapter $6$ of the thesis. A few examples of the schemes that don't require security are \acrfull{RIO} (covered in Chapter $4$), \acrfull{MQT} (covered in Chapter $2$) and \acrfull{QB} of known state (covered in Chapter $3$).
\section{Summary of the thesis}
The thesis covers different aspects of quantum computation and quantum communication. The major findings of the thesis are briefly summarized as follows:
\begin{enumerate}
	\item The resources used to accomplish the \acrshort{MQT} task reported earlier have been optimized and tested experimentally on IBM quantum computer.
	\item It is shown that earlier reported schemes for \acrshort{QB} \cite{LuoM2013QB,Giovannetti2008QB,YuY2017QBandMQT,ZhouY2018probableQB,YuY2018generalQB} were incorrectly referred as \acrshort{QB}. The existence of no-broadcasting theorem doesn't allow to broadcast an unknown quantum state. We refer these schemes as \acrshort{QB} of known state. We have shown that the existing schemes for \acrshort{QB} can be accomplished using multi-party \acrshort{RSP} and generalized it for arbitrary number of senders and receivers.
	\item Three protocols for variants of \acrshort{RIO} are proposed which has a direct application in either distributed photonic quantum computing or blind quantum computing. The proposed protocols are named as \acrfull{CJRIO}, \acrfull{RIHO} and \acrfull{RIPUO}. Additionally, effect of noise has been studied on them.
	\item A set of proposed protocols for \acrfull{QAV} have been experimentally verified on IBM quantum computer and it is found that the \acrshort{QAV} protocol which uses Bell state has achieved higher fidelity.
	\item The \acrfull{DPR} category \acrshort{QKD} protocols (\acrfull{DPS} and \acrfull{COW}) have been experimentally implemented. The parameters which are affecting the obtained \acrfull{KR} for both the protocols have been analyzed. We have studied the effect on obtained \acrshort{KR} with respect to \acrfull{DR}, \acrfull{CR} and detector's \acrfull{DT} over various distances $40$ km, $80$ km and $120$ km.	
\end{enumerate}
This thesis contains total seven chapters and each chapter has a message that can help to mature the field of quantum computing and quantum communication. The concluding remarks of each chapter are briefed as follows:
\begin{description}
	\item[Chapter~1] This chapter gives a background of various concepts used in the thesis. An introduction to quantum information, specially focused on quantum computation and quantum communication is provided. We have discussed some preliminary concepts like, qubits, measurement basis, tensor products, quantum gates and quantum entanglement. On computing side, we have discussed available platforms to access quantum computers on cloud and also briefed about quantum state tomography. However on communication side, we have discussed various facets quantum communication such as, quantum teleportation, remote state preparation and quantum cryptography. Additionally, we have shown how one can model a noisy quantum system. A chronological history for major advancement in the field of quantum computation and quantum communication is also provided. The concepts discussed in this chapter are mainly from the Reference \cite{Pathak2013book,Nielsen2010book}. 
	\item[Chapter~2] In this chapter, it is shown that the quantum resources used in the previously proposed schemes for \acrshort{MQT} \cite{YuY2021MQT,Bikash2020IBM} are higher than the required (optimal) amount of quantum resources. We have used two copies of a Bell state instead of a five-qubit cluster state to teleport $m$ and $(m+1)$-qubit GHZ type state to two receivers. The modified \acrshort{MQT} scheme proposed by us that uses an optimal amount of quantum resources (two copies of a Bell state) is experimentally verified on a quantum computer (ibmq\_casablanca) available at IBM quantum experience on the cloud. Additionally, the robustness of the modified \acrshort{MQT} scheme is examined under various noise conditions such as, phase damping, amplitude damping, depolarizing, and bit-flip noise. The results reported in this chapter are published as Reference \cite{Satish2023MQT}.
	\item[Chapter~3] Despite the existence of no-broadcasting theorem, several research papers reported protocols for \acrshort{QB} \cite{LuoM2013QB,Giovannetti2008QB,YuY2017QBandMQT,ZhouY2018probableQB,YuY2018generalQB}. We have observed that all the existing schemes for \acrshort{QB} are actually broadcasting a known quantum state which can easily be reduced to multi-party \acrshort{RSP}. Such a reduction of existing \acrshort{QB} schemes to multi-party \acrshort{RSP} have been discussed in this chapter. A proof of principle experiment is also performed to show that the \acrshort{QB} schemes work better with the technique we provided than the earlier reported techniques on a quantum IBM quantum computer. Performance of the resources used earlier and the resources used by us is checked in the presence of a various noisy environment and it is found that a pair of Bell state outperforms the cluster state. A possible generalization of the \acrshort{QB} of known information has also been done in this chapter. The results reported in this chapter are published as Reference \cite{Satish2024QB}.
	\item[Chapter~4] A technique to operate one or more qubits remotely is known as \acrshort{RIO}. In this chapter, three protocols for different variants of \acrshort{RIO} are proposed. These protocols are referred as \acrshort{CJRIO}, \acrshort{RIHO} and \acrshort{RIPUO} which uses different quantum resources. For example, \acrshort{CJRIO} uses a four-qubit hyperentangled state (entangled in two \acrfull{DOF}, \acrshort{S-DOF} and \acrshort{P-DOF}) while \acrshort{RIHO} and \acrshort{RIPUO} uses a two-qubit maximally entangled state (entangled in \acrshort{S-DOF}). The proposed protocols use photon as a qubit and due to the least interacting properties of photons, cross-Kerr interaction method is used to allow interaction between photons.\\
	The \acrshort{CJRIO} scheme involves total four parties, a receiver (Alice), two senders ($\text{Bob}^1$ \& $\text{Bob}^2$), and one controller (Charlie). The two senders jointly teleport an operator to the receiver under the presence of a controller Charlie. This scheme is also generalized for arbitrary senders and controllers. Efficiency of the proposed scheme is also estimated.\\
	The \acrshort{RIHO} and \acrshort{RIPUO} schemes involve two parties, one sender (Bob), and one receiver (Alice). In the \acrshort{RIHO} scheme, a hidden operator is teleported to Alice. The operator is hidden in the sense that intended operator is dissolved inside a super-operator. In the \acrshort{RIPUO} scheme, a partially unknown operator is teleported. The operator is partially known or unknown in the sense that its structure is known but not its values. Both the schemes are then extended to their other possible variants. A technique is discussed which gives controllers a power to control the protocols even without keeping qubits. Most interestingly, the dissipation of coherent state used for the cross-Kerr interaction is well studied. The success probability of both the schemes is estimated in terms of error occurs during the interaction.\\
	The results reported in this chapter are published as Reference \cite{Satish2024CJRIO,Satish2024RIHO}.
	\item[Chapter~5] In real-life scenarios, voting plays a significant role in the decision-making process. Several schemes where voting is mandatory needs the anonymity of voters and also its security. As quantum mechanical features guarantee unconditional security, researchers have proposed several protocols for quantum voting since 2006. A particular type of quantum voting scenario known as \acrshort{QAV} where voters are given right to perform veto has drawn a special attention. A set of protocols for the \acrshort{QAV} have been recently proposed \cite{SandeepM2021QV}. In this chapter, two efficient \acrshort{QAV} protocols (referred as Protocol A and B) have been implemented for four voters case using a quantum computer (IBMQ Manila) available on cloud at IBM Quantum platform. Protocol A is realized using the Bell state, whereas Protocol B is realized using the cluster state and three-qubit GHZ state. Both protocols were successfully implemented with high success probabilities. Additionally, it is noticed that Protocol A that uses Bell state outperforms Protocol B that uses higher qubit entangled state under both ideal and noisy conditions. The results reported in this chapter are published as Reference \cite{Satish2022QV}. 
	\item[Chapter~6] In the current era, quantum technology is considered as the major players for secure communication. The well established example of quantum communication is \acrshort{QKD} where a secure key is generated among two legitimate parties. Several protocols for \acrshort{QKD} have been proposed and analyzed, most use either single photon source or entangled photons for communication. But creating a single photon source with currently available technology is quite difficult and also maintaining entangled photons is very challenging. Alternatively, an attenuated laser source is used for \acrshort{QKD}. Such class of \acrshort{QKD} that uses attenuated coherent pulses is called as \acrshort{DPR} \acrshort{QKD}. Hence, it is easier to realize experimentally. \acrshort{DPS} and \acrshort{COW} are the well known examples of \acrshort{DPR} \acrshort{QKD}. In this chapter, we have reported experimental realization of both \acrshort{COW} and \acrshort{DPS} \acrshort{QKD} protocols over various distances. A \acrshort{KR} analysis has been done with respect to classical shifting parameters such as, \acrshort{DR}, \acrshort{CR} and detector's \acrshort{DT}. The \acrshort{KR} obtained aligns with the results reported by other groups for the same protocols. This study provides valuable insights into optimizing these shifting parameters to enhance key rates over longer distances, marking another step toward the realization of secure quantum communication networks. The results reported in this chapter are published as Reference \cite{Satish2024ExpDPR-QKD}.
\end{description}
\section{Limitations and scope for future work}
Limitation of the present works will lead to a scope of future work. Also, with gradual advancements in the quantum technologies, the present work can be extended in numerous ways. Several ideas on which one can think in the near future are listed as follows: 
\begin{enumerate}
	\item In this thesis, the resources for \acrshort{MQT} and \acrshort{QB} tasks used for quantum communication have been optimized in Chapter 2 and 3 respectively. Instead of complex entangled state, an adequate copies of the two-qubit maximally entangled Bell state have been used. Hence, these schemes can be realized experimentally by other research groups which have sufficient experimental facilities.  
	\item The \acrshort{CJRIO} protocol proposed in Chapter 4 uses hyperentangled state (entangled in \acrshort{S-DOF} and \acrshort{P-DOF}). But it is to be noted that during the execution of the protocol only one \acrshort{DOF} is used at a time. One can extend this work to simultaneously using all \acrshort{DOF} of the hyperentangled state and its generalization.
	\item The proposed protocols for \acrshort{RIHO} and \acrshort{RIPUO} is restricted between two nodes of a network that share an entangled state. These protocols can be extended between two parts of a network that do not share an entangled state known as quantum multihop network.
	\item The implementation of \acrshort{COW} and \acrshort{DPS} protocol is restricted to point-to-point communication. This can be extended to multi-node quantum networks, including trusted and non-trusted node configurations. This would be a step towards the realization of the metropolitan-scale \acrshort{QKD} networks.
	\item The security analysis of both the \acrshort{COW} and \acrshort{DPS} protocol in realistic scenarios is still an open problem. Finite key analysis for both protocols will be done in the near future with the help of the experimental parameters. Further, it may be noted that several attacks (including zero error attack) and countermeasures (theoretical in nature to date) applicable to COW and DPS protocols have been proposed in the recent past \cite{trenyi2021zero,curty2021zero,inoue2005dpsPNS}. Due to the limitations of available technologies, all such countermeasures cannot be implemented in this experimental works. In the future, an extension of the experiments to more secure implementations is planned.
\end{enumerate}  

\newpage


\addcontentsline{toc}{chapter}{\fontsize{12}{10}\selectfont REFERENCES}
\bibliographystyle{JIITREFStyle}
\renewcommand{\bibname}{\fontsize{16}{16} \selectfont REFERENCES}
\bibliography{References/references} 
\newpage


\addcontentsline{toc}{chapter}{\fontsize{12}{10}\selectfont LIST OF AUTHOR'S PUBLICATIONS}

\chapter*{\fontsize{16}{16} \selectfont List of Author's Publications}
\vspace{-0.25in}
\begin{itemize}
\item \textbf{International Journals}
\begin{enumerate}
\item Kumar S. and Pathak A., \enquote{Experimental realization of quantum anonymous veto protocols using IBM quantum computer,} Quantum Information Processing, vol. 21, no. 9, p. 311, 2022. 
\item Kumar S., \enquote{Multi-output quantum teleportation of different quantum information with optimal quantum resources,} Optical and Quantum Electronics, vol. 55, no. 4, p. 296, 2023. 
\item Kumar S. and Pathak A., \enquote{Many facets of multiparty broadcasting of known quantum information using optimal quantum resource,} Quantum Information Processing, vol. 23, no. 5, p. 158, 2024. 
\item Kumar S., Ba An N., and Pathak A., \enquote{Controlled-joint remote implementation of operators and its possible generalization,} The European Physical Journal D, vol. 78, no. 7, p. 90, 2024.
\item Kumar S., Malpani P., Britant, Mishra S., and Pathak A., \enquote{Experimental implementation of distributed phase reference quantum key distribution protocols,} Optical and Quantum Electronics, vol. 56, no. 7, p. 1190, 2024.
\item Malpani P., Kumar S., and Pathak A., \enquote{Implementation of coherent one way protocol for quantum key distribution up to an effective distance of 145 km,} Optical and Quantum Electronics, vol. 56, no. 8, p. 1369, 2024.
\item Kumar S., Gangwar K., and Pathak A., \enquote{Remote implementation of hidden or partially unknown quantum operators using optimal resources: A generalized view,} Physica Scripta, vol. 99, no. 11, p. 115106, 2024.
\end{enumerate}

\end{itemize}
\addtocontents{toc}{\protect\contentsline {chapter}{\fontsize{12}{10}\selectfont SYNOPSIS}{Synopsis-1}}
\newpage
\afterpage{\blankpage}
\newpage
\end{document}